\documentclass[twocolumn,tighten]{aastex62}%
\usepackage{graphicx}
\usepackage{graphicx}		
\usepackage{url}
\usepackage{epstopdf}
\usepackage{color}
\usepackage{textcomp}
\usepackage{gensymb}
\usepackage{multirow}
\usepackage{ragged2e}
\usepackage{hyperref}
\usepackage{url}
\usepackage{amsmath}
\usepackage{booktabs}
\usepackage{comment}
\usepackage{afterpage}
\usepackage{mathtools}
\usepackage{tabularx}
\usepackage{bold-extra}
\usepackage{xspace}
\usepackage{relsize}
\usepackage{newunicodechar}
\usepackage[encapsulated]{CJK}
\usepackage{ucs}
\usepackage{longtable}
\usepackage{enumerate}
\usepackage{amssymb}
\usepackage{natbib}
\usepackage{soul}
\usepackage[utf8]{inputenc}
\bibpunct{(}{)}{;}{a}{}{,}
\DeclareGraphicsExtensions{.pdf,.png,.jpeg}
\newcommand{\pas}{$\rlap{.}^{\prime\prime}$}
\newcommand{\as}{$^{\prime\prime}$}
\newcommand{\dg}{$^{\circ}$}

\newcommand{\radmsq}{rad~m$^{-2}$}
\newcommand{\radms}{rad~m$^{-2}$}
\newcommand{\msyr}{$M_{\odot}$~yr$^{-1}$}
\newcommand{\rs}{$R_{Sch}$}

\newcommand{\uvmf}{{\sc uvmf}}
\newcommand{\intf}{{\sc intf}}
\newcommand{\casa}{{\sc CASA}}

\newcommand{\txt}{{\sc 3x3}}
\newcommand{\uvmultifit}{{\sc uvmultifit}}

\newcommand{\cntext}[1]{\begin{CJK}{UTF8}{gbsn}#1\end{CJK}}

\newcommand{\newtext}[1]{\textcolor{black}{#1}}

\shorttitle{Polarimetric properties of Event Horizon Telescope targets from ALMA}
\shortauthors{Goddi, Mart\'i-Vidal, Messias, et al.}
\begin{document}

\title{Polarimetric properties of Event Horizon Telescope targets from ALMA\\
 }
\correspondingauthor{Ciriaco Goddi}
\email{cgoddi@gmail.com}

\author[0000-0002-2542-7743]{Ciriaco Goddi}
\affiliation{Department of Astrophysics, Institute for Mathematics, Astrophysics and Particle Physics (IMAPP), Radboud University, P.O. Box 9010, 6500 GL Nijmegen, The Netherlands}
\affiliation{Leiden Observatory---Allegro, Leiden University, P.O. Box 9513, 2300 RA Leiden, The Netherlands}

\author[0000-0003-3708-9611]{Iv\'an Martí-Vidal}
\affiliation{Departament d'Astronomia i Astrof\'{\i}sica, Universitat de Val\`encia, C. Dr. Moliner 50, E-46100 Burjassot, Val\`encia, Spain}
\affiliation{Observatori Astronòmic, Universitat de Val\`encia, C. Catedr\'atico Jos\'e Beltr\'an 2, E-46980 Paterna, Val\`encia, Spain}

 \author[0000-0002-2985-7994]{Hugo Messias}
\affil{Joint ALMA Observatory, Alonso de Cordova 3107, Vitacura 763-0355, Santiago de Chile, Chile}

\author[0000-0003-4056-9982]{Geoffrey C. Bower}
\affiliation{Institute of Astronomy and Astrophysics, Academia Sinica, 645 N. A'ohoku Place, Hilo, HI 96720, USA}

\author[0000-0002-3351-760X]{Avery E. Broderick}
\affiliation{Perimeter Institute for Theoretical Physics, 31 Caroline Street North, Waterloo, ON, N2L 2Y5, Canada}
\affiliation{Department of Physics and Astronomy, University of Waterloo, 200 University Avenue West, Waterloo, ON, N2L 3G1, Canada}
\affiliation{Waterloo Centre for Astrophysics, University of Waterloo, Waterloo, ON N2L 3G1 Canada}

\author[0000-0003-3903-0373]{Jason Dexter}
\affiliation{JILA and Department of Astrophysical and Planetary Sciences, University of Colorado, Boulder, CO 80309, USA}

\author[0000-0002-2367-1080]{Daniel P. Marrone}
\affiliation{Steward Observatory and Department of Astronomy, University of Arizona, 933 N. Cherry Ave., Tucson, AZ 85721, USA}

\author[0000-0002-4661-6332]{Monika Moscibrodzka}
\affiliation{Department of Astrophysics, Institute for Mathematics, Astrophysics and Particle Physics (IMAPP), Radboud University, P.O. Box 9010, 6500 GL Nijmegen, The Netherlands}

\author[0000-0003-0292-3645]{Hiroshi Nagai}
\affiliation{National Astronomical Observatory of Japan, 2-21-1 Osawa, Mitaka, Tokyo 181-8588, Japan}
\affiliation{Department of Astronomical Science, The Graduate University for Advanced Studies (SOKENDAI), 2-21-1 Osawa, Mitaka, Tokyo 181-8588, Japan}

\author[0000-0001-6993-1696]{Juan Carlos Algaba}
\affiliation{Department of Physics, Faculty of Science, University of Malaya, 50603 Kuala Lumpur, Malaysia}

% ...


\begin{abstract}
We present the results from a full polarization study carried out with the Atacama Large Millimeter/submillimeter Array (ALMA) during the first Very Long Baseline Interferometry (VLBI) campaign, which was conducted  in Apr 2017 in the $\lambda$3\,mm and $\lambda$1.3\,mm bands, in concert with the Global mm-VLBI Array (GMVA) and the Event Horizon Telescope (EHT), respectively.  
We determine the  polarization and  Faraday  properties  of all VLBI targets, including Sgr A*, M87, and a dozen radio-loud AGN, in the two bands at several epochs in a time window of ten days. 
We detect high linear polarization fractions (2--15\%) and large rotation measures (RM  $>10^{3.3}-10^{5.5}$~\radms), 
confirming the trends of previous AGN studies at mm wavelengths. 
We  find that blazars are more strongly polarized than other AGN in the sample, while exhibiting (on average) an order-of-magnitude  lower RM values, consistent with the AGN viewing angle unification scheme.  
  For Sgr~A* we report a mean RM  of $(-4.2\pm0.3) \times10^5$~\radms~at 1.3~mm,  consistent with measurements over the past decade, and, for the first time, a RM of $(-2.1\pm0.1) \times10^5$~\radms~at 3~mm, 
  suggesting that about half of  the Faraday rotation at 1.3~mm may occur between the 3~mm photosphere and the 1.3~mm source.  
   We also report the first  unambiguous measurement of RM toward the M87 nucleus at mm wavelengths, which
 undergoes significant  changes in magnitude and sign reversals on a one year time-scale, 
 spanning the range from --1.2 to 0.3 $\times\,10^5$~\radms~at 3~mm and --4.1 to 1.5 $\times\,10^5$~\radms~at 1.3~mm.
Given this time variability, we argue that, unlike the case of Sgr~A*,  the RM in M87 does not  provide an accurate estimate of the mass accretion rate onto the black hole. 
We put forward a two-component model, comprised of a variable compact region  and a static extended region, 
 that can  simultaneously explain  the polarimetric properties observed by both the EHT  (on horizon scales) and ALMA (which observes the combined emission from both  components).
These measurements provide critical constraints for the calibration, analysis, and interpretation of simultaneously obtained VLBI data with the EHT and GMVA.

\end{abstract}
\vspace{2.5cm}
\section{Introduction}
 Active galactic nuclei (AGN) are known to host supermassive black holes (SMBHs), which accrete gas through a disk and drive powerful relativistic jets that are observed on scales of parsecs to megaparsecs \citep{Blandford2019}.
Magnetic fields are believed to play a major role in the formation of such relativistic jets, 
by either extracting energy from a spinning SMBH via the Blandford–Znajek mechanism \citep{BlandfordZnajek1977} 
or by tapping into the rotational energy of a magnetized accretion flow via the Blandford–Payne mechanism \citep{BlandfordPayne1982}. 

Polarization observations are  a powerful tool to probe magnetic fields and to understand their role in black-hole mass-accretion and launching and acceleration of  relativistic AGN jets.  
In fact, the radio emission from AGN and their associated jets is thought to be produced by synchrotron processes, and thus they display high intrinsic linear polarization \citep[LP; e.g.,][]{Pacholczyk1970,Trippe2010,Agudo2018}. 
LP fractions and polarization vector orientations  can provide details on the magnetic field strength and topology.
  Besides LP,  circular polarization (CP) may also be present   as a consequence of Faraday conversion of the linearly-polarized synchrotron emission    \citep{Beckert2002}, and can also help constraining the magnetic field configuration \citep[e.g.,][]{Munoz2012}.  
 
As the linearly polarized radiation travels through  magnetized plasma, it experiences Faraday  rotation of the LP vectors. The externally magnetized plasma is also known as the ``Faraday screen'' and the amount of Faraday rotation is known as the ``rotation measure" (RM).  
 If the background source of polarized emission is entirely behind (and not intermixed with) the Faraday screen, the RM can be written as an integral of the product of the electron number density ($n_e$) and the magnetic field component along the line-of-sight ($B_{||}$) via:
 \begin{equation}\label{eq:RM}
\mathrm{RM} = 8.1 \times 10^5 \int n_e [\mathrm{cm^{-3}}]~ \mathbf{B} [\mathrm{G}] \cdot \mathbf{\mathrm{d}l}[\mathrm{pc}]  ~~\mathrm{rad\,m}^{-2} \,.
\end{equation}
Thus, by measuring the RM one can also constrain the electron density, $n_e$, and the magnetic field, $B_{||}$,   
in the plasma surrounding SMBHs.
Under the assumption that the polarized emission is produced close to the SMBH and then Faraday-rotated in the surrounding accretion flow, the RM has been used in some cases to infer the accretion rate onto SMBHs \citep[e.g,][]{Marrone2006,Marrone2007,Plambeck2014,Kuo2014,Bower2018}.  
Alternatively, the polarized emission may be Faraday-rotated along the jet boundary layers \citep[e.g.,][]{ZavalaTaylor2004,IMV2015}.
Therefore, Faraday rotation measurements can provide crucial constraints on magnetized accretion models and jet formation models.

RM studies are typically conducted at cm wavelengths
using the Very  Large Array (VLA) or the Very Long Baseline Array \citep[VLBA; e.g.,][]{ZavalaTaylor2004}.
However, cm wavelengths are strongly affected by synchrotron self-absorption (SSA) close to the central engines and can therefore only probe magnetized plasma in the optically thin regions at relatively larger distances (parsec scales) from the SMBH \citep{Gabuzda2017,Kravchenko2017}.
On the other hand, emission at mm wavelengths is optically thin from the innermost regions of the jet base (and accretion disc), enabling us to study the plasma and magnetic fields much closer to the SMBH. 
In addition, LP can be more easily detected at mm wavelengths because the mm emission region is smaller  \citep[e.g.,][]{Lobanov1998}, and so depolarization induced by RM variations across the source (e.g., owing to a tangled magnetic field) is less significant. 
Finally, since Faraday rotation is smaller at shorter wavelengths (with a typical $\lambda^2$ dependence), mm-wavelength measurements more clearly reflect the intrinsic LP properties, and therefore the magnetic field of the system.

Unfortunately, polarimetric measurements at mm wavelengths have so far been limited by sensitivity and instrumental systematics.
The first interferometric measurements of RM at (sub)mm wavelengths were conducted towards Sgr~A* with the Berkeley-Illinois-Maryland Association (BIMA) array    \citep[][]{Bower2003,Bower2005} and the Submillimeter Array  \citep[SMA][]{Marrone2006,Marrone2007}, which yielded a RM $\sim -5 \times10^5$~\radms. 
SMA measurements towards M87 provided an upper limit  $\left|{\rm RM}\right| < 7.5 \times10^5$~\radms\ \citep{Kuo2014}.  
Other AGN with RM detections with mm interferometers include 3C~84 with RM $=8 \times10^5$~\radms\ (\citealt{Plambeck2014};  
see also \citealt{Nagai2017} for a similarly high RM measured with the VLBA at 43~GHz), 
PKS 1830-211 (at a redshift $z=2.5$) with RM $ \sim 10^7$~\radms\   \citep{IMV2015}, 
and 3C~273 with RM $=5 \times10^5$~\radms\ \citep{Hovatta2019}. 
Additional examples of AGN RM studies with mm single-dish telescopes can be found in  \citet{Trippe2012} and  \citet{Agudo2018}.

In order to  progress in this field,  polarization interferometric studies at mm wavelengths should be extended to a larger sample of AGN and it will be important to investigate both time and frequency dependence effects, by carrying out observations at multiple frequency bands and epochs. 
Ultimately, observational studies should be conducted at the highest possible angular resolutions in order to resolve the innermost regions of the accretion flow and/or the  base of  relativistic jets. 

The advent of the Atacama Large Millimeter/ submillimeter Array (ALMA) as a phased array   \citep[hereafter phased-ALMA;][]{APPPaper,QA2Paper} as a new element to Very Long Baseline Interferometry (VLBI)  at millimeter (mm) wavelengths (hereafter mm-VLBI) has been a game charger in terms of sensitivity and polarimetric studies. 
In this paper, we present a complete polarimetric analysis of ALMA observations carried out  during the first VLBI campaign.

\subsection{mm-VLBI with ALMA}

 The first science observations with phased-ALMA were conducted in April 2017  \citep[][]{QA2Paper}, in concert with two different VLBI networks: the  Global mm-VLBI Array (GMVA) operating at 3\,mm wavelength \citep[e.g.,][]{GMVA_IMV} and the Event Horizon Telescope (EHT)  operating at 1.3\,mm wavelength  \citep{EHTC2019_2}. 
These observations had two ``key science'' targets,
the SMBH candidate at the Galactic center, Sgr A*, and  the nucleus of the giant elliptical galaxy M87 in the Virgo cluster, M87*, both enabling studies at horizon-scale resolution \citep[][]{Doeleman2008,Doeleman2012,Goddi2017,EHTC2019_1}. 
In addition to those targets, VLBI observations with phased-ALMA also targeted a sample of  a dozen radio-loud AGN, including the closest and most luminous quasar 3C~273, the bright $\gamma$-ray-emitting  blazar 3C~279, the closest radio-loud galaxy Centaurus~A (Cen~A), and the best supermassive binary black hole candidate OJ287.

In 2019, the first EHT observations with phased-ALMA yielded groundbreaking results, most notably 
 the first ever event-horizon-scale image of the M87* SMBH \citep{EHTC2019_1,EHTC2019_2,EHTC2019_3,EHTC2019_4,EHTC2019_5,EHTC2019_6}. 
Beyond this breakthrough, EHT observations have now  imaged polarized emission in the ring surrounding M87*, resolving for the first time the magnetic field structures within a few Schwarzschild radii (\rs)  of a SMBH \citep{EHTC2020_1}. In addition, these new polarization images enable us to place tight constraints on physical models of the magnetized accretion flow around the M87* SMBH and, in general, on relativistic jet launching theories \citep{EHTC2020_2}. 

Both the VLBI imaging and the theoretical modelling use constraints  from ALMA observations \citep{EHTC2020_1, EHTC2020_2}. In fact, besides providing a huge boost in sensitivity and uv-coverage \citep{EHTC2019_3,Goddi2019msng}, the inclusion of ALMA in a VLBI array provides another important advantage: 
standard interferometric visibilities among the ALMA antennas are computed by the ALMA
correlator and simultaneously stored in the  ALMA archive together with the VLBI recording of the phased signal   \citep{APPPaper,QA2Paper}.
Furthermore, VLBI observations are always performed in full-polarization mode in order to supply the inputs to the polarization conversion process (from linear to circular) at the VLBI correlators,  
which is carried out using the {\sc PolConvert} software \citep{PCPaper} after 
the "Level 2 Quality Assurance"  (QA2) process \citep{QA2Paper}.  
Therefore, VLBI observations with ALMA yield  a full-polarization interferometric dataset, 
which provides both source-integrated information for  refinement and validation of VLBI data calibration \citep{EHTC2020_1} as well as observational constraints to theoretical models  \citep{EHTC2020_2}. 
Besides these  applications, this dataset carries valuable scientific value on its own and can be used to derive  mm emission,  polarization, and Faraday properties of  a selected sample of  AGN  on arcsecond scales. 

\subsection{This paper}

In this paper, we present a full polarization study carried out with ALMA in the $\lambda$3\,mm and $\lambda$1\,mm bands towards Sgr A*, M87, and  a dozen radio-loud AGN, with particular emphasis on their polarization and Faraday properties. 
The current paper is structured as follows. 

Section \ref{sect:data} summarizes the 2017 VLBI observations (\S\ref{obs}), the procedures followed for the data calibration (\S\ref{sect:datacal}),  the details of the full-polarization image deconvolution (\S\ref{sect:dataimg}), and additional observations on M87 (\S\ref{m87_add_data}). 

Section~\ref{sect:analysis} describes the procedures of data analysis. After presenting some representative total-intensity images of Sgr~A* and M87   (\S\ref{Sgr_A_m87_im}),  two independent algorithms to estimate the Stokes parameters of the compact cores are described (\S\ref{allstokes}). 
 The Stokes parameters for each source and spectral window are then converted into fractional LP and EVPA (\S\ref{subsec:LP}),  
 and used to estimate Faraday rotation (\S\ref{subsec:RM}) and (de)polarization  effects (\S\ref{subsec:depol}). Finally, the CP analysis is summarised in \S\ref{subsec:CP}.

Section~\ref{sect:results}  reports the polarimetric and Faraday properties of all the GMVA and EHT target sources, with dedicated subsections on AGN, M87, and Sgr~A*.

In Section~\ref{sect:discussion}, the polarization properties presented in the previous sections are used to explore potential physical origins of the polarized emission and location of Faraday screens in the context of  SMBH accretion and  jet formation models. \S\ref{sect:LP_1mmvs3mm} presents a comparison between the $\lambda$3\,mm and $\lambda$1.3\,mm bands, including a discussion on the effects of synchrotron opacity and Faraday rotation; \S\ref{sect:blazarsVSothers} presents a comparison between the case of blazars and other AGN; \S\ref{cena_ngc1052} discusses about depolarization in radio galaxies and its possible connection to instrumental effects. 
\S\ref{sect:m87_disc} is devoted to the special case of M87, including a discussion about the origin of the Faraday screen (internal vs. external; \S\ref{int_vs_ext_FR}) as well as a simple two-component Faraday model (\S\ref{2polcomp}). 
Finally, \S\ref{sect:sgra_disc} is dedicated to the special case of Sgr~A*. 

Conclusions are drawn in Section~\ref{sect:conclusions}.

The paper is supplemented with a number of appendices including:  the list of ALMA projects observed during the VLBI campaign in April 2017 (Appendix~\ref{app:exp}), a full suite of polarimetric images  (\ref{appendix:maps})    for all the observed targets, 
comparisons between multiple flux-extraction methods (\ref{app:fluxext}) and between the polarimetry results obtained during the VLBI campaign and the monitoring programme with the Atacama Compact Array (\ref{app:amapola_comp}),   tables with polarimetric quantities  per ALMA spectral-window (\ref{app:stokes}),  
 Faraday RM plots  (\ref{app:RMplots}), 
quality assessment of the circular polarization estimates (\ref{app:stokesV}), and 
mm spectral indices  of all the observed targets (\ref{app:alpha}).
Finally, a two-component polarization model for M87, which combines constraints from ALMA and EHT observations, is presented in Appendix~\ref{app:twocomp}.

\section{Observations, data processing, and imaging}
\label{sect:data}

\subsection{2017 VLBI observations with ALMA}
\label{obs}
The observations with phased-ALMA were  conducted as part of Cycle~4 during the 2017 VLBI campaign 
in ALMA Band 3 (April 1-3) and Band 6 (April 5-11), respectively.  
The ALMA data were acquired simultaneously with the VLBI observations (in this sense they are a ``byproduct" of the VLBI operations).
 The ALMA array was in the compact configurations C40-1 (with 0.15~km longest baseline)  and, after Apr 6, C40-3 (with 0.46~km longest baseline).
Only antennas within a  radius of 180~m (from the array center) were used for phasing on all days. 
About 37 antennas were normally phased together, which is equivalent to a telescope of 73 m diameter\footnote{A few more antennas participated in the observations without being phased, the so-called “comparison” antennas, which are mostly used to provide feedback on the efficiency of the phasing process \citep[see][for details]{APPPaper,QA2Paper}.}. 
 In both Band 3 and 6, 
the spectral setup includes four spectral windows (SPWs) of 1875~MHz,
two in the lower and two in the upper sideband, 
correlated with 240 channels per SPW (corresponding to a spectral resolution of 7.8125 MHz\footnote{The recommended continuum setup for standard ALMA observations in full polarization mode is somewhat different and consists of 64 channels,  31.25\,MHz wide, per SPW.}).
In Band~3 the four SPWs are centered at 86.268, 88.268, 98.328, and 100.268 GHz\footnote{
The "uneven" frequency separation with SPW=2 is due to constraints on the first and second Local Oscillators  in the ALMA's tuning system.}
while in Band~6 they are centered at 213.100, 215.100, 227.100, and 229.100 GHz. 

Three projects were observed in Band 3 with the GMVA 
(science targets: OJ\,287, Sgr\,A*, 3C\,273)
and six projects were observed in Band 6 with the EHT (science targets: OJ~287, M87, 3C~279, Sgr~A*, NGC~1052, Cen~A). 
The projects were arranged and calibrated in ``tracks'' (where one track consists of the observations taken during the same day/session).
In Appendix~\ref{app:exp} we provide a list of the observed projects and targets on each day, 
with the underlying identifications of (calibration and science target) sources within each project  
(see Tables~\ref{table:gmva_exp} and \ref{table:eht_exp} in Appendix~\ref{app:exp}). 
 More details of the observation structure and calibration sources can be found in \citet{QA2Paper}.

\subsection{Data calibration and processing}
\label{sect:datacal}

During phased-array operations, the data path from the antennas to the ALMA correlator is different with respect to standard interferometric operations \citep{APPPaper,QA2Paper}. 
This makes the  calibration of VLBI observations within the Common Astronomy Software Applications ({\sc casa}) package intrinsically different and some essential modification in the procedures is required with respect to ALMA standard  observations. 
The special steps added to the standard ALMA  polarization calibration procedures \citep[e.g.,][]{Nagai2016} are described in detail in \citet{QA2Paper}. 
The latter focus mostly on the LP calibration and the polarization conversion at the VLBI correlators \citep{PCPaper}. In this paper we extend the data analysis also to the CP.

Only sources observed in VLBI mode were 
calibrated in polarization \citep[see Section 5 in][]{QA2Paper}. 
Therefore the sources exclusively observed for ordinary ALMA calibration during the VLBI schedule gaps  (i.e., \texttt{Flux} and \texttt{Gain} calibrators) are excluded from this analysis (compare the source list in 
Tables~\ref{table:gmva_exp} and \ref{table:eht_exp} in Appendix~\ref{app:exp} 
with Tables~4 and 6 in \citealt{QA2Paper}). 
Two additional sources observed on Apr 7, 3C~84 and J0006-0623, are also excluded from the following analysis. These sources are in fact flagged in a final flagging step (run on the fully calibrated $uv$-data before imaging and data analysis), which  
 removes visibility data points having amplitudes outside  a certain range (set  by  three times the RMS from the median of the  data) and
a source elevation below 25\dg. 
Finally,  the two weakest targets observed at 1.3~mm, NCG~1052 and J0132-1654, 
were found to fall below the flux threshold 
(correlated flux density of $>$0.5~Jy on intra-ALMA baselines)
required to enable on-source phasing of the array as commissioned  \citep{APPPaper}.
Despite   these two sources are detected with high signal-to-noise ratio (SNR) in total intensity (SNR$>1000$) and polarized flux (SNR$>50$ for J0132-1654), we recommend extra care in interpreting these source measurements owing to lower data quality.

\subsection{Full-stokes imaging} 
\label{sect:dataimg}

All targets observed in Band 3 and Band 6 are imaged using the {\sc casa} task \texttt{tclean} in all Stokes parameters: $I$, $Q$, $U$, $V$. 
A Briggs weighting scheme \citep{Briggs1995} is adopted with a robust parameter of 0.5, and a cleaning gain of 0.1. 
A first quick cleaning (100 iterations over all four Stokes parameters) is done in the inner 10\arcsec\ and 4\arcsec\, in bands 3 and 6, respectively. 
Providing there is still significant emission ($>7\sigma$) in the residual maps (e.g., in M87 and Sgr~A*), an automatic script changes the cleaning mask  accordingly, and a second, deeper cleaning is done down to 2$\sigma$ (these two clean steps are run with parameter \texttt{interactive=False}). 
A final interactive clean step (with \texttt{interactive=True}) is run to adjust the mask to include real emission which was missed by the automatic masking and to clean deeper sources with complex structure and high-signal residuals (this step was essential for proper cleaning of Sgr A*). 
No self-calibration was attempted during the imaging stage (the default calibration scheme for ALMA-VLBI data already relies on self-calibration; see \citealt{QA2Paper} for details). 

We produced maps of size 256$\times$256 pixels, with a pixel size of 0\pas5 and 0\pas2 in Band~3 and Band~6, respectively, resulting in maps with a field-of-view (FOV) of 128\as $\times$ 128\as\ and 51\as $\times$ 51\as, respectively, thereby comfortably covering the primary beams of ALMA Band~3 (60\as) and Band~6 (27\as) antennas. 
We produced maps for individual SPWs and by combining SPWs in each sideband (SPW=0,1 and SPW=2,3), setting the \texttt{tclean} parameters \texttt{deconvolver=`hogbom'} and \texttt{nterms=1}, as well as by combining all four SPWs, using  \texttt{deconvolver=`mtmfs'} and \texttt{nterms=2}.  
The latter achieved better sensitivity and yielded higher quality images\footnote{The \texttt{deconvolver=`mtmfs'} performed best when combining all four SPW, yielding on average 30--40\% better sensitivity than \texttt{deconvolver=`hogbom'} combining two SPW at a time, as expected for RMS~$\propto 1/\sqrt{\Delta \nu}$. However, \texttt{deconvolver=`hogbom'} performed poorly when combining all four SPW, especially for steep spectral index sources, yielding up to 50\% worse RMS than \texttt{deconvolver=`mtmfs'}.}, so we used the combined SPW images for the imaging analysis presented in this paper (except for the per-SPW analysis).

Representative total-intensity images in Band~3  and  Band~6 are shown in Figures~\ref{fig:m87+sgra_I_maps} (Stokes\,$I$) and \ref{fig:m87+sgra_polimage} (Stokes\,$I$ + polarized intensity),  whereas the full suite of images including each source observed in Band~3  and  Band~6 on each day of the 2017 VLBI campaign is reported in Appendix~\ref{appendix:maps} (Figures~\ref{fig:polimages_sgra_1mm}--\ref{fig:polimages_3mm}). 

The array  configurations employed during phased-array observations yielded synthesized beams in the range 
[4\pas7--6\pas1] $\times$ [2\pas4--3\pas4]  in Band~3 
 and 
[1\pas2--3\pas0] $\times$ [0\pas7--1\pas5]   
 in Band~6 (depending on the day and the target). 
 Images on different days achieve different sensitivities and angular resolutions, depending on the time on-source and baseline lengths of the phased-array. 
In particular, the relatively large range of beamsizes in Band~6  is due to the fact that, 
during the EHT campaign, progressively more antennas were moved out from the ``central cluster'' (with a diameter $<$ 150~m). 
As a consequence,  in the last day of the campaign (Apr 11) the observations were carried out with a more extended array, yielding a beamsize in the range [1\pas2--1\pas5] $\times$ [0\pas7--0\pas9] 
(i.e., an angular resolution  roughly two times better than that of other tracks). 
Tables~\ref{tab:GMVA_im_rms} and~\ref{tab:EHT_im_rms} in Appendix~\ref{appendix:maps}     report the synthesized beamsize and the RMS achieved in the images of each Stokes parameter for each source observed in Band~3  and  Band~6 on each day.

\subsection{Additional ALMA polarization datasets on M87}
\label{m87_add_data}

In addition to the April 2017 data,  we have also analysed  ALMA data acquired during the 2018 VLBI campaign as well as ALMA archival polarimetric experiments targeting M87. 

The  2018 VLBI campaign was conducted as part of Cycle 5 in Band 3 (April 12-17) and Band 6 (April 18-29), respectively. 
The observational setup  was the same as in Cycle 4, as outlined in \S\ref{obs} (a full description of the 2018 VLBI campaign will be reported elsewhere). 
Three observations of M87 at $\lambda$1.3~mm were conducted on Apr 21, 22, and 25 under the  project 2017.1.00841.V.
For the data processing and calibration, we followed the same procedure used for the 2017 observations, as outlined in \S\ref{sect:datacal}.

The archival experiments include three observations at $\lambda$3mm carried out on Sep and Nov 2015 (project codes: 2013.1.01022.S and 2015.1.01170.S, respectively) and Oct 2016 (project code: 2016.1.00415.S), and one observation at $\lambda$1.3mm from Sep 2018 (project code: 2017.1.00608.S).    
For projects 2013.1.01022.S and  2015.1.01170.S,  we used directly the imaging products released with the standard QA2 process and publicly available for download from the ALMA archive. 
For  projects 2016.1.00415.S and  2017.1.00608.S, 
we downloaded the raw visibility data and the QA2 calibration products from the ALMA archive, and we revised the polarization calibration after additional data flagging, following the procedures outlined in \citet{Nagai2016}. 

The data imaging was performed following the same procedures outlined in \S\ref{sect:dataimg}.  
After imaging,  we found that in 2017.1.00608.S,
Stokes $I$, $Q$, and $U$ are not co-located: $U$ is shifted $\sim$0.07\as\ to the East, while $Q$ is shifted $\sim$0.13\as\ West and $\sim$0.07\as\ north, with respect to $I$, respectively. 
This shift (whose origin is unknown) prevents us to assess reliably the polarimetric properties of M87. Therefore, we will not use 2017.1.00608.S in the analysis presented in this paper.
The analysis and results of the other datasets will be presented in \S\ref{sect:results_m87}. 
\section{Data analysis}
\label{sect:analysis}
\begin{figure*}
\includegraphics[width=0.5\textwidth]{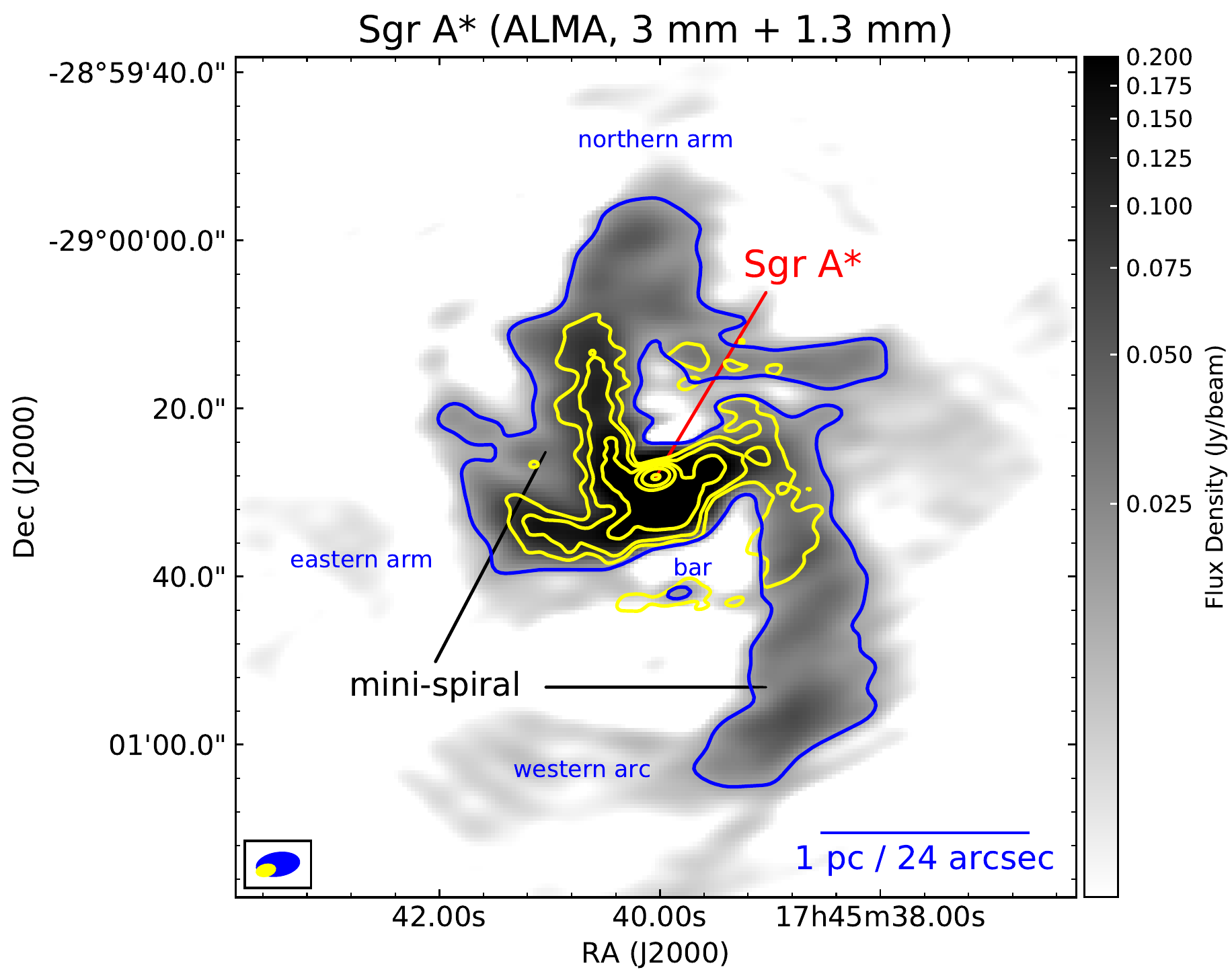}
\includegraphics[width=0.5\textwidth]{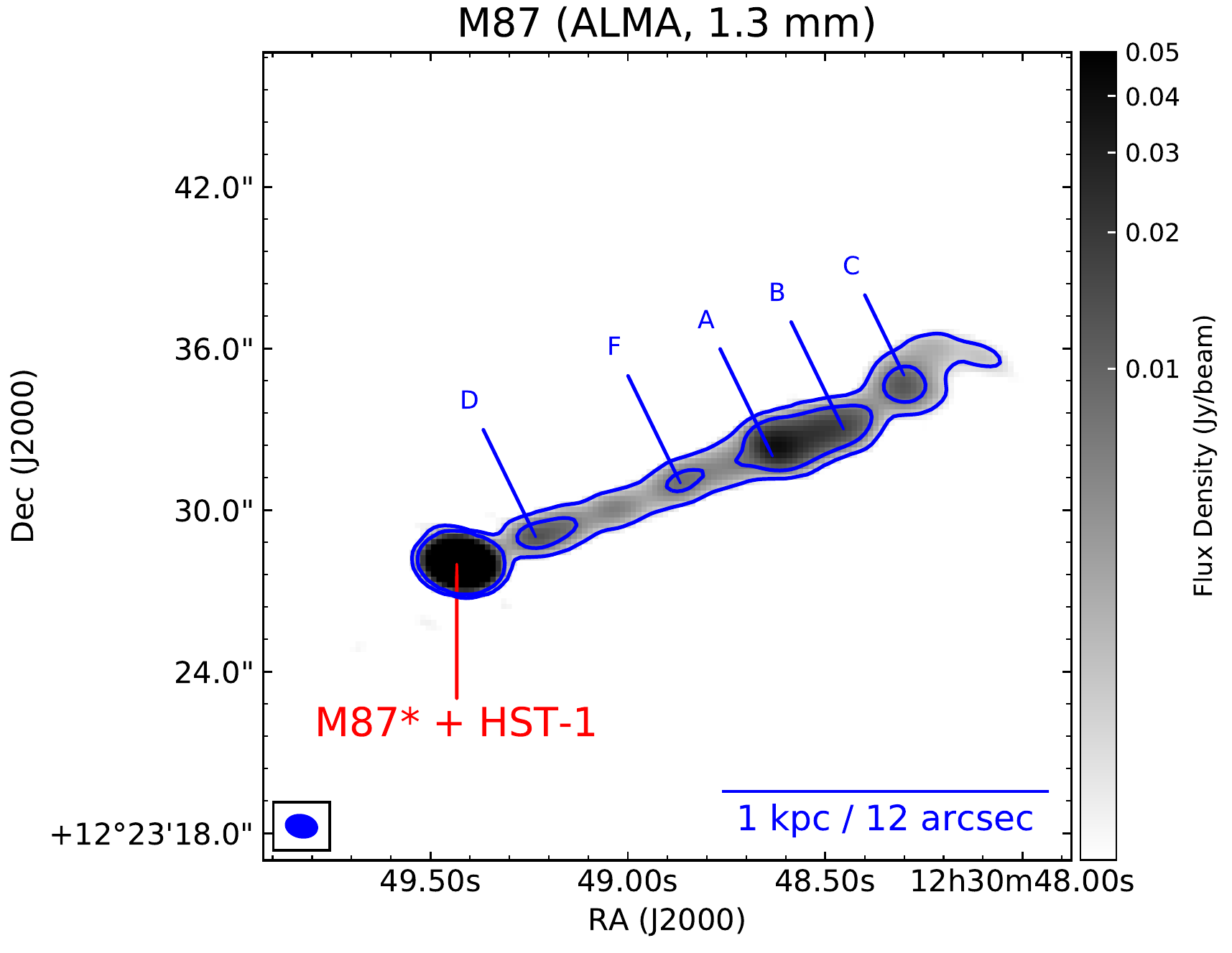}
\caption{
Representative total intensity images. 
Left panel: 
Image of Sgr A* at  3~mm on Apr 3 (grey-scale and blue contour) and at  1.3~mm (yellow contours) on Apr 6 2017. 
The image showcases the well-known ``mini-spiral" structure surrounding the central compact core, including  the eastern and northern arms, the western arc, and the bar at the center.
The contour levels at 1.3~mm are $5\sigma \times 2^n$ where $\sigma=$0.44~mJy~beam$^{-1}$ 
and n$=0\,,1\,,2\,,3\,\ldots$ up to the peak flux-density; the contour level at 3~mm corresponds to  $20\sigma$ ($\sigma=$0.8~mJy~beam$^{-1}$). 
The peak flux-density is $2.5\,\pm\,0.1$ ($2.6\,\pm\,0.3$) Jy~beam$^{-1}$ and the integrated flux density across the entire source is $9.9\,\pm\,0.5$ ($4.9\,\pm\,0.5$) Jy at a representative frequency of 93 (221)~GHz. 
The FOV is given by the primary beam in Band 3 ($\sim$60\arcsec) and 1 pc corresponds to 24\as. 
The beamsizes are 5\pas0 $\times$ 2\pas7 (P.A.= --81.1\dg) in Band~3 and 2\pas2 $\times$ 1\pas3 (P.A.= -77.5\dg) in Band~6, shown as a blue and yellow ovals, respectively, in the lower left corner. 
Right  panel: 
Image of M87  at 1.3~mm  on Apr 11 2017. The image showcases the structure of the kpc-scale  relativistic  jet comprised of a  bright core at the nucleus and the  knots along the jet labeled as D, F, A, B, C; HST-1 is not resolved from the nucleus in these images.
The RMS noise level is 0.16~mJy~beam$^{-1}$, and the contour levels are a factor of 10 and 40 of the RMS. 
The peak flux-density is 1.34~Jy~beam$^{-1}$ and the integrated flux-density is 1.57~Jy at a representative frequency of 221~GHz.  
The FOV is given by the primary beam in Band 6 ($\sim$ 27\arcsec\ at 1.3~mm). 
1 kpc  corresponds to 12\as.
This observation was conducted with the most extended array during the VLBI campaign, yielding the highest angular resolution (beam size = 1\pas2 $\times$ 0\pas8,  P.A.= 79.3\dg, shown in the lower left corner). 
In both panels the four observing SPWs (see \S~\ref{obs}) were used together for imaging. The intensity brightness is plotted using  a logarithmic weighting function (starting from the $5\sigma$-level), in order to highlight the full extent of both the mini-spiral (in Sgr~A*)  and the jet (in M87).  
}
\label{fig:m87+sgra_I_maps}
\end{figure*}
\subsection{Representative total intensity images}
\label{Sgr_A_m87_im}
The sources targeted by the GMVA and EHT are generally unresolved  at 
arcsecond scales and their images are mostly consistent with point sources (see images displayed in Appendix~\ref{appendix:maps}). 
 The EHT key science targets, Sgr~A* and M87, are clear  exceptions, and show  complex/extended structures  across  tens of arcseconds. 
We show representative images of  Sgr~A* (3~mm, Apr 3; 1.3~mm, Apr 6) and M87 (1.3~mm, Apr 11)  in Figure~\ref{fig:m87+sgra_I_maps}. 
The images displayed cover an area corresponding to the primary beam of the ALMA antennas (27\arcsec\ in Band 6 and 60\arcsec\ in Band 3; the correction for the attenuation of the primary beam is not  applied to these maps).

The Sgr~A* images clearly depict the well-known ``mini-spiral'' structure which traces ionized gas streams surrounding the central compact source; the mini-spiral has been studied in a wide range of wavelengths  \citep[e.g.][]{Zhao2009,Irons2012,Roche2018}. 
 The  ``eastern arm'', the ``northern arm'', and the ``bar'' are clearly seen in both Band 3 and Band 6, while the ``western arc'' is clearly traced only in the Band 3 image (it falls mostly outside of the antenna primary beam for Band 6). 
Similar images were obtained in the 100, 250, and 340~GHz bands in ALMA Cycle~0 by \citet[][see their Fig.~1]{Tsuboi2016}.
Since Sgr~A* shows considerable variability in its core at  mm-wavelengths \citep[e.g.][]{Bower2018}, the displayed maps and quoted flux values throughout this paper should be considered as time-averaged images/values at the given epoch.

The M87 jet has been observed across the entire electromagnetic spectrum \citep[e.g.,][]{Prieto2016}, and imaged in detail at radio wavelengths from $\lambda$1\,metre \citep[with LOFAR:][]{deGasperin2012} through $\lambda$[15--0.7]\,cm  \citep[with the VLA and the VLBA: e.g.,][]{Hada2013,Walker2018} up to $\lambda$\,3\,mm  \citep[with the GMVA: e.g.,][]{Kim2018}.
VLA images at lower radio frequencies \citep[e.g.,][]{Biretta1995} showcase a bright component at the nucleus and a kiloparsec-scale (kpc-scale) relativistic jet, extending across  approximately 25\arcsec\ ($\sim 2$~kpc) from  the central core. 
Images of the kpc-scale relativistic jet were also produced with ALMA Cycle-0 observations at $\lambda$\,3\,mm \citep{Doi2013} and with the SMA at $\lambda$\,1\,mm \citep{Tan2008,Kuo2014}, but could only recover the bright central core and the strongest knots along the jet.  

Our $\lambda$1.3\,mm ALMA image showcases a similar structure, but the higher dynamic range (when compared with these earlier studies) allows us to recover the continuous  structure of the   straight and narrow kpc-scale  jet across approximately 25\as\ from the nucleus, including knots D, F, A, B, C, at increasing distance from the central core (HST-1 is not resolved from the nucleus in these images).
The jet structure at larger radii ($\gtrsim 2$~kpc) as well as the jet-inflated radio lobes, imaged in great detail with observations at lower frequencies, are not recovered in our images (see for example the NRAO 20 cm VLA image).

\subsection{Extracting Stokes parameters in the compact cores}
\label{allstokes}

We extract flux values for Stokes $I$, $Q$, $U$, and $V$ in the compact cores of each target observed in Band 3 and Band 6. 
We employ three different methods which use 
both the visibility data and the full-Stokes images. 
In the $uv$-plane analysis, we use the external {\sc casa} library {\sc uvmultifit}
\citep{UVMULTIFIT}. 
To reduce its processing time, we first average all (240) frequency channels to obtain  one-channel four-SPW visibility $uv$-files. 
We assume that the emission is dominated by a central point source at the phase centre and  we  fit a \texttt{delta} function to the visibilities to obtain  Stokes $I$, $Q$, $U$, and $V$ parameters in each individual SPW. 
Uncertainties are assessed with Monte Carlo (MC) simulations, as the standard deviation of 1000 MC simulations for each Stokes parameter.  
For the image-based values, we take the sum of the central nine pixels of the  \texttt{CLEAN} model component map (an area of $3 \times 3$ pixels, where the pixel size is 0\pas2 in Band~6 and 0\pas5 in Band~3).  
Summing only the central pixels in the model maps allows one to isolate the core emission from the surroundings in sources with extended structure. 
A third independent method  provides the integrated flux by fitting a Gaussian model to the compact source at the phase-center in each image with the  {\sc casa} task \texttt{IMFIT}. 
In the remaining of this paper, we will indicate these three methods as {\sc uvmf}, {\sc 3x3}, and {\sc intf}. 

From a statistical perspective, any fitting method in the visibility domain should be statistically  more reliable  than a $\chi^2$-based fitting analysis in the image-plane (whose pixels have correlated noise), and should therefore be preferred to image-based methods. 
However, we have two reasons for considering both approaches in this study: 
(i) some of our targets exhibit prominent emission structure at arc-second scales (see Figure~\ref{fig:m87+sgra_I_maps}, and the maps in Appendix~\ref{appendix:maps}), 
(ii) the observations are carried out 
with various array configurations, resulting in a different degree of filtering of the source extended emission. Both elements can  potentially bias  the flux values of  the compact cores extracted in the visibility domain vs. the image domain.

In Appendix~\ref{app:fluxext} we  present a comparative analysis of three flux-extraction methods to assess the magnitude of such systematic biases (reported in Table~\ref{tab:methodcomp}). 
The statistical analysis shows that the Stokes~$I$ values estimated  with {\sc uvmf} are consistent with those estimated from the images, with a median absolute deviation (MAD) $\leq 0.07$\%  and individual offsets $<$1\% (for both  point sources and  extended sources) in the case of the {\sc 3x3} method (the agreement is slightly worse for the {\sc intf} method). 
These deviations are negligible when compared to the absolute uncertainty of ALMA's flux calibration (10\% in Band~6).
 This consistency generally holds also for  Stokes $Q$ and $U$ (with MAD $< 1$\%) and other derived parameters within their  uncertainties (see Table~\ref{tab:methodcomp}).
We therefore conclude that, for the purpose of the polarimetric analysis conducted in this paper, the {\it uv}-fitting method  {\sc uvmf} provides  sufficiently precise flux values  for the Stokes parameters (but see Appendix~\ref{app:fluxext}  for details on M87 and  Sgr~A*). 

\citet{QA2Paper} report the Stokes $I$ flux values per source  estimated in the {\it uv}-plane from amplitude gains using the {\sc casa} task \texttt{fluxscale}. 
We assess that the Stokes $I$ estimated from the visibilities with {\sc uvmf} are consistent with those estimated with \texttt{fluxscale} generally within 1\%.  
In addition, \citet{QA2Paper} compared the \texttt{fluxscale} flux values (after opacity correction) with the predicted values from the regular flux monitoring programme with the ALMA Compact Array (ACA), showing that these values are generally within 10\% (see their Appendix B and their Fig.~16). 
In Appendix~\ref{app:amapola_comp} we perform a similar comparative analysis for the sources commonly observed in the ALMA-VLBI campaign and the AMAPOLA polarimetric Grid Survey, concluding that our polarimetric measurements are generally consistent with historic trends of grid sources (see Appendix~\ref{app:amapola_comp} for more details and comparison plots).
\begin{table*}
\caption{Frequency-averaged polarization properties of GMVA targets (at a representative frequency of 93 GHz). }
\label{tab:GMVA_uvmf_RM}
\begin{tabular}{ccccccccc} 
\hline\hline 
Source & Day & I & Spectral Index & LP & EVPA$^{a}$ & $\chi_0$ & RM & Depol.\\
  &  [2017]  & [Jy]&  $\alpha$  & [\%] & [deg] & [deg] & [$10^5$ rad m$^{-2}$] & [$10^{-4}$  GHz$^{-1}$] \\
\hline\hline 
               OJ287             &  Apr 2        &   5.97$\pm$0.30            &      -0.619$\pm$0.029              &      8.811$\pm$0.030              &      -70.02$\pm$0.10         &      -71.85$\pm$0.37           &      0.0305$\pm$0.0062&      2.244$\pm$0.071 \\
          J0510+1800             &  Apr 2        &   3.11$\pm$0.16            &      -0.6360$\pm$0.0059              &      4.173$\pm$0.030              &      81.86$\pm$0.21         &      65.49$\pm$0.81           &      0.273$\pm$0.013&      2.639$\pm$0.078 \\
            4C 01.28             &  Apr 2        &   4.86$\pm$0.24            &      -0.480$\pm$0.033              &      4.421$\pm$0.030              &      -32.27$\pm$0.19         &      -31.73$\pm$0.74           &      -0.009$\pm$0.012&      2.117$\pm$0.054 \\
              Sgr A*             &  Apr 3        &   2.52$\pm$0.13            &      -0.08$\pm$0.13              &      0.735$\pm$0.030              &      8.1$\pm$1.4         &      135.4$\pm$5.3           &      -2.13$\pm$0.10&      4.72$\pm$0.13 \\
          J1924-2914             &  Apr 3        &   5.11$\pm$0.26            &      -0.462$\pm$0.026              &      4.841$\pm$0.030              &      -46.38$\pm$0.18         &      -46.68$\pm$0.70           &      0.005$\pm$0.012&      2.34$\pm$0.22 \\
            NRAO 530             &  Apr 3        &   2.74$\pm$0.14            &      -0.588$\pm$0.010              &      0.921$\pm$0.030              &      38.8$\pm$1.0         &      51.5$\pm$3.7           &      -0.213$\pm$0.061&      0.4372$\pm$0.0034 \\
            4C 09.57             &  Apr 3        &   2.85$\pm$0.14            &      -0.3056$\pm$0.0057              &      4.069$\pm$0.030              &      -28.47$\pm$0.21         &      -31.15$\pm$0.83           &      0.045$\pm$0.014&      0.43$\pm$0.11 \\
               3C279             &  Apr 4        &   12.93$\pm$0.65            &      -0.3703$\pm$0.0087              &      12.159$\pm$0.030              &      43.906$\pm$0.070         &      44.98$\pm$0.27           &      -0.0179$\pm$0.0045&      0.456$\pm$0.041 \\
               3C273             &  Apr 4        &   9.86$\pm$0.49            &      -0.2887$\pm$0.0049              &      3.984$\pm$0.030              &      -45.45$\pm$0.22         &      -41.87$\pm$0.85           &      -0.060$\pm$0.014&      -2.06$\pm$0.38 \\
\hline\hline
\end{tabular} 
\tablenotetext{a}{The EVPAs are the frequency-averaged $\bar{\chi}$ (as defined in Eq.~\ref{eq:rm}).}
\end{table*} 
\begin{table*}
\caption{Frequency-averaged polarization properties of EHT targets (at a representative frequency of 221 GHz).} \label{tab:EHT_uvmf_RM}
\begin{tabular}{ccccccccc} 
\hline\hline 
Source & Day & I & Spectral Index  & LP & EVPA$^{a}$ & $\chi_0$ & RM  & Depol. \\
  &  [2017]  & [Jy]&  $\alpha$ & [\%] & [deg] & [deg] & [$10^5$ rad m$^{-2}$] & [$10^{-4}$  GHz$^{-1}$] \\
\hline 
\hline &&&&&&&& \\
               3C279             &  Apr 5        &   8.99$\pm$0.90            &      -0.642$\pm$0.019              &      13.210$\pm$0.030              &      45.180$\pm$0.060         &      45.20$\pm$0.51           &      -0.002$\pm$0.048&      0.242$\pm$0.051 \\
               3C279             &  Apr 6        &   9.36$\pm$0.94            &      -0.619$\pm$0.033              &      13.010$\pm$0.030              &      43.340$\pm$0.070         &      43.41$\pm$0.52           &      -0.007$\pm$0.049&      0.303$\pm$0.018 \\
               3C279             &  Apr 10        &   8.56$\pm$0.86            &      -0.6090$\pm$0.0030              &      14.690$\pm$0.030              &      40.140$\pm$0.060         &      40.10$\pm$0.46           &      0.004$\pm$0.043&      0.473$\pm$0.033 \\
               3C279             &  Apr 11        &   8.16$\pm$0.82            &      -0.683$\pm$0.019              &      14.910$\pm$0.030              &      40.160$\pm$0.060         &      40.15$\pm$0.46           &      0.001$\pm$0.043&      1.027$\pm$0.015 \\
\hline &&&&&&&& \\
                 M87             &  Apr 5        &   1.28$\pm$0.13            &      -1.212$\pm$0.038              &      2.420$\pm$0.030              &      -7.79$\pm$0.36         &      -14.6$\pm$2.8           &      0.64$\pm$0.27&      1.318$\pm$0.031 \\
                 M87             &  Apr 6        &   1.31$\pm$0.13            &      -1.112$\pm$0.011              &      2.160$\pm$0.030              &      -7.60$\pm$0.40         &      -23.6$\pm$3.1           &      1.51$\pm$0.30&      0.888$\pm$0.046 \\
                 M87             &  Apr 10        &   1.33$\pm$0.13            &      -1.171$\pm$0.023              &      2.740$\pm$0.030              &      0.03$\pm$0.31         &      2.5$\pm$2.5           &      -0.24$\pm$0.23&      0.540$\pm$0.048 \\
                 M87             &  Apr 11        &   1.34$\pm$0.13            &      -1.208$\pm$0.019              &      2.710$\pm$0.030              &      -0.64$\pm$0.32         &      3.5$\pm$2.5           &      -0.39$\pm$0.24&      1.553$\pm$0.064 \\
\hline &&&&&&&& \\
              Sgr A*             &  Apr 6        &   2.63$\pm$0.26            &      -0.0270$\pm$0.0030              &      6.870$\pm$0.030              &      -65.83$\pm$0.13         &      -14.7$\pm$1.0           &      -4.84$\pm$0.10&      3.75$\pm$0.10 \\
              Sgr A*             &  Apr 7        &   2.41$\pm$0.24            &      -0.057$\pm$0.059              &      7.230$\pm$0.030              &      -65.38$\pm$0.12         &      -18.77$\pm$0.93           &      -4.412$\pm$0.088&      3.33$\pm$0.12 \\
              Sgr A*             &  Apr 11        &   2.38$\pm$0.24            &      -0.1450$\pm$0.0080              &      7.470$\pm$0.030              &      -49.33$\pm$0.12         &      -14.66$\pm$0.92           &      -3.281$\pm$0.087&      2.52$\pm$0.32 \\
\hline &&&&&&&& \\
          J1924-2914             &  Apr 6        &   3.25$\pm$0.32            &      -0.780$\pm$0.012              &      6.090$\pm$0.030              &      -49.28$\pm$0.14         &      -53.6$\pm$1.1           &      0.41$\pm$0.10&      0.13$\pm$0.20 \\
          J1924-2914             &  Apr 7        &   3.15$\pm$0.31            &      -0.8510$\pm$0.0070              &      5.970$\pm$0.030              &      -49.22$\pm$0.15         &      -52.1$\pm$1.2           &      0.27$\pm$0.11&      0.1470$\pm$0.0080 \\
          J1924-2914             &  Apr 11        &   3.22$\pm$0.32            &      -0.677$\pm$0.031              &      4.870$\pm$0.030              &      -51.82$\pm$0.18         &      -56.2$\pm$1.4           &      0.42$\pm$0.13&      0.16$\pm$0.21 \\
\hline &&&&&&&& \\
               OJ287             &  Apr 5        &   4.34$\pm$0.43            &      -0.91$\pm$0.10              &      9.020$\pm$0.030              &      -61.190$\pm$0.090         &      -62.32$\pm$0.73           &      0.108$\pm$0.069&      0.11$\pm$0.63 \\
               OJ287             &  Apr 10        &   4.22$\pm$0.42            &      -0.781$\pm$0.088              &      7.000$\pm$0.030              &      -61.81$\pm$0.12         &      -62.6$\pm$1.0           &      0.077$\pm$0.091&      0.09$\pm$0.61 \\
               OJ287             &  Apr 11        &   4.26$\pm$0.43            &      -0.715$\pm$0.043              &      7.150$\pm$0.030              &      -59.61$\pm$0.12         &      -62.97$\pm$0.92           &      0.317$\pm$0.087&      0.110$\pm$0.049 \\
\hline &&&&&&&& \\
            4C 01.28             &  Apr 5        &   3.51$\pm$0.35            &      -0.73$\pm$0.16              &      5.900$\pm$0.030              &      -23.18$\pm$0.15         &      -22.5$\pm$1.1           &      -0.06$\pm$0.11&      0.58$\pm$0.20 \\
            4C 01.28             &  Apr 10        &   3.59$\pm$0.36            &      -0.679$\pm$0.079              &      5.080$\pm$0.030              &      -16.82$\pm$0.17         &      -16.3$\pm$1.3           &      -0.05$\pm$0.12&      0.68$\pm$0.26 \\
            4C 01.28             &  Apr 11        &   3.57$\pm$0.36            &      -0.630$\pm$0.024              &      5.000$\pm$0.030              &      -14.74$\pm$0.18         &      -18.2$\pm$1.4           &      0.33$\pm$0.13&      0.416$\pm$0.054 \\
\hline &&&&&&&& \\
            NRAO 530             &  Apr 6        &   1.61$\pm$0.16            &      -0.96$\pm$0.14              &      2.350$\pm$0.030              &      51.59$\pm$0.37         &      51.7$\pm$2.9           &      -0.01$\pm$0.28&      0.940$\pm$0.062 \\
            NRAO 530             &  Apr 7        &   1.57$\pm$0.16            &      -0.812$\pm$0.017              &      2.430$\pm$0.030              &      50.67$\pm$0.36         &      51.1$\pm$2.8           &      -0.04$\pm$0.27&      0.82$\pm$0.15 \\
\hline &&&&&&&& \\
          J0132-1654             &  Apr 6        &   0.420$\pm$0.040            &      -0.625$\pm$0.086              &      1.990$\pm$0.050              &      15.54$\pm$0.67         &      23.4$\pm$5.3           &      -0.74$\pm$0.50&      0.04$\pm$0.40 \\
          J0132-1654             &  Apr 7        &   0.410$\pm$0.040            &      -0.75$\pm$0.10              &      2.010$\pm$0.050              &      17.85$\pm$0.78         &      14.3$\pm$6.2           &      0.34$\pm$0.58&      -0.18$\pm$0.21 \\
\hline &&&&&&&& \\
            NGC 1052             &  Apr 6        &   0.430$\pm$0.040            &      -0.83$\pm$0.11              &      0.120$\pm$0.030              &      - - - -         &     - - - -                      &         - - - -          &      - - - - \\
            NGC 1052             &  Apr 7        &   0.380$\pm$0.040            &      -1.33$\pm$0.16              &      0.160$\pm$0.040              &      - - - -         &     - - - -                      &         - - - -          &      - - - - \\
\hline &&&&&&&& \\
               Cen A             &  Apr 10        &   5.66$\pm$0.57            &      -0.197$\pm$0.038              &      0.070$\pm$0.030              &      - - - -         &     - - - -                      &         - - - -          &      - - - - \\
\hline &&&&&&&& \\
               3C273             &  Apr 6        &   7.56$\pm$0.76            &      -0.705$\pm$0.024              &      2.390$\pm$0.030              &      -55.50$\pm$0.36         &      -82.2$\pm$2.8           &      2.52$\pm$0.27&      -2.54$\pm$0.11 \\
\hline &&&&&&&& \\
          J0006-0623             &  Apr 6        &   1.99$\pm$0.20            &      -0.789$\pm$0.059              &      12.530$\pm$0.030              &      16.480$\pm$0.070         &      15.83$\pm$0.57           &      0.061$\pm$0.054&      0.78$\pm$0.27 \\
\hline\hline 
\end{tabular} 
\tablenotetext{a}{The EVPAs are the frequency-averaged $\bar{\chi}$ (as defined in Eq.~\ref{eq:rm}).} 
\end{table*} 
\begin{figure*}
\includegraphics[width=0.5\textwidth]{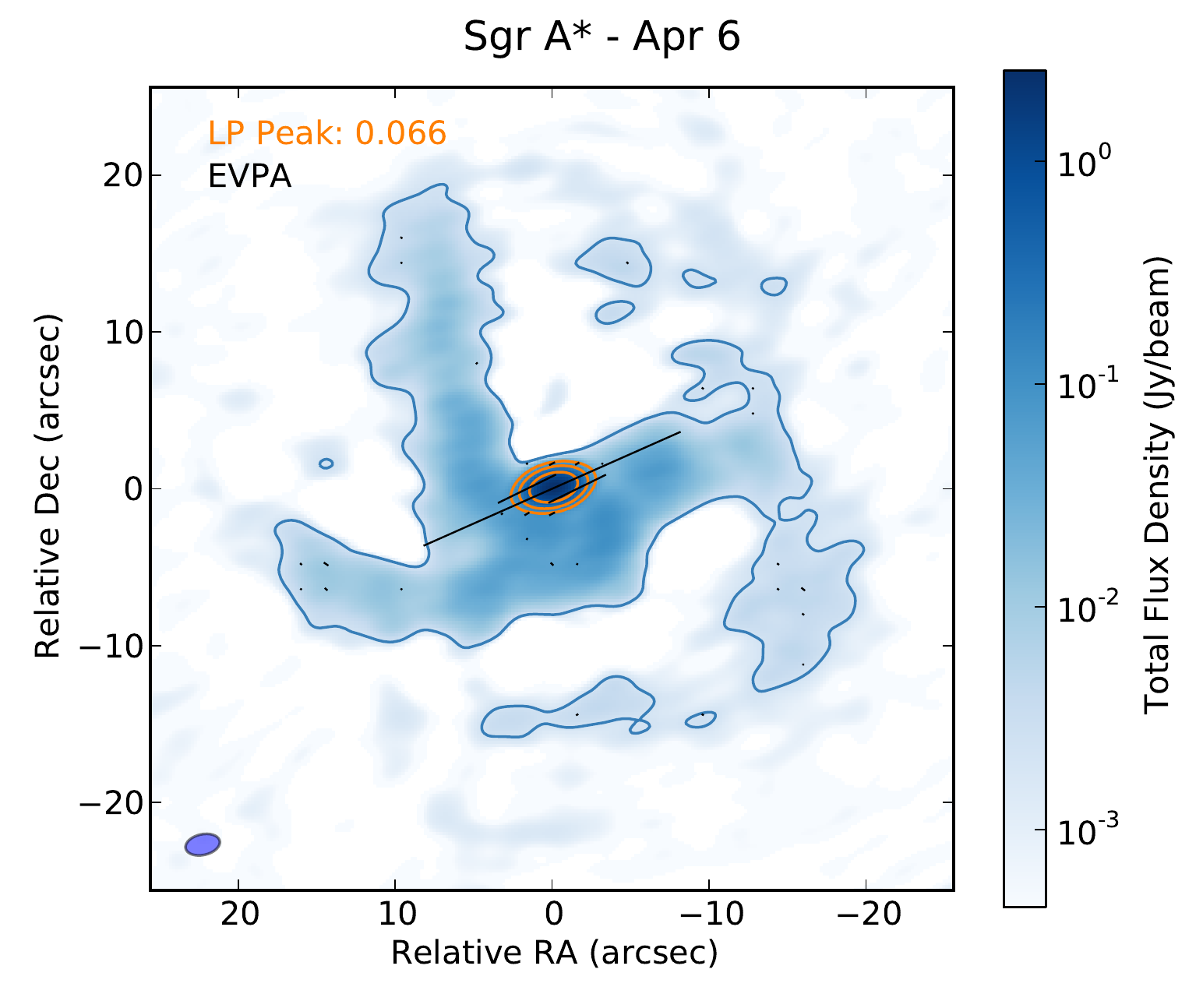} 
\includegraphics[width=0.5\textwidth]{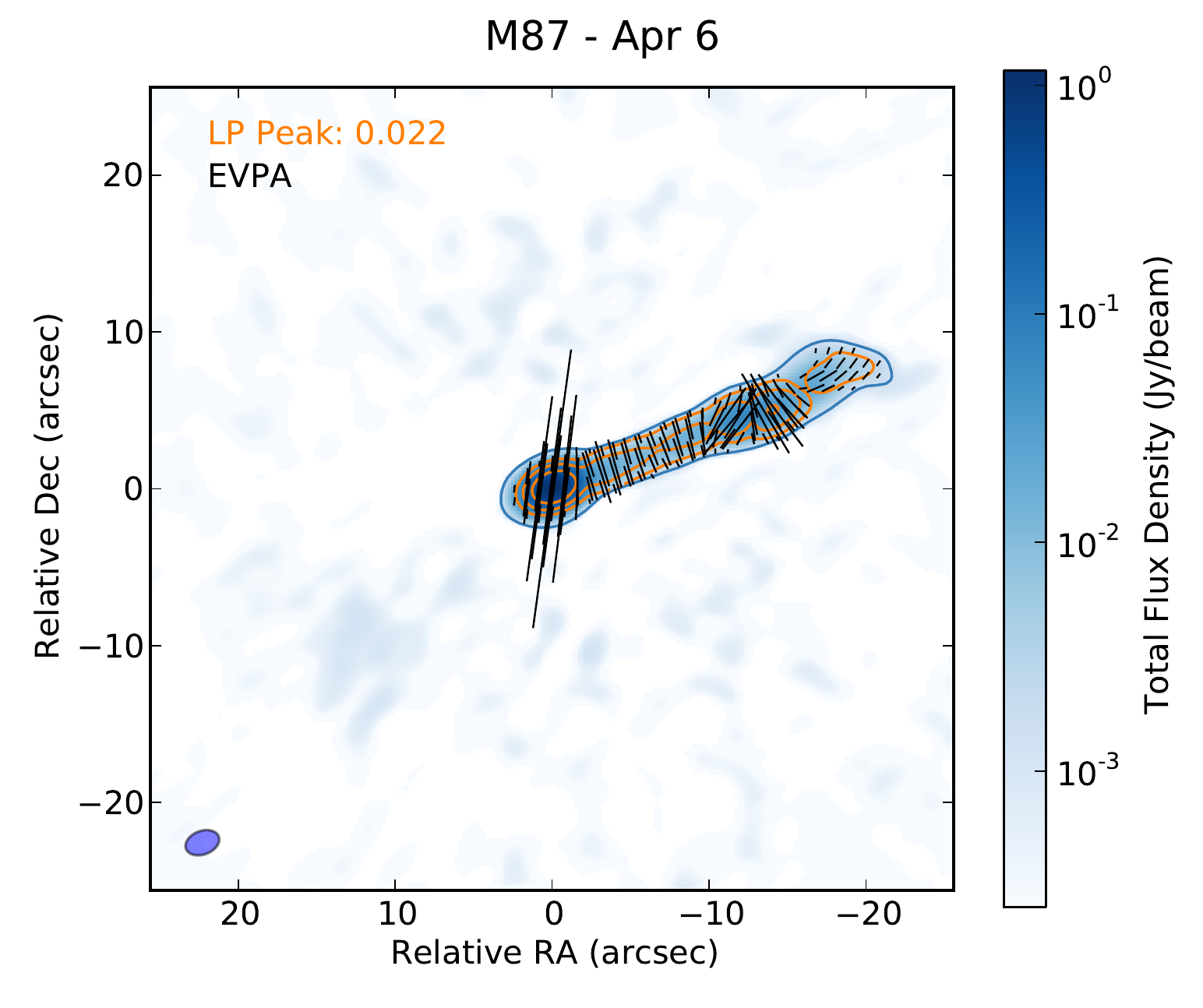}  
\caption{
Polarization images of Sgr A*  (left panel) and M87 (right panel) at 1.3~mm  on Apr 6 2017. 
The raster image and blue contour   show the total intensity emission, the orange contours show the linearly polarized emission, and the black vectors showcase the orientation of the EVPAs  (their length is linearly proportional to the polarized flux).  
 The total intensity brightness is plotted using  a logarithmic weighting function (starting from the $1\sigma$-level), the blue contour corresponds to  $5\sigma$ (where $\sigma$ is the Stokes I map RMS), 
 while the orange contour levels are $5\sigma \times 2^n$ (where $\sigma$ is the LP map RMS and n=0,1,2,3... up to the peak in the image). The LP fraction  at the peak of the compact core is reported in the  upper left corner in each panel. 
The EVPAs are plotted every 8 pixels (1\pas6 or about 1 per beam) for Sgr A* and every 4 pixels (0\pas8 or about 2 per beam) for M87 (in order to sample more uniformly the jet). 
According to the measured RM, the EVPAs towards the compact core should be rotated by $-23$\dg\ (east of north) in Sgr A* and by $-16$\dg\   in M87. 
The beamsizes (shown as an oval in the lower left corner) are  2\pas2 $\times$ 1\pas3 (P.A. --77\dg) and  2\pas2$\times$1\pas5 (P.A. --69\dg) in the left and right panels, respectively. 
Note that there are several tiny EVPAs plotted across the mini-spiral, apparently locating regions with polarized flux above the  image RMS noise cutoff ($5\sigma$). The LP and EVPA errors are however dominated by the systematic leakage (0.03\% of I onto QU), which is not added to the images.  Once these systematic errors are added, the LP flux in those points falls below the $3\sigma$ measurement threshold. Therefore we do not claim detection of polarized emission outside of the central core in Sgr A*. 
Besides, only the polarisation within the inner 1/3 of the primary beam is guaranteed by ALMA. 
The full set of 1.3~mm observations of Sgr A*  and M87  are reported in Figures~\ref{fig:polimages_sgra_1mm} and \ref{fig:polimages_m87}, respectively. 
}
\label{fig:m87+sgra_polimage}
\end{figure*}
\begin{figure*}
\centering
\includegraphics[width=4cm]{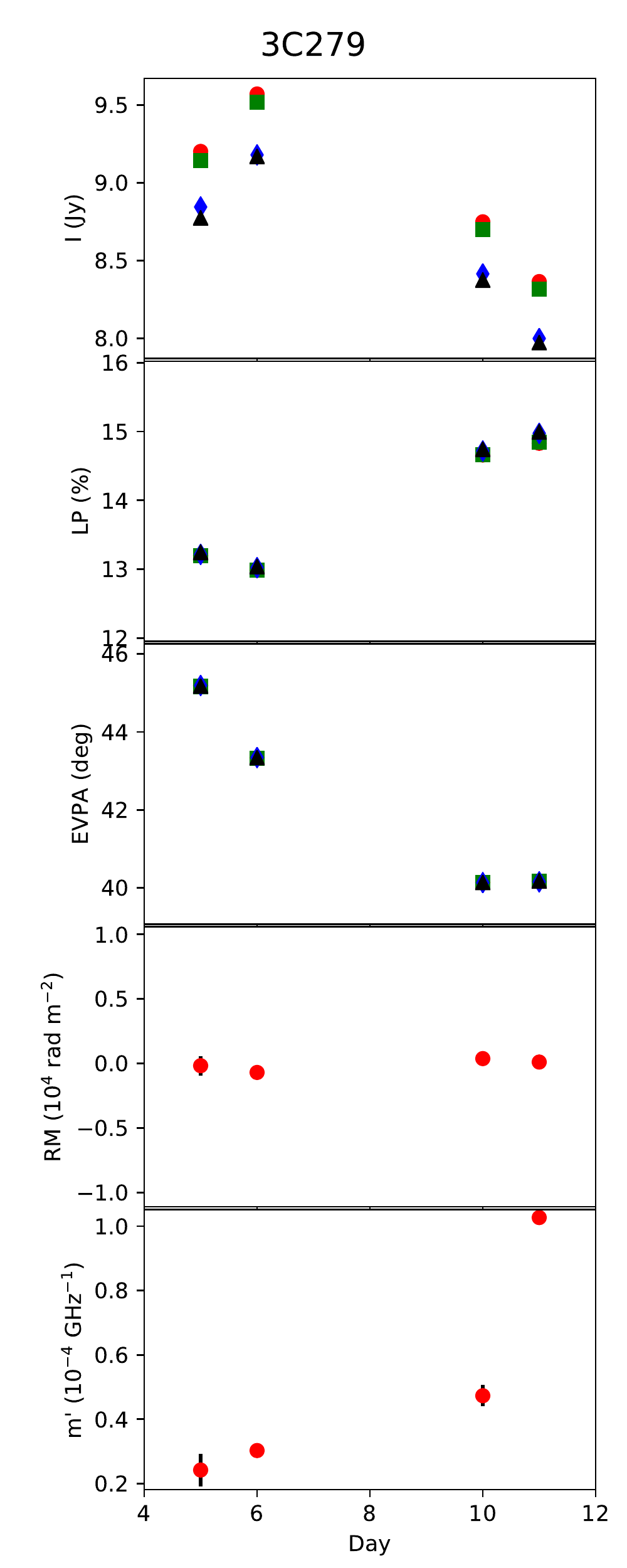} \hspace{-0.35cm}
\includegraphics[width=4cm]{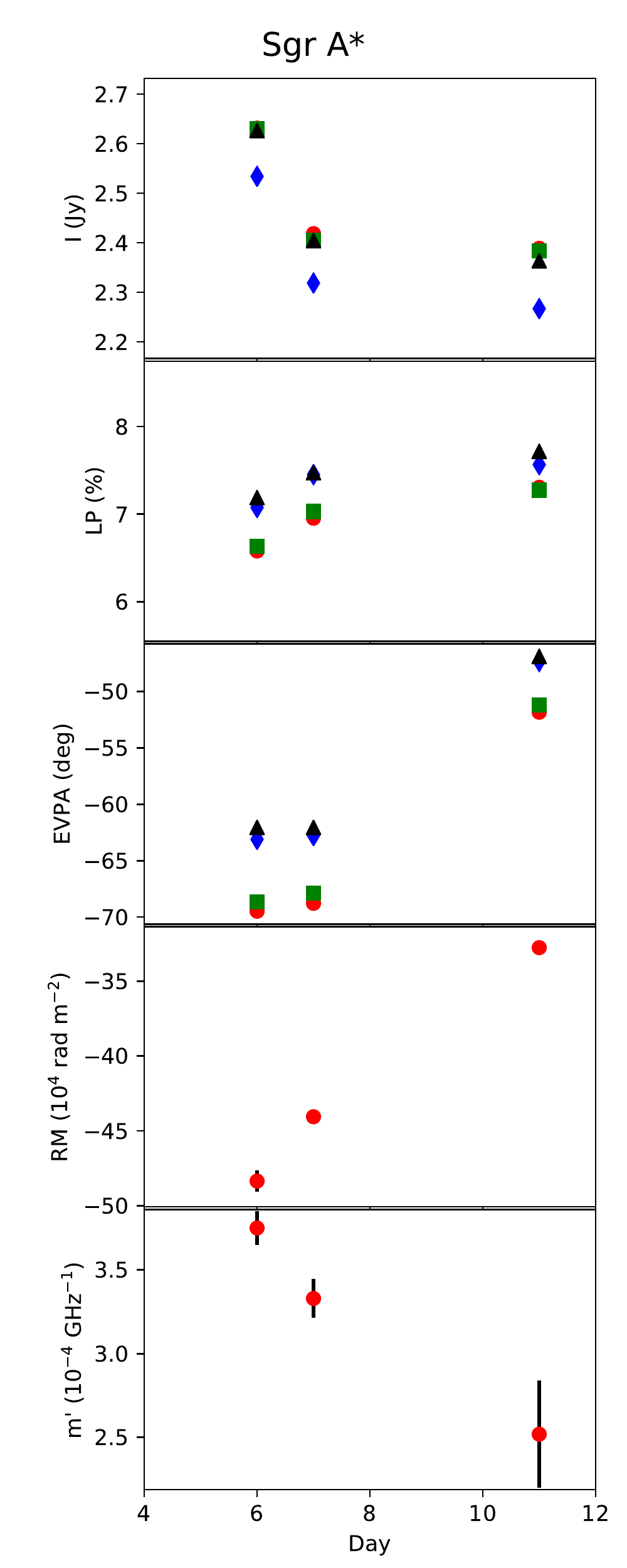}  \hspace{-0.35cm}
\includegraphics[width=4cm]{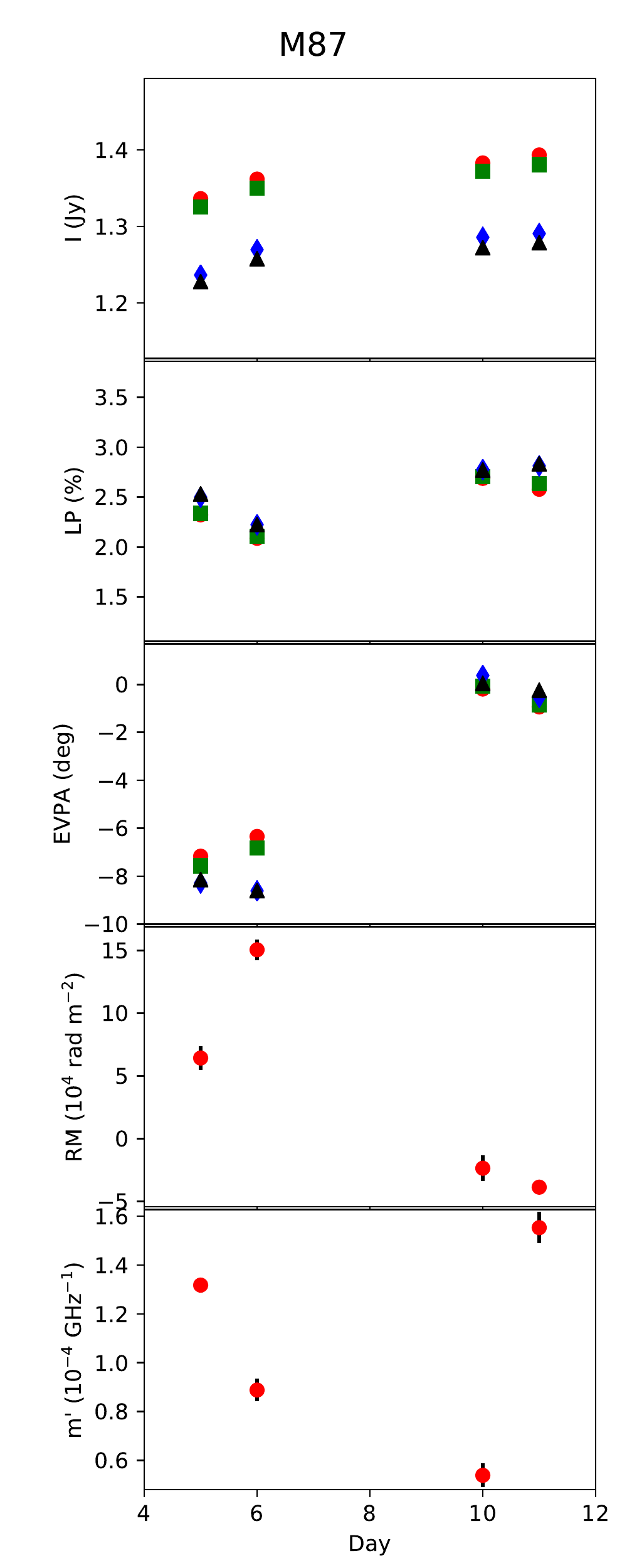}  \hspace{-0.35cm}
\includegraphics[width=4cm]{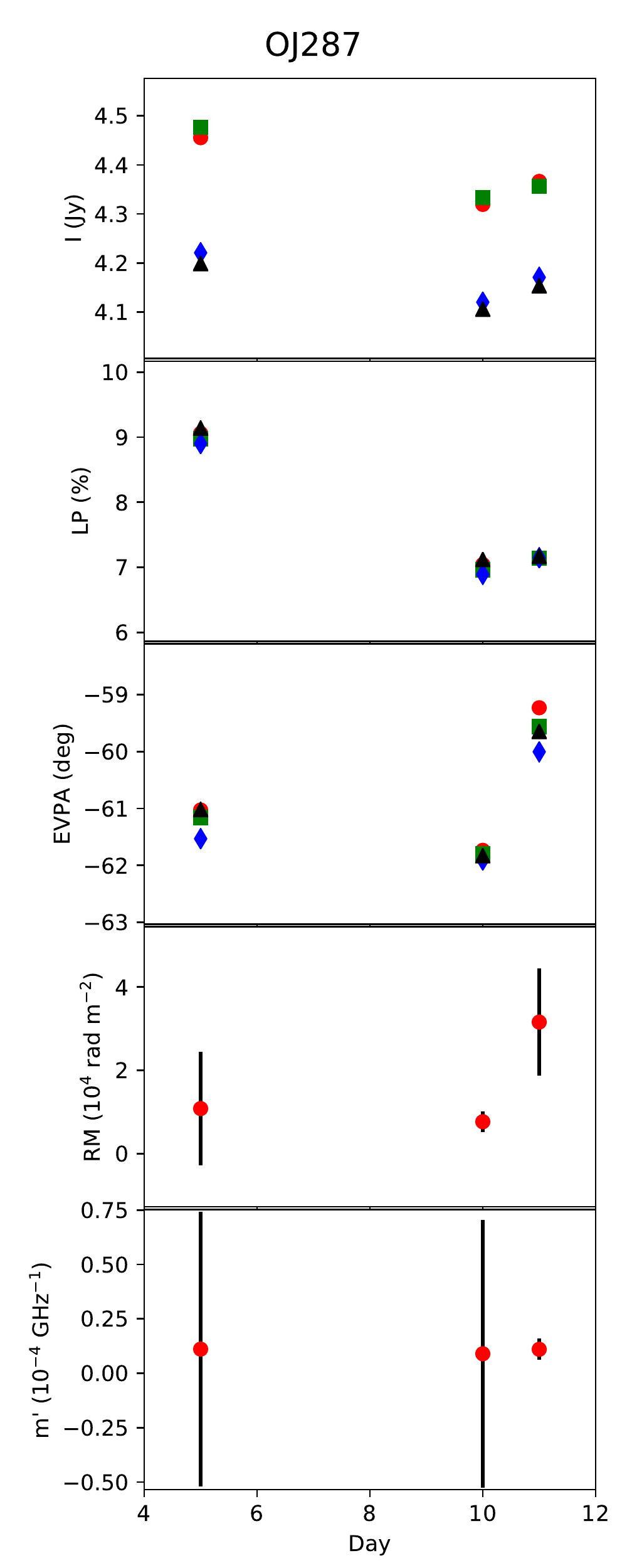}  \hspace{-0.35cm}
\includegraphics[width=4cm]{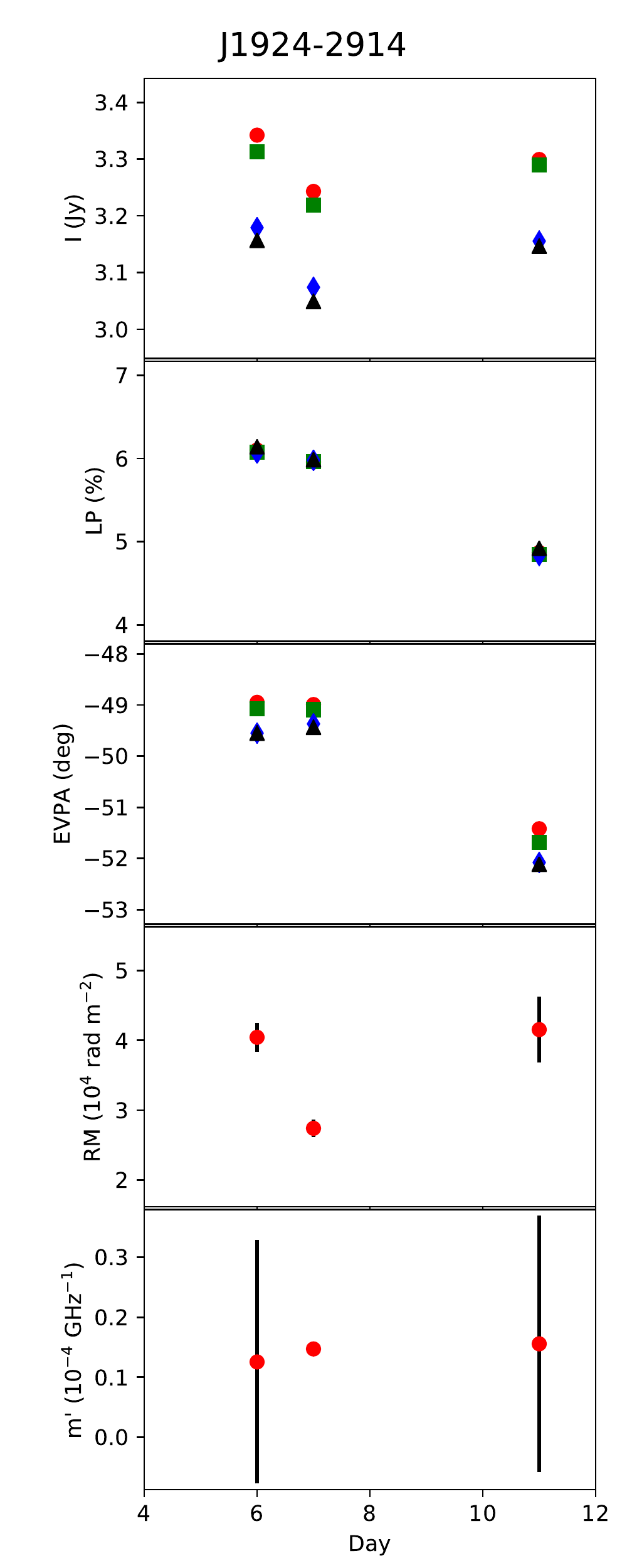}  \hspace{-0.35cm}
\includegraphics[width=4cm]{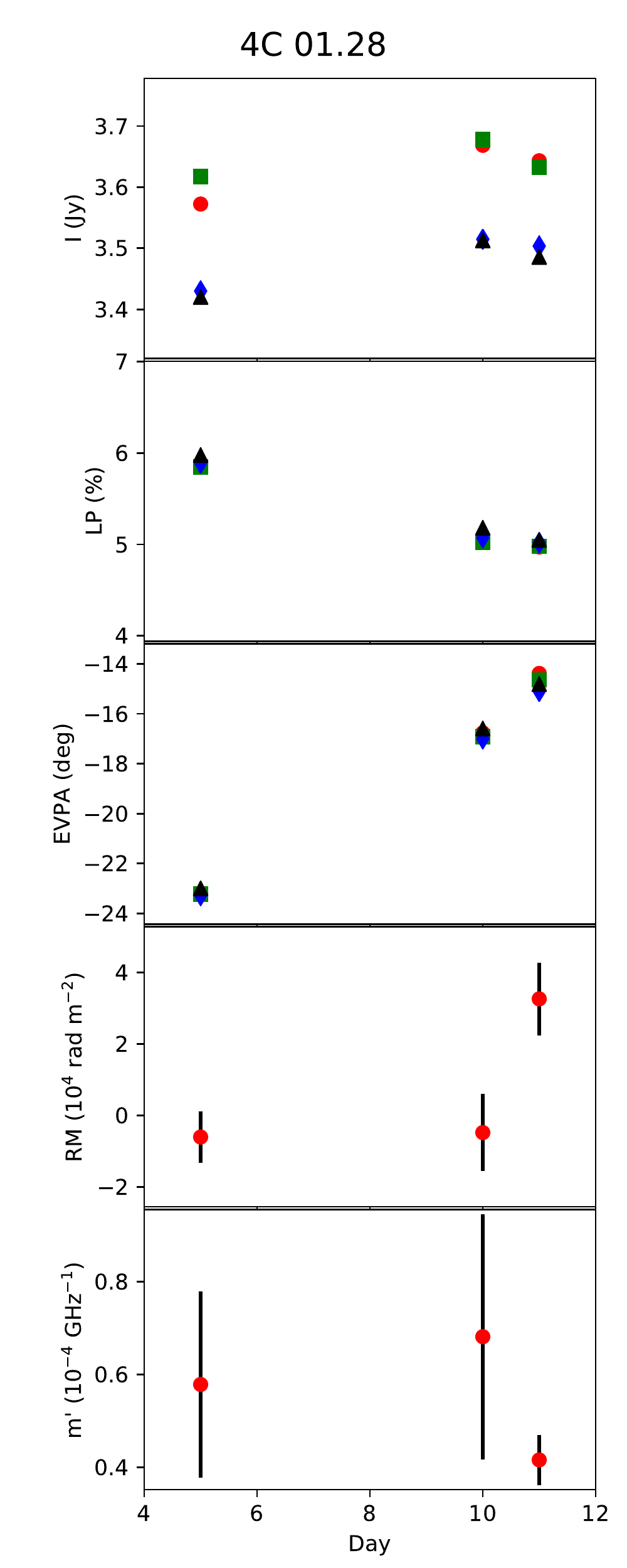}  \hspace{-0.35cm}
\includegraphics[width=4cm]{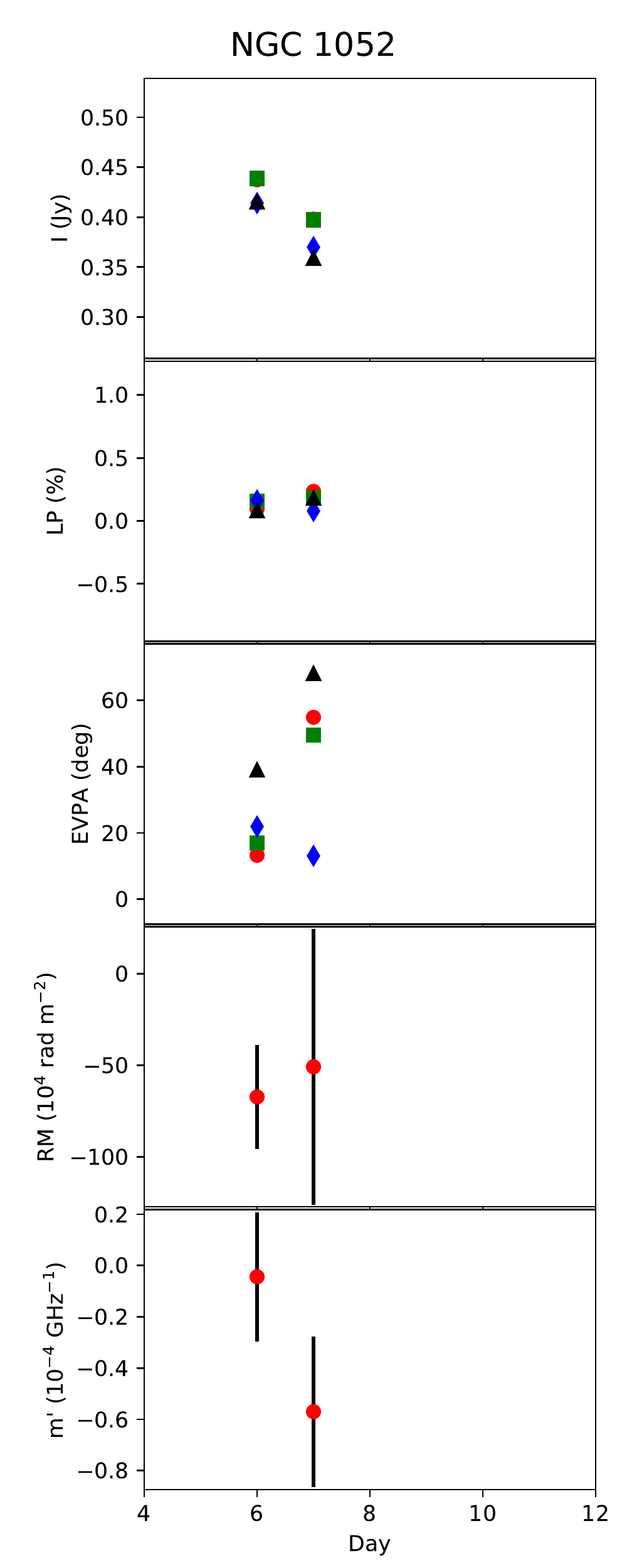} \hspace{-0.35cm}
\includegraphics[width=4cm]{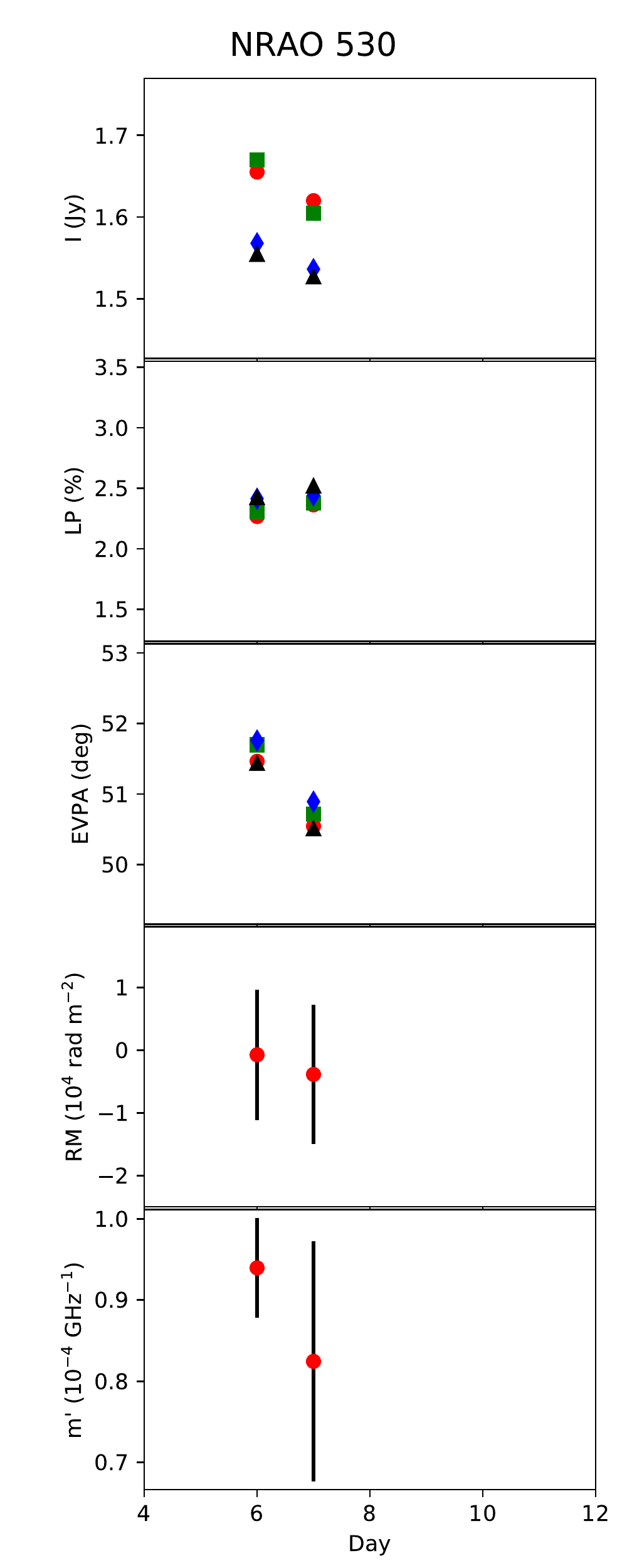}  
\caption{
Polarization properties of selected EHT targets observed during the 2017 VLBI campaign as a function of observing day. 
For each source (labeled at the top of each column), the top  panel shows Stokes I (in Jy), the second panel shows the LP degree (in \%), the third  panel shows the EVPA (in degrees), the fourth  panel shows the RM (in units of 10$^4$~rad~m$^{-2}$), and the bottom panel shows the depolarization (in units of 10$^{-4}$~GHz$^{-1}$). 
The different symbols and colors in the upper three panels indicate 4 different observing bands centered at 213 GHz  (black), 215 GHz  (blue), 227 GHz  (green), and 229 GHz (red), corresponding to SPW=0,1,2,3. 3C\,279 was used as polarization calibrator in all days, except on Apr 7  when J1924-2914 was used. 
}
\label{fig:stokes}
\end{figure*}
\subsection{Polarimetric data analysis}
\label{subsec:polan}

In this section we use the measured values of  the Stokes parameters 
to determine the  polarization properties for all targets, including the fractional LP (\S~\ref{subsec:LP}), the electric vector position angle (EVPA) and its variation as a function of frequency or Faraday Rotation (\S~\ref{subsec:RM}), the degree of depolarization (\S~\ref{subsec:depol}), 
and the fractional CP (\S~\ref{subsec:CP}).  
These polarization quantities, averaged across the four SPWs, are reported in
Tables~\ref{tab:GMVA_uvmf_RM} and \ref{tab:EHT_uvmf_RM} for each target observed with the GMVA and the EHT, respectively (while Table~\ref{tab:M87_RM} summarizes all the ALMA polarimetric observations towards M87 analysed in this paper).
For selected EHT targets, the polarization properties (per SPW and per day)  are displayed in Figure~\ref{fig:stokes}.

\subsubsection{Linear polarization and EVPA}
\label{subsec:LP}
The values estimated for Stokes {\it Q} and {\it U} can be combined to directly provide the fractional LP  in the form  $\sqrt{Q^{2}+U^{2}}/I$, as well as the EVPA, $\chi$, via the equation  $2\chi=\mathrm{arctan}(U/Q)$. 
Tables~\ref{tab:GMVA_uvmf_spw} and  \ref{tab:EHT_uvmf_spw} in Appendix~\ref{app:stokes} report Stokes parameters, LP, and EVPA, for each SPW. The  LP has been {\it debiased} in order to correct for the LP bias in the low SNR regime (this correction is especially relevant for low polarization sources; see Appendix~\ref{app:stokes} for the debiased LP derivation).

The estimated LP fractions  range from $\lesssim 0.1-0.2$\% for the most weakly polarized targets (Cen A and NGC~1052) to 15\% for the most strongly polarized target (3C279), consistent with previous measurements (see Appendix~\ref{app:amapola_comp}). 
The uncertainties in LP include the fitting (thermal) error of Stokes $Q$ and $U$ and 
the (systematic) Stokes $I$ leakage onto Stokes $Q$ and $U$ (0.03\% of Stokes~$I$) added in quadrature.
This  analysis yields LP uncertainties $< 0.1\%$, similar to those quoted in previous studies \citep{Nagai2016,Bower2018}. 

 Figure~\ref{fig:m87+sgra_polimage} showcases representative polarization images of Sgr A*  (left panel) and M87 (right panel) as observed at 1.3~mm  on Apr 6. The  individual images display the measured EVPAs overlaid on  the polarized flux contour images and the total intensity images. 
Note that the EVPAs are not Faraday-corrected and that the  measured\footnote{The actual magnetic field in the source may be different from the measured one, which can be affected by Lorentz transformation and light aberration.} magnetic field orientations should be rotated by 90\dg. 
In Sgr~A*,  polarized emission is present only towards the compact core, while none is observed from the mini-spiral. 
In M87, the EVPA distribution appears  quite smooth along the jet, with no evident large fluctuations of the EVPAs  in nearby regions, except between Knots A and B.
For a negligible RM along the jet, one can infer that the magnetic field  orientation is first parallel to the jet axis, then in Knot A it changes direction (tending to be perpendicular to the jet), and then turns back to be parallel in Knot B, and finally becomes perpendicular to the jet axis further downstream (Knot C). 
This behaviour  can be explained if Knot A is a standing or recollimation shock: 
if multiple standing shocks with different magnetic field configurations  form along the jet and the latter is  threaded with a helical magnetic field, its helicity (or magnetic pitch) would be different before and after the  shock  owing to a different radial dependence of the poloidal and toroidal components of the magnetic field \citep[e.g.,][]{Mizuno2015}. The EVPA distribution is also in good agreement with the polarization characteristics derived from observations at centimeter wavelengths with the VLA \citep[e.g.,][]{Algaba2016}. 
We nevertheless explicitly  note that only the polarisation within the inner third of the primary beam is guaranteed by the ALMA observatory. Since we focus on the polarization properties in the core, the analysis presented in this paper is not affected by this systematics.  

\subsubsection{Rotation measure}
\label{subsec:RM}

Measuring the EVPA for each SPW (i.e., at four different frequencies) enables us to estimate the  RM  
in the 3~mm band (spanning a 16 GHz frequency range of 85--101 GHz) and in the 1.3~mm band (spanning a 18~GHz frequency range of 212--230 GHz), respectively.
In the simplest assumption that the Faraday rotation is caused by a single external Faraday screen (i.e., it occurs outside of the plasma responsible for the polarized emission), a
linear dependence is expected between the EVPA and the wavelength squared. 
In particular, we fit the RM and the mean-wavelength ($\bar{\lambda}$) EVPA ($\bar{\chi}$) following the relation:
\begin{equation}
\chi = \bar{\chi} + RM (\lambda^2 - \bar{\lambda}^2) \,,
\label{eq:rm}
\end{equation}
where $\chi$ is the observed EVPA at  wavelength $\lambda$ and $\bar{\chi}$ is the EVPA at wavelength $\bar{\lambda}$. The EVPA extrapolated to zero wavelength (assuming that the $\lambda^2$ relation holds) is: 
\begin{equation}
\chi_{0} = \bar{\chi} - RM \bar{\lambda}^2 \,.
\label{eq:chi0}
\end{equation}
The RM fitting  is done using a weighted least-squares method of $\chi$ against $\lambda^2$.   
The $\bar{\chi}, \ \chi_{0}$, and the fitted RM values are reported in the sixth, seventh, and eighth    columns of Tables~\ref{tab:GMVA_uvmf_RM} and \ref{tab:EHT_uvmf_RM}, respectively.   

The EVPA uncertainties quoted in Tables~\ref{tab:GMVA_uvmf_RM},  \ref{tab:EHT_uvmf_RM}, \ref{tab:M87_RM},  \ref{tab:GMVA_uvmf_spw},   \ref{tab:EHT_uvmf_spw}, are typically dominated by the systematic leakage of 0.03\% of Stokes $I$ into Stokes $Q$ and $U$.  
At 1.3~mm, this results in estimated errors between 0.06\dg~for the most strongly polarized   source (3C279) and 0.8\dg\ for the weakest  source (J0132-1654), with most sources in the range $0.1$\dg -- $0.4$\dg. These EVPA uncertainties imply RM propagated errors between $0.4 \times 10^4$~\radmsq\ and $6 \times 10^4$~\radmsq, with most sources in the range $(1-3) \times 10^4$~\radmsq. 
Similarly, at 3~mm we find  EVPA uncertainties of 0.07\dg--1.4\dg, with a typical value of 0.2\dg, and RM uncertainties in the range $(0.06-1.0) \times 10^4$~\radmsq, with a typical value of $\sim 0.13 \times 10^4$~\radmsq.

In Appendix~\ref{app:RMplots}, we present plots of the measured EVPAs at the four ALMA SPWs and their RM fitted models as a function of $\lambda^2$ (displayed  in  Figures~\ref{fig:RM1},  \ref{fig:RM3}, \ref{fig:RM_3mm}, and \ref{fig:RM_M87_3+1mm})
 and 
we discuss the  magnitude of thermal and  systematic errors in the  RM analysis. 

\subsubsection{Bandwidth depolarization}
\label{subsec:depol}
In the presence of high RM, 
the large EVPA rotation within the observing frequency bandwidth will decrease the measured fractional polarization owing to Faraday frequency or ``bandwidth" depolarization, which depends on the observing frequency band. The RM values inferred in this study (e.g., Table \ref{tab:EHT_uvmf_RM}) introduce an EVPA rotation of less than one degree within each 2\,GHz spectral window, indicating that the bandwidth depolarization in these data should be very low ($<$0.005\%). However, if there is an internal component of Faraday rotation (i.e., the emitting plasma is itself causing the RM), there will be much higher frequency-dependent (de)polarization effects (the ``differential" Faraday rotation), which will be related to the structure of the Faraday depth across the source \cite[e.g.][]{Cioffi1980, sokoloff98}.

We have modeled the frequency dependence of LP using a simple linear model:
\begin{equation}
m = \bar{m} + m'(\nu - \bar{\nu}) \,,
\label{eq:mnu}
\end{equation}
where 
$m$ is the observed LP at frequency $\nu$, 
$\bar{m}$ is the LP at the mean frequency $\bar{\nu}$, and $m'$ is the change of LP per unit frequency (in GHz$^{-1}$). Given the relatively narrow fractional bandwidth ($\lesssim$2~GHz), the linear approximation given in Eq. \ref{eq:mnu} should suffice to model the frequency depolarization (multifrequency broadband single-dish studies fit more complex models; see for example  \citealt{Pasetto2016,Pasetto2018}).  
We have fitted the values of $m'$ from a least-squares fit of Eq. \ref{eq:mnu} to all sources and epochs, using   
LP estimates for each spectral window from Table~\ref{tab:EHT_uvmf_spw}. 
We show  the fitting results for selected sources in Fig. \ref{fig:stokes} (lower panels). There are clear detections of $m'$ for 3C279, Sgr~A*, and M87; these detections also differ between epochs.
Such complex time-dependent frequency effects in the polarization intensity may be indicative of an internal contribution to the Faraday effects observed at mm wavelengths. 

\subsubsection{Circular polarization}
\label{subsec:CP}

Measuring Stokes {\it V} provides, in principle, a direct estimate of the fractional CP as $V/I$.  
In practice, the polarization calibration for ALMA data in \casa\ is done by solving the polarization equations in the linear approximation, where parallel-hands and cross-hands visibilities are expressed as a linear function of $I$, $Q$, $U$, while it is assumed $V=0$ \citep[e.g.,][]{Nagai2016,QA2Paper}. 
A non-negligible Stokes $V$ in the polarization calibrator will introduce a spurious instrumental Stokes $V$ into the visibilities of all the other sources. 
Moreover, such a Stokes $V$ introduces a bias in the estimate of the cross-polarization phase, $\beta$, at the reference antenna (see Appendix~\ref{app:stokesV}), which translates into a leakage-like effect in the polconverted VLBI visibilities \citep[see Eq.~13 in][]{QA2Paper}. 
 The magnitude of such a bias may depend on the fractional CP of the polarization calibrator, the parallactic-angle coverage of the calibrator, and the specifics of the calibration algorithm. 
In Appendix~\ref{app:stokesV} we attempt to estimate such a spurious contribution to Stokes $V$ by computing the cross-hands visibilities of the polarization calibrator  as a function of parallactic angle (see Figures~\ref{VFigure} and \ref{AllVStokes}). 
This information can then be used to assess Stokes $V$ and CP for all sources in all days (reported in Tables~\ref{tab:GMVA_uvmf_CP} and~\ref{tab:EHT_uvmf_CP} for  GMVA and EHT sources, respectively). 

We stress two main points here. 
First, the reconstructed Stokes $V$ values of the polarization calibrators are non-negligible and are therefore expected to introduce a residual instrumental X--Y phase difference in all other sources, after QA2 calibration.
This can be  seen in the dependence of the reconstructed Stokes $V$ with feed angle in almost all the observed sources (displayed in Figure~\ref{AllVStokes}). 
The  estimated X--Y residual phase offsets are of the order of $0.5^{\circ}$, but they can be as high as $2^{\circ}$ (e.g., on Apr 5). These values would translate to a (purely imaginary) leakage term of the order of a few \% in the polconverted VLBI visibilities. 

The second point is that there is a significant variation in the estimated values of reconstructed stokes $V$  across the observing week. 
In particular, on Apr 5, 3C279 shows a much higher value, indicating either an intrinsic change in the source, or systematic errors induced by either the instrument or the calibration.  
In either case, this anomalously large Stokes $V$ in the polarization calibrator introduces a large X--Y phase difference in all other sources. 
 This can be seen in the strong dependence of reconstructed Stokes $V$  on feed angle for sources OJ287 and 4C 01.28 (displayed in Figure~\ref{AllVStokes}, upper left panel) and in their relatively high Stokes $V$ when compared to the following days (see Table~\ref{tab:EHT_uvmf_CP}). 
Besides the anomalous value in Apr 5, it is interesting to note that the data depart from the sinusoidal model described by Eq.~\ref{SpuriousVEq}, for observations far from transit, especially on Apr 11. These deviations may be related to other instrumental effects which however we are not able to precisely quantify. 
For these reasons, we cannot precisely estimate the magnitude of the true CP  fractions for the observed sources (see Appendix~\ref{app:stokesV} for details). 
 Nevertheless, our analysis still enables us to obtain order-of-magnitude values of CP. 
 In particular, excluding the anomalous Apr 5, we report CP =[--1.0,--1.5] \%  in Sgr~A*,
 CP$\sim$ 0.3\% in M87, and possibly a lower CP level ($\sim 0.1-0.2$\%) in a few other AGNs (3C273, OJ287, 4C 01.28, J0132-1654, J0006-0623; see Table~\ref{tab:EHT_uvmf_CP}).  
In the 3~mm band, we do not detect appreciable CP above 0.1\%, except for 4C 09.57 (--0.34\%), J0510+1800 (--0.14\%) and 3C273 (0.14\%). 
We however note that the official accuracy of CP guaranteed by the ALMA observatory is $<$ 0.6\% ($1\sigma$) or 1.8\% ($3\sigma$), and therefore all of these CP measurements should be regarded as tentative detections.

\section{Results}
\label{sect:results}

\subsection{AGN}

We observed a dozen AGN, eight at 3~mm and ten at 1.3~mm (with six observed in both bands), in addition to M87.
Following the  most prevalent  classification scheme found in the literature \citep[e.g.][]{Lister2005,Veron-Cetty2010}, 
our sample includes three radio galaxies (M87, NGC 1052, Cen A), 
three BL Lacs (OJ 287, J0006-0623, 4C 09.57), and 
seven additional QSOs (3C 273, 3C 279, NRAO 530, 4C 01.28,  J1924-2914, J0132-1654, and J0510+1800).
Following the standard definition of {\it blazar} (i.e. an AGN with a relativistic jet nearly directed towards the L.O.S.), we can further combine the last two categories into seven blazars (3C 279, OJ 287, J1924-2914,  4C 01.28, 4C 09.57, J0006-0623, J0510+1800) and three additional QSOs (3C 273, NRAO 530, J0132-1654). 
The observed radio galaxies have a core that is considered as a LLAGN \citep[e.g.,][]{Ho2008}.

Their polarimetric quantities at 3~mm and  1.3~mm  are reported in Tables~\ref{tab:GMVA_uvmf_RM} and~\ref{tab:EHT_uvmf_RM}, respectively, and displayed  in Figure~\ref{fig:stokes}. 
Overall, we find LP fractions in the range 0.1–15\% (with $SNR \sim 3-500\sigma$) and RM in the range $10^{3.3}-10^{5.5}$~rad/m$^2$ (with $SNR \sim 3-50\sigma$), in line with previous studies at mm-wavelengths with single-dish telescopes \citep[e.g.,][]{Trippe2010,Agudo2018} and interferometers  \citep[e.g.,][]{Plambeck2014,IMV2015,Hovatta2019}. 
We also constrain CP to $<$0.3\% in all the observed AGN, consistent with previous single-dish  \citep[e.g.,][]{Thum2018} and VLBI \citep[e.g.,][]{HomanLister2006} studies, suggesting that at mm-wavelengths  AGN are  not strongly circularly polarized and/or that Faraday conversion of the linearly polarized synchrotron emission is not an efficient process \citep[but see][]{Vitrishchak2008}.

In Appendix~\ref{appendix:maps}, we also report maps  of all the AGN targets observed at 1.3~mm (Figures~\ref{fig:polimages_3c279}, \ref{fig:polimages_agns_3days}, \ref{fig:polimages_agns_1_2days}), and  at 3~mm (Figure~\ref{fig:polimages_3mm}), showcasing their arcsecond-structure at mm wavelengths. 

In the rest of this section, we briefly comment on the properties of selected AGN.

\paragraph{3C~279} 
3C 279 is a bright and highly magnetized gamma-ray emitting blazar, whose jet is inclined at a very small viewing angle ($\lesssim3$\dg). At its distance  (z=0.5362), 1 arcsecond subtends 6.5~kpc. 3C~279 was observed on four days at 1.3~mm and one day at 3~mm.
It is remarkably highly polarized both at 1.3~mm and 3~mm. 
At 1.3~mm, LP varies from 13.2\% on Apr 5 to 14.9\% in Apr 11, while the EVPA goes from 45\dg\ down to 40\dg.  
At 3~mm, LP is slightly lower ($\sim12.9$\%) and the EVPA is 44\dg. 

While at 1.3~mm we can only place a $1\sigma$ upper limit of $<$5000~rad/m$^2$, 
at 3~mm we measure a RM$=1790 \pm 460$~rad/m$^2$ (with a $\sim4\sigma$ significance).
\citet{Lee2015} used the Korean VLBI Network to measure the LP at 13, 7, and 3.5 mm, finding  RM values in the range  -650  to -2700~\radmsq, which appear to scale as a function of wavelength as $\lambda^{-2.2}$. 
These VLBI measurements are not inconsistent with  our 3~mm measurement and our upper limits at 1.3~mm, but more accurate measurements at higher frequencies are needed to confirm an  increase of the RM with frequency. 

The total intensity images at 1.3~mm reveal, besides the bright core, a jet-like feature extending approximately 5\as\ towards south-west (SW)  (Fig.~\ref{fig:polimages_3c279}); such a feature is not discernible in the lower-resolution 3~mm image (Fig.~\ref{fig:polimages_3mm}).
The  jet-like feature is oriented at approximately 40\dg, i.e. is roughly aligned with the EVPA in the core. 
Ultra-high resolution images with the EHT reveal a jet component
approximately along the same PA but on angular scales $10^5$  times
smaller \citep{Kim2020}. 

\paragraph{3C~273} 
3C 273 was the first discovered quasar \citep{Schmidt1963}, and is one of the closest (z=0.158, 1 arcsec = 2.8~kpc)  and brightest radio-loud quasars.
3C~273 was observed both at 1.3~mm and 3~mm (2 days apart). 
Total intensity and LP are higher in the lower frequency band: F=9.9 Jy and LP=4.0\% (at 3~mm) vs. F=7.6 Jy and LP=2.4\% (1.3~mm). 
We estimate a RM = $(2.52 \pm 0.27) \times 10^5$~rad/m$^2$ at 1.3~mm, confirming the high RM revealed in previous
ALMA observations (conducted in Dec 2016 with 0\pas8 angular resolution) by \citet{Hovatta2019} who report LP=1.8\% and a  (twice as large) RM$=(5.0 \pm 0.3) \times 10^5$~rad/m$^2$.  
We also report  for the first time a RM measurement at 3~mm,  RM = $(-0.60 \pm 0.14) \times 10^4$~rad/m$^2$, about 40 times lower and with opposite sign with respect to  the higher frequency band. 
The $\chi_0$ changes from $-82\pm3$\dg\ at 1.3~mm to $-41.9\pm0.8$\dg\ at 3~mm. These large differences may be explained with opacity effects \citep[\S~\ref{sect:LP_1mmvs3mm}; see also][]{Hovatta2019}. 
The EVPAs  measured at 3~mm and 1.3~mm are in excellent agreement with predictions from the AMAPOLA survey (which however over-predicts LP$\sim$3.5\% at 1.3~mm; see Fig.~\ref{fig:stokescomp_gs_2}).

The total intensity images both at 1.3~mm and 3~mm display, besides  the bright core, a bright, one-sided jet extending approximately 20\as\ (54 kpc) to the SW. 
In the higher resolution 1.3~mm image (Fig.~\ref{fig:polimages_agns_1_2days}), the bright component of the jet is narrow and nearly straight,  
starts  at a separation of  $\sim$10\as\ from the core and has a length of  $\sim$10\as. We also detect (at the 3$\sigma$ level) two weak components of the inner jet (within $\sim$10\as\ from the core) joining the bright nucleus to the outer jet.  The  jet structure is qualitatively similar to previous $\lambda$\,cm images made with the VLA at several frequencies between 1.3 and 43~GHz \citep[e.g.,][]{Perley2017}, where the   outer jet appears highly linearly polarized\footnote{ \citet{Perley2017} report an LP as high as 55\% in their at 15 GHz map along the jet boundaries (although in the central regions LP is much lower).}.
We do not detect LP in the jet feature. 
 
\paragraph{OJ~287} 
The bright blazar OJ 287 (z=0.306,  1~arcsec = 4.7~kpc) is among the best candidates for hosting a compact supermassive binary black hole \citep[e.g.][]{Valtonen2008}.
OJ287 was observed on three days at 1.3~mm and one day at 3~mm\footnote{These ALMA observations of OJ 287 in April 2017 were preceded by a major X-ray--optical outburst in late 2016 to early 2017 \citep{Komossa2020}.}.
OJ287 is one of the most highly polarized targets both at 1.3~mm (LP$\sim 7-9$\%) and 3~mm (LP = 8.8\%). LP drops from 9\% on Apr 5 down to 7\% on Apr 10, while 
the EVPA is stable around [-59.6\dg,-61.8\dg]  at  1.3~mm and -70\dg\  at  3~mm.
The LP variation and stable EVPA are consistent with the historical trends derived from the AMAPOLA survey (see Fig.~\ref{fig:stokescomp_gs_1}).  
Its flux density is also stable.  
At 1.3mm, the EVPA either does not follow a $\lambda^2$-law  (Apr 5 and 11) or the formal fit is consistent with RM = 0 (Apr 10). 
  Although we do not have a RM detection at 1.3~mm, we  measure a RM $=3050 \pm 620$~rad/m$^2$ at 3~mm. 
A 30 years monitoring of  the radio jet in OJ287 has revealed that its (sky-projected) PA varies  both at cm and mm wavelengths and follows the modulations of the EVPA at optical wavelengths \citep{Valtonen2012}. 
The observed EVPA/jet-PA trend can be explained with a jet precessing model from the binary black hole which successfully predict an optical EVPA = --66.5\dg\ in 2017 (Dey et al., submitted), consistent with actual measurements from optical  polarimetric observations during 2016/17  \citep{Valtonen2017} 
and close to the EVPA measured at 3~mm and 1.3~mm with ALMA.   

\paragraph{NRAO~530}
J1733-1304 (alias NRAO~530) is a highly variable QSO (at z = 0.902; 1 arcsec = 8~kpc) that exhibits strong gamma-ray flares. 
It was observed on two consecutive days at 1.3~mm and one day at 3~mm.
It is linearly polarized at a $\sim$2.4\% level at 1.3~mm but only 0.9\%  at 3~mm. 
 The EVPA goes from $\sim51$\dg\ at 1.3~mm to 39\dg\ at 3~mm, while $\chi_0$ is stable around 51--52\dg. 
At  3~mm, we estimate RM = $-0.21\pm0.06 \times 10^5$~rad/m$^2$ at a $3.6\sigma$ significance, which is comparable to the inter-band RM 
between 1 and 3\,mm ($-0.33 \times 10^5$~rad/m$^2$).
These RM values  are in agreement with those reported by \citet{Bower2018} at 1.3\,mm.

The arcsecond-scale  structure at 1.3~mm is dominated by a compact core with a second weaker component at a separation of approximately 10\as\ from the core towards west (Fig.~\ref{fig:polimages_agns_1_2days}). At 3~mm, there is another  
feature in opposite direction (to the east), which could be a counter jet component (Fig.~\ref{fig:polimages_3mm}). 
This geometry is apparently inconsistent with the north-south elongation of the jet revealed on scales $<100$~pc by recent VLBI multi-frequency  (22, 43 and 86 GHz) imaging  \citep[e.g.,][]{Lu2011}, although the Boston University Blazar monitoring program\footnote{\url{https://www.bu.edu/blazars/VLBA_GLAST/1730.html}} conducted with the  VLBA at 7~mm has revealed significant changes in the jet position angle over the years, and possibly jet bending. 

\paragraph{J1924-2914} 
J1924-2914 
is  a  radio-loud  blazar at z=0.352 (1 arcsec = 5.1~kpc),  which shows  strong  variability  from  radio  to  X-ray.
It was observed on three days at 1.3~mm and one day at 3~mm.
J1924-2914  appears strongly polarized with LP varying from 6.1\% (on Apr 6) to 4.9\% (on Apr 11) at 1.3~mm, and LP=4.8\% (on Apr 4) at 3~mm. 
The EVPA is stable around [-49.2\dg,-51.8\dg]  at  1.3~mm and -46.4\dg\  at  3~mm.
We report a RM $\sim ([0.3-0.4] \pm 0.1) \times 10^5$~rad/m$^2$ at 1.3~mm and a $ 3\sigma$ upper limit of 3600~\radms\  at 3~mm (approximately an order of magnitude lower). 
\citet{Bower2018} report a higher RM value of $(-0.71\pm0.06)\times10^5$~\radmsq\ at 1.3\,mm from ALMA observations carried out in August 2016, when the source LP was considerably lower ($\lesssim2\%$). 
The  AMAPOLA monitoring revealed a considerable variation in the source EVPA  during Mar--Dec 2016\footnote{\url{www.alma.cl/$\sim$skameno/AMAPOLA/J1924-2914.flux.html}}, likely due to a period of low LP. We therefore ascribe the difference with the \citet{Bower2018} measurement  to source variability. 

J1924-2914  is completely unresolved on arcsecond scales both at 1.3~mm and 3~mm (see Figures~\ref{fig:polimages_agns_3days} and \ref{fig:polimages_3mm}), a result consistent with images at cm-wavelengths made with the VLA   \citep[e.g.,][]{Perley1982}.

\paragraph{4C~01.28} 
J1058+0133 (alias 4C~01.28) is a blazar at z= 0.888 (1 arcsec = 8~kpc). 
It was observed on three days at 1.3~mm and one day at 3~mm. 
The source is  strongly polarized with a mean LP of 5.5\% at 1.3~mm  and 4.4\% at 3~mm. 
At 1.3~mm, the LP  varies by $<$15\% while the EVPA changes from $\sim-23$\dg\ (Apr 5) to $\sim-15$\dg\ (Apr 11); the EVPA at 3~mm, measured on Apr~2, is --32\dg, apparently consistent with the trend at 1.3~mm. 
Both the measured EVPA and LP values at 1.3~mm and 3~mm follow very closely the time evolution predicted in the AMAPOLA survey (see Fig.~\ref{fig:stokescomp_gs_1}), 
where the  LP and  EVPA  follow a  trend parallel to the Stokes $I$ evolution. 
On Apr 11, we tentatively detect  RM $\sim (0.33\pm0.13)  \times 10^5$~\radmsq\ at the $\sim 3\sigma$ level;  we however caution that on Apr 5 and 10 the EVPAs  do not follow the $\lambda^2$ trend  (Fig.~\ref{fig:RM1}), and  we do not have a RM detection at 3~mm (with a $3\sigma$ upper limit of 3600~\radmsq). 
  
\paragraph{Cen~A} 
Centaurus A (Cen~A) is the closest radio-loud AGN (at a distance of 3.8 Mpc, 1 arcsec = 18~pc).  
Although it is a bright mm source (with F=5.7~Jy), it is unpolarized  at 1.3~mm (with a $3\sigma$ LP upper limit of 0.09\%). 
We find a spectral index of --0.2 in the central core, consistent with a flat spectrum, 
as also measured between 350 and 698 GHz with (non-simultaneous) ALMA observations \citep{Espada2017}.  

The total intensity images reveal a diffuse emission component around the central bright core, extending   across 12\arcsec\ and mostly elongated north-south, and two additional compact components towards north-east (NE) separated by roughly 14\arcsec\ and 18\arcsec\ from the central core and aligned  at P.A.$\sim$50\dg\ (see Fig.~\ref{fig:polimages_agns_1_2days}, bottom-right panel).
The first component could be associated with the inner circumnuclear disk, mapped in CO with the SMA  \citep{Espada2009} and ALMA \citep{Espada2017}, and may  indicate the presence of a dusty torus.
The two additional components correspond to two knots of the northern lobe of the relativistic jet,  labelled as A1 (inner) and A2 (outer) in a VLA study by \citet{Clarke1992}; no portion of the southern jet is seen,  consistent with previous observations \citep{McCoy2017}.

\paragraph{NGC~1052} 
NGC~1052 is a  nearby (19.7~Mpc; 1 arcsec = 95~pc)  radio-galaxy  that showcases an exceptionally bright twin-jet  system with a large viewing angle close to 90 degrees \citep[e.g.,][]{Baczko2016}. 
 With  F$\sim$0.4 Jy and LP$<$0.15\%, it is the  weakest mm source (along with J0132-1654) and the second least polarized AGN in  our sample. 
The apparent discrepancy in flux-density and spectral index between Apr 6 and 7 is most likely a consequence of the low flux-density (below the threshold required by the  commissioned on-source phasing mode; see \S~\ref{sect:datacal}) and the much poorer data quality on Apr 7, rather than  time-variability of the source.

\paragraph{Remaining AGN} 
J0006-0623  is the most highly polarized blazar (after 3C279) observed at 1.3~mm, with LP = 12.5\%.  
J0132-1654 is the weakest QSO observed at 1.3~mm ($F\sim$0.4 Jy) and has LP$\sim2$\%. 
The blazar J0510+1800 has an LP $\sim$ 4\% at 3~mm and shows  indication of a  large RM ($\sim 0.27\times10^5\,$rad/m$^2$), although the EVPA distribution does not follow a $\lambda^2-$dependence (see Fig.~\ref{fig:RM_3mm}, upper-right panel). 

\subsection{M87}
\label{sect:results_m87}

We report the first unambiguous measurement of RM toward the M87 nucleus at mm wavelengths (Table~\ref{tab:EHT_uvmf_RM}; Figure~\ref{fig:RM1}, middle panels). 
We measure
$(1.51\pm0.30) \times 10^5$~\radms\ (with a $5\sigma$ significance) on Apr 6
and tentatively $(0.64\pm0.27) \times 10^5$~\radms\ (with a $2.4\sigma$ significance) on Apr 5. 
On the last two days we can only report best-fit values of 
$(-0.24\pm0.23) \times 10^5$~\radms\ (with a $3\sigma$ confidence level range $ [-0.93,0.45] $) on Apr 10 and
$(-0.39\pm0.24) \times 10^5$~\radms\ (with a $3\sigma$ confidence level range $ [-1.11,0.33] $) on Apr 11. 
Although we cannot determine precisely the RM value on all days, we can conclude that the RM appears to vary substantially across days and there is marginal evidence of sign reversal.

Before this study, the only RM measurement was done with the SMA at 230 GHz by \citet{Kuo2014}, who reported a best-fit RM = $ (-2.1\pm1.8) \times 10^5$~\radmsq\, ($1\sigma$ uncertainty) and could therefore only provide an upper limit. 
In order to better constrain the RM amplitude and its time variability, in addition to the 2017 VLBI observations (which are the focus of this paper), 
we have also analysed the ALMA data acquired during the Apr 2018 VLBI campaign as well as additional ALMA archival polarimetric experiments (these are introduced in \S\ref{m87_add_data}).  
For two projects (2016.1.00415.S and 2017.1.00608.S) we produced fully-calibrated $uv$-files and then used the \uvmf\ flux extraction method with \uvmultifit\ to determine the M87 Stokes parameters. 
For the remaining two projects (2013.1.01022.S and 2015.1.01170.S), we used the full-Stokes images released with QA2. Since these images do not include clean component models, we used the \intf\   method  to extract the Stokes parameters in the compact core directly in the images\footnote{Based on the analysis of the 2017 datasets, we have  assessed that 
\intf\  yields consistent polarimetric parameters with respect to \uvmf\ and \txt\ (see  \S~\ref{allstokes} and Appendix~\ref{app:fluxext}).}. 

\begin{figure}
\includegraphics[width=0.5\textwidth]{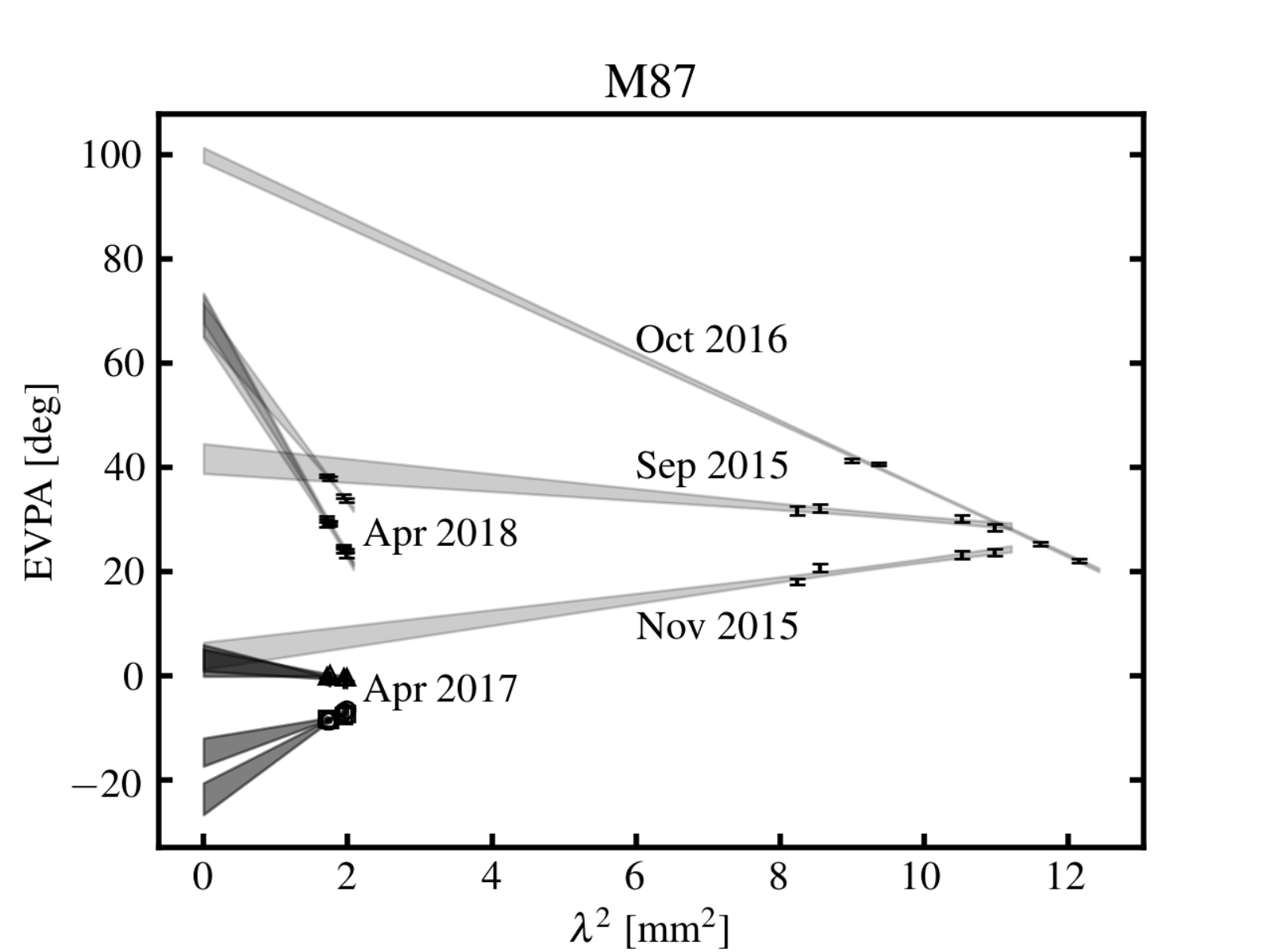}
\caption{M87 EVPA as a function of $\lambda^2$ observed in multiple epochs  at 3~mm (from Sep 2015 to Nov 2016) and 1.3~mm  (from Apr 2017 to Apr 2018).   
Each grey line is a linear fit to the EVPAs measured at the four ALMA Band~3 and Band~6 SPW, yielding the RM in each epoch, and the extrapolated intercept at the Y-axis is $\chi_0$. Note the large offset in  $\chi_0$ between the 3~mm and 1.3~mm bands. 
}
\label{fig:m87_rm_var}
\end{figure}

Table~\ref{tab:M87_RM} reports the full list of ALMA observations,  project codes, and derived polarimetric parameters. 
In total, we have collected data from three and eight different observations at 3~mm and 1.3~mm, respectively, spanning three years (from Sep 2015 to Sep 2018). 
The main findings revealed by the analysis of the full dataset are the following:

\begin{enumerate}[1.]
\item
The total flux density is quite stable on a week timescale, varying by $\lesssim$5\% in both Apr 2017 and Apr 2018, and exhibiting total excursion of about 15-20\% across one year both at 1.3~mm (decreasing from Apr 2017 to Apr 2018) and 3~mm (increasing from Sep 2015 to Oct 2016).

\item
We detect LP $\sim$ 1.7--2.7\% (2.3\% mean; Apr 2017 -- Apr 2018) at 1.3~mm and LP $\sim$ 1.3--2.4\% (1.7\% mean; Sep 2015 -- Oct 2016) at 3~mm. 

\item
The EVPA  distributions across the four ALMA SPWs clearly display  a $\lambda^2$ dependence, on specific dates, within both the 1.3~mm and 3~mm bands (e.g., see Figures~\ref{fig:RM1} and \ref{fig:RM_M87_3+1mm}).

\item
The magnitude of the RM varies both at 3~mm (range $| \rm{RM} | =   [ 0.2-1.2 ] \times 10^5$~\radms) and 1.3~mm (range $| \rm{RM} | =   [ 1.5-4.1 ] \times 10^5$~\radms, including $<3\sigma$ non-detections).

\item
The RM can either be positive or negative in both bands (with a preference for a negative sign), indicating that sign flips are present both at 3~mm and 1.3~mm.

\item
In Apr 2017, the RM magnitude appears to vary significantly (from non-detection  up to 1.5 $\times$ $10^5$~\radms) in just 4--5 days. 

\item
In Apr 2017, $\chi_0$  varies substantially across a week, being 
 [$-14.6\pm2.8,\ -23.6\pm3.1,\ 2.5\pm2.5,\ 3.5\pm2.5$] in Apr 5, 6, 10, and 11, respectively. Therefore, although the EVPA at 1.3~mm changes only by $\sim$+8\dg\ during the observing week, the $\chi_0$ varies by -9\dg\ in the first two days, and +27\dg\ between the second day and the last two days. 
In Apr 2018 $\chi_0$ appears instead to be consistently around 68.4\dg--70.6\dg\footnote{The change of about +10\dg\ in the EVPA at 1.3~mm between Apr 21 and Apr 25 2018 can be completely explained with a decrease in RM $\sim - 1 \times 10^5$~\radms.}.  
The  $\chi_0$ derived from  the three 3~mm experiments (Sep, Nov 2015 and Oct 2016) spans a range from $\sim$4\dg\ to 107\dg\ (see Fig.~\ref{fig:m87_rm_var} for a summary plot of RM+$\chi_0$ in all the available M87 observations). 

\item
The EVPAs measured  at 1.3 mm in the 2017 campaign ($\sim [-8,0]$\dg) are significantly different to the ones measured in the 2018 campaign ($\sim [26,36]$\dg), which are instead consistent with the ones measured in 2015--2016  at 3~mm ($\sim [21,33]$\dg).

\item
We find hints of CP at 1.3~mm at the $[-0.3 \pm 0.6,-0.4\pm 0.6]$\% level, but these should be regarded as tentative measurements (see also  Appendix~\ref{app:stokesV} for caveats on the CP estimates).

\end{enumerate}

We will interpret these findings in Section~\ref{sect:m87_disc}.

\subsection{Sgr~A*}
\label{sect:results_sgra}

In this section, we analyse the polarimetric properties of Sgr~A* and its variability on a week timescale based on  the ALMA observations at 1.3~mm and 3~mm.   

  \paragraph{LP.}
We measured LP between 6.9\% and 7.5\% across one week at 1.3~mm (Table~\ref{tab:EHT_uvmf_RM}). 
These values are broadly consistent with historic  measurements using BIMA on several epochs in the time span 2002--2004 at 227 GHz \citep[7.8--9.4\%; ][]{Bower2003,Bower2005},  
 SMA on several days in Jun--Jul 2005  
 \citep[4.5--6.9\% at 225 GHz;][]{Marrone2007}, 
and more recently  with ALMA in Mar--Aug 2016 at 225 GHz 
\citep[3.7--6.3\%, 5.9\% mean;][who also report intra-day variability]{Bower2018}. 
Besides observations at 1.3~mm, LP variability  has been reported also at 3.5~mm with BIMA \citep[on a timescale of days--][]{Macquart2006} and at 0.85~mm with the SMA \citep[on a timescale from hours to days--][]{Marrone2006}. 
All together, these measurements imply significant time-variability of LP across timescales of hours/days to months/years.

While at 1.3~mm LP $\sim$ 7\%, at 3~mm we detect  LP $\lesssim$ 1\% (Table~\ref{tab:GMVA_uvmf_RM}). 
It is interesting to note that the LP fraction increases from $\sim$ 0.5\% at $\sim$ 86\,GHz (our SPW=0,1) up to $\sim$ 1\% at $\sim$ 100\,GHz (our SPW=2,3; see Table~\ref{tab:GMVA_uvmf_spw}). 
This trend is  consistent with earlier measurements  at 22~GHz and 43~GHz with the VLA, and at 86~GHz and 112--115~GHz with BIMA, yielding upper limits of LP$\sim$ 0.2, 0.4\%, 1\% \citep{Bower1999a}, and 1.8\% \citep{Bower2001}, respectively \citep[but see][who report LP $\sim$ 2\% at 85~GHz with BIMA observations in Mar 2004]{Macquart2006}.

\paragraph{RM.}

We report a mean RM of $-4.2 \times10^5$~\radmsq\, at 1.3~mm with a significance of $\sim 50\sigma$ (Table~\ref{tab:EHT_uvmf_RM}; Figure~\ref{fig:RM1}, upper second to fourth panels), consistent with measurements over the past 15 years since the first measurements with BIMA+JCMT \citep{Bower2003}, BIMA+JCMT+SMA \citep[][]{Macquart2006}, and the SMA alone \citep{Marrone2007}\footnote{Both \citet{Bower2003} and \citet[][]{Macquart2006} used non-simultaneous EVPA measurements in the frequency range 150--400~GHz and 83--400~GHz, respectively. \citet{Marrone2007} determined for the first time the RM  comparing EVPAs 
measured simultaneously at each (1.3 and 0.85~mm) band.}.
 Across the observing week,  we see a change in RM from $-4.84 \pm 0.1\times10^5$~\radmsq\, (on Apr 6) to $-3.28 \pm 0.09 \times10^5$~\radmsq\, (on Apr 11), corresponding to a change of $\sim -1.5 \times10^5$~\radmsq\  ($\sim$30\%), detected with a significance of $\sim 20\sigma$. 
 This RM change  can completely explain the EVPA variation from  --65.8\dg$\pm$0.1\dg\ to  --49.3\dg$\pm$0.1\dg\ (or a $\sim$16\dg\ change across 5 days), given the consistency in $\chi_0$ between Apr 6  and Apr 11 ($\sim -14.7$\dg$\pm1.0$\dg; see Table~\ref{tab:EHT_uvmf_RM}). 
\citet{Marrone2007} find a comparable dispersion based on six measurements in the time period Jun--Jul 2005 ($\Delta |{\rm RM}|= 1.3  \times 10^5$~\radmsq\, excluding their most discrepant point, or $\Delta |{\rm RM}|= 3.8  \times 10^5$~\radmsq\, including all 6 measurements spanning almost 2 months). 
\citet{Bower2018} find an even larger  $\Delta |{\rm RM}|=-4.9 \times10^5$~\radmsq\, across 5 months; they also report intra-day variability in a similar range on timescales of several hours.
 
 Variations in RM appear to be coupled with LP fraction: the lower the polarization flux density, the higher the absolute value of the RM. In particular, we find $\Delta {\rm LP} \sim +5\%$ ($\Delta {\rm RM}\sim -9\%$) and $\Delta {\rm LP} \sim +9$\% ($\Delta {\rm RM}\sim -32\%$) in Apr 7 and 11, respectively, with respect to Apr 6. 
This can be understood if a larger RM scrambles more effectively the polarization vector fields resulting in lower net polarization. 
Although with only three data points we cannot draw a statistically  significant conclusion, we note that the same trend was also seen by \citet{Bower2018} on shorter (intra-day) timescales.

\begin{figure}
\includegraphics[width=0.5\textwidth]{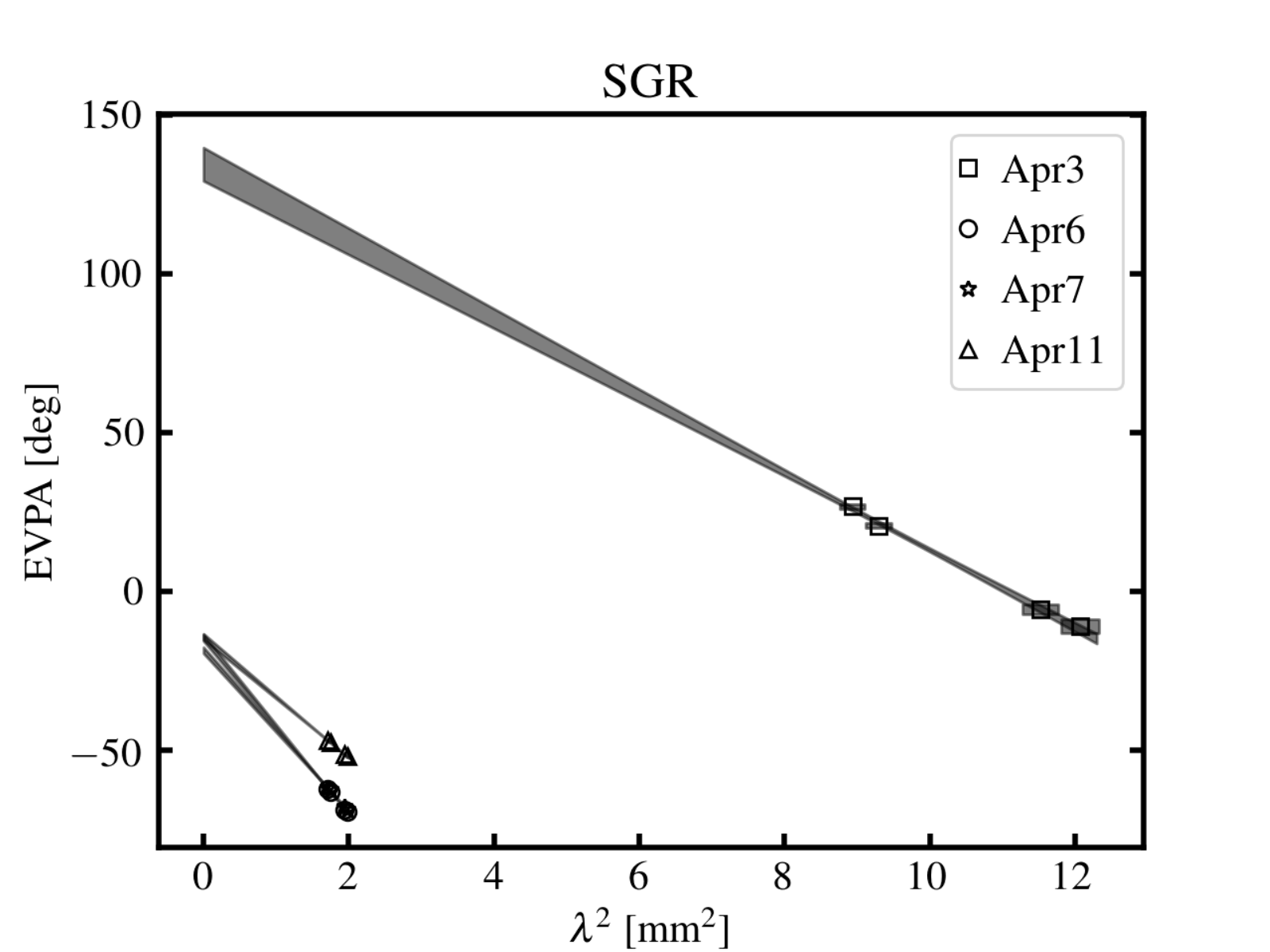}
\caption{Sgr~A* EVPA as a function of $\lambda^2$ observed in 2017 at 3~mm (Apr 3) and 1.3~mm (Apr 6, 7, 11). 
Each grey line is a linear fit to the EVPAs measured at the four ALMA Band~3 and Band~6 SPWs, yielding the RM in each epoch, and the extrapolated intercept at the Y-axis is $\chi_0$. Note the large offset in $\chi_0$ between the 3~mm and 1.3~mm bands, despite the consistency of $\chi_0$ at 1.3~mm across 6 days. 
}
\label{fig:sgra_rm_var}
\end{figure}

We report for the first time a measurement of  RM at 3~mm, with a magnitude of $(-2.1\pm0.1) \times10^5$~\radmsq\,(Table~\ref{tab:GMVA_uvmf_RM}; Figure~\ref{fig:RM1}, upper-left panel). 
The RM magnitude at 3~mm (measured on Apr 3) is  a factor of 2.3 (2.1) smaller  than the RM value measured at 1.3~mm on Apr 6 (Apr 7).  
Furthermore, we note a large offset in  $\chi_0$ between the 3~mm (+135\dg\  or --45\dg\ for a full 180\dg\ wrap) and the 1.3~mm bands ($\sim [-14.7,-18.8]$\dg), which is unlikely a consequence of time variability  (given the $\chi_0$ consistency  on Apr~6--11).
The comparison of  RM and $\chi_0$ in the two frequency bands (showcased in Fig.~\ref{fig:sgra_rm_var}) indicates the presence of both Faraday and intrinsic changes of the source.
We will provide an interpretation of the differences observed between the two frequency bands in \S~\ref{sect:sgra_disc}.

\paragraph{CP.}

We report a tentative detection of  CP at 1.3~mm in the range  $(-1.0 \pm 0.6)$\% to $(-1.5 \pm 0.6)$\%.
This is consistent with the first detection with the SMA  from observations carried out in 2005--2007 \citep{Munoz2012} and with a more recent ALMA study based on 2016 observations \citep{Bower2018}.  
This result suggests that the handedness of the mm-wavelength CP is stable over timescales larger than 12 yr. 
Interestingly, historical VLA data (from 1981 to 1999) between 1.4 and 15 GHz  show that  the emission is circularly polarized at the 0.3\% level and is consistently left-handed  \citep{Bower1999b,Bower2002}, possibly extending the stability of the CP sign to 40 years. 
Such a remarkable consistency of the sign of CP over (potentially four) decades suggests a stable magnetic field configuration (in the emission and conversion region).

Similarly to the RM, we also note a weak anti-correlation  between LP and CP (although more observations are needed to confirm it).  

We do not detect  CP at 3~mm ($<0.06\% $, $3\sigma$ upper limit). 
\citet{Munoz2012} and  \citet{Bower2018} find that, once one combines the cm and mm measurements,  the CP fraction as a function of frequency should be characterized by a power law with $\nu^{0.35}$. Using the measurements at 1.3~mm, this shallow power-law would imply a CP fraction at the level of $\sim 0.7-1.1$\% at 3~mm,  which would have been readily observable. 
 The non detection of CP at 3~mm suggests that the CP spectrum may not be monotonic.

Although the origin of the CP is not well understood, 
since a relativistic synchrotron plasma is expected to produce little CP, \citet{Munoz2012} suggest that the observed CP is likely generated close to the event horizon by the Faraday conversion which transforms LP into CP via thermal electrons that are mixed with the relativistic electrons responsible for the linearly polarized synchrotron emission \citep{Beckert2002}.  
In this scenario, while the high degree of order in the magnetic field necessary to produce LP $\sim$ 7\% at 1.3~mm naturally leads to a high CP  in a synchrotron source, the absence of CP at 3~mm is consistent with the low LP measured. 
See  \citet{Munoz2012} for a detailed discussion of potential origins for the CP emission. 

A final caveat is that based on the analysis presented in \S\ref{subsec:CP} and Appendix~\ref{app:stokesV}, the physical interpretations above should be considered as tentative. 

\paragraph{Flux-density variability}
We do not report significant variability  in total intensity and polarized intensity,  
which is about 10\% in six days (comparable to the absolute flux-scale uncertainty in ALMA Band~6). 
\citet{Marrone2006} and \citet{Bower2018} report more  significant variability in all polarization parameters based on {\it intra-day} light curves in all four Stokes parameters. This type of analysis is beyond the scope of this paper, and will be investigated elsewhere. 

\begin{table*}
 \caption{M87 RM measurements with ALMA. }
\label{tab:M87_RM}
\centering
\begin{tabular}{cccccccc}
\hline\hline 
Date &  I     &LP & EVPA & $\chi_0$ & RM & Beamsize & Project Code \\
         & [Jy] &    [\%] & [deg] & [deg] & [$10^5$ rad m$^{-2}$]  & \\
\hline\hline 
\multicolumn{7}{c}{3~mm}\\

              2015/09/19        &  2.17$\pm$0.11    &   1.37$\pm$ 0.03            &        30.68$\pm$0.74     &  41.7$\pm$3.1             &    -0.201$\pm$0.054&  0\pas53      &    2013.1.01022.S\footnote{Stokes I, Q, and U were extracted from the images using the CASA task \texttt{IMFIT}. \texttt{UVMULTIFIT} was used for all the other experiments.}
 \\
              2015/11/11        &  1.93$\pm$0.10    &   1.30$\pm$ 0.03            &        21.47$\pm$0.69      &  3.9$\pm$2.7           &     0.318$\pm$0.049  &  0\pas15   &    2015.1.01170.S$^{\rm a}$ \\
              2016/10/04        &  1.85$\pm$0.10    &   2.39$\pm$0.03            &        33.35$\pm$0.36      &  107.4$\pm$1.4       &     -1.227$\pm$0.023      &    0\pas 43   &    2016.1.00415.S  \\       
 \hline 
\multicolumn{7}{c}{1.3~mm}\\


              2017/04/05        &   1.28$\pm$0.13   &   2.42$\pm$ 0.03            &         -7.78$\pm$0.37       &      -14.6$\pm$2.9        &     0.64$\pm$0.27      &    1\pas5     &    2016.1.01154.V \\
              2017/04/06        &   1.31$\pm$0.13   &   2.16$\pm$ 0.03            &        -7.61$\pm$0.39        &      -23.6$\pm$3.1        &     1.51$\pm$0.29       &    1\pas8     &    2016.1.01154.V  \\
              2017/04/10        &   1.33$\pm$0.13   &   2.74$\pm$ 0.03            &        0.11$\pm$0.32         &      3.5$\pm$2.5       &     -0.32$\pm$0.24     &    1\pas5     &    2016.1.01154.V  \\
              2017/04/11        &   1.31$\pm$0.13   &   2.71$\pm$ 0.03            &        -0.63$\pm$0.31        &      3.7$\pm$2.4        &     -0.41$\pm$0.23       &    1\pas0    &    2016.1.01154.V  \\
              2018/04/21        &   1.11$\pm$0.11   &   2.29$\pm$ 0.03            &        27.18$\pm$0.38         &     70.6$\pm$3.0              &     -4.11$\pm$0.28       &    0\pas9    &    2017.1.00841.V \\
              2018/04/22        &   1.18$\pm$0.12   &   1.71$\pm$ 0.03            &       26.42$\pm$0.52         &      68.9$\pm$4.0              &     -4.02$\pm$0.39       &    0\pas9    &    2017.1.00841.V \\
              2018/04/25        &   1.14$\pm$0.11   &   2.21$\pm$ 0.03            &       36.12$\pm$0.39          &      68.4$\pm$3.0        &     -3.05$\pm$0.29       &    0\pas9    &    2017.1.00841.V \\
              
           2018/09/25        &   1.16$\pm$0.12   &   0.78$\pm$ 0.04            &        --             &     --       &  --  & 0\pas35    &    2017.1.00608.S\footnote{\newtext{The lower LP estimated for this project is likely caused by a systematic offset between Stokes $QU$ and $I$ (see \S\ref{m87_add_data}). 2017.1.00608.S was not used in the analysis.}} \\
\hline\hline
\end{tabular}
\end{table*}
 
\section{Discussion}
\label{sect:discussion}

In this section, we review general polarization properties of AGN comparing the two (1.3~mm and 3~mm) frequency bands, different AGN classes, and depolarization mechanisms   (\S~\ref{sect:LP_sample}); then we 
interpret the Faraday properties  derived for M87 in the context of existing accretion and jet models as well as a new two-component  polarization model (\S~\ref{sect:m87_disc}); 
and finally we discuss additional constraints on the Sgr~A* polarization model from a comparison of 1.3~mm and 3~mm observations (\S~\ref{sect:sgra_disc}). 
\subsection{Polarization degree and Faraday rotation in AGN}
\label{sect:LP_sample}

\subsubsection{1.3~mm vs. 3~mm}
\label{sect:LP_1mmvs3mm}

\paragraph{Synchrotron emission opacity} 
The total intensity spectral indexes  for  the AGN sources in the sample vary in the range $\alpha$=[--0.7,--0.3] at 3~mm and $\alpha$=[--1.3,--0.6] at 1.3~mm, Cen~A being the only exception, with $\alpha$=--0.2 (see Tables~\ref{tab:GMVA_uvmf_RM} and \ref{tab:EHT_uvmf_RM} and Appendix~\ref{app:alpha}). 
This contrasts with the flat spectra  ($\alpha$=0) typically found at longer cm wavelengths in AGN cores \citep[e.g.,][]{Hovatta2014}, corresponding to optically-thick emission. 
In addition, we  observe a spectral  steepening  (with $\Delta \alpha=[-0.2,-0.4]$) between 3~mm and 1.3~mm; although one should keep in mind the caveat of time variability, 
since the observations in the two frequency bands were close in time (within ten days) but not simultaneous.
Such spectral steepening can naturally be explained by decreased opacity of the synchrotron emission  at higher frequencies in a standard jet model \citep[e.g.,][]{BlandfordKonigl1979,Lobanov1998}.

\paragraph{LP degree}  
We detect LP in the range 0.9–13\% at 3~mm and 2--15\% at 1.3~mm (excluding the unpolarized sources NGC~1052 and Cen~A). 
At 1.3 mm, the median fractional polarization is 5.1\%, just slightly higher than the median LP at 3~mm, 4.2\%, yielding a ratio of 1.2. 
If we consider only the sources observed in both bands, then the ratio goes slightly up to 1.3 (or 1.6 including also Sgr~A*). 
Despite the low statistics, these trends are marginally consistent with results from previous single-dish surveys with the IRAM 30-m telescope \citep{Agudo2014,Agudo2018} and  the Plateau de Bure Interferometer or PdBI \citep{Trippe2010}. In particular, \citet{Agudo2014} find an LP ratio of 1.6 between 1~mm and 3~mm based on  simultaneous, single-epoch observations of a sample of 22 radio-loud ($F>1$~Jy) AGN, 
while  \citet{Agudo2018} find an LP ratio of 2.6 based on long-term  monitoring, non-simultaneous observations of 29 AGN. 
\citet{Trippe2010} find similar numbers from a sample of 73 AGN observed
as part of the IRAM/PdBI calibration measurements  during   standard interferometer science operations\footnote{The polarimetric data analysis is based on Earth rotation polarimetry and is antenna-based, i.e. executed for each antenna separately. Therefore, no interferometric polarization images are available from this study.}.
The comparison of these statistics at both  wavelengths suggests  a general higher degree of polarization at 1 mm as compared to 3 mm. 
This finding can be related either to a smaller size of the emitting region   and/or to a higher ordering  of  the  magnetic-field  configuration \citep[e.g., see discussion in][]{HUGHES1991}. 
In fact, according to the standard jet model, the size of the core region decreases as a power-law of the observing frequency, which could help explain the higher LP observed at 1\,mm.  Alternatively, the more ordered magnetic-field  configuration  could be related  to a large-scale (helical) magnetic-field structure along the jets.  

\paragraph{Faraday RM} 
Among the six sources observed both  at 3~mm and 1.3~mm, we have RM detections at the two  bands only in  3C273, where 
the estimated value at 3~mm is significantly lower than at 1.3~mm. 
For the remaining sources with RM
detections at 3~mm (NRAO 530, OJ~287, and 3C279) 
and  at 1.3~mm (J1924-2914 and 4C~01.28), their $3\sigma$ upper limits, respectively at 1.3~mm and 3~mm,  still allow a larger RM at the higher frequency band.

A different  'in-band' RM  in the 3~mm and 1.3~mm bands can be explained either with  
(i)  the  presence of an internal Faraday screen or multiple external  screens in the beam; or with 
(ii)  a different opacity of the synchrotron emission  between the two bands.
Case (i) will cause a non-$\lambda^2$ behaviour of the EVPA and a non-trivial relation between the  'in-band' RM determined at only two narrow radio bands. 
Evidence for non-$\lambda^2$ behaviour of the EVPA can be possibly seen in OJ~287, 4C~01.28, and J0006-0623 at 1.3~mm (Fig.~\ref{fig:RM3}) and J0510+1800 at 3~mm (Fig.~\ref{fig:RM_3mm}). 
In order to estimate $B$ or $n_e$ from the RM (see Eq.~\ref{eq:RM}), one would need to sample densely the EVPA over a broader frequency range and perform a more sophisticated  analysis, using techniques like  
the Faraday RM synthesis or Faraday tomography \citep[e.g.][]{BrentjensdeBruyn2005}. 
This type of analysis is beyond the scope of this paper and can be investigated in a future study (we refer to  \S~\ref{int_vs_ext_FR} for evidence of internal Faraday rotation and    \S~\ref{2polcomp} for an example of a multiple component Faraday model for the case of M87).
Since the spectral index analysis shows that  the AGN in the sample become more optically thin at 1~mm, the  observed  differences in the 'in-band' RM  at 3~mm and 1.3~mm  can be likely explained with synchrotron opacity effects alone (with the caveat of time variability since the observations are near-in-time but not simultaneous). 

It is also interesting to note that we also see a sign reversal  between the RM measured at 3~mm and 1.3~mm for 3C~273. 
 RM sign reversals require  reversals in $B_{||}$ either over time  (the observations in the two bands were not simultaneous) or across the emitting region (the orientation of the magnetic field is different in the 3~mm and 1~mm regions). 
 With the data in hand we cannot distinguish between time variability
or spatial incoherence of the magnetic field (we refer to \S~\ref{sect:m87_disc} for a discussion on possible origins of RM sign reversals in AGN). 

\subsubsection{Blazars vs. other AGN}
\label{sect:blazarsVSothers}

We find that blazars are more strongly polarized than other AGN in our sample,  with a median LP $\sim$7.1\% vs. 2.4\% at 1.3 mm, respectively.
Furthermore, blazars have approximately an order-of-magnitude  lower RM values  (on average)   than other AGN, with a median value of 
$\sim 0.07 \times 10^5$~\radms\  at 1.3~mm (with the highest values of $\sim 0.4 \times 10^5$~\radms\ exhibited by J1924-2914), whereas for other AGN we find a median value of $\sim 0.4 \times 10^5$ at 1.3~mm\footnote{In computing the median we exclude the unpolarized Cen~A and NGC~1052 for which we cannot measure a RM.} (with the highest values $> 10^5$~\radms\  exhibited by M87 and 3C~273). 

\citet{Bower2017} used the Combined Array for Millimeter Astronomy (CARMA) and  the SMA to observe at 1.3~mm two low-luminosity AGN (LLAGN), M81 and M84, finding upper limits to LP of 1\%--2\%.  Similarly,  \citet[][]{Plambeck2014} used CARMA to observe the LLAGN 3C~84 at 1.3 and 0.9~mm, measuring an LP in the 1\%--2\% range, and a very high RM of $\sim (9\pm2) \times 10^5$. 
These low values of LP (and high values of RM) are comparable to what we find in M87, which is also classified as a LLAGN \citep[e.g.][]{dimatteo2003}.  

When put together, these results suggest that blazars have different polarization properties at mm wavelengths from all other AGN, including LLAGN, radio galaxies, or regular QSOs\footnote{Similar conclusions were reached from VLBI imaging studies of large AGN samples at cm wavelengths \citep[e.g.,][]{Hodge2018}.}. 
These mm polarization differences can be understood in the context of the viewing angle unification scheme of AGN. 
A smaller  viewing angle implies a stronger Doppler-boosting  of the synchrotron emitting plasma in the jet, which in turn  implies a  higher polarization fraction for  blazars. Furthermore, their face-on geometry    allows the observer to reach the innermost radii of the nucleus/jet and reduces the impact of  the ‘scrambling’ of  linearly polarized radiation  by averaging different polarization components within the source (e.g. Faraday and beam depolarization -- see next section), also resulting in  higher LP (and lower RM).

\subsubsection{Depolarization in radio galaxies}
\label{cena_ngc1052}

In the previous section we point out that radio galaxies and LLAGN exhibit  lower polarization degree than blazars. 
In particular, the radio galaxies Cen~A and NGC~1052 do not show appreciable polarized intensity (LP$<$0.2\%) at 1.3~mm. 
We suggest several  depolarization mechanisms that may be at play in these radio galaxies (and potentially other LLAGN):
\begin{enumerate} [a.]
\item Faraday depolarization due to a thick torus or a dense accretion flow.
\item Bandwidth depolarization due to a strong  magnetic field. 
\item Beam depolarization due to a disordered magnetic field.
\item Thermal (non-synchrotron) emission.
\end{enumerate} 
In the following, we elaborate on these mechanisms. 

\paragraph{a. Faraday depolarization due to a thick torus or a dense accretion flow} 
Radio galaxies are often surrounded by an obscuring torus. The cold gas in the torus can be photo-ionized by UV photons from the inner accretion disk  and  the mixture of thermal and non-thermal material    could be responsible for the strong depolarization of the inner regions  via Faraday rotation -- and one speaks of   {\it Faraday} depolarization. 
 In the case of NGC~1052, 
\citet{Fromm2018,Fromm2019} created synthetic radio maps of the jets using  special-relativistic hydrodynamic (SRHD) simulations and suggested that an obscuring torus can explain some of the observed properties of these jets. 
In fact, the presence of a massive ($\sim 10^7 \, M_\odot$) and dense ($> 10^7$~cm$^{-3}$) molecular torus has been recently demonstrated with ALMA observations \citep{Kameno2020}. 
A clumpy  torus is also known to surround the nucleus of Cen~A \citep[e.g.][]{Espada2017}, as also suggested by our 1.3~mm map (see Fig.~\ref{fig:polimages_agns_1_2days}, bottom-right panel). 
Therefore the presence of a thick torus of cold gas could naturally explain the low polarization degree in both radio galaxies. 
A similar mechanism can be at play  in LLAGN whose radio emission is thought to be powered by synchrotron radiation from a geometrically thick, hot accretion flow \citep[e.g.,][]{narayan1994,blandford1999,quataert2000}. 

\paragraph{b. Bandwidth depolarization due to a strong magnetic field} 
A large homogeneous magnetic field implies an intrinsically large homogeneous RM (see Eq.~\ref{eq:RM}), 
resulting in the source to appear unpolarized in broadband observations -- and one speaks of {\it bandwidth} depolarization. 
 In  NGC~1052, GMVA imaging at 86 GHz helped constrain the magnetic field at Schwarzschild radius (\rs) scales in the range  360--70000~G \citep{Baczko2016},  providing evidence of an extremely high magnetic field near the SMBH. Coupled with its high inclination \citep[e.g.][]{Kadler2004}, such a strong magnetic field would result in  the source to appear unpolarized in ALMA broadband observations.

\paragraph{c. Beam depolarization due to a tangled magnetic field} 
If the magnetic field in the emitting regions or in a foreground Faraday screen is tangled or generally disordered on scales much smaller than the observing beam, magnetic field regions with similar  polarization degrees but opposite signs will cancel out and the net observed polarization degree would be  significantly decreased -- and one speaks of {\it beam} depolarization. 
We do not have evidence of such tangled magnetic field for any of the low LP sources in our sample, which will require high-resolution polarization imaging with the GMVA or the EHT. 

\paragraph{d. Thermal (non-synchrotron) emission} 
Multi-wavelength (MWL) studies in Cen~A show that its SED is moderately inverted up to the infrared, possibly indicating a dust contribution at mm wavelengths \citep[e.g.][] {Espada2017}.
Using VLBI imaging at 229~GHz with the EHT, Janssen et al. (submitted) measured a flux of  $\sim 2$~Jy in the VLBI core, indicating that the EHT filters out $\sim 65$\% of the emission seen by ALMA. 
While we cannot exclude contribution from thermal emission to the total flux measured by ALMA, the flux measured with the EHT must necessarily be associated with non-thermal emission. We therefore conclude that dust emission is an unlikely explanation for the lack of LP at 1.3~mm.  
\vspace{0.2cm}

An improved data analysis including spectro-polarimetry could be helpful to measure the actual RM in both NGC~1052 and Cen~A and thus assess which is the dominant depolarization mechanism among the ones discussed above (this is however beyond  the scope of this paper). 
As a final note, an interesting insight may come from a comparison between mm and IR wavelengths, where both  NGC 1052 \citep{Fernandez-Ontiveros2019} and Cen~A \citep{Jones2000, Lopez-Rodriguez2021} are highly polarized. These characteristic are similar to Cygnus A, where the low polarized core at mm-wavelengths and the high polarized core at IR wavelengths may be the signature of an ordered magnetic field in the plane of the accretion disk supporting the accretion flow and/or jet formation \citep{Lopez-Rodriguez2018}. 
\subsection{Physical origin of the rotation measure in M87}
\label{sect:m87_disc}
 We can now use   the polarimetric and Faraday properties of the mm emission from M87  reported in \S~\ref{sect:results_m87},  to constrain properties of accretion models onto the M87 SMBH.
Models aiming to explain the origin of the RM in M87 should match the following key observed features (see \S~\ref{sect:results_m87} for a full list of findings):

\begin{enumerate}[i.]

\item 
{\it Low LP and high RM.} 
 M87 has a rather low LP ($\sim 2.3\%$ mean at 1.3~mm) when compared to Sgr A* and other blazars in the sample (see \S\ref{sect:blazarsVSothers}), while the measured RM can be as high as a few times $10^5$~\radms. 

\item 
{\it RM sign reversals.}  
Observations on different dates yield large differences in the measured RM values, which can be either positive or negative  (in both the 3~mm and 1.3~mm bands). 
This requires sign flips in $B_{||}$ over time and/or across the emitting region. 

\item 
{\it Rapid RM time variability.} 
In Apr 2017, the RM magnitude appears to vary significantly (from non-detection  up to $1.5 \times 10^5$~\radms) in just 4-5 days. 
This suggests the presence of small-scale fluctuations in the emitting source and/or the Faraday screen.  

\item 
{\it $\lambda^2$ scaling.}  
Plots in Figures~\ref{fig:RM1} and~\ref{fig:RM_M87_3+1mm} clearly display a $\lambda^2$ dependence of the EVPA at 1.3~mm and 3~mm on specific days (although this is not always the case). 

\end{enumerate}

The MWL spectral energy distribution (SED) of the M87 core is best explained
by emission from advection dominated/radiatively inefficient accretion flow 
 \citep[ADAF/RIAF --][]{reynolds1996,dimatteo2003} or from a jet \citep[e.g.,][]{Dexter2012,Prieto2016,Moscibrodzka2019}, 
or emission from a combination of these two components \citep[e.g.,][]{BroderickLoeb2009,Nemmen2014,Moscibrodzka2016,Feng2016,Moscibrodzka2017,Davelaar2019,EHTC2019_5}. In the hybrid models, the low-frequency radio emission is produced by the jet while the optically-thin mm/submm (and X-ray) emission can either come from the jet base or the inner accretion flow.

The traditional  approach  adopted  in  previous  studies was to  assume that 
the large scale ($r \sim~100$~\rs) accretion flow itself may act as a Faraday screen and that the  core emission region lies entirely behind the same portion of the Faraday screen
(e.g., the core emission size is small compared to the scale of any fluctuations in large scale flow).  
In the framework of semi-analytic RIAF/ADAF models, the RM magnitude 
 has then been used  to estimate mass accretion rates onto black holes in Sgr~A*  \citep{Marrone2006,Marrone2007} and in 3C~84 \citep{Plambeck2014}.
 A similar approach has been used in M87, yielding estimates of the accretion rate in the range from $\dot{M} < 9 \times 10^{-4}$ \citep{Kuo2014} to $\dot{M}\sim [0.2,1] \times 10^{-3}$\,\msyr \citep{Feng2016}, where the quoted values are either upper limits or depend on specific model assumptions (e.g. black hole spin or the exact location of the Faraday screen).
From the largest RM that we measured, 
$ 4 \times 10^5$~\radmsq\ (on Apr 2018), using Equation (9) in \citet{Marrone2006}, we would infer a mass accretion rate of $\dot{M} = 7.7 \times 10^{-8} |RM|^{2/3} \sim 4 \times 10^{-4}$ \msyr\ assuming an inner boundary to the Faraday screen of  $R_{\rm RM, in}=21$~\rs\ (as suggested by \citealt{Kuo2014}).

While our estimates of mass accretion rates are consistent with previous similar estimates and upper limits from the ADAF/RIAF models, the observed properties listed above, especially the time variable RM and its sign reversals, provide new constraints. 
In particular, the timescale of the RM variability can be set by  the rotating medium  dynamical time ($\propto \sqrt{R^3/GM}$), thus 
constraining the radius at which the RM originates. 
The rapid variability observed in Apr 2017 implies that the RM should occur much closer to the SMBH (within a few \rs) than  assumed in previous mass accretion models, which in turn suggests the possibility of a co-location of the emitting and rotating medium. 
In the alternative, the Faraday screen could be at further distance and  the observed variability could be ascribed to rapid fluctuations in the emitting source. 
Therefore, both  a turbulent accretion flow acting  as a Faraday screen or a varying compact source with an external screen can  explain the observed time-variability. 
 Finally, the accretion flow is not the only possible source of Faraday rotation. 
 Since  simulations show that relativistic jets can have a ``spine-sheath'' structure \citep[e.g.,][]{McKinney2006}, the jet sheath can provide a magnetized screen surrounding the jet, and indeed it has been also suggested as a plausible source of Faraday rotation in AGN  \citep[e.g.,][]{ZavalaTaylor2004}. 
Therefore, either (or both) the accretion flow and/or the jet can in principle be the sources of the mm emission and/or  the Faraday rotation. 

All the scenarios described above imply a more complicated physical origin of the Faraday rotation than is usually assumed in traditional semi-analytic models that use the RM to infer a mass accretion rate. We conclude that, unlike the case of Sgr~A*, the RM in M87 may not provide an accurate estimate of the mass accretion rate onto the black hole.

In what follows, we review clues on the location
of the Faraday screen using observational constrains from  ALMA    (\S~\ref{int_vs_ext_FR})
as well as information on horizon scales from the EHT 
(\S~\ref{2polcomp}).  

\subsubsection{Location of the Faraday screen: internal vs. external}
\label{int_vs_ext_FR}

We distinguish between two general cases:  internal and  external Faraday rotation. 

\begin{enumerate}[I.]

\item
Internal Faraday rotation: 
the accretion flow or jet can simultaneously be the source of synchrotron radiation and the Faraday screen.  

\begin{enumerate}[a.]

\item {\it RM rapid time variability and sign reversals.} 
Recent time-dependent GRMHD simulations of the M87 core \citep{Ricarte2020}  
show that turbulence within the accretion flow is able to change $B_{||}$ in both amplitude and orientation, resulting in significant RM fluctuations and sign reversals on the dynamical time at $R \simeq 2.5-5$~\rs, corresponding to short timescales of a few days for M87 (consistent  with properties \#ii and \#iii). 

\item {\it Beam Depolarization.} 
An internal Faraday screen could cause 
beam depolarization of the synchrotron emission. This has been theoretically predicted by GRMHD simulations of the M87 core emission,  which yield  low  values of LP (typically  in the range 1\%—3\%) and  large Faraday RM ($\gtrsim 10^5$~\radms)  \citep{Moscibrodzka2017,EHTC2020_2}, broadly consistent with the  observed feature \#i. 
We however note that the beam depolarization could be also caused by external Faraday rotation in an  inhomogeneous screen.

\end{enumerate}

\item
External Faraday rotation: 
 the emission region lies entirely behind (and it is not inter-mixed with) the Faraday screen. 

\begin{enumerate}[a.]

\item {\it $\lambda^2$ scaling of the EVPA.} 
A $\lambda^2$ dependence is typically adopted as observational evidence of the fundamental assumption on the location of the Faraday screen as external relative to the background source.
Although it can be argued that EVPA  variations in a narrow frequency range could  be  approximated  to  be linear 
  (at  1.3/3~mm the fractional $\lambda^2$ bandwidth is only 16/32\% of the central wavelength), 
good linear fits of EVPA  vs.  $\lambda^2$  are also obtained from lower-frequency observations, including the range 2, 5 and 8 GHz  \citep{Park2019}, 8, 15, 22 and 43 GHz \citep{Algaba2016}, and 24, 43 and 86 GHz \citep{Kravchenko2020}.

\item {\it RM sign reversals and helical magnetic fields.}
Polarimetric images with the VLBA at 43~GHz have revealed magnetic field vectors wrapped around the core \citep{Walker2018}, suggesting that toroidal fields might be dominant on scales of hundreds of \rs. 
Helical magnetic fields threading the jet may be produced by the differential rotation either in the BH ergosphere or in the innermost regions of the accretion disk  \citep[e.g.,][]{BroderickLoeb2009,BroderickMcKinney2010,Tchekhovskoy2011}.
 If toroidal fields are dominant in the sheath, one would expect transverse RM gradients across the jet, with opposite signs of the RM from one edge to the other, as   shown in a handful cases where VLBI images resolve the jet width \citep[ e.g.,][]{Asada2002,Gomez2008,Gabuzda2014}.
Systematic changes in the signs of these gradients, leading to RM sign reversals in unresolved measurements, 
can be explained with a number of models,   
including the magnetic "tower" model \citep{Lynden-Bell1996,Contopoulos1998,Lico2017}, 
or  the "striped" jet model \citep{Parfrey2015,Mahlmann2020,Nathanail2020}. 
 Nevertheless, it remains difficult to explain the rapid fluctuations observed in Apr 2017 with these models.  

\end{enumerate}

\end{enumerate}

A long-term  monitoring   with beam-matched simultaneous observations at multiple frequency bands would be required to assess the frequency-dependence of the RM and to conclusively discriminate between internal and external Faraday rotation. 
Clear evidence of $\lambda^2$ scaling in a wider frequency range   would be evidence of the external scenario, while deviations from $\lambda^2$ would support the internal scenario.
 A time cadence from a few days to a few months would allow us to assess whether the RM sign flips are stochastic (favoring the internal scenario), or systematic (favoring the external scenario).

\subsubsection{Two-component polarization model for M87}
\label{2polcomp}

\begin{figure}
\begin{center}
\includegraphics[width=\columnwidth]{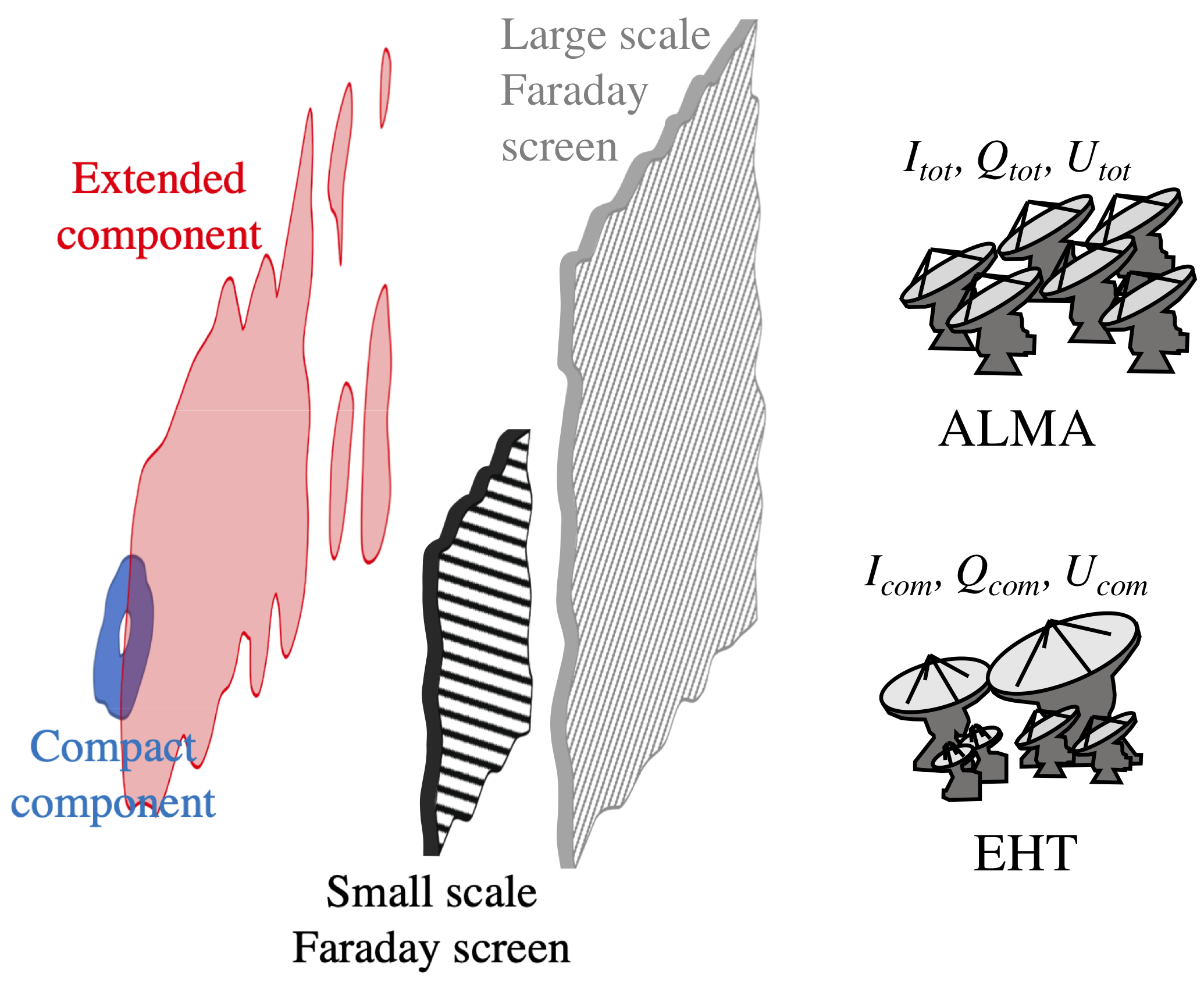}
\end{center}
\caption{Diagram of the two-component model, comprised of a compact (blue) and extended (red) polarized emission regions with corresponding small-scale and large-scale Faraday screens.  The small-screen, which may be external (shown) or internal (not shown) acts only on the compact component, which is observed by the EHT.  ALMA observes the combined emission from both components.  
}
\label{fig:twocomp_cartoon}
\end{figure}

EHT estimates of the  flux in the compact core, i.e., that arising on horizon scales, can at most account for $50\%$ of that measured by ALMA, both in total intensity \citep{EHTC2019_4} and polarized intensity \citep{EHTC2020_1} emission.  Natural origins for the additional emission are the larger-scale structures in the jet.  
The additional components may encompass many scales, be discrete features (e.g., HST-1), or be a combination thereof.  
In order to interpret these differences revealed by the EHT and ALMA, we adopt the simplest version of a multi-scale model permissible -- a two-component model comprised of a variable compact region and static extended region (see Figure~\ref{fig:twocomp_cartoon}).  We find that this is sufficient to harmonize the polarimetric properties observed by both the EHT and ALMA in Apr 2017, including the interday variability in the ALMA RMs   and the  EVPA variation of the compact core as observed by the EHT.

Both the compact and extended components of the two-component model consist of total intensity, spectral index, linearly polarized flux, and polarization angle.  We consider both internal and external Faraday screen models for the compact component.  In both cases, the Faraday screen for the extended component is assumed to be external.  A model likelihood is constructed using the integrated EHT Stokes $I$, $Q$, and $U$ ranges presented in Table~7 in Appendix~H2 of \citet{EHTC2020_1}, and the ALMA core Stokes $I$, $Q$, and $U$ values for the individual SPWs in Table~\ref{tab:EHT_uvmf_spw}, assuming Gaussian errors.  This likelihood is then sampled with the EMCEE python package \citep{emcee2013} to obtain posterior probability distributions for the model components.  For more details regarding the model, priors, and fit results, see Appendix~\ref{app:twocomp}.

Across days, only the LP and EVPA of the compact component is permitted to vary.  This is consistent with the extended component being associated with much larger physical structures and required by the polarimetric variability observed by the EHT \citep{EHTC2020_1}.  There is no evidence that variability in any other component of the two-component model is required: despite static Faraday screens, permitting the polarization of the compact component to vary is capable of reproducing the rapid changes in the ALMA RMs.  In this picture, the observed RMs arise in part from the wavelength-dependent competition between the two components, and thus are not directly indicative of the properties of either Faraday screen.

\begin{figure}
\begin{center}
\includegraphics[width=\columnwidth]{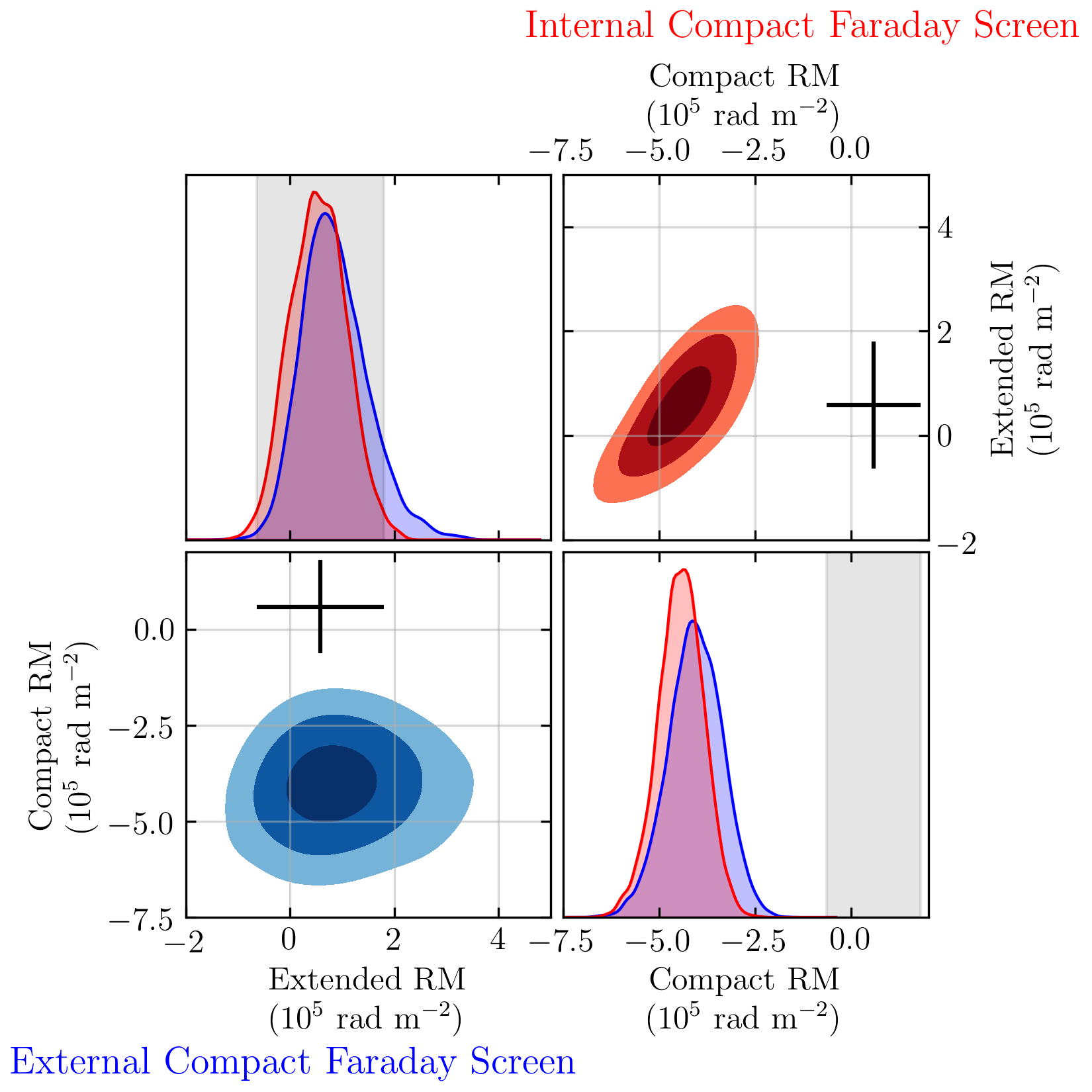}
\end{center}
\caption{Posteriors implied by the April 5, 6, 10, and 11, 2017 ALMA and EHT observations for the   RMs measured on ALMA (extended) and EHT (compact) scales  in the two component model when it is assumed that the compact Faraday screen is external (lower left, blue) and internal (upper right, red) to the emission region.  Contours are shown for the 50\%, 90\%, and 99\% quantiles.  For comparison, the $2\sigma$ range of April 5, 6, 10, and 11, 2017 ALMA RMs reported in Table~\ref{tab:M87_RM} are shown by the black crosses and gray bands.}\label{fig:twocomp_RMsummary}
\end{figure}

Nevertheless, via this model we are able to separately constrain the RMs that are observed on ALMA and EHT scales; these are shown in Figure~\ref{fig:twocomp_RMsummary}.
Specifically, while the RM of the large-scale Faraday screen is comparable to the Apr 2017 values reported in Table~\ref{tab:M87_RM}, 
that associated with the compact component 
 is not directly constrained by the ALMA measurements and can be factors of many larger: at 95\% confidence, the compact RM is between $-5.4\times10^5~{\rm rad~m^{-2}}$ and $-2.9\times10^5~{\rm rad~m^{-2}}$. 
 Interestingly, the estimated range is consistent with 
 the RM inferred from low-inclination GRMHD models of the M87 core \citep{Ricarte2020,EHTC2020_2}  
 and comparable to  the Apr 2018 values reported in Table~\ref{tab:M87_RM}.   
 This consistency  suggests the possibility that in Apr 2018 ALMA may be   seeing the core RM (e.g., as a consequence of a different beating of the  two components). 
 This hypothesis can be directly tested with the 2018 EHT observations which, unlike the 2017 ones, covered the full frequency spacing of ALMA (212--230~GHZ; see \S~\ref{obs}), and  are therefore expected to directly measure the resolved RM of the core. This will in turn allow to quantify the interplay between compact and extended components, 
 and potentially explain the time variability observed with ALMA.  

\subsection{Constraints on Sgr~A*  model  from  polarization and Faraday properties at 3~mm and 1.3~mm}
\label{sect:sgra_disc}

Measurements of Faraday rotation at radio/mm wavelengths, either towards Sgr~A* itself \citep[e.g.,][]{Marrone2007, Bower2018} or the nearby pulsar PSR J1745-2900 \citep[e.g.,][]{Eatough2013, Kravchenko2016}, have been used to probe the magnetized accretion flow in Sgr~A* on scales from tens of \rs\ out to the Bondi radius ($\sim10^5$ $R_{Sch}$). 
Using the same semi-analytic RIAF/ADAF models introduced in \S~\ref{sect:m87_disc} \citep{Marrone2006}, from the measured RM values at 1.3~mm we obtain an accretion rate of order $10^{-8}$~\msyr\ (assuming $R_{\rm RM, in}=10$~\rs), with a maximum variation of approximately 20\% across the observing week. 

We have derived for the first time the polarization and Faraday properties of Sgr~A* both at 3~mm and 1.3~mm in a time-window of ten days. 
Since the synchrotron photosphere in the accretion flow  moves outwards with decreasing frequency (because of increased opacity), 
the polarized emission at 3~mm and 1.3~mm is expected to arise from different locations with potentially different magnetic field structures. 
Any variation of the intrinsic EVPA or RM with frequency  can therefore provide interesting insights on the polarized source and magnetic field structure.  
In \S~\ref{sect:results_sgra} we infer that the intrinsic polarization vector is rotated  between 
the 3~mm ($\chi_o \sim +135$\dg\   or --45\dg\ assuming a full 180\dg\ wrap) and the 1.3~mm  ($\chi_o \sim$ [--15\dg,--19\dg]) bands 
and that the RM magnitude at 3~mm is about half of the RM value measured at 1.3~mm over a three-days separation. 
From earlier VLBI measurements, we know that the emission at mm wavelengths must come from very closely situated regions of the black hole, with an intrinsic (i.e. unscattered) size of  $\sim$120~$\mu$as (or 12 \rs) at 3.5~mm \citep{Issaoun2019} and $\sim$50~$\mu$as (or 5~\rs) at 1.3~mm \citep{Lu2018}. Therefore the radius of the 3.5~mm source is 2.4 times larger than the 1.3~mm source, i.e. very close to the ratio of the RM values measured with ALMA at the two wavelengths  (see RM paragraph in \S~\ref{sect:results_sgra}). 
 Taken at face value, 
 this result suggests that about half of  the Faraday rotation at 1.3~mm may occur between the 3~mm photosphere and the 1.3~mm source.

Although this result would be extremely constraining for the model of Sgr~A*, we should point out two caveats: 
i) possible presence of multiple components;  
ii) potential RM time variability.   
We explain these caveats below. 

 \paragraph{Multiple components}   
 In addition to the  mini-spiral,  which is however unpolarized (see \S~\ref{subsec:LP})  and thus should not contribute significantly to the polarized flux, 
 the presence of a relativistic compact jet has been proposed based on theoretical modeling of the source SED, in particular to explain the radio emission in Sgr~A* \citep[e.g.,][]{FalckeMarkoff2000}. 
In addition,  the only available VLBI polarimetry study at 1.3~mm has shown that the polarization structure of Sgr~A* is complex \citep{Johnson2015}  
and can be in principle different at the two wavelengths. 
Therefore, we cannot completely exclude the presence of an additional jet component to the accretion flow or a more complex morphology for the intrinsic polarization. 
Nevertheless, we argue that  the stability of LP, RM, and CP (including their sign) observed in Sgr~A* over more than a decade, unlike the case of M87, favors the presence of one single dominating polarized component.

 \paragraph{RM time variability} 
At 1.3~mm, the RM changes by $+1.5 \times10^5$~rad/m$^2$ ($\sim$ 30\%) between the Apr 6/7 and Apr 11. 
Assuming the same rate of time variability at 3~mm and 1.3~mm, such variability  is likely not  responsible for a factor of two difference over three days. 
Likewise, the large offset in $\chi_o$ observed at 3~mm and 1.3~mm  is unlikely a consequence of time variability, given the $\chi_0$ consistency  on Apr~6--11.
Larger variations in RM were however recorded by \citet{Marrone2007} and \citet{Bower2018} on timescales from hours to months (see for example Figure~12 in  \citealt{Bower2018}). 
Since the observations in the two frequency bands  were close in time but not simultaneous,  we cannot definitely exclude time variability as the origin of the observed difference in RM magnitude  at 3~mm and 1.3~mm. \\

Future simultaneous measurements over a wider wavelength range (including  3~mm and 1.3~mm)  will allow us to separate time variability and source structure effects. 
\section{Conclusions}
\label{sect:conclusions}

We have determined and analysed the polarization and Faraday properties of Sgr~A*, the nucleus of M87, and  a dozen radio-loud AGN, 
observed with ALMA during the 2017 VLBI campaign in the 3~mm and 1~mm bands in concert with the GMVA and the  EHT, respectively. 

Our main findings can be summarised as follows: 

\begin{enumerate}
\item 
The AGN sources in our sample are highly polarized, with linear polarization degrees in the range 2--15\% at 1.3~mm and 0.9–13\% at 3~mm. 
The radio galaxies NGC~1052 and Cen~A are the only exceptions with LP$<$0.2\%.

    \item 
The AGN sources have negative spectral indexes varying in the range $\alpha=[-1.3,-0.2]$,  in contrast with the flat spectra  ($\alpha$=0) typically found at longer cm wavelengths in AGN cores. 
We also observe  a spectral steepening between the 3~mm and the 1.3~mm bands, which  
can naturally be explained by decreased opacity of the synchrotron emission  at higher frequencies in a standard jet model \citep[e.g.,][]{Lobanov1998}. 

    \item 
 We find marginal evidence for a general higher degree of polarization and RM magnitude 
 in the 1 mm band as compared to the 3 mm band (a 
trend which is consistent with  single-dish surveys). 
The increase of polarized intensity at higher frequency may be the result of an increased magnetic-field order in the inner portions of jets and/or to the smaller size of the high-frequency emitting regions. 
The increase of RM with frequency can be explained by opacity effects: emission at higher frequencies is generated in, and propagates along, regions with higher magnetic fields and plasma densities \citep[e.g.][]{Hovatta2014}. 
 Given the small-number statistics (eight AGN observed at 3~mm, eleven   at 1.3~mm, and six  in both bands) and the caveat of time variability (in a time window of ten days), simultaneous observations of a larger AGN sample at multiple frequency bands are needed  to confirm these results. 
 
    \item 
The blazars  (seven in our sample) have on average the highest level of polarization (LP $\sim$7.1\% at 1.3~mm) and an order of magnitude lower RM ($\sim 0.07 \times 10^5$~\radms\  at 1.3~mm) when compared with other AGN in our sample (with  LP $\sim$2.4\% and RM $\sim 0.4 \times 10^5$~\radms, respectively).  These mm polarization differences can be understood in the context of the viewing angle unification scheme of AGN:  blazars'  face-on geometry  implies a stronger Doppler-boosting of the synchrotron emitting plasma in the jet and reduces the effect of Faraday and beam depolarization in the accretion flow, resulting in higher LP (and lower RM). 
Future observations of a broader sample of sources are necessary for assessing the statistical significance of these trends.

    \item 
 We constrain the  circular polarization fraction in the observed AGN to $<$0.3\%. For Sgr A* we report CP = [-1.0,-1.5]\%, consistent with previous SMA and ALMA studies. 
 However, we explicitly note that the ALMA observatory does not guarantee a CP level $<$0.6\% ($1\sigma$), therefore these measurements should be regarded as tentative detections.

    \item 
We derive for the first time 
 the polarization and Faraday properties of Sgr~A* both at 3~mm and 1.3~mm in a time-window of ten days. 
 The RM magnitude at 3~mm, $(-2.1\pm0.1) \times10^5$~rad/m$^2$, is about half of the RM value measured at 1.3~mm over a three-days separation, suggesting that about half of  the Faraday rotation at 1.3~mm may occur between the 3~mm photosphere and the 1.3~mm source (although we cannot exclude effects related to time variability). 
 
    \item 
 We report the first unambiguous measurement of Faraday rotation   toward the M87 nucleus at mm wavelengths. At variance with Sgr~A*, the M87 RM exhibits significant changes in magnitude and sign reversals.
 At 1.3~mm, it spans from positive values 
($+1.5\times 10^5$~\radms\ at a $5\sigma$ level), 
to $<3\sigma$ non-detections  in Apr 2017, 
to negative values ($-3$ to $-4\times 10^5$~\radms\ at a $10\sigma$ level) in Apr 2018. 
At 3~mm the RM measured values span the range from $-1.2 \rm \ to\  0.3 \times 10^5$~\radms\ from Sep 2015 to Oct 2016. 
 The large scatter and time variability revealed by the ALMA measurements 
suggest a more complicated physical origin of the Faraday rotation than is usually assumed in models using the RM to infer a mass accretion rate. 
 We conclude that, unlike the case of Sgr~A*, the RM in M87 may not  provide an accurate estimate of the mass accretion rate onto the black hole. 

    \item 
 The observed RM in M87 may result from Faraday rotation internal to the emission region, as commonly found in GRMHD models of turbulent accretion flows or expected in a structured jet, or from a time-varying helical magnetic field threading the jet boundary layer acting as an external Faraday screen.
As an alternative, we put forward a two-component model comprised of a variable compact region  and static extended region.    We find that this simple model is able to simultaneously explain the polarimetric properties observed  in Apr 2017 by both the EHT (on horizon scales) and ALMA (which observes the combined emission from both components).

\end{enumerate}

 The ALMA measurements presented in this work provide critical constraints for the calibration and analysis of simultaneously obtained VLBI data. 
This is an essential resource for two instruments like the EHT and the GMVA which have the resolving power to reveal polarization structures and measure magnetic field strengths and particle densities on horizon scales (in the case of M87 and Sgr A*) and/or in the inner few parsecs for the AGN.

\acknowledgments{
ALMA is a partnership
of the European Southern Observatory (ESO;
Europe, representing its member states), NSF, and
National Institutes of Natural Sciences of Japan, together
with National Research Council (Canada), Ministry
of Science and Technology (MOST; Taiwan),
Academia Sinica Institute of Astronomy and Astrophysics
(ASIAA; Taiwan), and Korea Astronomy and
Space Science Institute (KASI; Republic of Korea), in
cooperation with the Republic of Chile. The Joint
ALMA Observatory is operated by ESO, Associated
Universities, Inc. (AUI)/NRAO, and the National Astronomical
Observatory of Japan (NAOJ). The NRAO
is a facility of the NSF operated under cooperative agreement
by AUI. 
This paper makes use of the following ALMA data:\\
ADS/JAO.ALMA\#2016.1.00413.V \\ 
ADS/JAO.ALMA\#2016.1.01116.V \\
ADS/JAO.ALMA\#2016.1.01216.V \\
ADS/JAO.ALMA\#2016.1.01114.V \\
ADS/JAO.ALMA\#2016.1.01154.V \\
ADS/JAO.ALMA\#2016.1.01176.V \\
ADS/JAO.ALMA\#2016.1.01198.V \\
ADS/JAO.ALMA\#2016.1.01290.V \\
ADS/JAO.ALMA\#2016.1.01404.V \\
ADS/JAO.ALMA\#2017.1.00841.V \\ 
ADS/JAO.ALMA\#2013.1.01022.S \\ 
ADS/JAO.ALMA\#2015.1.01170.S \\ 
ADS/JAO.ALMA\#2016.1.00415.S \\ 
ADS/JAO.ALMA\#2017.1.00608.S \\ 
The authors of the present paper thank the following
organizations and programs: the Academy
of Finland (projects 274477, 284495, 312496, 315721); the
Advanced European Network of E-infrastructures for
Astronomy with the SKA (AENEAS) project, supported
by the European Commission Framework Programme
Horizon 2020 Research and Innovation action
under grant agreement 731016; the Alexander
von Humboldt Stiftung; an Alfred P. Sloan Research Fellowship;
Allegro, the European ALMA Regional Centre node in the Netherlands, the NL astronomy
research network NOVA and the astronomy institutes of the University of Amsterdam, Leiden University and Radboud University;
the Black Hole Initiative at
Harvard University, through a grant (60477) from
the John Templeton Foundation; the China Scholarship
Council; 
Agencia Nacional de Investigación y Desarrollo (ANID), Chile via NCN19-058 (TITANs) and Fondecyt 3190878; 
Consejo Nacional de Ciencia y Tecnología (CONACYT, Mexico, projects  U0004-246083, U0004-259839, F0003-272050, M0037-279006, F0003-281692,
104497, 275201, 263356);
the Delaney Family via the Delaney Family John A.
Wheeler Chair at Perimeter Institute; Dirección General
de Asuntos del Personal Académico--Universidad
Nacional Autónoma de México (DGAPA--UNAM,
projects IN112417 and IN112820); the European Research Council Synergy
Grant "BlackHoleCam: Imaging the Event Horizon
of Black Holes" (grant 610058); the Generalitat
Valenciana postdoctoral grant APOSTD/2018/177 and
GenT Program (project CIDEGENT/2018/021); MICINN Research Project PID2019-108995GB-C22;
the
Gordon and Betty Moore Foundation (grants GBMF-
3561, GBMF-5278); the Istituto Nazionale di Fisica
Nucleare (INFN) sezione di Napoli, iniziative specifiche
TEONGRAV; the International Max Planck Research
School for Astronomy and Astrophysics at the
Universities of Bonn and Cologne; the Jansky Fellowship
program of the National Radio Astronomy Observatory
(NRAO); 
Joint Princeton/Flatiron and Joint Columbia/Flatiron Postdoctoral Fellowships, research at the Flatiron Institute is supported by the Simons Foundation; 
the Japanese Government (Monbukagakusho:
MEXT) Scholarship; the Japan Society for
the Promotion of Science (JSPS) Grant-in-Aid for JSPS
Research Fellowship (JP17J08829); the Key Research
Program of Frontier Sciences, Chinese Academy of
Sciences (CAS, grants QYZDJ-SSW-SLH057, QYZDJSSW-
SYS008, ZDBS-LY-SLH011); the Leverhulme Trust Early Career Research
Fellowship; the Max-Planck-Gesellschaft (MPG);
the Max Planck Partner Group of the MPG and the
CAS; the MEXT/JSPS KAKENHI (grants 18KK0090,
JP18K13594, JP18K03656, JP18H03721, 18K03709,
18H01245, 25120007); the Malaysian Fundamental Research Grant Scheme (FRGS) FRGS/1/2019/STG02/ UM/02/6; the MIT International Science
and Technology Initiatives (MISTI) Funds; the Ministry
of Science and Technology (MOST) of Taiwan (105-
2112-M-001-025-MY3, 106-2112-M-001-011, 106-2119-
M-001-027, 107-2119-M-001-017, 107-2119-M-001-020,
and 107-2119-M-110-005, MOST 108-2112-M-001-048 and MOST 109-2124-M-001-005); the National Aeronautics and
Space Administration (NASA, Fermi Guest Investigator
grant 80NSSC20K1567, NASA Astrophysics Theory Program grant 80NSSC20K0527, and Hubble Fellowship grant
HST-HF2-51431.001-A awarded by the Space Telescope
Science Institute, which is operated by the Association
of Universities for Research in Astronomy, Inc.,
for NASA, under contract NAS5-26555, ); the National
Institute of Natural Sciences (NINS) of Japan; the National
Key Research and Development Program of China
(grant 2016YFA0400704, 2016YFA0400702); the National
Science Foundation (NSF, grants AST-0096454,
AST-0352953, AST-0521233, AST-0705062, AST-0905844, AST-0922984, AST-1126433, AST-1140030,
DGE-1144085, AST-1207704, AST-1207730, AST-1207752, MRI-1228509, OPP-1248097, AST-1310896, AST-1337663, AST-1440254, AST-1555365, AST-1615796, AST-1715061, AST-1716327, AST-1716536,
OISE-1743747, AST-1816420, AST-1903847,
AST-1935980, AST-2034306); the Natural Science
Foundation of China (grants 11573051, 11633006,
11650110427, 10625314, 11721303, 11725312, 11933007, 11991052, 11991053); a fellowship of China Postdoctoral Science Foundation (2020M671266); the Natural
Sciences and Engineering Research Council of
Canada (NSERC, including a Discovery Grant and
the NSERC Alexander Graham Bell Canada Graduate
Scholarships-Doctoral Program); the National Youth
Thousand Talents Program of China; the National Research
Foundation of Korea (the Global PhD Fellowship
Grant: grants NRF-2015H1A2A1033752, 2015-
R1D1A1A01056807, the Korea Research Fellowship Program:
NRF-2015H1D3A1066561, Basic Research Support Grant 2019R1F1A1059721); the Netherlands Organization
for Scientific Research (NWO) VICI award
(grant 639.043.513) and Spinoza Prize SPI 78-409; the
New Scientific Frontiers with Precision Radio Interferometry
Fellowship awarded by the South African Radio
Astronomy Observatory (SARAO), which is a facility
of the National Research Foundation (NRF), an
agency of the Department of Science and Technology
(DST) of South Africa; the Onsala Space Observatory
(OSO) national infrastructure, for the provisioning
of its facilities/observational support (OSO receives
funding through the Swedish Research Council under
grant 2017-00648) the Perimeter Institute for Theoretical
Physics (research at Perimeter Institute is supported
by the Government of Canada through the Department
of Innovation, Science and Economic Development
and by the Province of Ontario through the
Ministry of Research, Innovation and Science); the Spanish
Ministerio de Economía y Competitividad (grants
PGC2018-098915-B-C21, AYA2016-80889-P, PID2019-108995GB-C21); the State
Agency for Research of the Spanish MCIU through
the "Center of Excellence Severo Ochoa" award for
the Instituto de Astrofísica de Andalucía (SEV-2017-
0709); the Toray Science Foundation; the Consejería de Economía, Conocimiento, Empresas y Universidad of the Junta de Andalucía (grant P18-FR-1769), the Consejo Superior de Investigaciones Científicas (grant 2019AEP112);
the US Department
of Energy (USDOE) through the Los Alamos National
Laboratory (operated by Triad National Security,
LLC, for the National Nuclear Security Administration
of the USDOE (Contract 89233218CNA000001);
the Italian Ministero dell’Istruzione Università e Ricerca
through the grant Progetti Premiali 2012-iALMA (CUP
C52I13000140001); the European Union’s Horizon 2020
research and innovation programme under grant agreement
No 730562 RadioNet; ALMA North America Development
Fund; the Academia Sinica; Chandra TM6-
17006X; the GenT Program (Generalitat Valenciana)
Project CIDEGENT/2018/021; NASA NuSTAR award 80NSSC20K0645; Chandra award DD7-18089X. 
This work used the
Extreme Science and Engineering Discovery Environment
(XSEDE), supported by NSF grant ACI-1548562,
and CyVerse, supported by NSF grants DBI-0735191,
DBI-1265383, and DBI-1743442. XSEDE Stampede2 resource
at TACC was allocated through TG-AST170024
and TG-AST080026N. XSEDE JetStream resource at
PTI and TACC was allocated through AST170028.
The simulations were performed in part on the SuperMUC
cluster at the LRZ in Garching, on the
LOEWE cluster in CSC in Frankfurt, and on the
HazelHen cluster at the HLRS in Stuttgart. This
research was enabled in part by support provided
by Compute Ontario (http://computeontario.ca), Calcul
Quebec (http://www.calculquebec.ca) and Compute
Canada (http://www.computecanada.ca). 
We thank
the staff at the participating observatories in the GMVA  and EHT, correlation
centers, and institutions for their enthusiastic support.
The GMVA is coordinated by the VLBI group at the Max-Planck-Institut f{\"u}r Radioastronomie (MPIfR) and consists of telescopes operated by MPIfR, IRAM, Onsala, Metsahovi, Yebes, the Korean VLBI Network, the Green Bank Observatory and the VLBA.
APEX is a collaboration between the MPIfR (Germany),
ESO, and the Onsala Space Observatory (Sweden). The
SMA is a joint project between the SAO and ASIAA
and is funded by the Smithsonian Institution and the
Academia Sinica. The JCMT is operated by the East
Asian Observatory on behalf of the NAOJ, ASIAA, and
KASI, as well as the Ministry of Finance of China, Chinese
Academy of Sciences, and the National Key R\&D
Program (No. 2017YFA0402700) of China. Additional
funding support for the JCMT is provided by the Science
and Technologies Facility Council (UK) and participating
universities in the UK and Canada. The
LMT is a project operated by the Instituto Nacional
de Astrófisica, Óptica, y Electrónica (Mexico) and the
University of Massachusetts at Amherst (USA). The
IRAM 30-m telescope on Pico Veleta, Spain is operated
by IRAM and supported by CNRS (Centre National de
la Recherche Scientifique, France), MPG (Max-Planck-
Gesellschaft, Germany) and IGN (Instituto Geográfico
Nacional, Spain). The SMT is operated by the Arizona
Radio Observatory, a part of the Steward Observatory
of the University of Arizona, with financial support of
operations from the State of Arizona and financial support
for instrumentation development from the NSF.
The SPT is supported by the National Science Foundation
through grant PLR- 1248097. Partial support is
also provided by the NSF Physics Frontier Center grant
PHY-1125897 to the Kavli Institute of Cosmological
Physics at the University of Chicago, the Kavli Foundation
and the Gordon and Betty Moore Foundation grant
GBMF 947. The SPT hydrogen maser was provided on
loan from the GLT, courtesy of ASIAA. The EHTC has
received generous donations of FPGA chips from Xilinx
Inc., under the Xilinx University Program. The EHTC
has benefited from technology shared under open-source
license by the Collaboration for Astronomy Signal Processing
and Electronics Research (CASPER). The EHT
project is grateful to T4Science and Microsemi for their
assistance with Hydrogen Masers. This research has
made use of NASA’s Astrophysics Data System. We
gratefully acknowledge the support provided by the extended
staff of the ALMA, both from the inception of
the ALMA Phasing Project through the observational
campaigns of 2017 and 2018. We would like to thank
A. Deller and W. Brisken for EHT-specific support with
the use of DiFX. 
We thank Sergio Martin for the help interpreting the Sgr~A* spectrum in SPW=2.
We acknowledge the significance that
Maunakea, where the SMA and JCMT EHT stations
are located, has for the indigenous Hawaiian people.
MP acknowledges support by the Spanish Ministry of Science through Grants PID2019-105510GB-C31, and PID2019-107427GB-C33, and from the Generalitat Valenciana through grant PROMETEU/2019/071.}

\appendix

\begin{table*}
\caption{Projects and sources observed in  the 3~mm Band.}
\centering  
\small
\begin{tabular}{cccccc}
\hline\hline                  
\noalign{\smallskip}
Date  & Project & Target &  Pol. Cal. & Other sources & UT range \\
\noalign{\smallskip}
\hline
\noalign{\smallskip}  
 2017 Apr 02   &2016.1.01116.V     & OJ287  & 4C 01.28  & J0510+1800 & 06:55:08 --  15:19:43     \\  
 2017 Apr 03   &2016.1.00413.V     & Sgr A*  & NRAO~530  &J1924--2914, 4C 09.57& 20:52:28  --  04:43:54    \\
 2017 Apr 04   &2016.1.01216.V     & 3C273  & 3C279  & --- &00:24:57  --  05:32:46     \\  
\noalign{\smallskip}
\hline   
\end{tabular} 
\label{table:gmva_exp}
\end{table*}

\begin{table*}
\caption{Projects and sources observed in the 1.3~mm Band.}
\centering  
\small
\begin{tabular}{cccccc}
\hline\hline                  
\noalign{\smallskip}
Date  & Project & Target &  Pol. Cal. & Other sources & UT range \\
\noalign{\smallskip}
\hline
\noalign{\smallskip}  
 2017 Apr 05   &&&&& 04/22:12 --  05/09:13  \\
& 2016.1.01114.V  & OJ287   &  3C279 & 4C 01.28, M87 &04/22:12 --  05/03:22  \\
& 2016.1.01154.V  & M87      &  3C279 & 4C 01.28, OJ287&  05/03:24 --  05/07:18    \\
& 2016.1.01176.V  & 3C279  &  3C279 & M87 &  05/07:19 --  05/09:13   \\
\noalign{\smallskip}
\hline   
\noalign{\smallskip}
  2017 Apr 06   &&&&& 06/00:18 --  06/16:19  \\
& 2016.1.01154.V  & M87      &  3C279 & 3C273 &  06/00:18 --  06/08:02    \\
& 2016.1.01404.V  & Sgr\,A*  &  3C279 &NRAO~530, J1924--2914 &  06/08:03 --  06/14:40     \\
& 2016.1.01290.V  & NGC~1052  &  3C279 & J0132-1654, J0006-0623&  06/14:51 --  06/16:19    \\
\noalign{\smallskip}
\hline   
\noalign{\smallskip}
  2017 Apr 07   &&&&& 07/03:45  --  07/20:47  \\
& 2016.1.01404.V  & Sgr\,A*  &  J1924--2914 & NRAO~530 & 07/03:45 --  07/14:31     \\
& 2016.1.01290.V  & NGC~1052  &  J1924--2914 &  J0132-1654, 3C 84$^a$  & 07/19:23 --  07/20:47    \\
&&&& J0006-0623$^a$ & \\
\noalign{\smallskip}
\hline   
\noalign{\smallskip}
 2017 Apr 10  &&& && 09/23:02 --  10/10:02  \\
& 2016.1.01114.V  & OJ287   &  3C279 & 4C 01.28, M87 &  09/23:02 --  10/03:49  \\
& 2016.1.01176.V  & 3C279  &  3C279 & Cen\,A, M87  &  10/03:51 --  10/06:21  \\
& 2016.1.01198.V  & Cen\,A  &  3C279 &--- &  10/06:23 --  10/10:02  \\
\noalign{\smallskip}
\hline   
\noalign{\smallskip}
  2017 Apr 11   &&&&& 10/21:44 --  11/10:31  \\
& 2016.1.01114.V  & OJ287   &  3C279 & 4C 01.28&  10/21:44 --  11/00:22  \\
& 2016.1.01154.V  & M87      &  3C279 & --- &  11/00:23 --  11/05:03    \\
& 2016.1.01176.V  & 3C279  &  3C279 &  M87 &  11/05:05 --  11/08:45   \\
& 2016.1.01404.V  & Sgr\,A* &  3C279 & J1924--2914 &  11/08:46 --  11/14:03     \\
\noalign{\smallskip}
\hline\hline  
\end{tabular}
 \tablenotetext{a}{Flagged before data analysis (see text in Appendix~\ref{app:exp} and \S\ref{sect:datacal}). }
\label{table:eht_exp}
\end{table*}
\section{Projects observed  during the 2017 VLBI campaign}
\label{app:exp}

\newtext{
Tables~\ref{table:gmva_exp} and \ref{table:eht_exp}  list the projects observed in the 3~mm and 1.3~mm bands, ordered by date of execution.  
Each row reports the observing date, the ALMA project code, the science target, the source used as polarization calibrator, other sources observed in the project, and the duration of each observation. 
In Table~\ref{table:eht_exp}, each row-group refers to an individual VLBI run or ''Track'', which includes observations of different projects carried out during the same night. 
The calibration of EHT projects was done per  track (and not per project; see \citealt{QA2Paper}). 
Two sources listed in Table~\ref{table:eht_exp}, 3C 84 and  J0006-0623, both observed on Apr 07, were excluded from the analysis presented in this paper: 
3C 84 was observed with an elevation below 25\dg, while J0006-0623 was observed for just $\sim2$~min  close to an elevation of 25\dg; 
 the resulting calibrated data display critical phase and amplitude scatter and hence were flagged before data analysis (see \S~\ref{sect:datacal}). 
}

\begin{table*}
\caption{Frequency-averaged imaging parameters of GMVA sources (at a representative frequency of 93 GHz). }
\label{tab:GMVA_im_rms}
\centering \begin{tabular}{ccccccc}
\hline\hline 
Source & T$_{\rm on \ source}$ & RMS(I) & RMS(Q) & RMS(U) & RMS(V) & Synthesized Beam\\
 & [h]& [mJy] & [mJy] & [mJy] & [mJy] & \colhead{$\theta_M$[\as] $\times \theta_m$[\as] (P.A.[$^{\circ}$])}\\
\hline
\multicolumn{7}{c}{Apr 2}\\
     OJ287                           &    2.584     &   0.31   &    0.34    &    0.46   &   0.17  & 4.7 \as~$\times$ 2.7\as~(-86.2$^{\circ}$) \\
     4C 01.28                        &     0.269    &   0.17   &    0.30    &    0.36   &   0.14  & 4.9 \as~$\times$ 2.4\as~(-86.8$^{\circ}$) \\
     J0510+1800                      &    0.363     &   0.31   &    0.29    &    0.52   &   0.12  & 5.8 \as~$\times$ 2.5\as~(-70.1$^{\circ}$) \\
\hline
\multicolumn{7}{c}{Apr 3}\\
     J1924-2914                      &     0.270    &   0.16   &    0.65    &    0.13   &   0.13  & 5.5 \as~$\times$ 2.5\as~(-75.2$^{\circ}$) \\
     NRAO 530                        &  0.479       &   0.02   &    0.02    &    0.02   &   0.03  & 4.8 \as~$\times$ 2.4\as~(-83.5$^{\circ}$) \\
     Sgr A*                          &    2.643     &   0.80   &    0.09    &    0.08   &   0.04  & 5.0 \as~$\times$ 2.7\as~(-81.1$^{\circ}$) \\
     4C 09.57                        &   0.133      &   0.09   &    0.08    &    0.14   &   0.07  & 6.1 \as~$\times$ 2.7\as~(72.0$^{\circ}$) \\
\hline
\multicolumn{7}{c}{Apr 4}\\
     3C273                           &      1.396   &   0.48   &    0.26    &    0.54   &   0.13  & 5.0 \as~$\times$ 3.4\as~(-86.7$^{\circ}$) \\
     3C279                           &   0.215      &   0.37   &    0.20    &    0.15   &   0.18  & 5.0 \as~$\times$ 3.4\as~(-85.8$^{\circ}$) \\
\hline\hline
\end{tabular}\end{table*}
\begin{table*}
\caption{Frequency-averaged imaging parameters of EHT sources (at a representative frequency of 221 GHz). }
\label{tab:EHT_im_rms}
\centering \begin{tabular}{ccccccc}
\hline\hline 
Source & T$_{\rm on-source}$ & RMS(I) & RMS(Q) & RMS(U) & RMS(V) & Synthesized Beam\\
 & [h]& [mJy] & [mJy] & [mJy] & [mJy] & \colhead{$\theta_M$[\as] $\times \theta_m$[\as] (P.A.[$^{\circ}$])}\\
\hline
\multicolumn{7}{c}{Apr 5}\\
     M87                             &     1.645    &   0.18   &    0.05    &    0.05   &   0.02  & 2.0 \as~$\times$ 1.0\as~(-85.5$^{\circ}$) \\
     3C279                           &    1.068     &   0.24   &    0.09    &    0.07   &   0.16  & 2.2 \as~$\times$ 0.9\as~(-80.9$^{\circ}$) \\
     OJ287                           &     1.406      &   0.13   &    0.16    &    0.15   &   0.11  & 2.0 \as~$\times$ 1.1\as~(88.7$^{\circ}$) \\
     4C 01.28                        &    0.230     &   0.12   &    0.11    &    0.08   &   0.12  & 2.0 \as~$\times$ 0.9\as~(87.0$^{\circ}$) \\
\hline
\multicolumn{7}{c}{Apr 6}\\
     J0006-0623                      &   0.045      &   0.47   &    0.12    &    0.15   &   0.13  & 2.2 \as~$\times$ 1.4\as~(-81.1$^{\circ}$) \\
     J0132-1654                      &   0.059      &   0.06   &    0.06    &    0.06   &   0.06  & 2.3 \as~$\times$ 1.4\as~(87.8$^{\circ}$) \\
     NGC 1052                        &   0.373      &   0.05   &    0.03    &    0.03   &   0.03  & 2.7 \as~$\times$ 1.3\as~(80.3$^{\circ}$) \\
     Sgr A*                          &      2.529   &   0.44   &    0.18    &    0.33   &   0.08  & 2.2 \as~$\times$ 1.3\as~(-77.5$^{\circ}$) \\
     J1924-2914                      &   0.269      &   0.10   &    0.07    &    0.08   &   0.11  & 2.2 \as~$\times$ 1.3\as~(-82.5$^{\circ}$) \\
     NRAO 530                        &   0.269      &   0.04   &    0.06    &    0.05   &   0.05  & 2.2 \as~$\times$ 1.3\as~(-76.4$^{\circ}$) \\
     M87                             &       1.613  &   0.24   &    0.06    &    0.06   &   0.02  & 2.2 \as~$\times$ 1.5\as~(-69.4$^{\circ}$) \\
     3C279                           &     0.430    &   0.30   &    0.13    &    0.08   &   0.13  & 2.2 \as~$\times$ 1.3\as~(-78.4$^{\circ}$) \\
     3C273                           &    0.403     &   0.19   &    0.10    &    0.09   &   0.10  & 2.3 \as~$\times$ 1.4\as~(-75.0$^{\circ}$) \\
\hline
\multicolumn{7}{c}{Apr 7}\\
     NGC 1052                        &    0.200     &   0.11   &    0.04    &    0.04   &   0.04  & 2.6 \as~$\times$ 1.0\as~(-76.3$^{\circ}$) \\
     J0132-1654                      &    0.056     &   0.08   &    0.07    &    0.07   &   0.06  & 3.0 \as~$\times$ 1.0\as~(-72.8$^{\circ}$) \\
     NRAO 530                        &   0.403      &   0.08   &    0.03    &    0.03   &   0.05  & 2.1 \as~$\times$ 0.9\as~(-89.6$^{\circ}$) \\
     J1924-2914                      &   0.312      &   0.18   &    0.04    &    0.04   &   0.08  & 2.1 \as~$\times$ 0.9\as~(89.7$^{\circ}$) \\
     Sgr A*                        &     4.109    &   0.25   &    0.14    &    0.13   &   0.04  & 2.1 \as~$\times$ 0.9\as~(-88.6$^{\circ}$) \\
\hline
\multicolumn{7}{c}{Apr 10}\\
     Cen A                           &     1.401    &   0.11   &    0.07    &    0.07   &   0.12  & 2.3 \as~$\times$ 0.9\as~(-79.0$^{\circ}$) \\
     M87                             &     0.454    &   0.25   &    0.06    &    0.07   &   0.04  & 2.0 \as~$\times$ 1.0\as~(-88.9$^{\circ}$) \\
     OJ287                           &   1.083      &   0.15   &    0.14    &    0.18   &   0.08  & 2.0 \as~$\times$ 1.1\as~(-82.5$^{\circ}$) \\
     3C279                           &   1.120     &   0.29   &    0.12    &    0.08   &   0.16  & 2.1 \as~$\times$ 0.8\as~(-85.2$^{\circ}$) \\
     4C 01.28                        &   0.289      &   0.15   &    0.14    &    0.10   &   0.09  & 2.2 \as~$\times$ 0.9\as~(80.0$^{\circ}$) \\
\hline
\multicolumn{7}{c}{Apr 11}\\
     Sgr A*                          &       1.934  &   0.35   &    0.24    &    0.20   &   0.06  & 1.2 \as~$\times$ 0.7\as~(-85.1$^{\circ}$) \\
     J1924-2914                      &   0.244      &   0.29   &    0.10    &    0.10   &   0.14  & 1.2 \as~$\times$ 0.7\as~(89.9$^{\circ}$) \\
     3C279                           &    1.705      &   0.23   &    0.11    &    0.08   &   0.11  & 1.2 \as~$\times$ 0.7\as~(-86.6$^{\circ}$) \\
     M87                             &     1.831    &   0.16   &    0.04    &    0.04   &   0.03  & 1.2 \as~$\times$ 0.8\as~(79.3$^{\circ}$) \\
     OJ287                           &    0.804     &   0.21   &    0.15    &    0.11   &   0.17  & 1.2 \as~$\times$ 0.9\as~(59.6$^{\circ}$) \\
     4C 01.28                        &   0.110      &   0.26   &    0.17    &    0.14   &   0.23  & 1.5 \as~$\times$ 0.8\as~(67.9$^{\circ}$) \\
\hline\hline
\end{tabular}\end{table*}

\section{Polarimetric Images} 
\label{appendix:maps}

Tables~\ref{tab:GMVA_im_rms}  and~\ref{tab:EHT_im_rms} report the main imaging parameters for each source observed on each day of the 2017 VLBI campaign  in Band~3  and  Band~6, respectively.
These parameters include the on-source time, the RMS achieved in each Stokes parameter, and the synthesized beamsize. 
The on-source time is computed after full calibration and flagging of bad data (see \S\ref{sect:datacal}). 
The RMS does not simply scale as $\sqrt{T_{\rm on-source}}$ but depends on several parameters such as source structure, number of observing antennas, array configuration, weather, and details of the VLBI scheduling blocks (e.g. low-elevation scans).  
The synthesized beamsize changes by a factor of two  at 1.3~mm due to the changing array configuration during the observing week
(see \S\ref{obs}).
The resulting images are  dynamic range limited and showcase different structures on different days (depending on the array beamsize). 

\begin{figure*}[ht!]
\centering
\includegraphics[width=6cm]{Sgr_A_spw0123_TRACK_B_polimage.pdf} \hspace{-0.3cm}
\includegraphics[width=6cm]{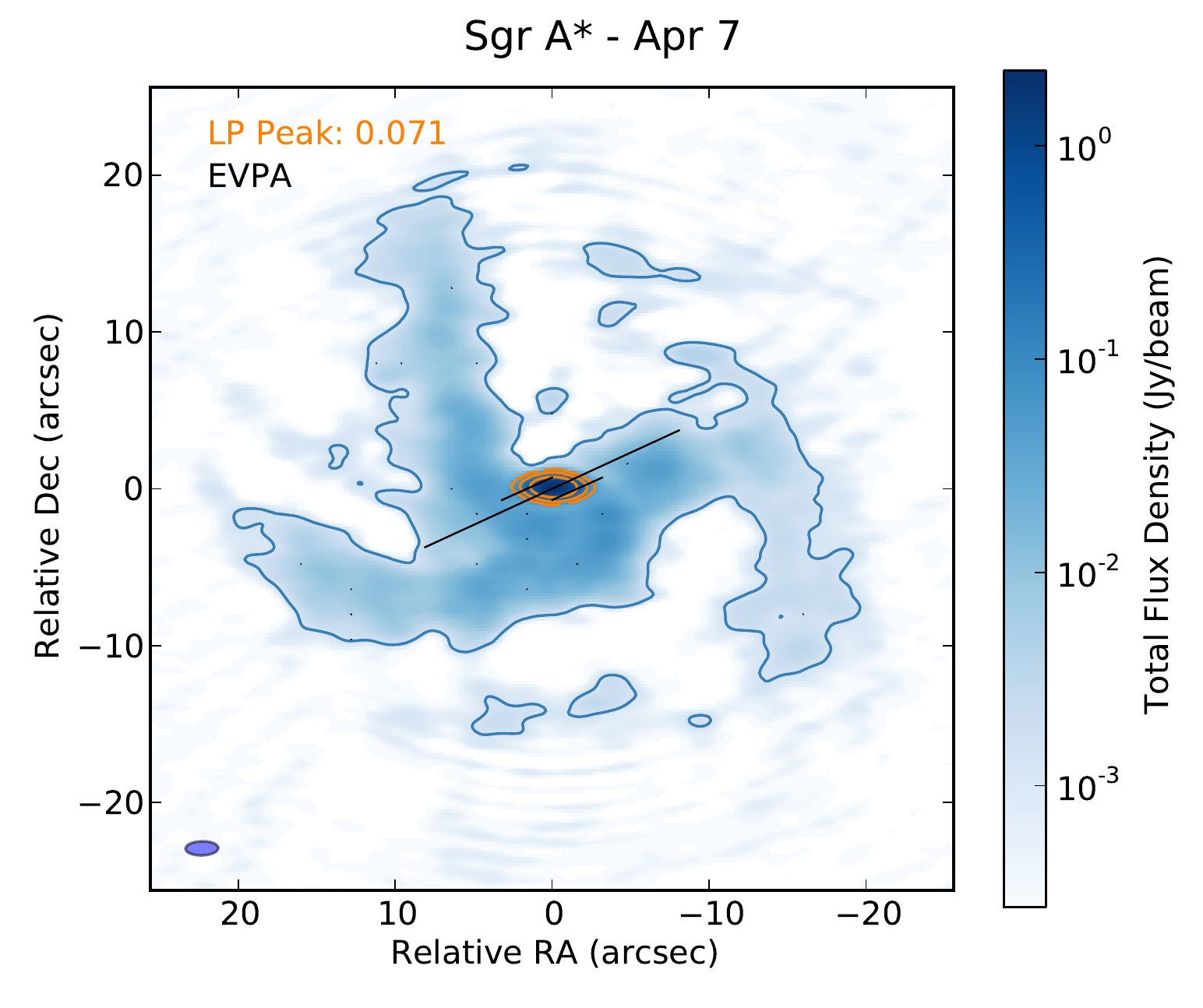}  \hspace{-0.3cm}
\includegraphics[width=6cm]{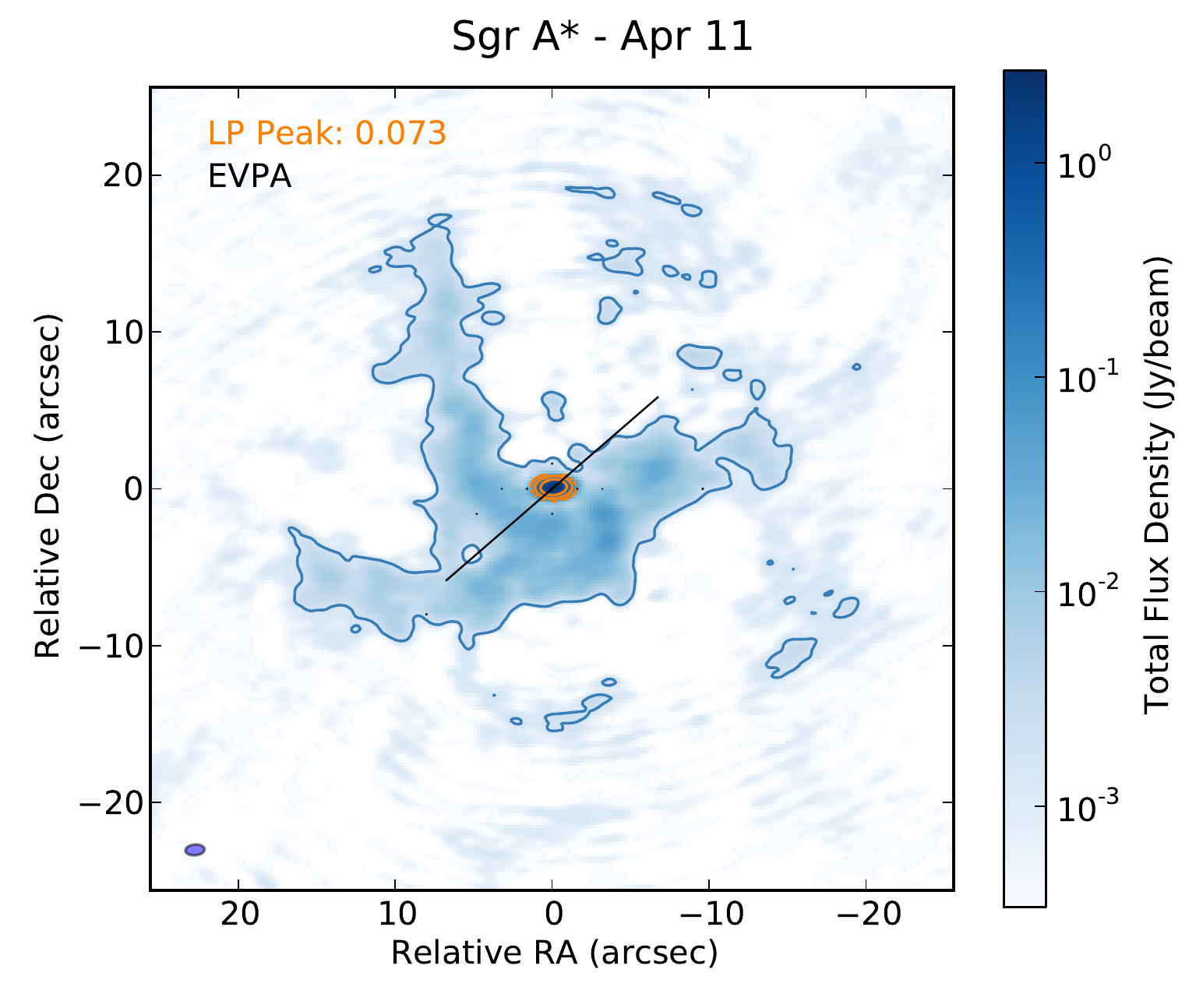}  \hspace{-0.3cm}
\caption{Polarization images of Sgr~A* at 1.3~mm (see Figure~\ref{fig:m87+sgra_polimage} for a description of the plotted quantities). 
The apparently lower quality of the image in the right panel is due to the fact that the observations on Apr 11 have about half the beam size, and  are therefore less sensitive to the extended emission in the mini-spiral, when compared to the other two days (the beamsizes are shown as  ovals in the lower left corner of each panel; see values reported in Table~\ref{tab:EHT_im_rms}). 
Note that there are several tiny EVPAs plotted across the mini-spiral, apparently locating regions with polarized flux above the  image RMS noise cutoff ($5\sigma$). The LP and EVPA errors are however dominated by the systematic leakage (0.03\% of I onto QU), which is not added to the images.  Once these systematic errors are added, the LP flux in those points falls below the $3\sigma$ detection threshold. Therefore we do not claim detection of polarized emission outside of the central core in Sgr A*.
}
\label{fig:polimages_sgra_1mm}
\end{figure*}

\begin{figure*}[ht!]
\centering
\includegraphics[width=9cm]{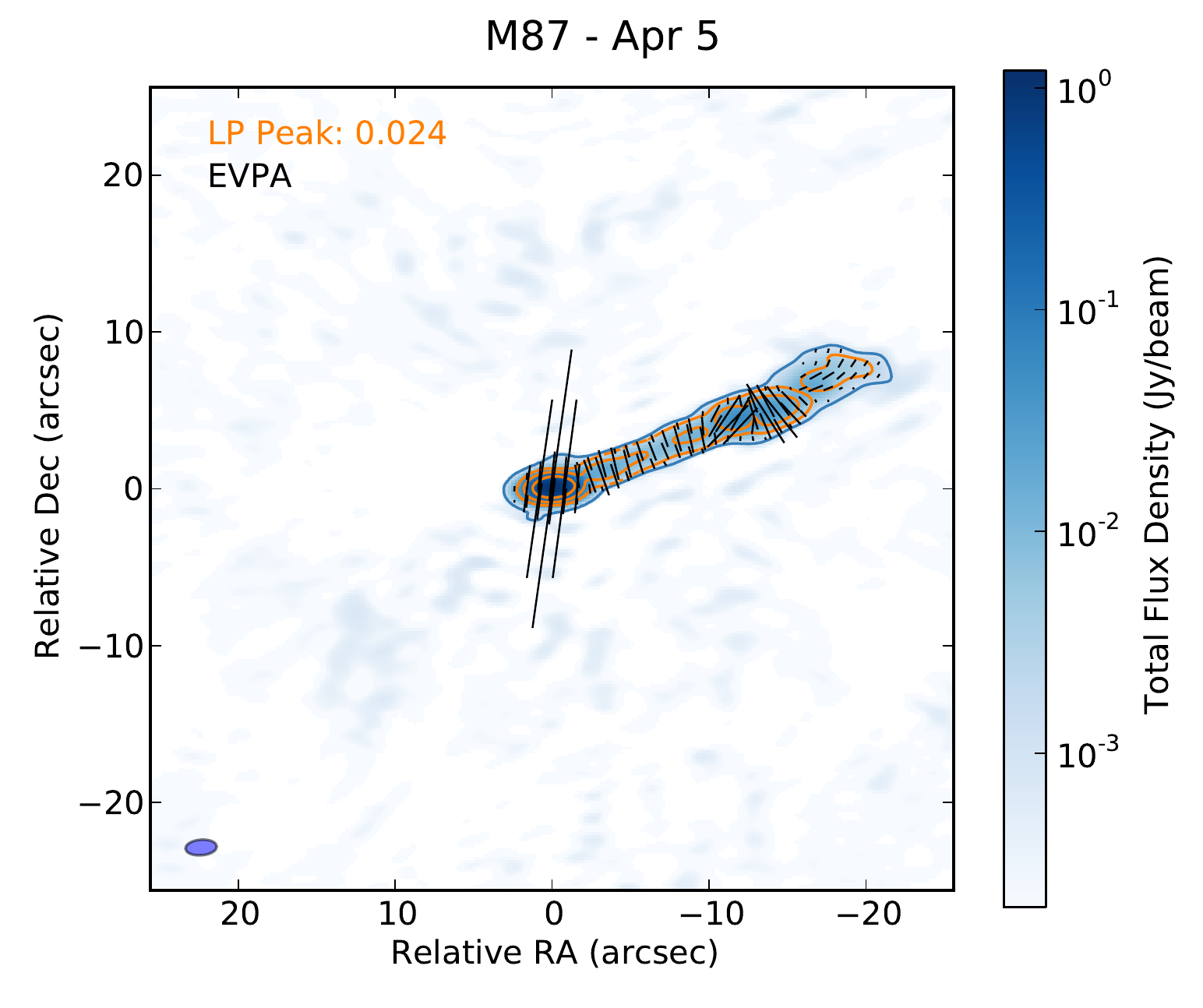} \hspace{-0.3cm}
\includegraphics[width=9cm]{M87_spw0123_TRACK_B_polimage.pdf} \hspace{-0.3cm}
\includegraphics[width=9cm]{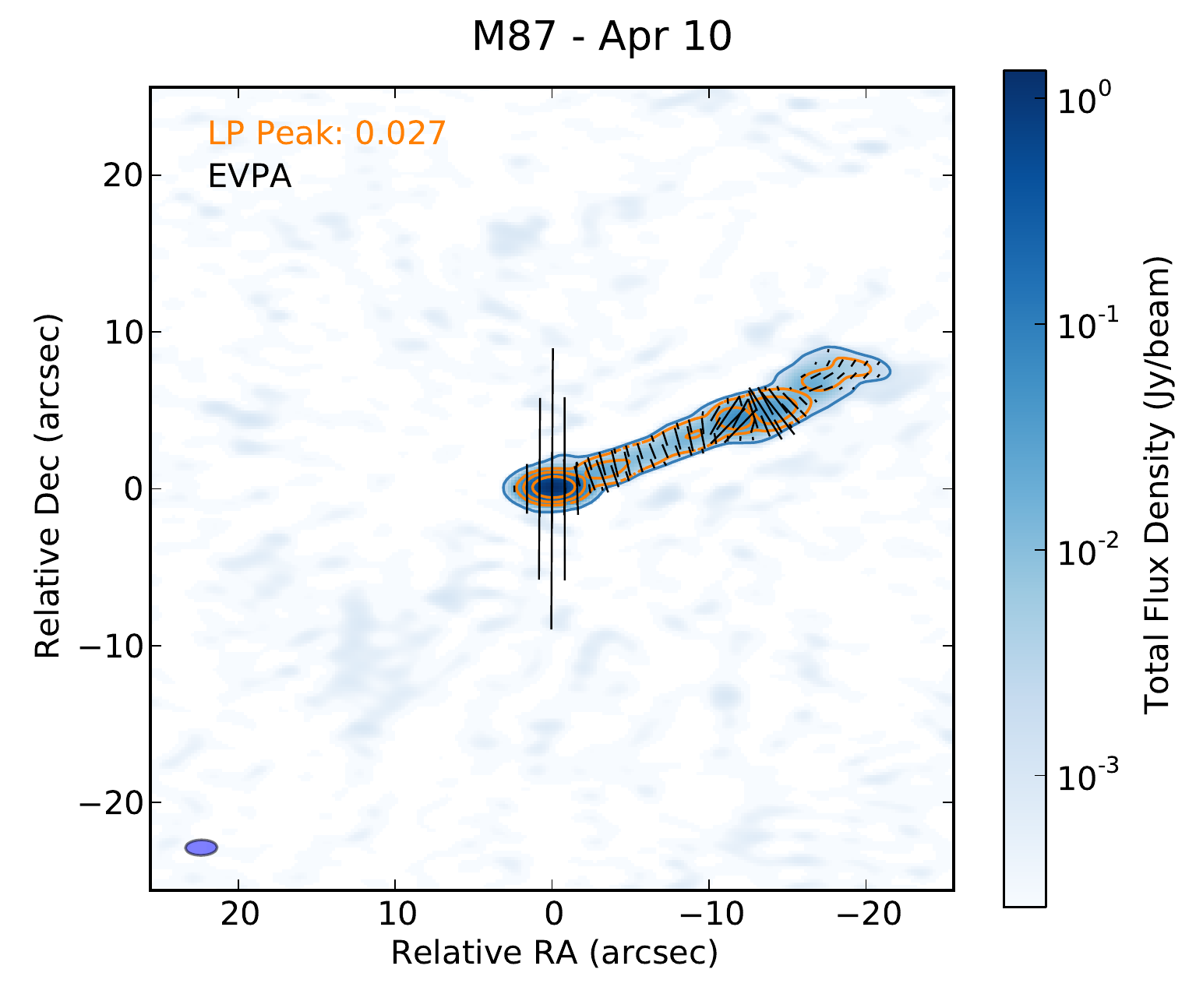}  \hspace{-0.3cm}
\includegraphics[width=9cm]{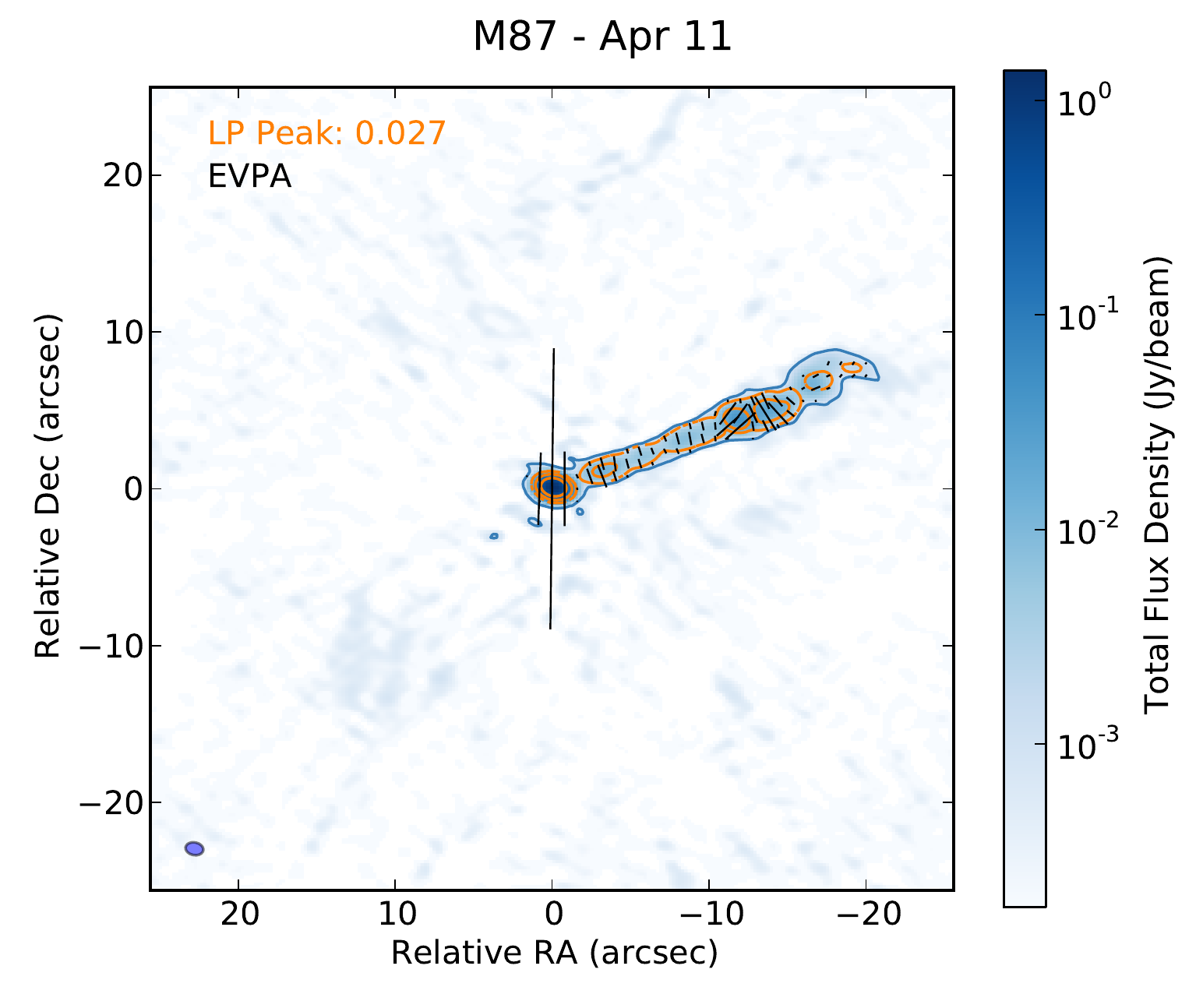}  \hspace{-0.3cm}
\caption{Polarization images of M87 at 1.3~mm (see Figure~\ref{fig:m87+sgra_polimage} for a description of the plotted quantities). 
 The apparently different jet structures (in total and polarized intensity) across days  are due to different beamsizes (shown as  ovals in the lower left corner of each panel; see values reported in Table~\ref{tab:EHT_im_rms}).
}
\label{fig:polimages_m87}
\end{figure*}

\begin{figure*}[ht!]
\centering
\includegraphics[width=9cm]{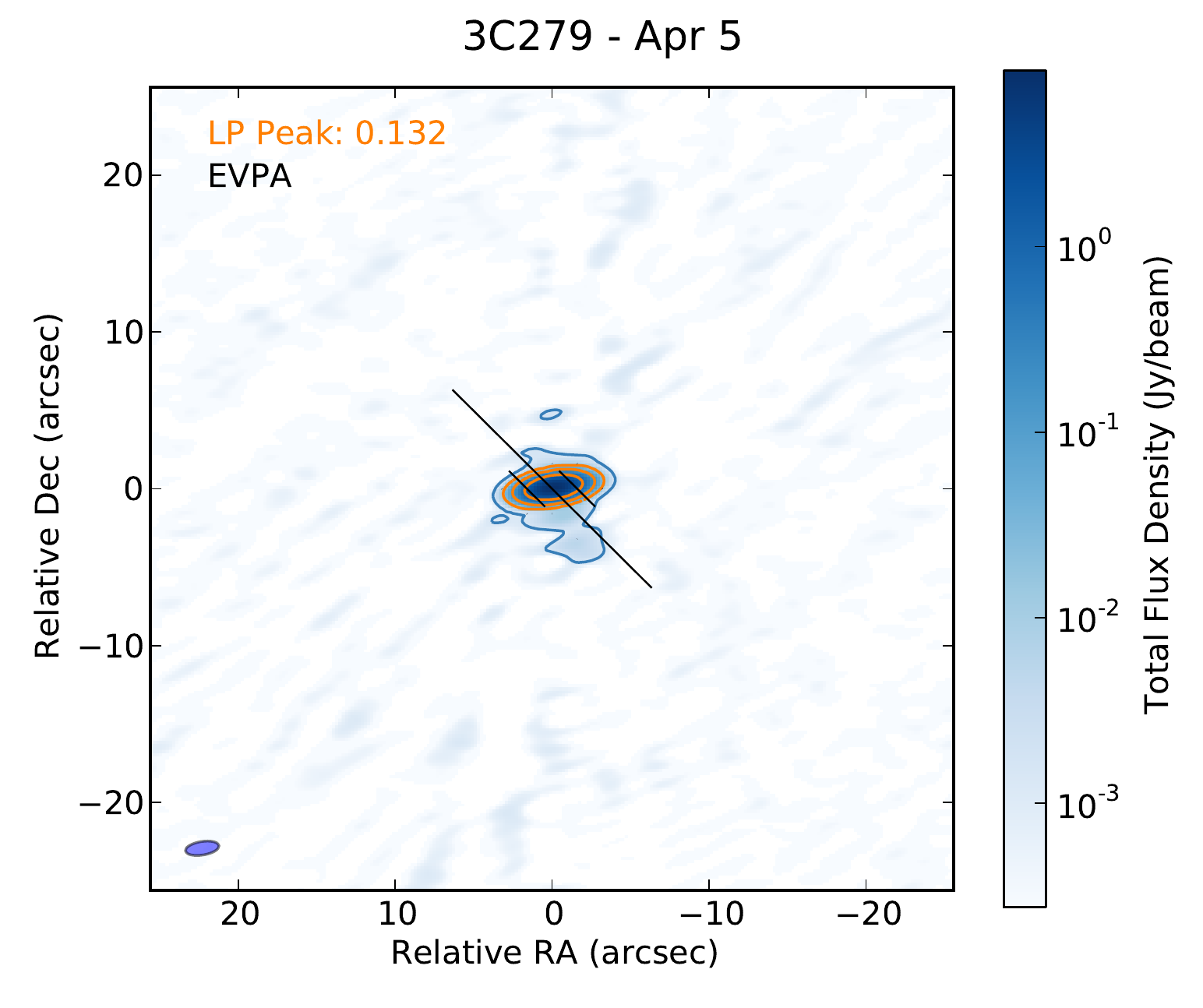} \hspace{-0.3cm}
\includegraphics[width=9cm]{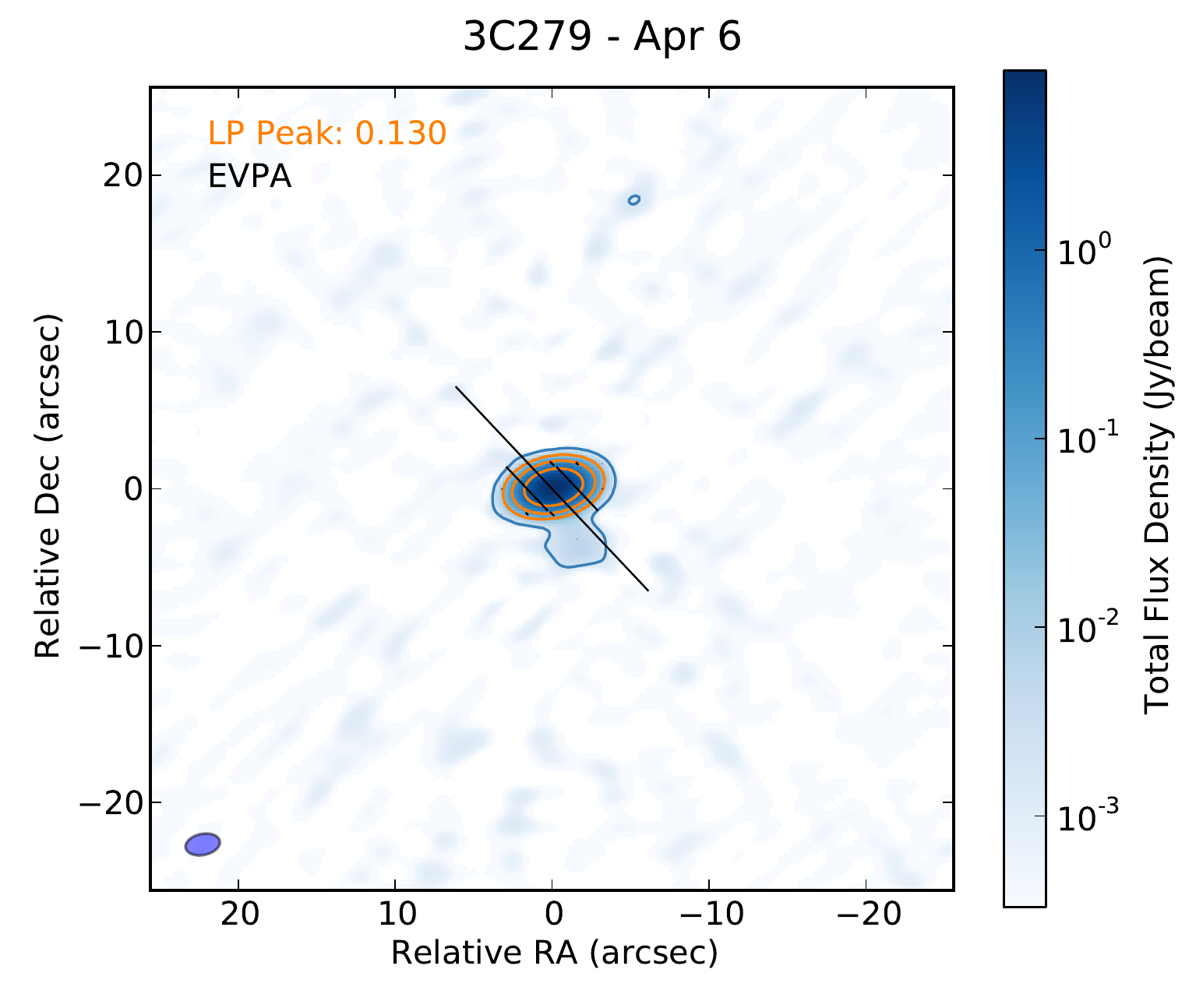} \hspace{-0.3cm}
\includegraphics[width=9cm]{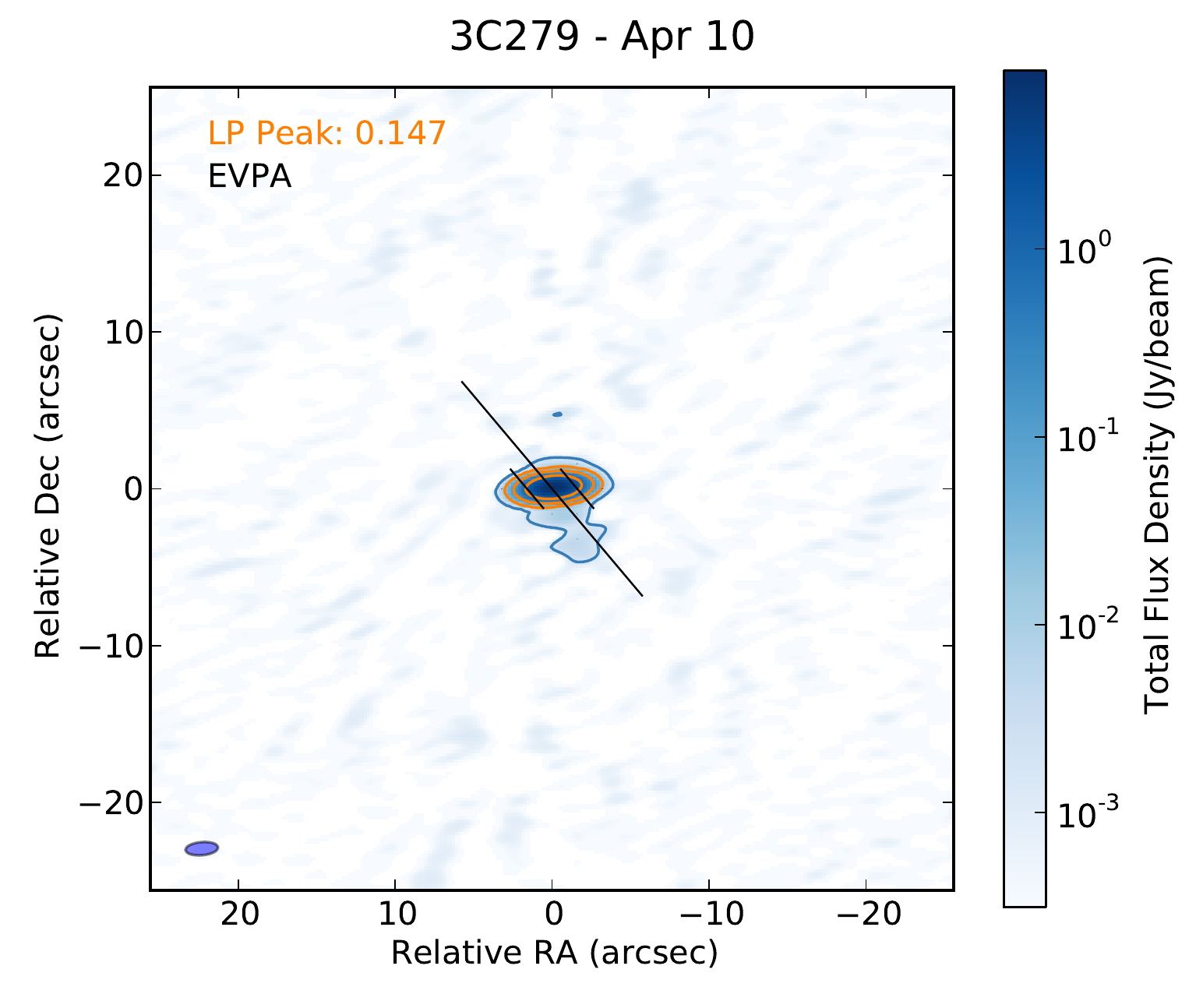}  \hspace{-0.3cm}
\includegraphics[width=9cm]{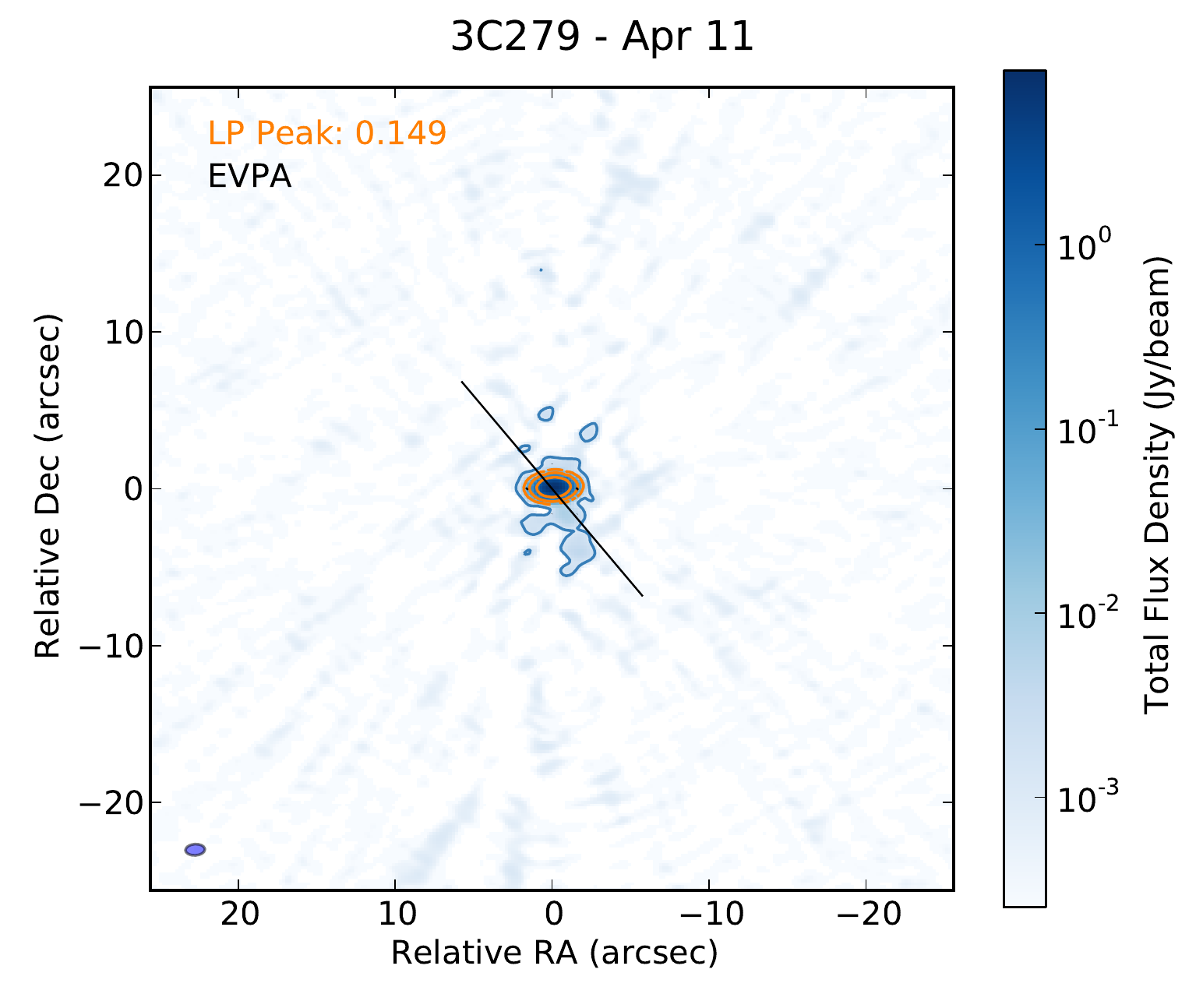}  \hspace{-0.3cm}
\caption{Polarization images of 3C279 at 1.3~mm (see Figure~\ref{fig:m87+sgra_polimage} for a description of the plotted quantities).
}
\label{fig:polimages_3c279}
\end{figure*}

\begin{figure*}[ht!]
\centering
\includegraphics[width=6cm]{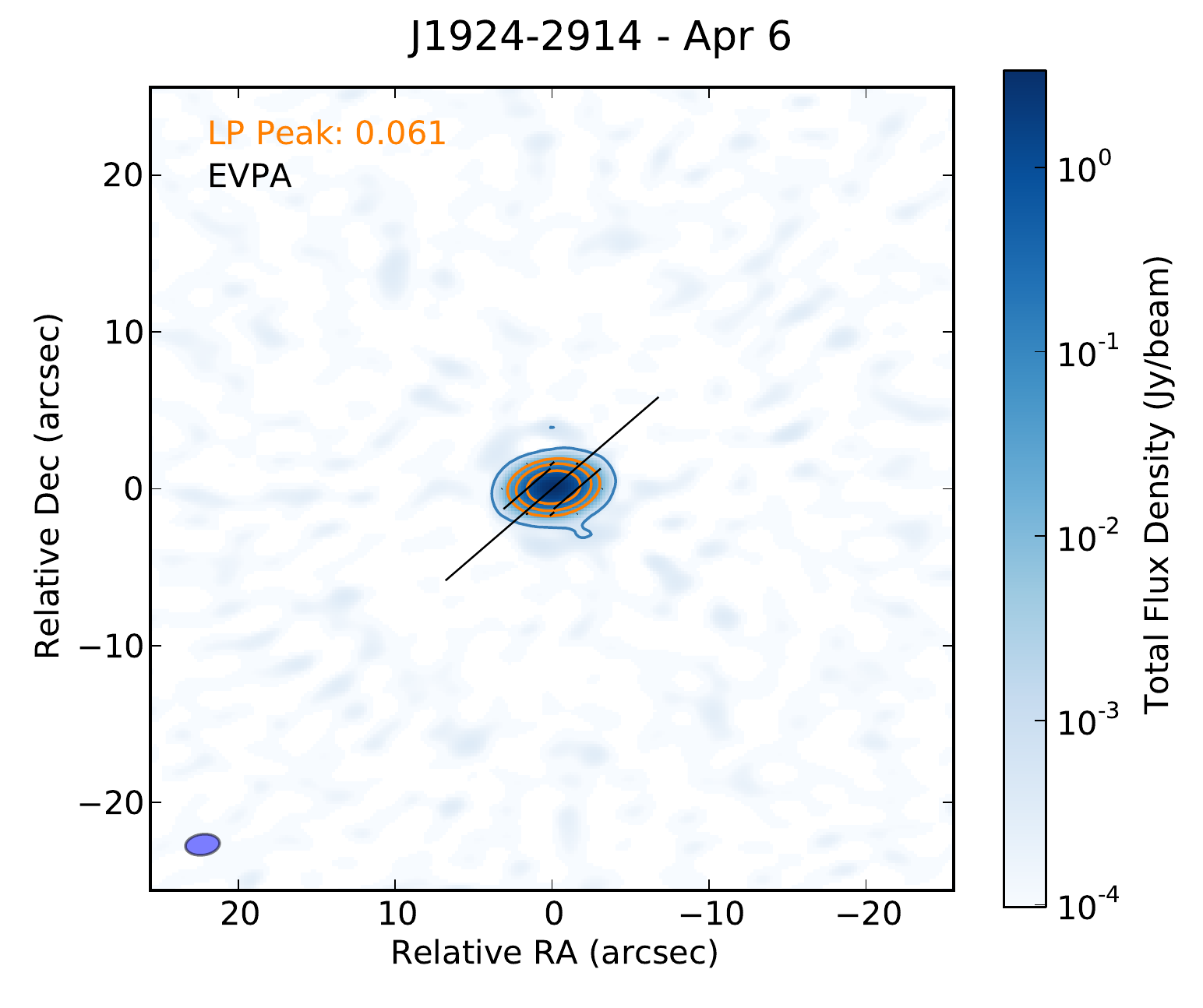} \hspace{-0.3cm}
\includegraphics[width=6cm]{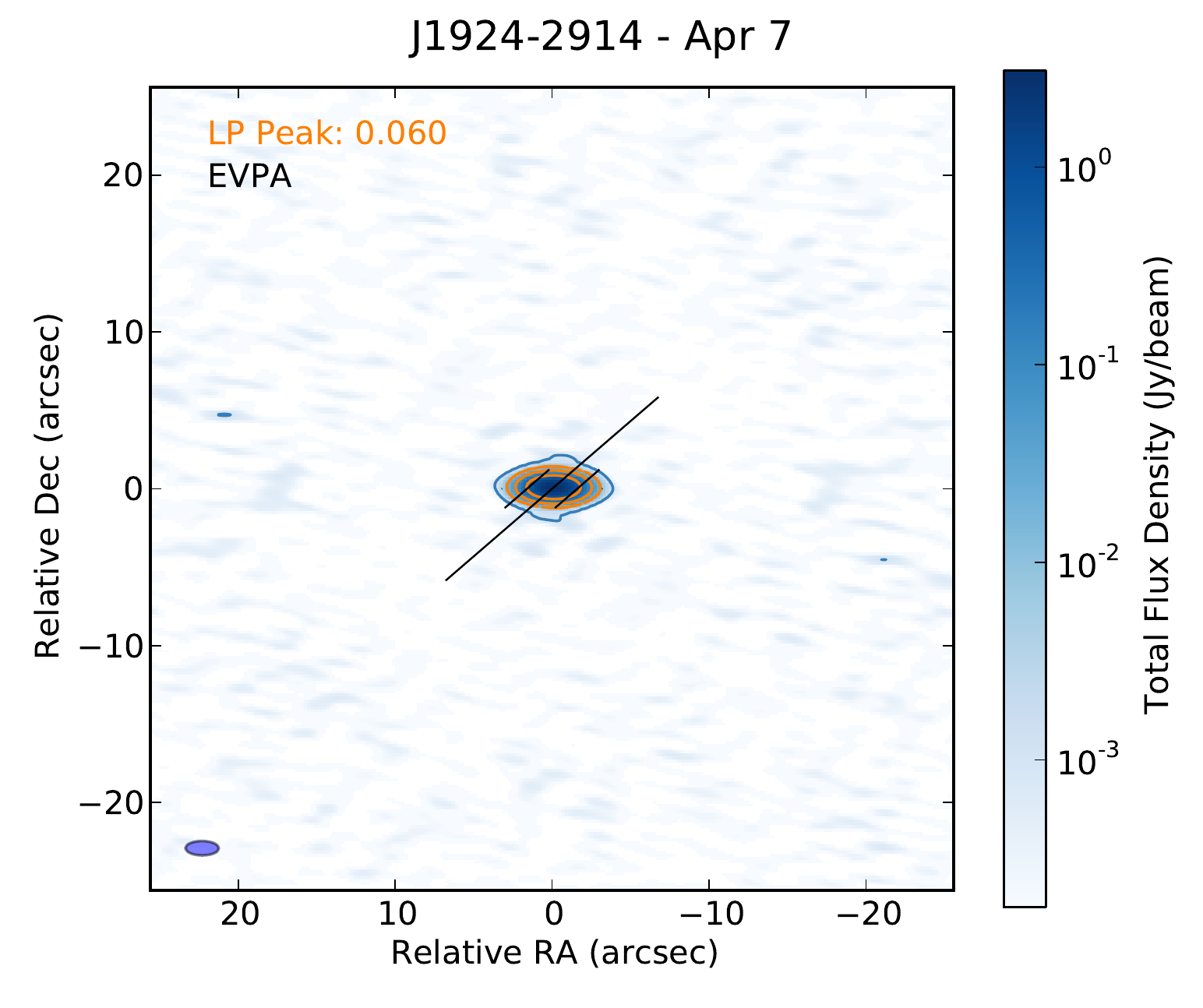}  \hspace{-0.3cm}
\includegraphics[width=6cm]{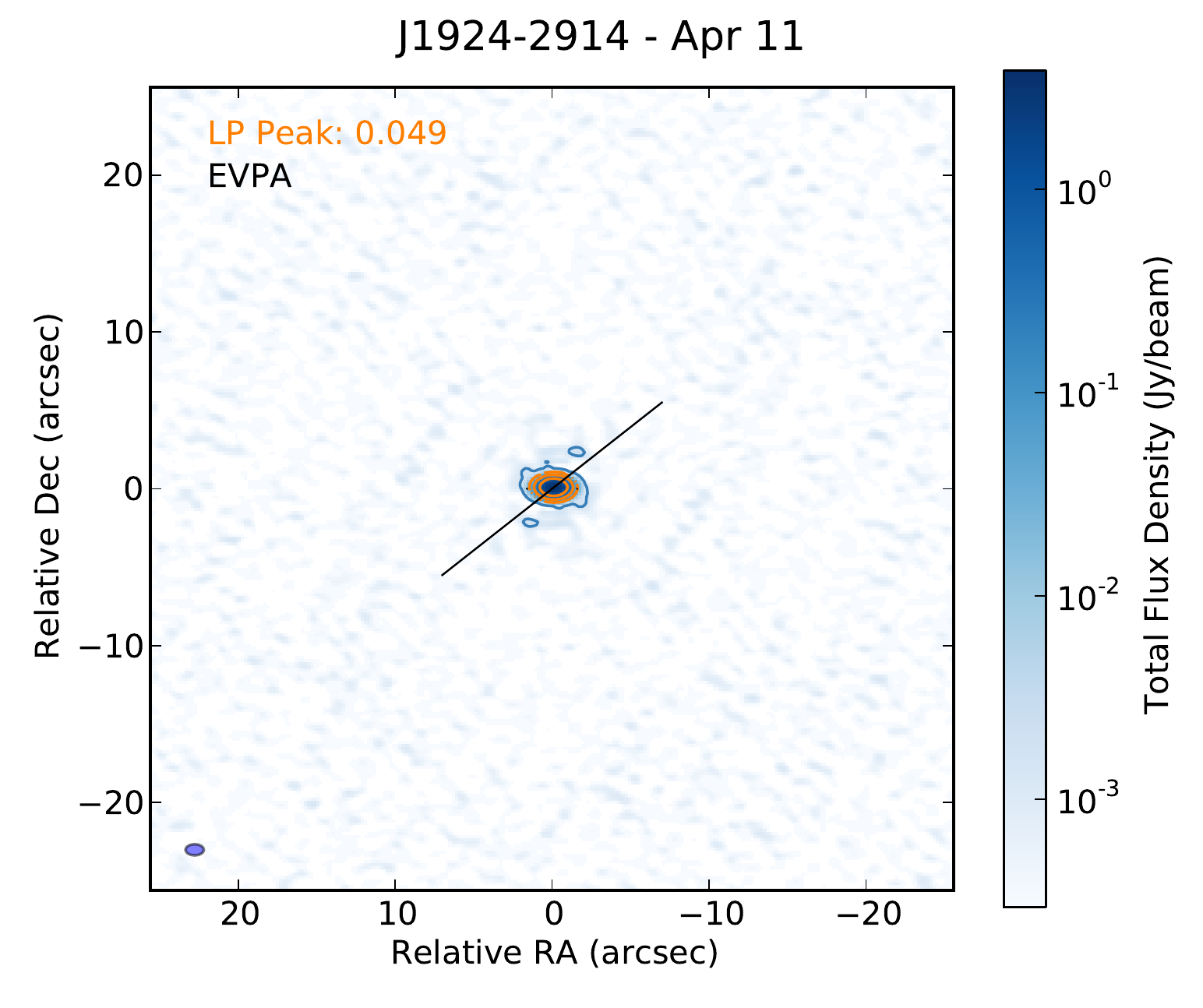}  \hspace{-0.3cm}
\includegraphics[width=6cm]{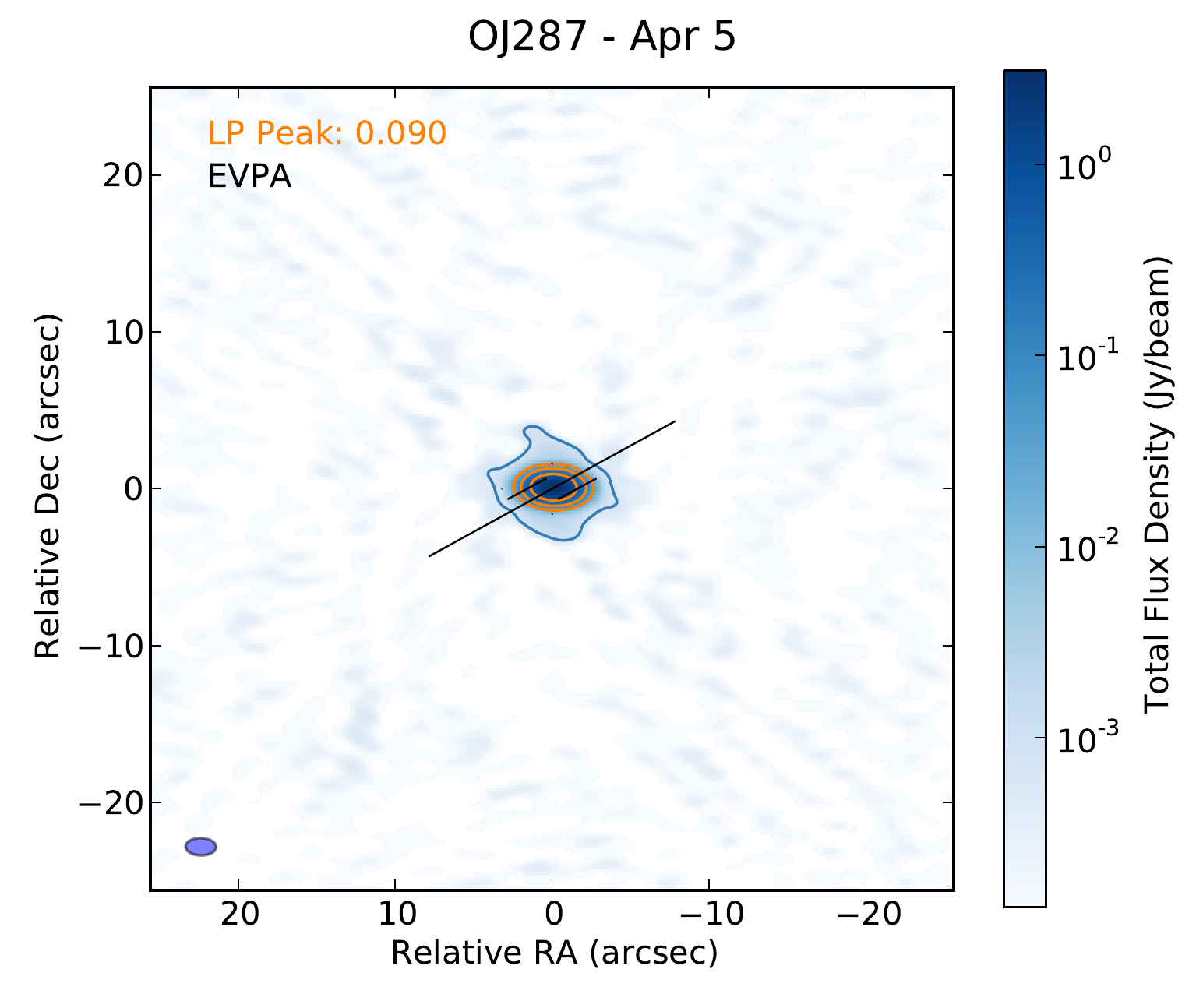} \hspace{-0.3cm}
\includegraphics[width=6cm]{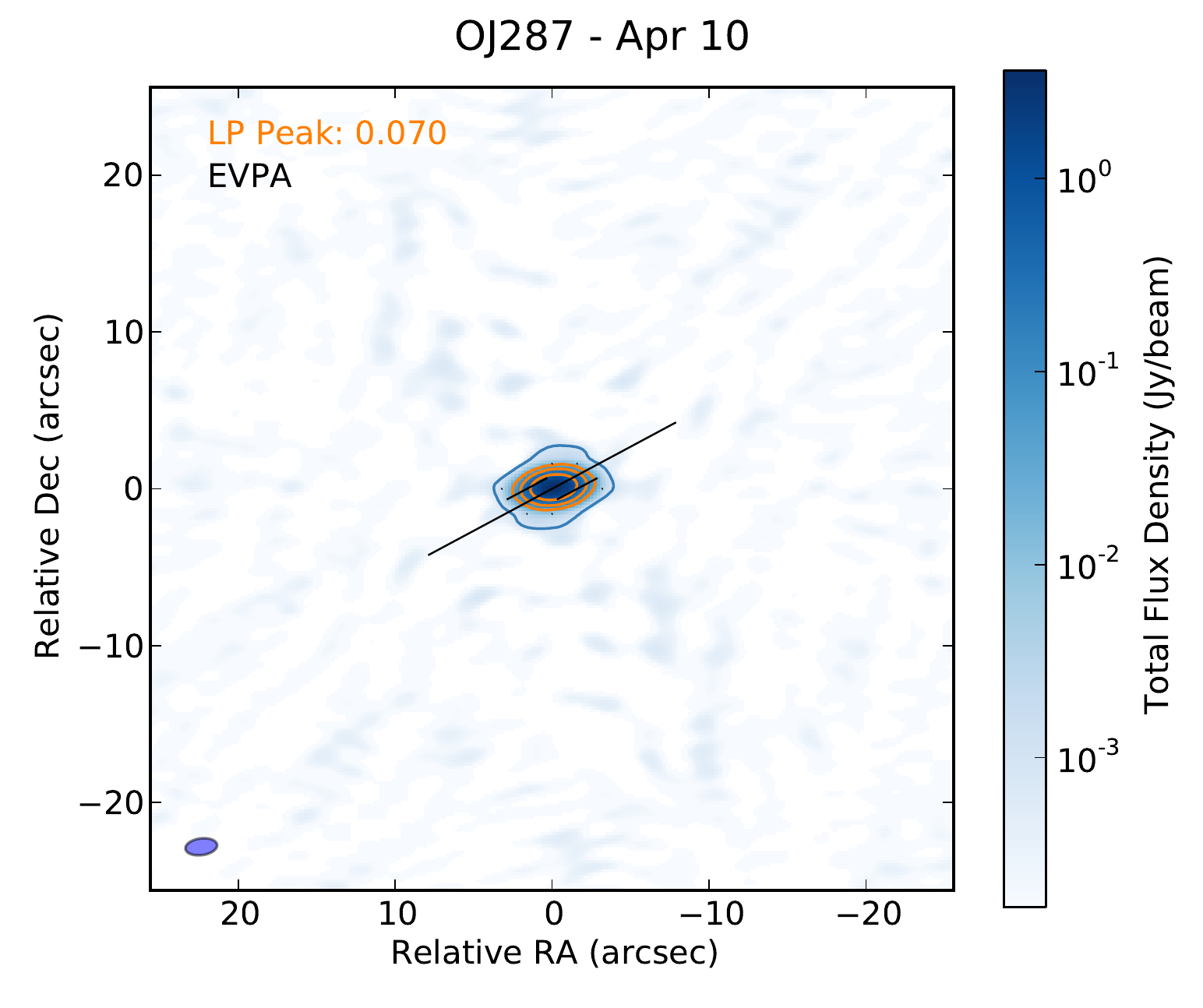}  \hspace{-0.3cm}
\includegraphics[width=6cm]{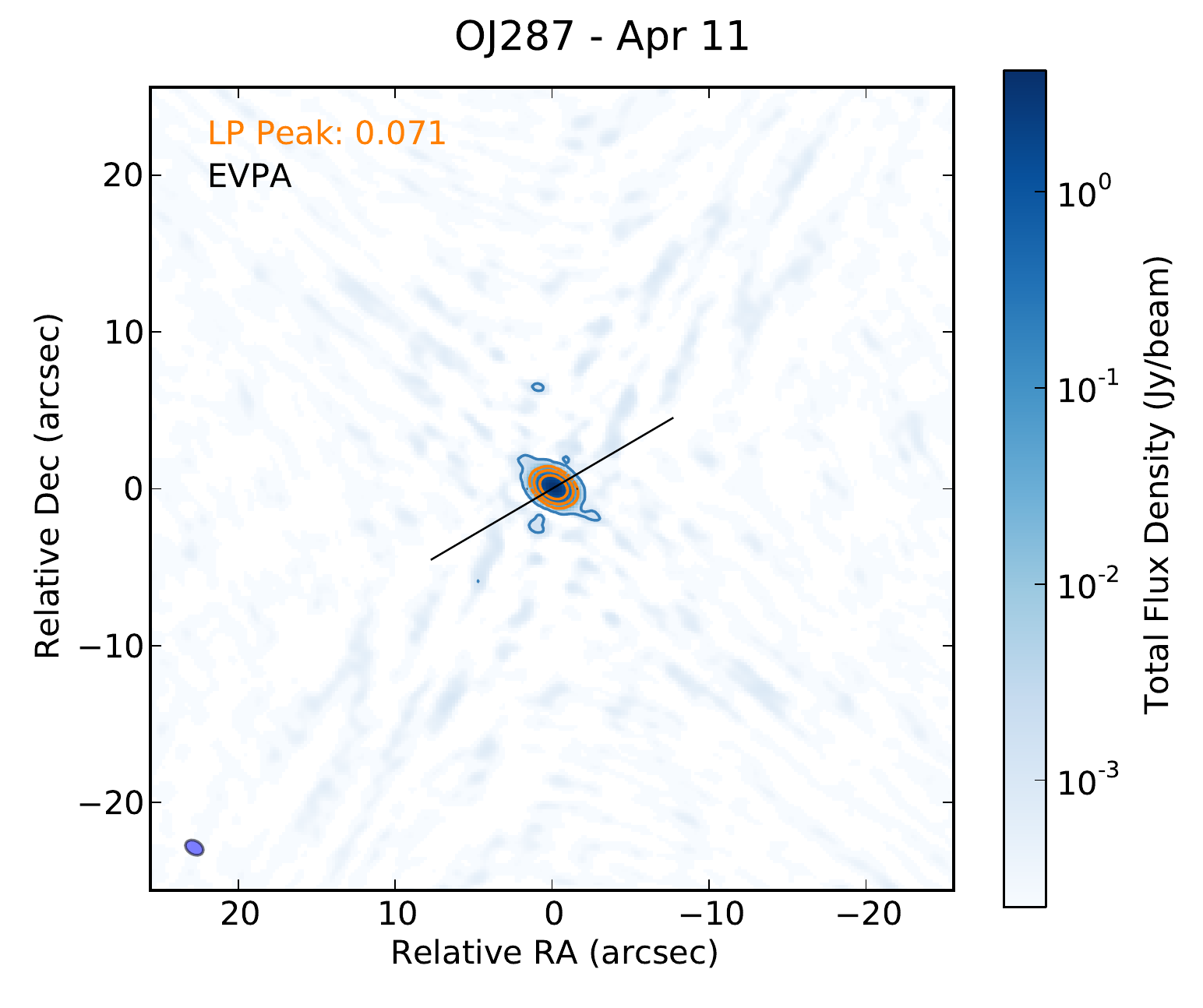}  \hspace{-0.3cm}
\includegraphics[width=6cm]{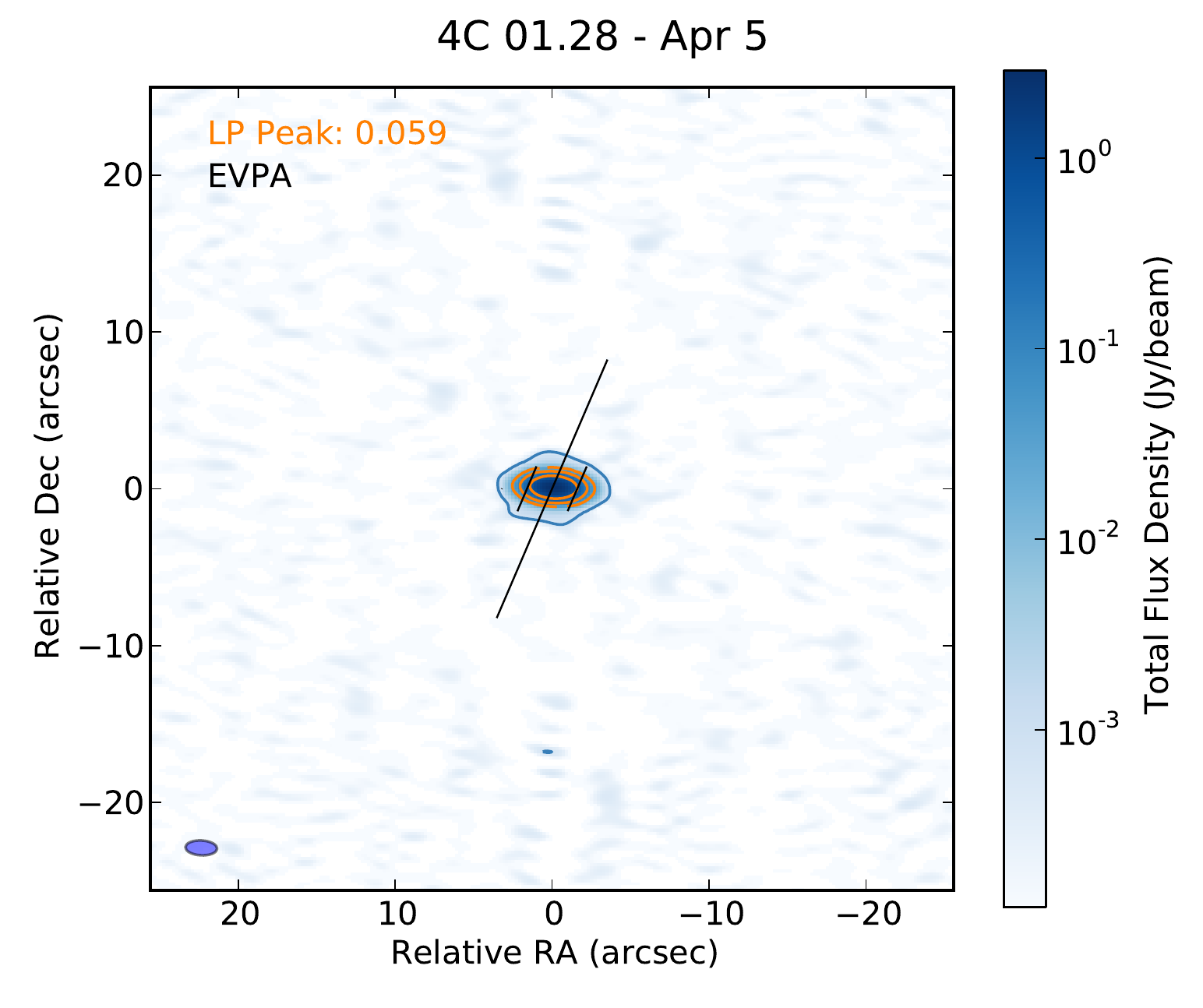} \hspace{-0.3cm}
\includegraphics[width=6cm]{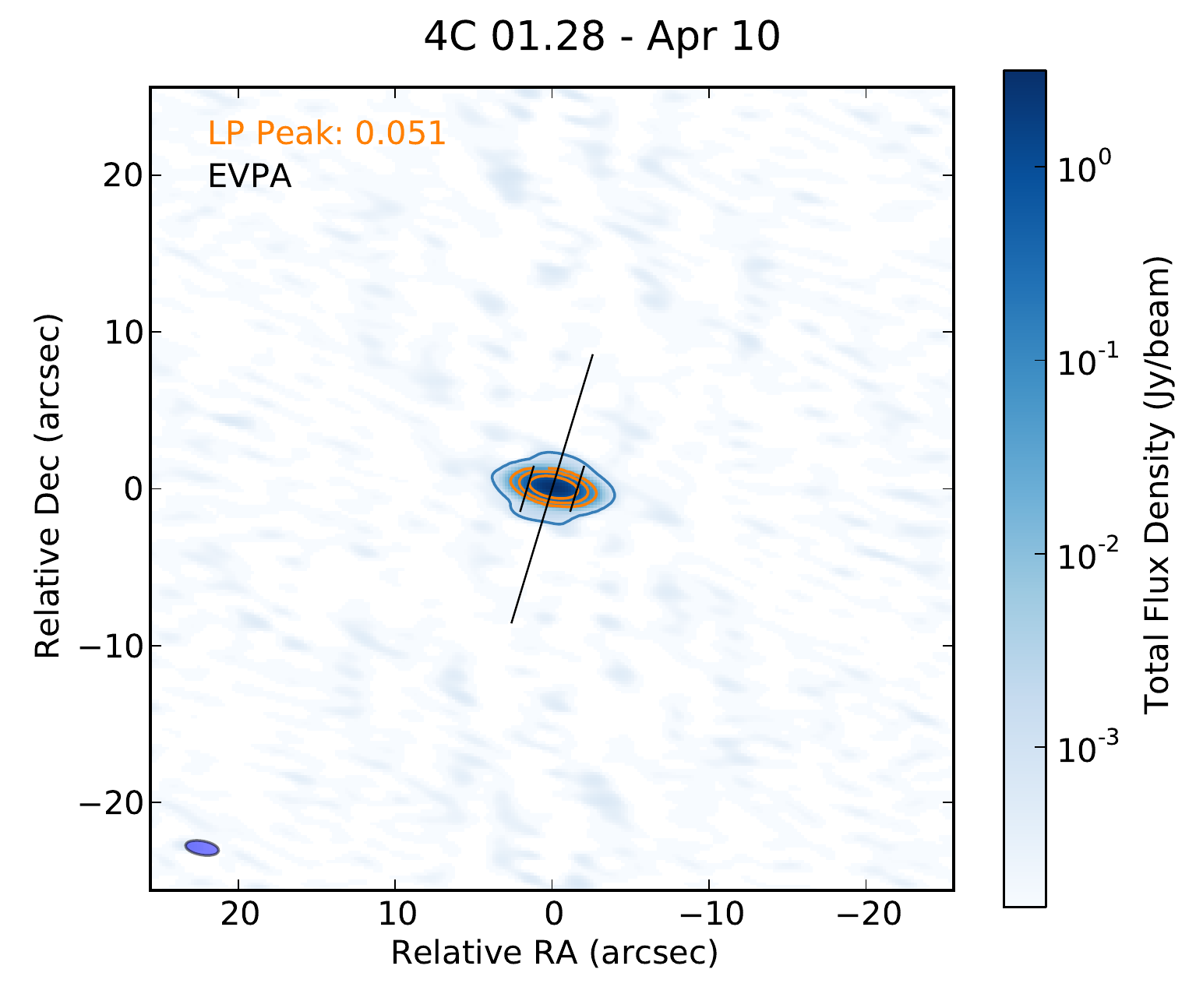}  \hspace{-0.3cm}
\includegraphics[width=6cm]{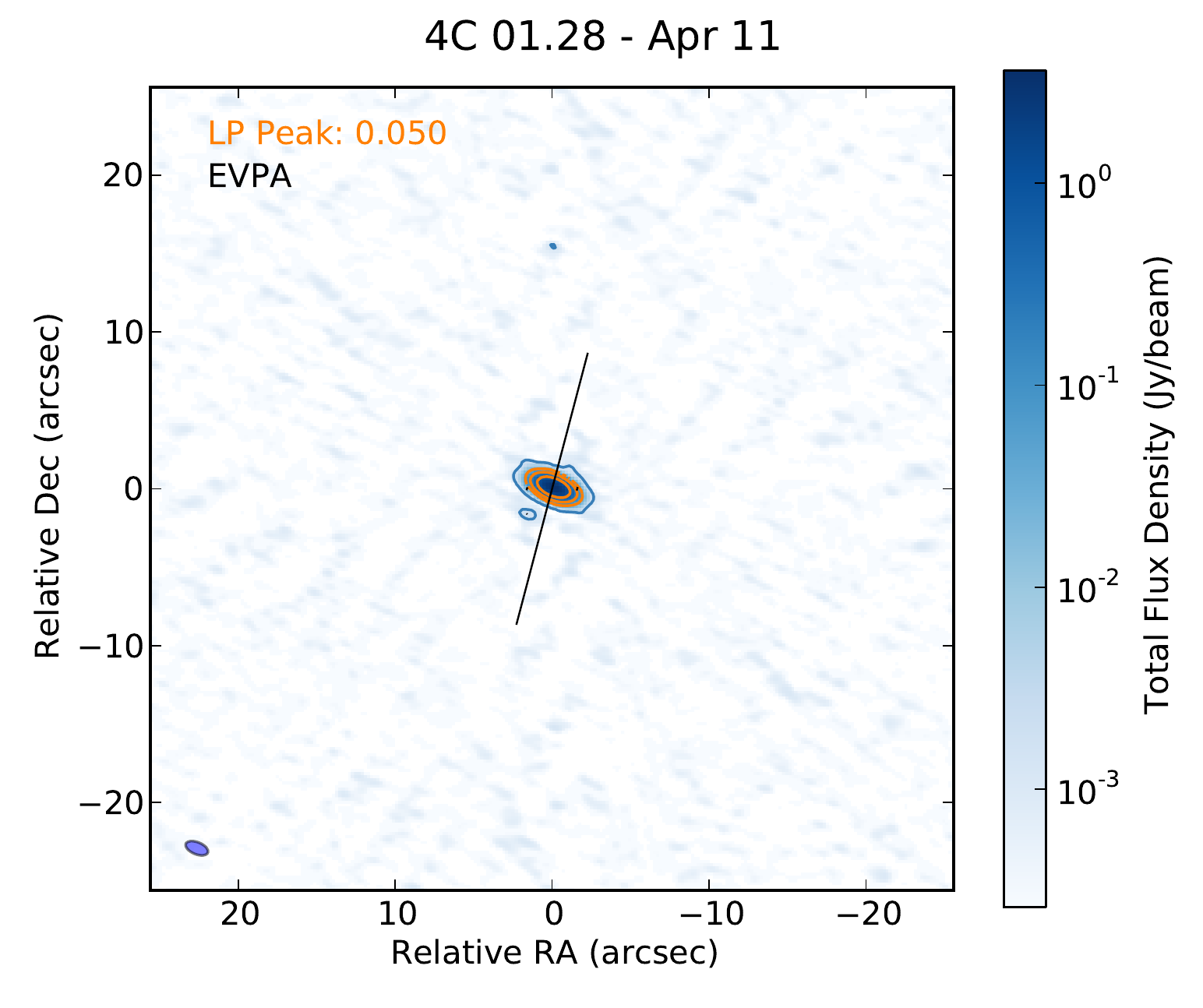}  \hspace{-0.3cm}
\caption{Polarization images of selected AGNs at 1.3~mm (see Figure~\ref{fig:m87+sgra_polimage} for a description of the plotted quantities). 
}
\label{fig:polimages_agns_3days}
\end{figure*}

\begin{figure*}[ht!]
\centering
\includegraphics[width=6cm]{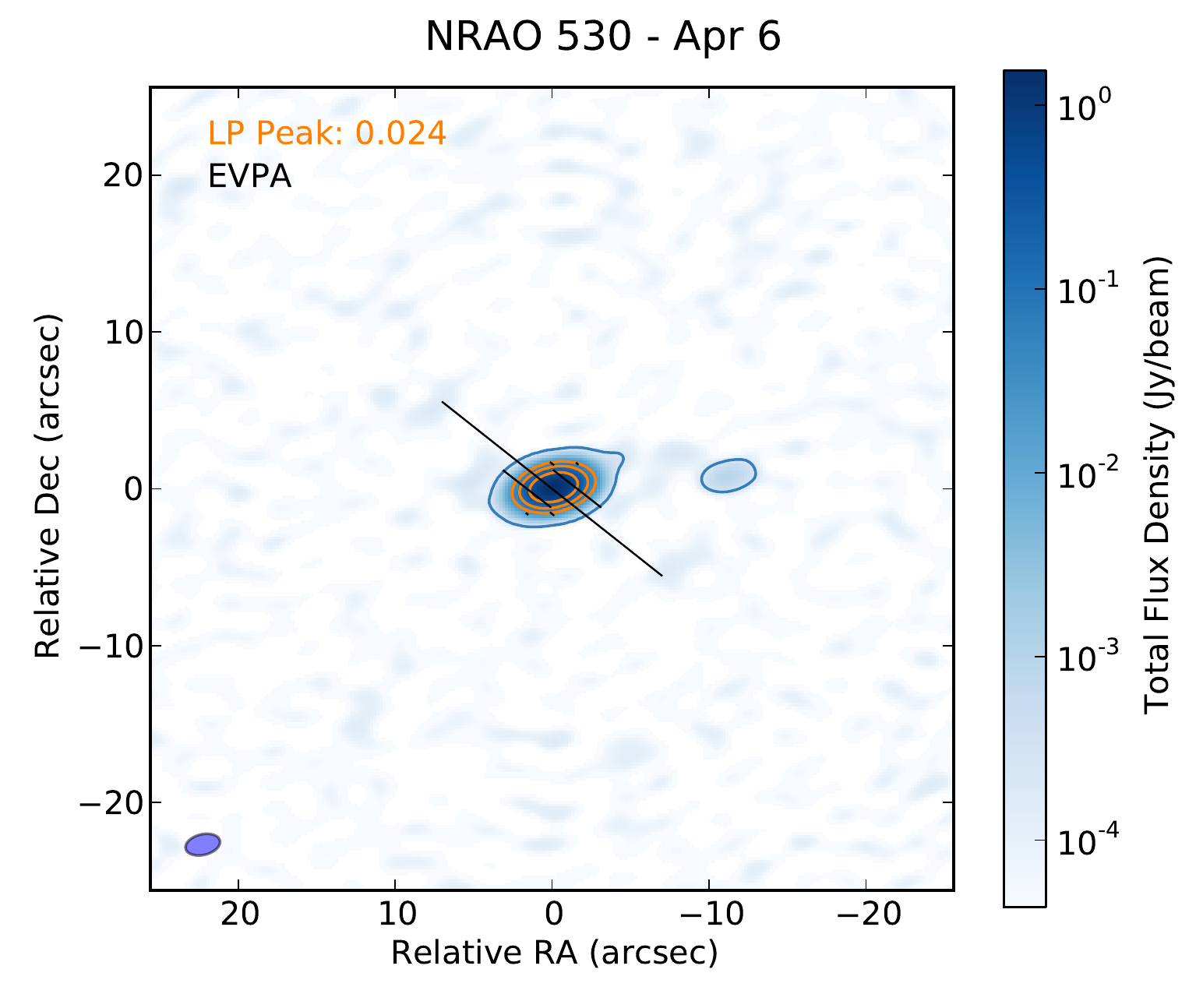} \hspace{-0.3cm}
\includegraphics[width=6cm]{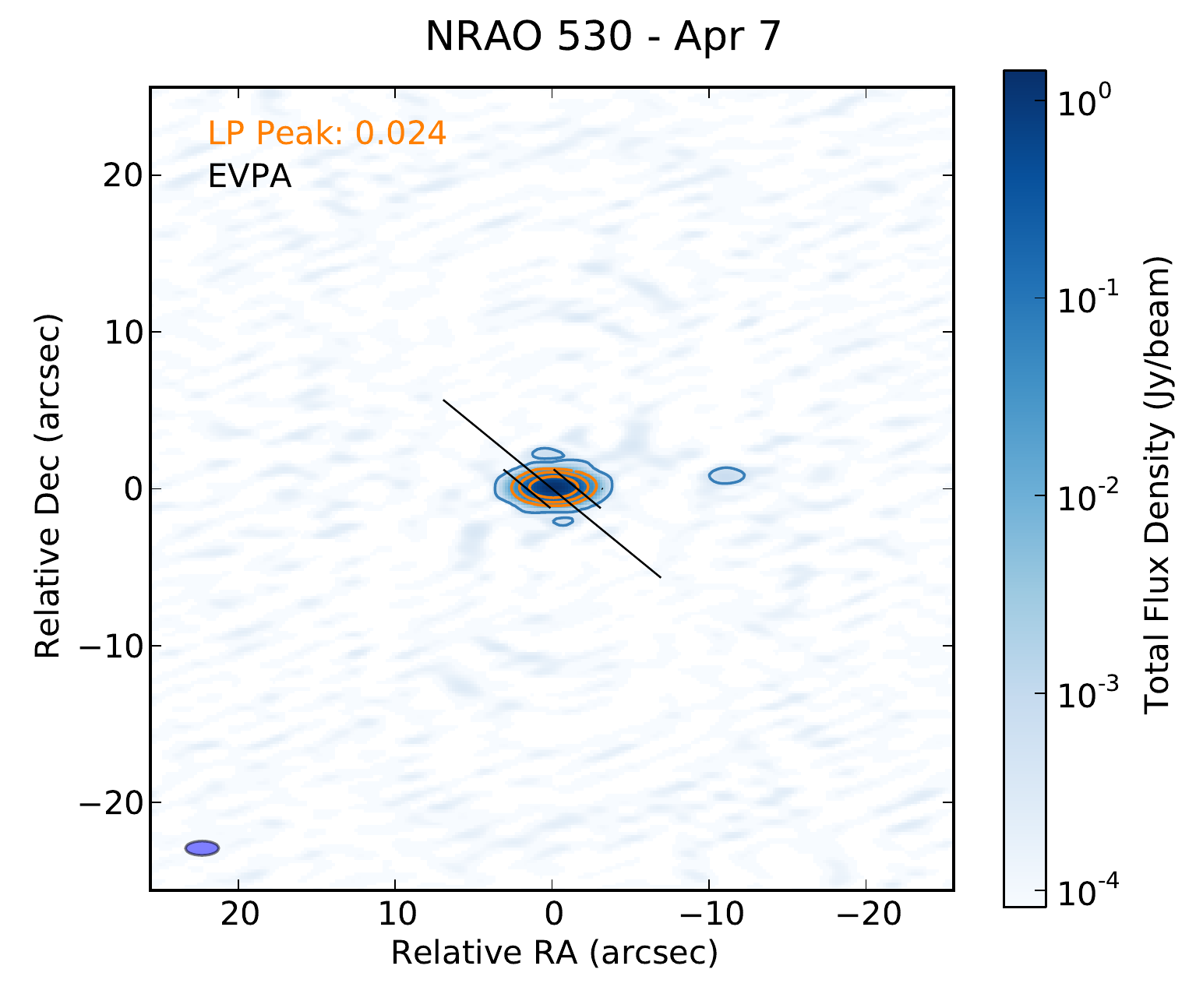}  \hspace{-0.3cm}
\includegraphics[width=6cm]{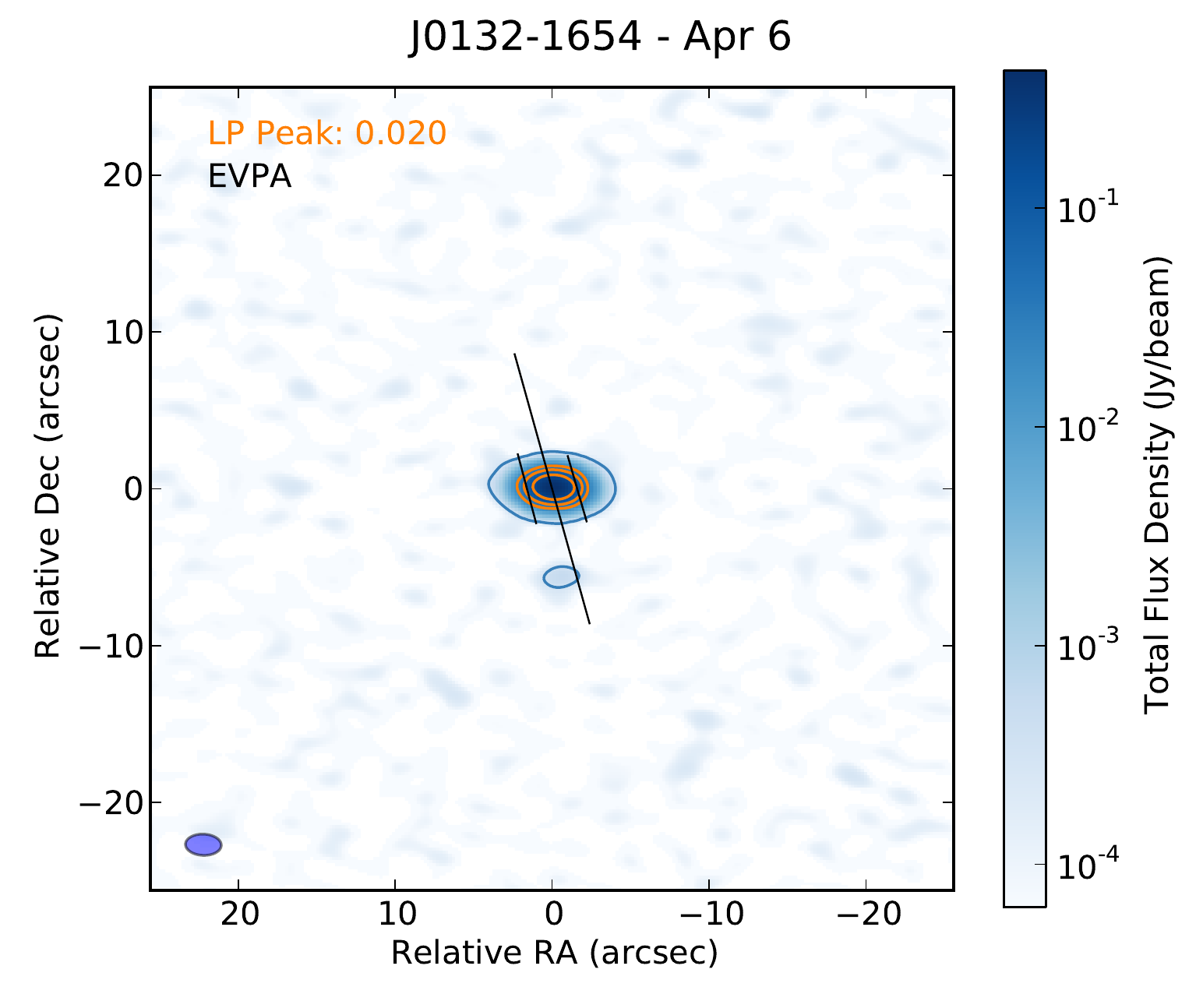} \hspace{-0.3cm}
\includegraphics[width=6cm]{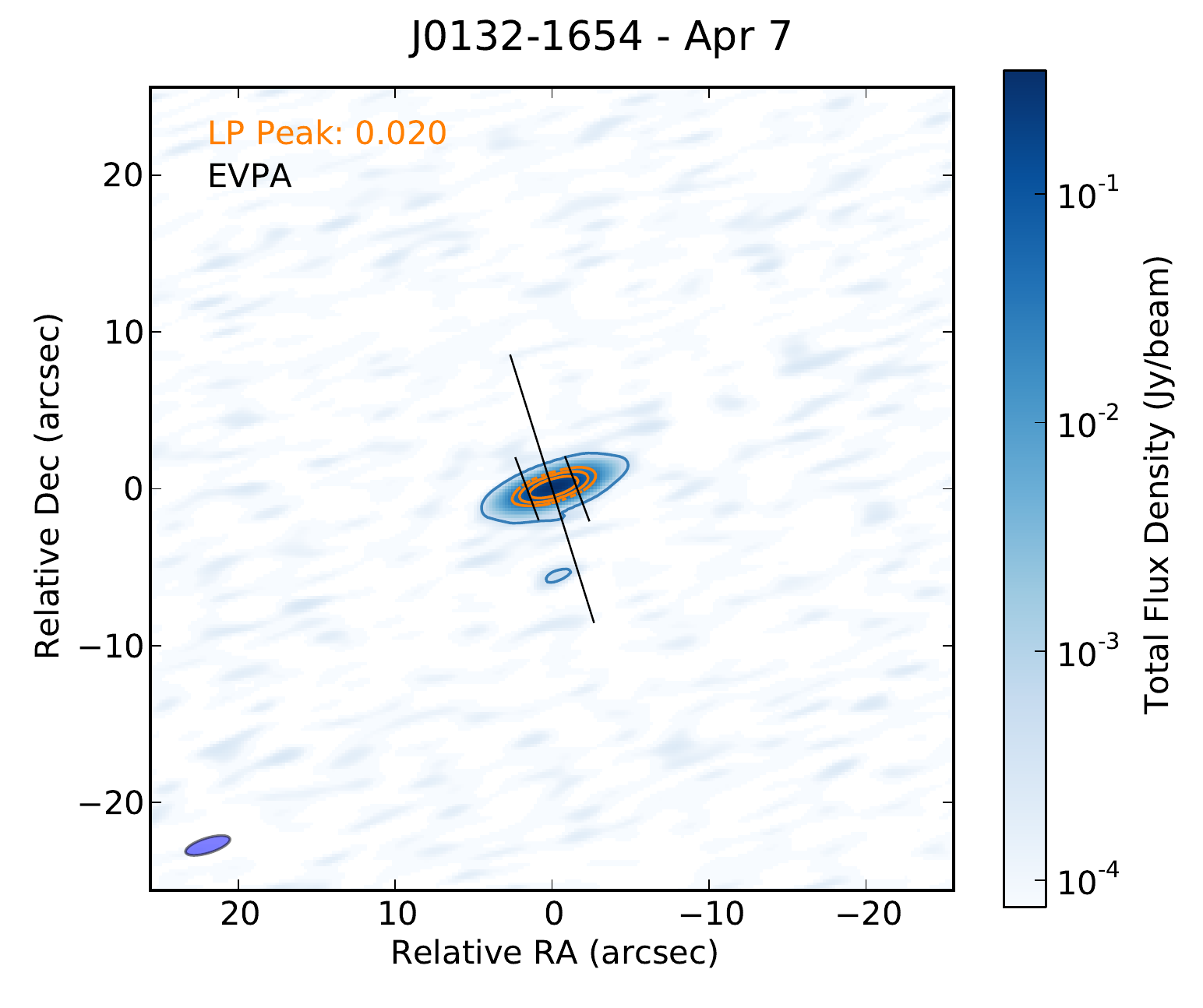}  \hspace{-0.3cm}
\includegraphics[width=6cm]{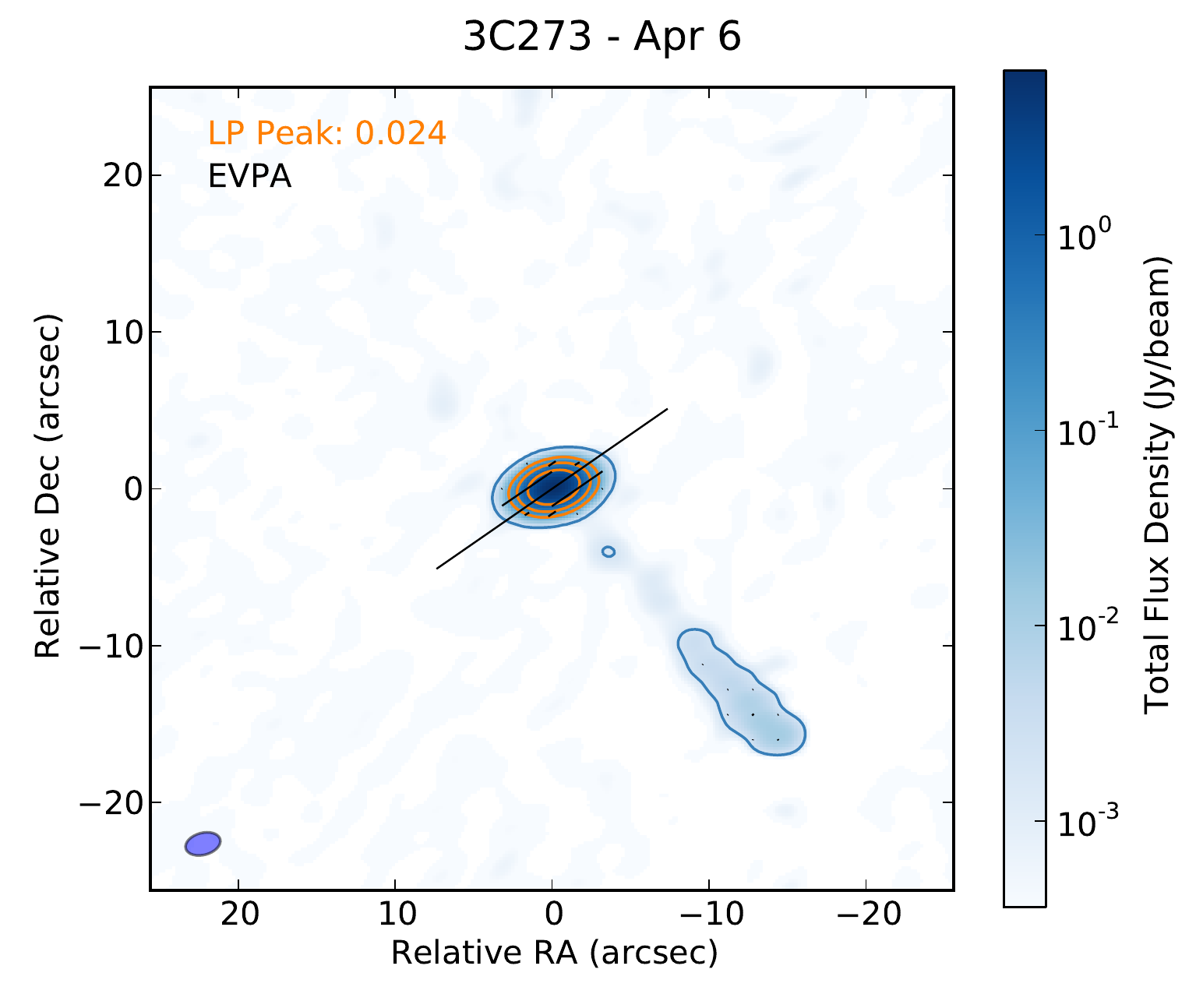} \hspace{-0.3cm}
\includegraphics[width=6cm]{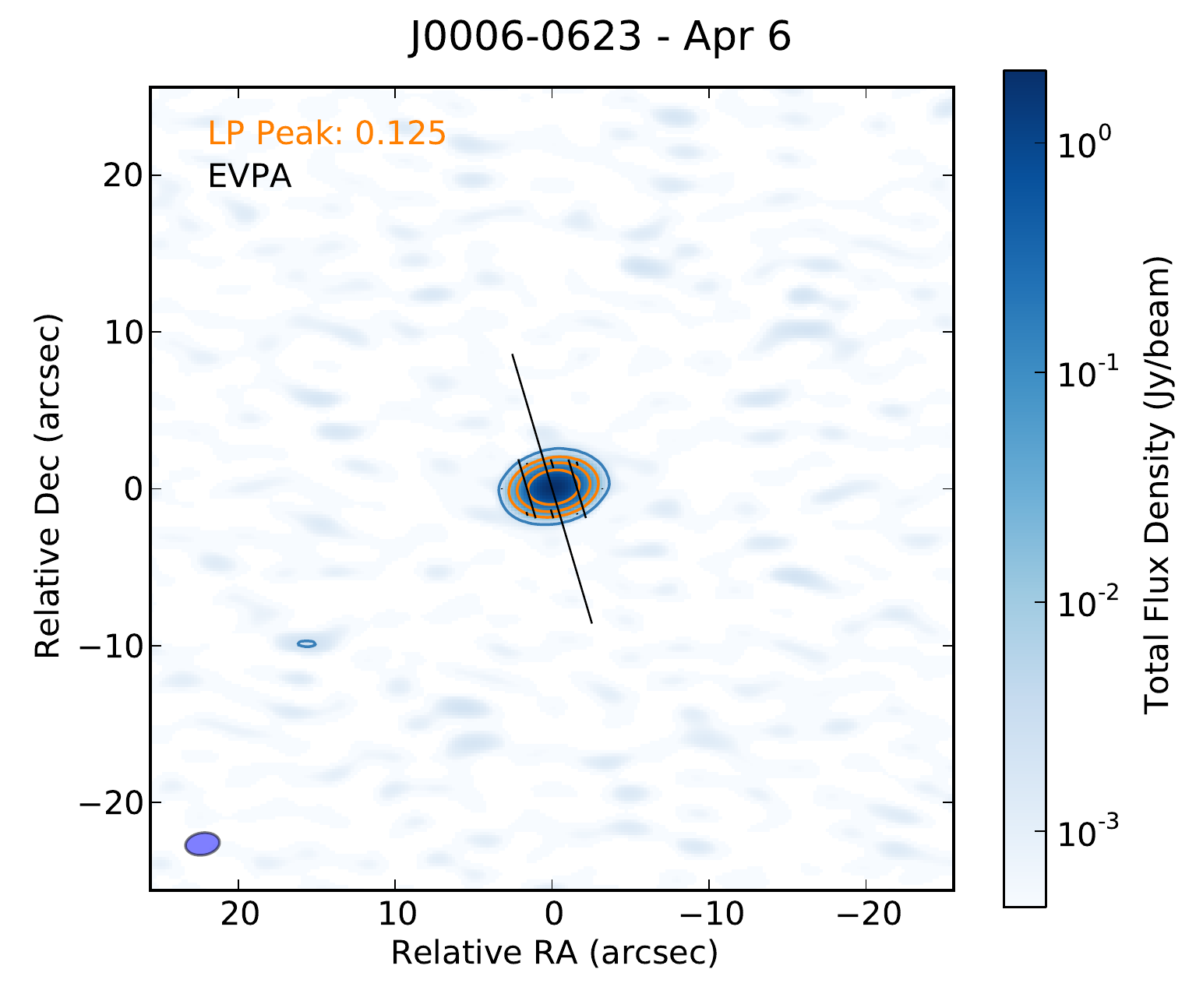}  \hspace{-0.3cm}
\includegraphics[width=6cm]{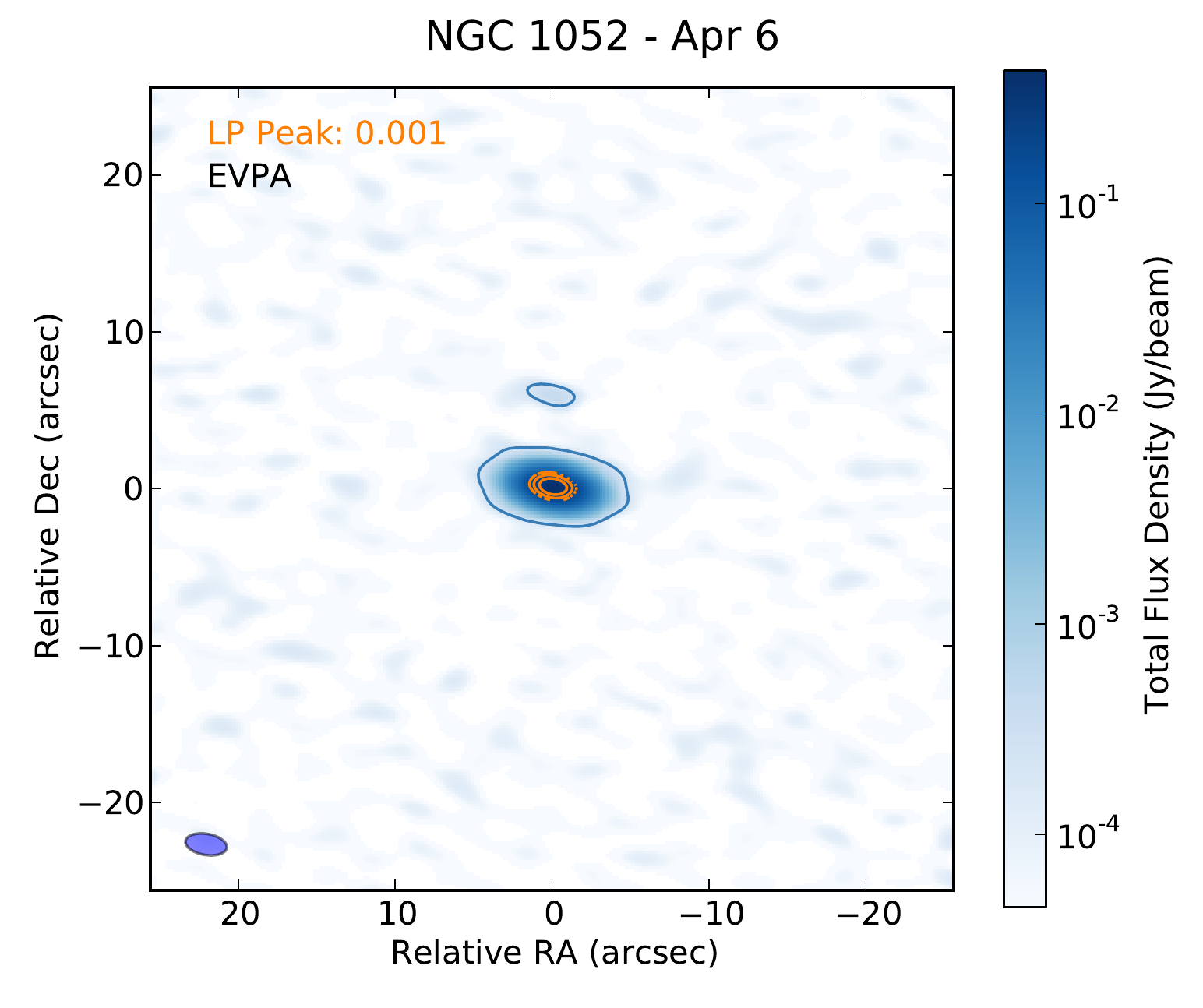}  \hspace{-0.3cm}
\includegraphics[width=6cm]{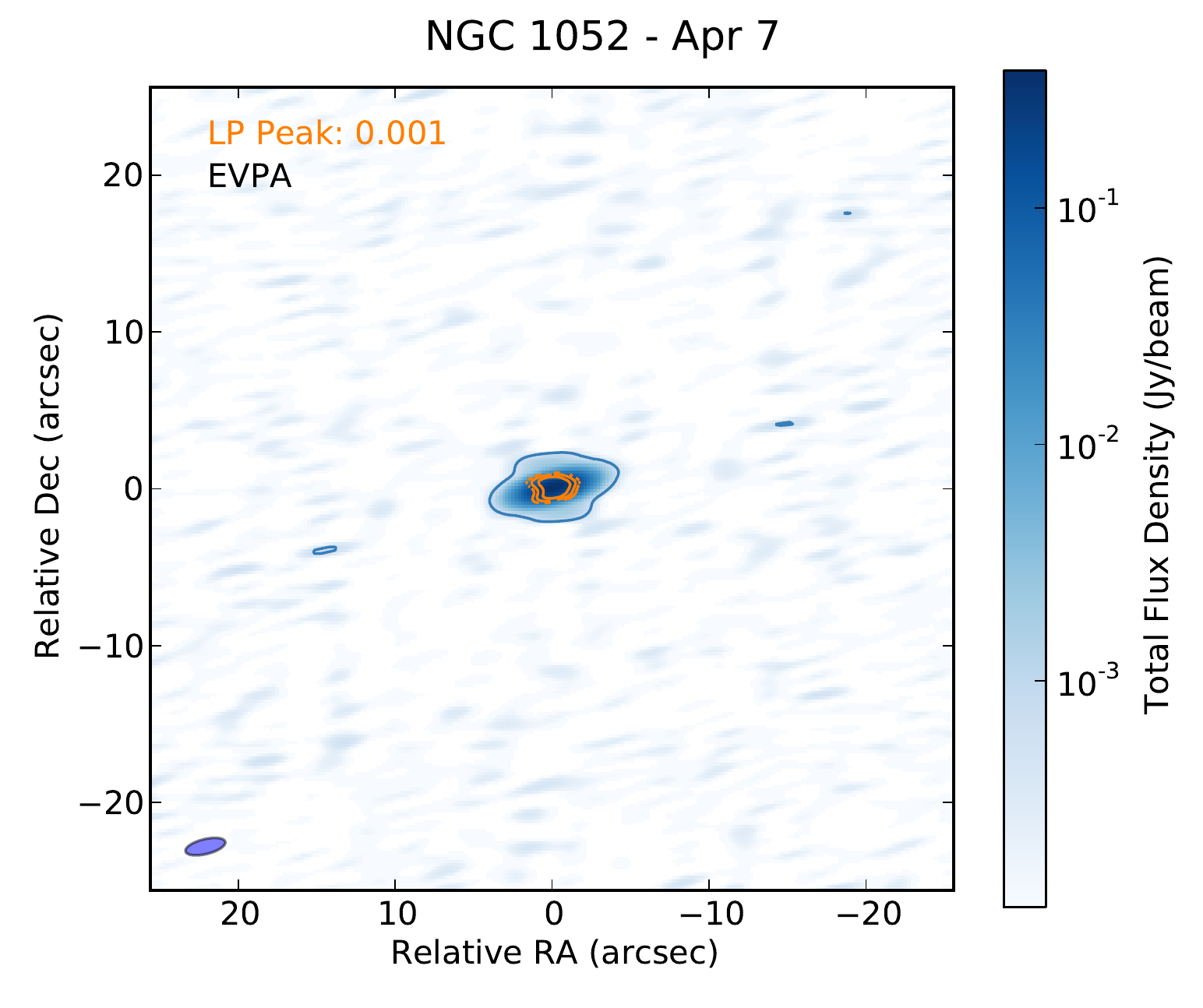}  \hspace{-0.3cm}
\includegraphics[width=6cm]{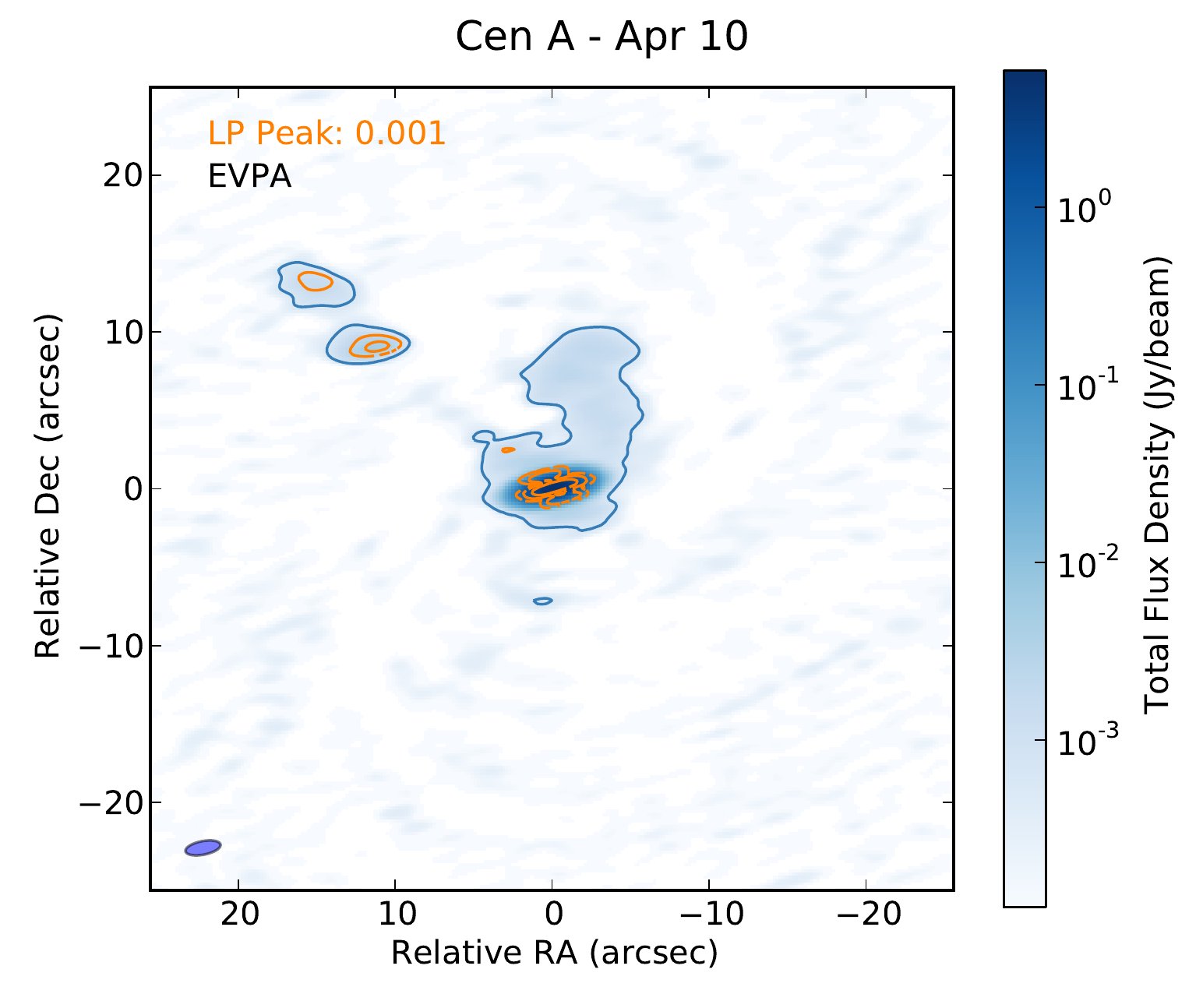}  \hspace{-0.3cm}
\caption{Polarization images of selected AGNs at 1.3~mm (see Figure~\ref{fig:m87+sgra_polimage} for a description of the plotted quantities). 
Note that for NGC~1052 and Cen~A no EVPA could be reliably derived owing to their low level of LP. 
We also note that although we detect  polarized flux  in Cen~A above the  image RMS noise cutoff ($5\sigma$), once  the systematic leakage (0.03\% of I onto QU) is added to thermal noise, the LP flux would fall below the $3\sigma$ detection threshold. Therefore we do not claim detection of polarized emission in Cen~A.}
\label{fig:polimages_agns_1_2days}
\end{figure*}

\begin{figure*}[ht!]
\centering                                     \includegraphics[width=6cm]{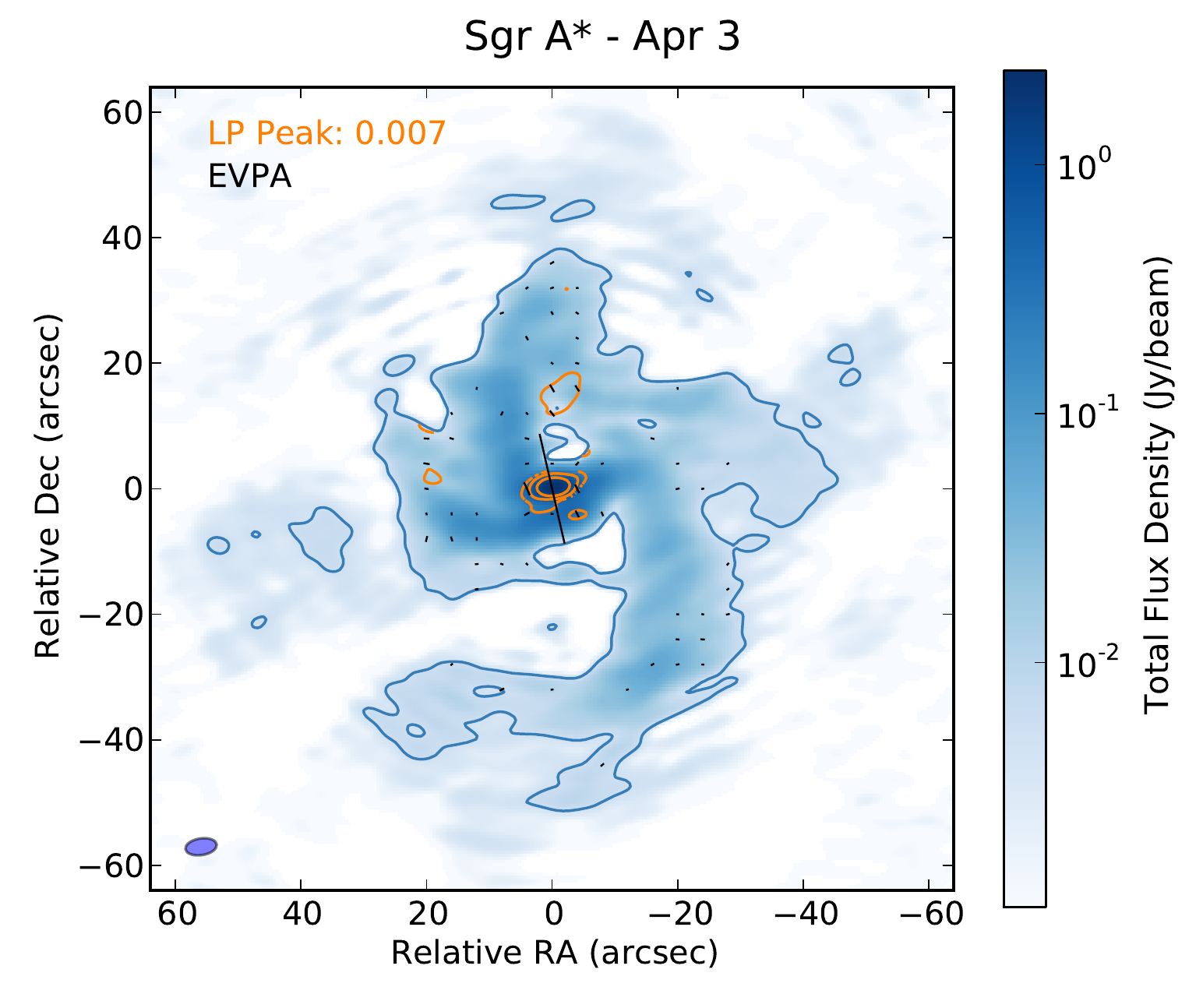} \hspace{-0.3cm}
\includegraphics[width=6cm]{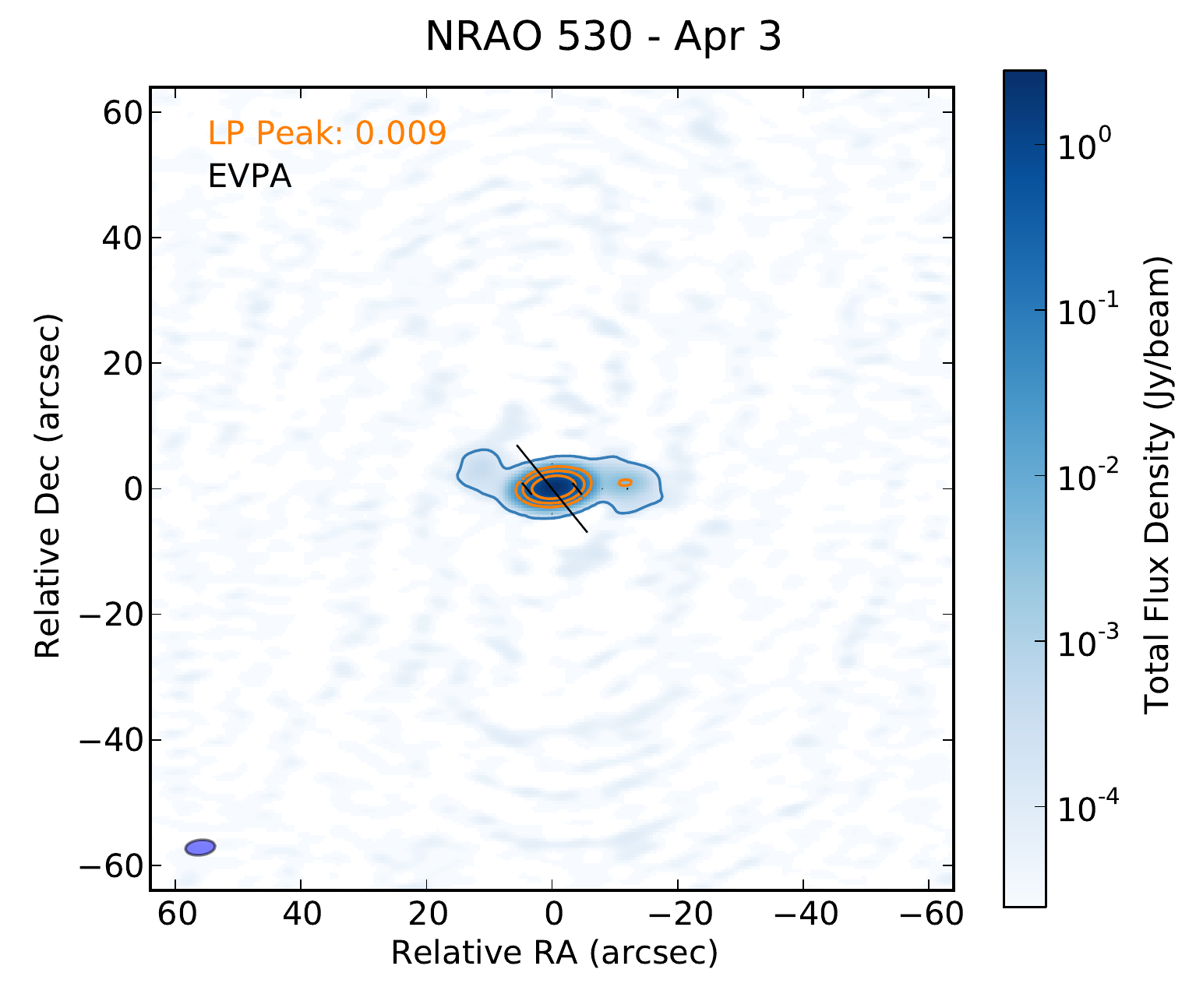} \hspace{-0.3cm}
\includegraphics[width=6cm]{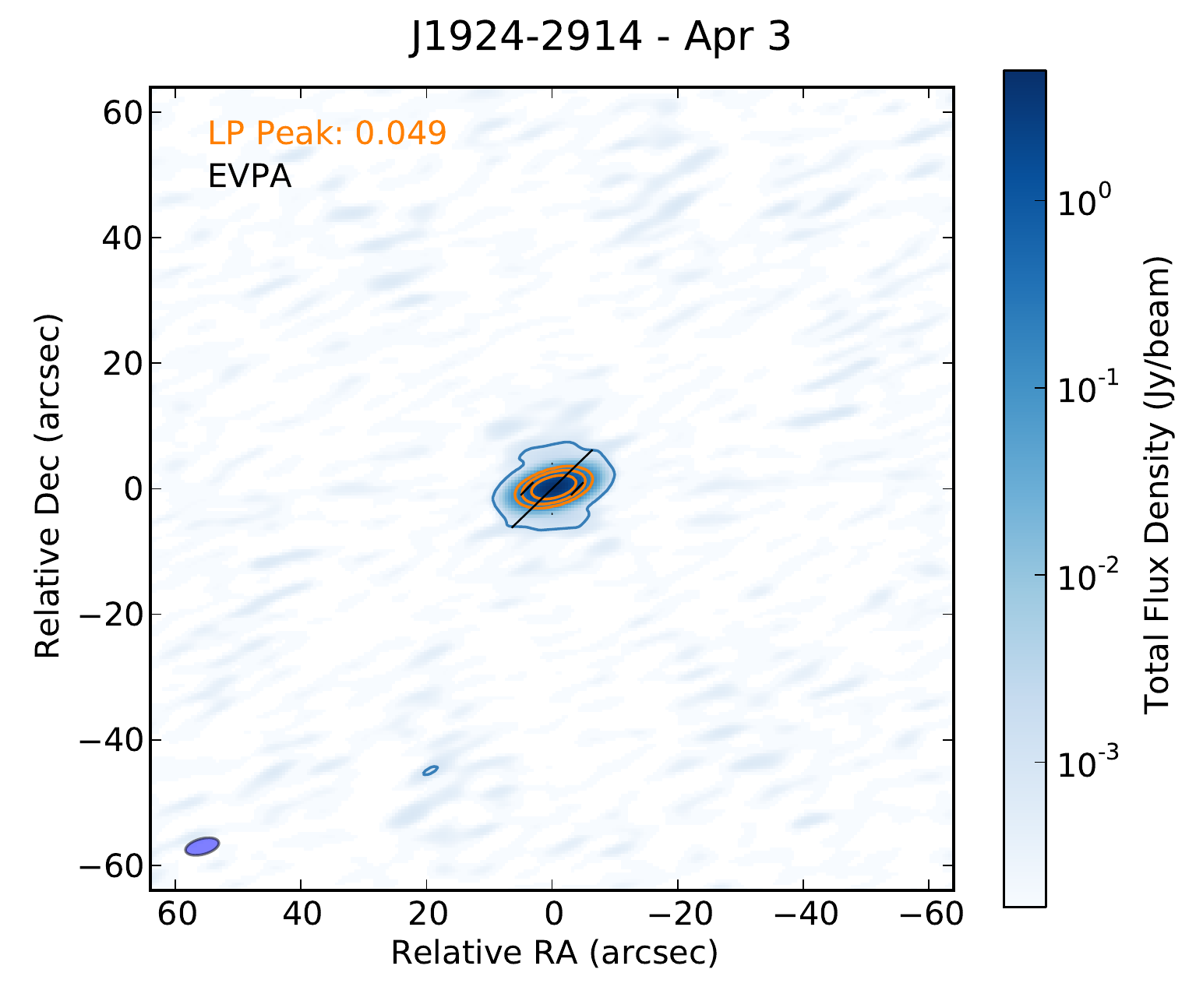} \hspace{-0.3cm}
\includegraphics[width=6cm]{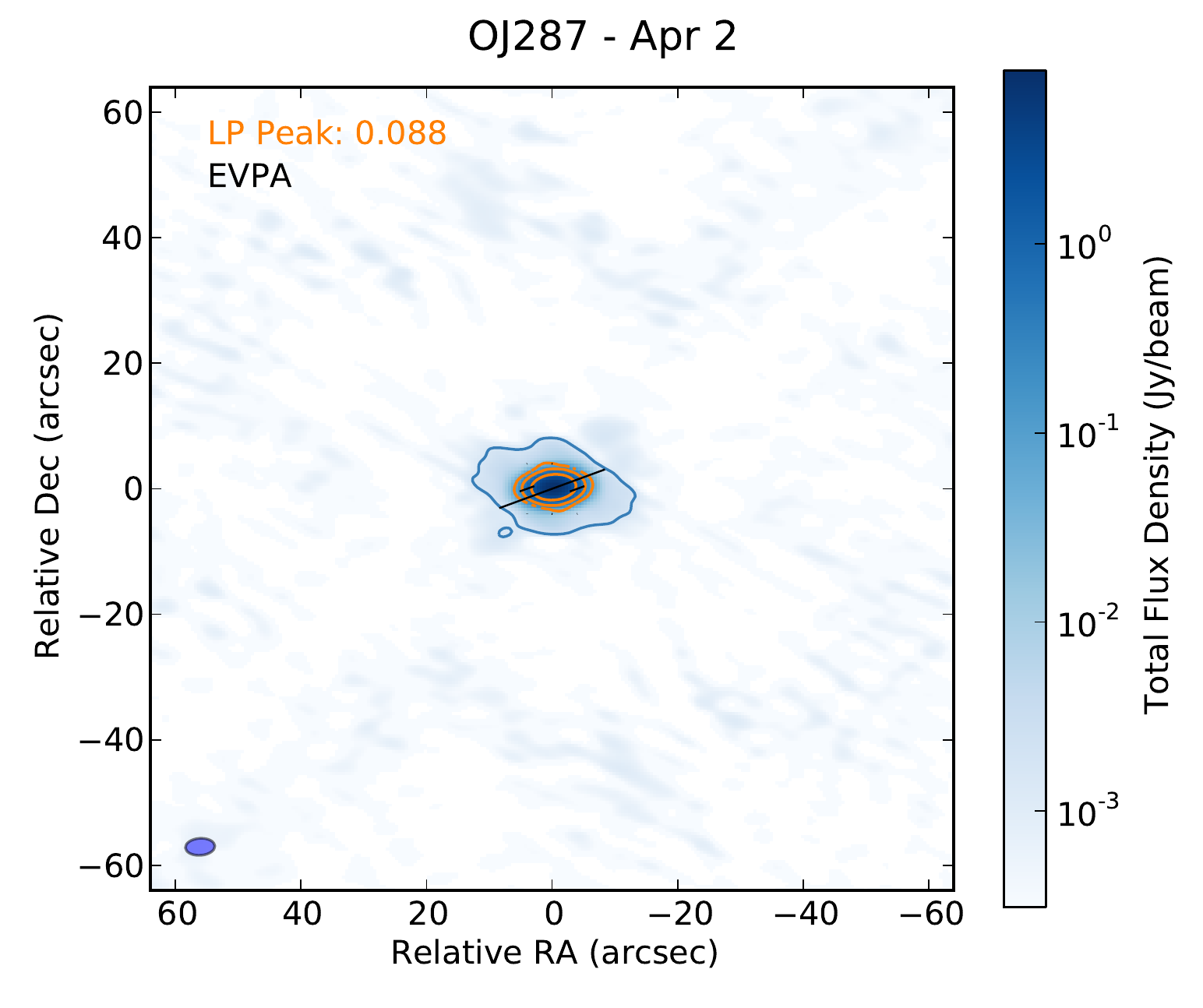} \hspace{-0.3cm}
\includegraphics[width=6cm]{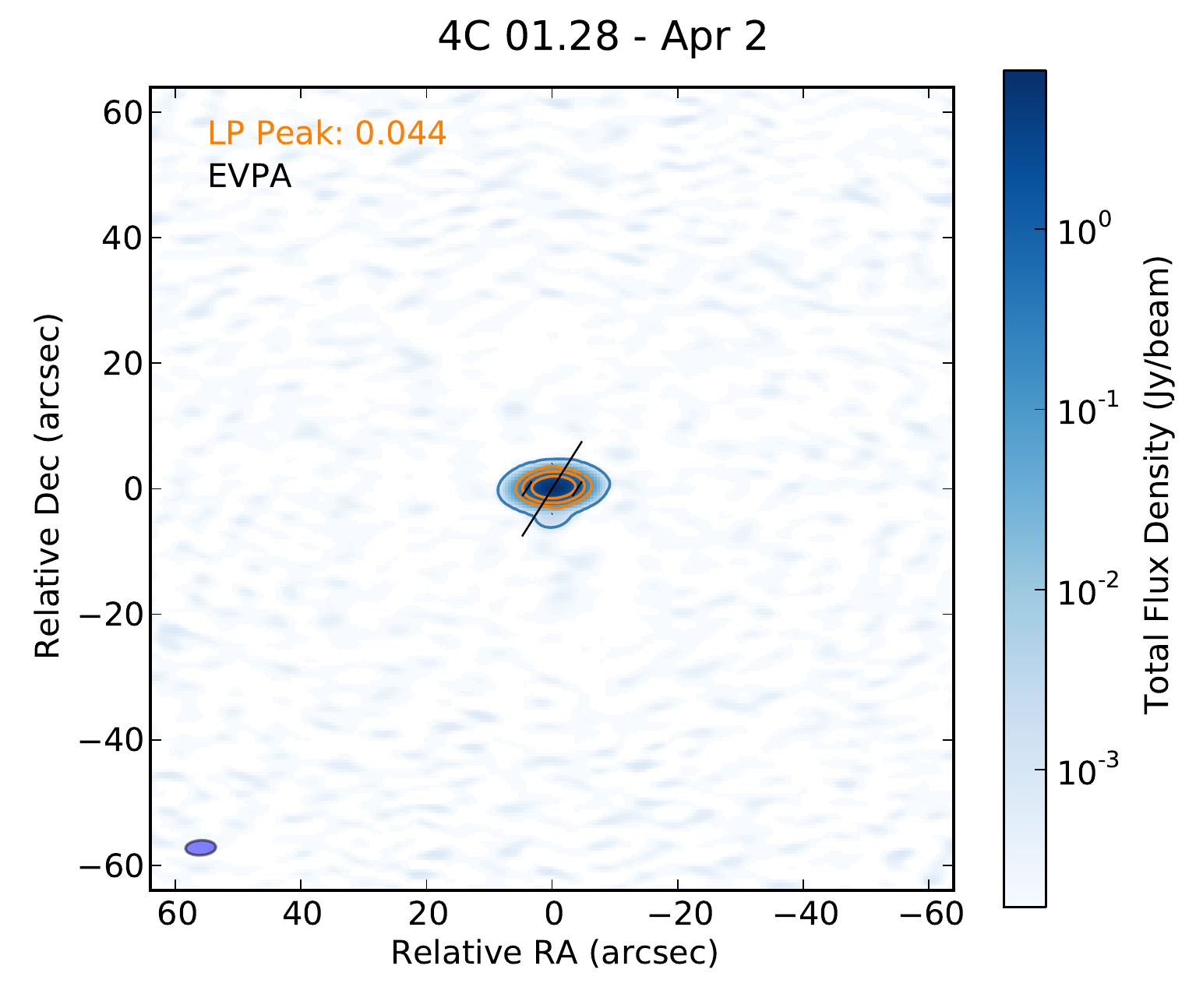} \hspace{-0.3cm}
\includegraphics[width=6cm]{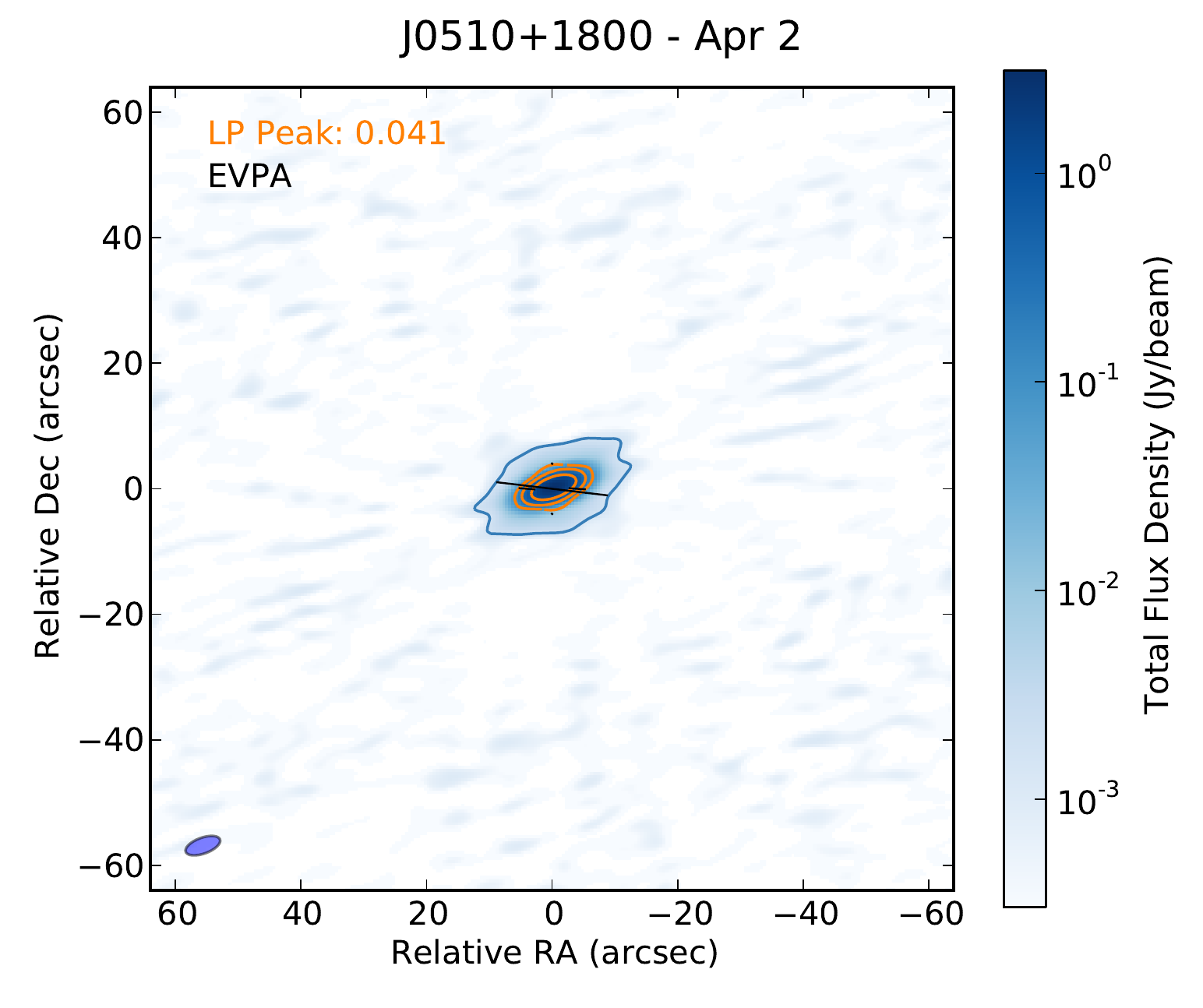} \hspace{-0.3cm}
\includegraphics[width=6cm]{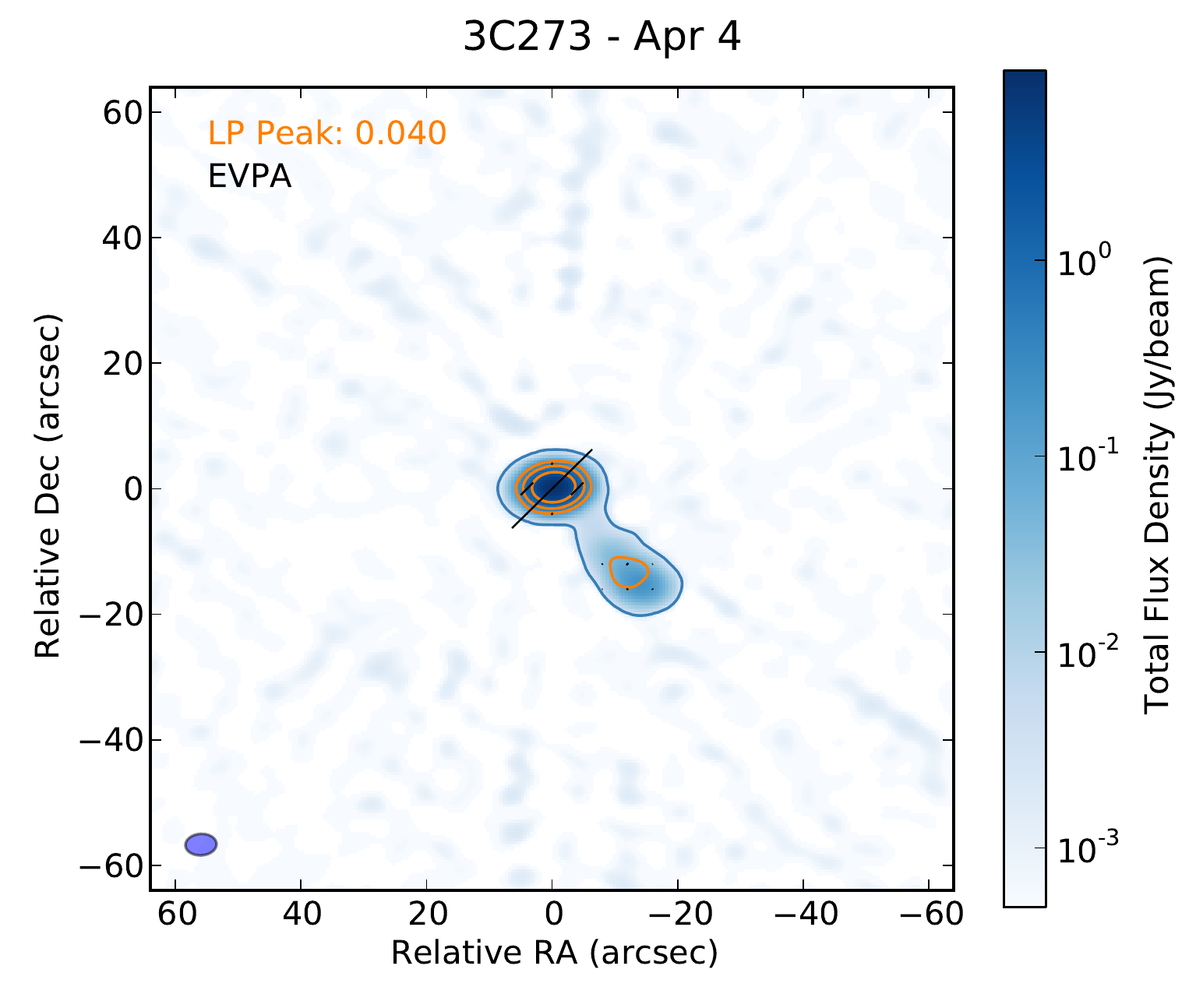} \hspace{-0.3cm}
\includegraphics[width=6cm]{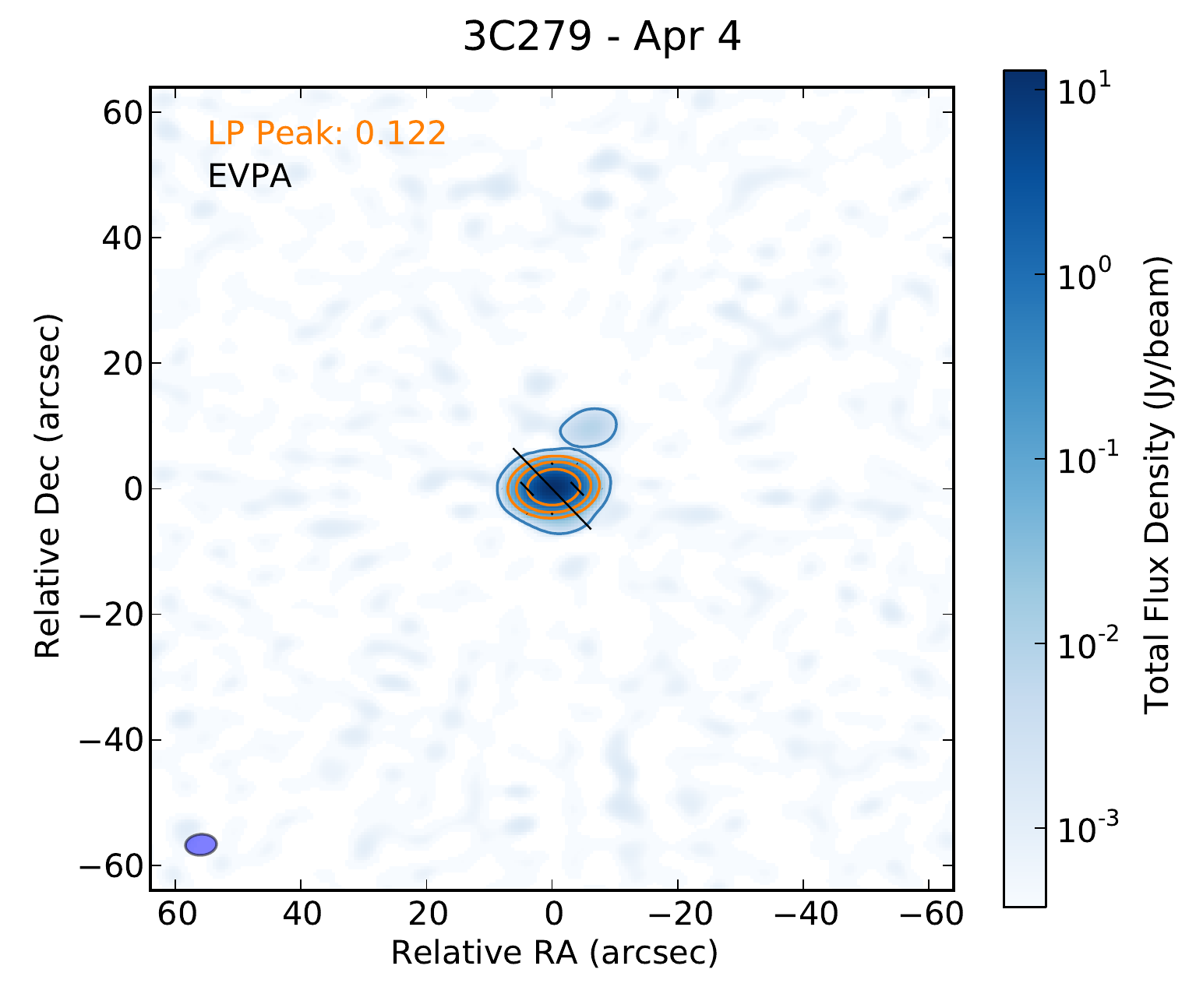} \hspace{-0.3cm}
\includegraphics[width=6cm]{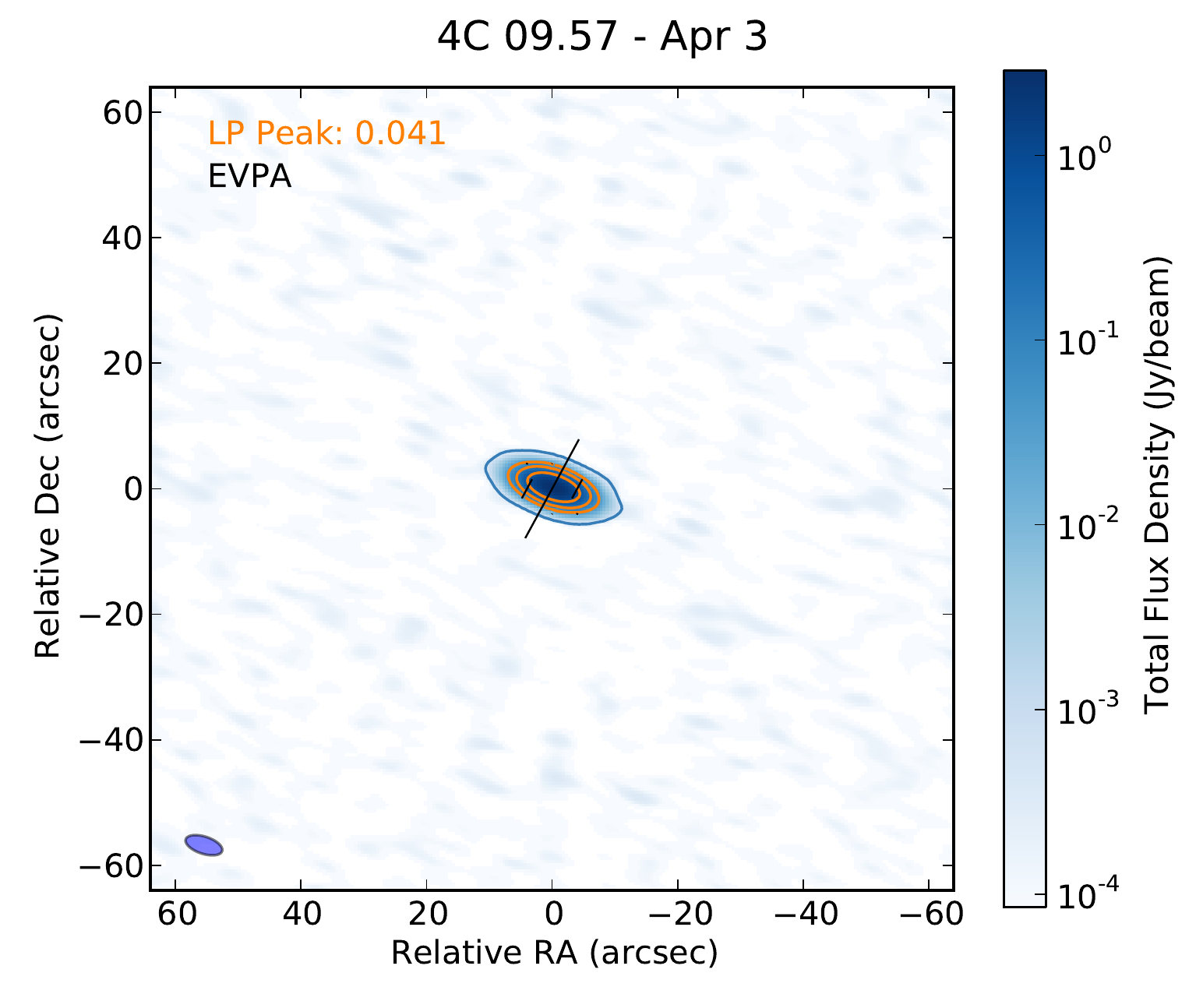} \hspace{-0.3cm}
\caption{Polarization images of all targets observed at 3~mm (at a central frequency of 93 GHz). 
The beamsizes are shown as an oval in the lower left corner of each panel (see values reported in table~\ref{tab:GMVA_im_rms}). 
See Figure~\ref{fig:m87+sgra_polimage} for a description of the plotted quantities. 
}
\label{fig:polimages_3mm}
\end{figure*}

The full suite of polarization images for all the sources observed in the VLBI campaign are shown in Figures~\ref{fig:polimages_sgra_1mm}--\ref{fig:polimages_3mm}. 
In particular, Figure~\ref{fig:polimages_sgra_1mm} displays 1.3~mm images of Sgr~A* in the three days of the EHT observations, Figures~\ref{fig:polimages_m87} and \ref{fig:polimages_3c279} display the 1.3~mm images of M87 and 3C279 on the four days of the EHT observations.  Figures~\ref{fig:polimages_agns_3days} and \ref{fig:polimages_agns_1_2days} report 1.3~mm maps for all the other AGN sources observed with the EHT for three days and one/two days, respectively. Finally, Figure~\ref{fig:polimages_3mm} shows 3~mm maps of the sources observed with the GMVA. 
In each plot,   the black vectors showcase the orientation of the EVPAs, while their length is linearly proportional to the polarized flux. The EVPAs are plotted every 8 pixels (i.e. are spaced by 1\pas6 at 1.3~mm and 4\arcsec\ at 3~mm) for all sources, except for M87, where the EVPAs are plotted  every 4 pixels (i.e. are spaced by 0\pas8), in order to sample more uniformly the jet. 
 Note that the EVPAs are not Faraday-corrected and that the magnetic field vectors should be rotated by 90\dg. 


\section{Comparative analysis across multiple flux-extraction methods}
  \label{app:fluxext}

Some of our targets (chiefly Sgr~A*  and  M87, but also a few other AGNs; see Appendix~\ref{appendix:maps})  reveal extended emission at arc-seconds scales, which is differently   resolved out by the different observing array configurations. Here we assess the impact of such extended emission in the flux values  extracted in the visibility domain for  the compact cores. 
For this purpose,  
we compare the parameters derived with {\sc uvmultifit} with those derived with two  imaged-based methods:  the sum of the nine central pixels  in the {\sc clean} model image (\txt) and  the integrated flux ({\sc intf})  from Gaussian fitting with the {\sc casa} task \texttt{IMFIT}\footnote{\texttt{IMFIT}  provides in output also the peak flux, but the integrated flux was preferred because its values were more consistent with the estimates from the other two methods.} (see \S\ref{allstokes}). Table~\ref{tab:methodcomp} reports the results from this comparative analysis between the image-based and the $uv$-based methods,   
showing the ratio of the four Stokes parameters, the LP, and the RM ({\sc 3x3/uvmf} and {\sc intf/uvmf}, respectively), 
and the difference of the EVPA in degrees ({\sc  3x3-uvmf} and {\sc intf-uvmf}, respectively). Three sources (Cen~A, NGC~1052, and J0132-1654) are excluded from this statistics due to their weak polarized signal.

Overall, the analysis clearly shows that  Flux({\sc 3x3}) $<$ Flux({\sc uvmf}) $<$ Flux({\sc intf}) 
when considering the median of the results. On the one hand, one can interpret this as the 3$\times$3 method being the least affected and {\sc intf} being the most affected by extended emission (with a weak dependency on the  array-configuration resolution). On the other hand, the 3$\times$3 method can also be more affected by phase and amplitude calibration errors (which will remove flux from the phase-center into the sidelobes), resulting in less recovered flux.
Despite this systematic deviation, the median offsets are compatible with no significant difference within the MAD values. In Stokes\,$I$, both the offset and MAD are $\leq 0.4$\% across the methods, which is negligible when compared to the absolute uncertainty of ALMA's flux calibration (10\% in Band~6). In the case of Stokes $Q$ and $U$, and resulting  LP, the systematic offset and MAD between methods can be up to 1\%.  With the LP measured in our sample, these differences are generally comparable (in absolute terms) to the 0.03\% of Stokes\,$I$ leakage onto $Q$ and $U$. 
 The EVPA shows a MAD value of  $\sim0.1\,$deg, which results in a MAD value of up to 10\% in the RM (note these are comparable or better than the observed statistical uncertainties -- see Tables~\ref{tab:GMVA_uvmf_spw} and \ref{tab:EHT_uvmf_spw}).  
The poorer results for Stokes\,$V$ are induced by the low level of CP in the sample (although the MAD is still comfortably around 4\%).

\begin{figure*}
\centering
\includegraphics[width=\textwidth]{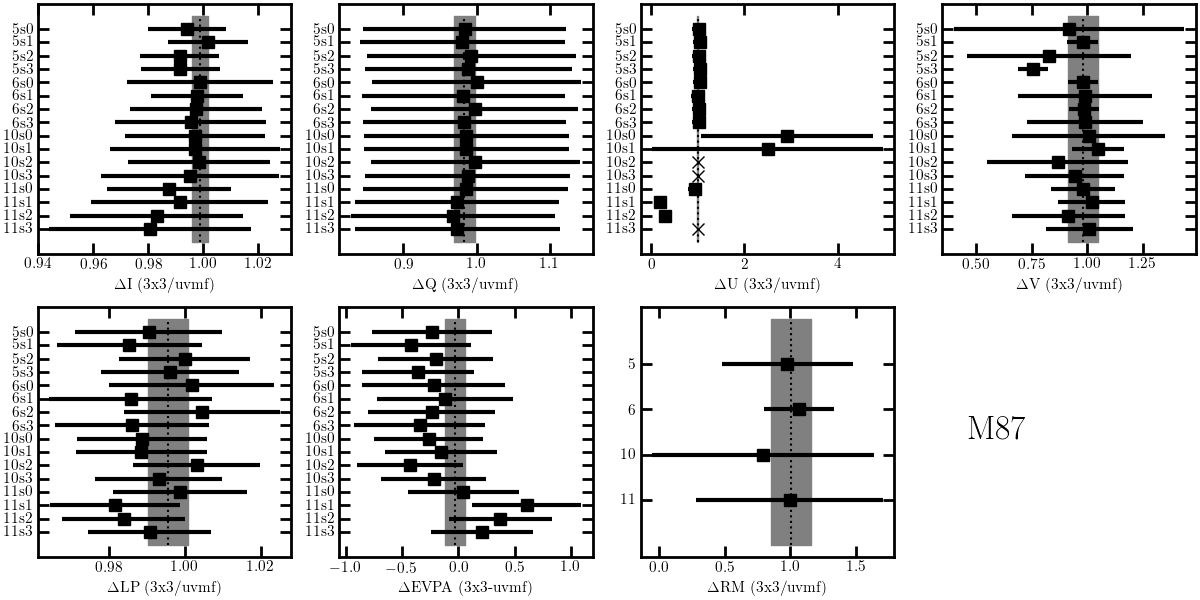}\\
\vspace{1cm}
\includegraphics[width=\textwidth]{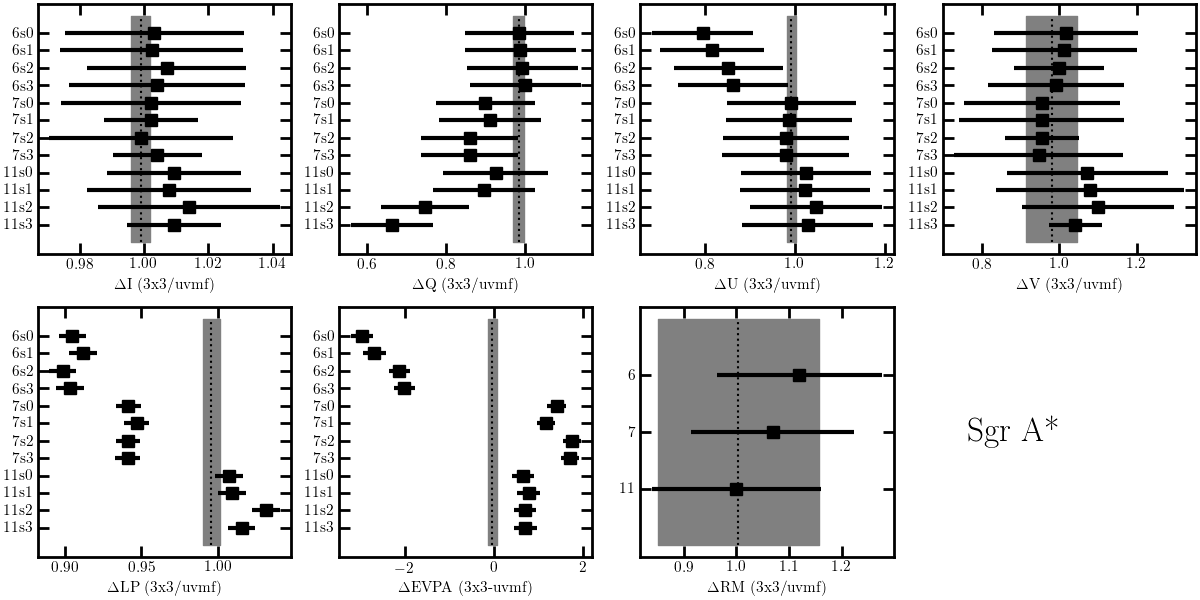} 
\caption{Comparison between image-based and $uv$-based flux-extraction methods for the two sources with the most prominent  extended emission in our sample: M87 (top panels), and Sgr~A* (bottom panels). The two methods being compared are: {\sc uvmultifit} ({\sc uvmf}) and the sum of the nine central pixels ({\sc 3x3}) of the model image. 
All parameters are compared via ratios, except  for the EVPA (showing the difference in degrees). The errorbar in each data-point is the combination in quadrature of the statistical error from each flux extraction method. 
The vertical dotted line and the shaded region show the median and MAD from Table~\ref{tab:methodcomp}. The labels in the Y-axis indicate  the observing day in April 2017 (5, 6, 7, 10, 11) and the observing SPW (s1, s2, s3, s4); the RM panel shows only the days.  
Non-detections in the images are indicated with a cross.
Note that the plotted uncertainties are not those of individual measurements but ratios or differences between the two methods. Therefore the errors on LP and EVPA for M87 on Apr 10/11 appear comparable to other days despite the large error bars and/or non detections in Stokes\,$U$ (i.e. the errors in U displayed do not propagate in the LP and EVPA plots).
}
\label{fig:uvmfcomp}
\end{figure*}

Table~\ref{tab:methodcomp} reports a fraction of ``outliers'' in the distributions (those cases that are 5$\sigma$ away from the sample median). These are generally associated with the sources with the most prominent extended emission (other outliers are associated to parameters estimated at low-significance values). In order to better assess the magnitude of these outliers, in Figure~\ref{fig:uvmfcomp} we show a detailed comparison between the $uv$-based and the {\sc 3x3} image-based methods for M87 (upper panels) and Sgr~A* (lower panels), respectively. 
Analogously to Table~\ref{tab:methodcomp}, individual panels   
show the ratio of the four Stokes parameters, the LP, and the RM ({\sc 3x3/uvmf}), 
and the difference of the EVPA in degrees ({\sc  3x3-uvmf}); 
the results are reported per day and SPW. 

In the case of M87, the two methods exhibit an excellent agreement, except for Stokes\,$U$. This is due to a combination of faint $U$ emission at the core  and appreciable emission from knots A and B. 
The discrepancy in Stokes\,$U$ results generally in $\Delta \rm EVPA \lesssim 0.4$\dg\, comparable with the uncertainties quoted in Table~\ref{tab:EHT_uvmf_spw}, with one single case of $\Delta \rm EVPA \sim 0.6$\dg\ (on Apr 11 SPW=1). 
Given the consistency in RM (within $1\sigma$), this worst-case scenario discrepancies in EVPA do not affect the results of the analysis of this paper. 

In the case of Sgr~A*, the discrepancies are much more pronounced in most parameters. 
Despite the prominence of the mini-spiral (see \S\ref{Sgr_A_m87_im} and Fig.~\ref{fig:polimages_sgra_1mm}), 
Stokes\,$I$ values actually agree within less than 1\% between the two methods. It is however noteworthy  that Flux({\sc 3x3})$>$Flux({\sc uvmf}) (i.e., the opposite trend with respect to Table~\ref{tab:methodcomp}). 
This inverted trend is due to a flux-decrement in the visibility amplitudes at around 25--30\,k$\lambda$ affecting most prominently  {\sc uvmfit}  on Apr 11 (this can be assessed by inspecting amplitude vs. baseline-length {\it uv}-plots -- not shown here -- for the parallel hands on all days). 
Unlike Stokes\,$I$, Stokes\,$QU$ are heavily affected, up to 20--30\% in the worst cases,  
 resulting in LP differences up to 10\% and EVPA offsets up to 2\dg. 
 These large deviations in Stokes\,$QU$ are systematic since all four SPW appear to deviate in each day by an approximate amount. This in turn results in 
consistent RM values within $1\sigma$. We note that we do not observe the same systematic offset either in M87 or J1924-2914, which were also calibrated using 3C279 as the polarization-calibrator (on Apr 6 and 11), hence we discard calibration issues as being responsible for these systematics.
One possible explanation is differential Stokes\,$I$ leakage onto $Q$ and $U$ from extended unpolarized emission (see the spurious EVPA vectors matching the spiral-arms in Fig.~\ref{fig:polimages_sgra_1mm}). Note also that such unpolarized emission extends farther than the inner third of the image field of view, where beam-squash \citep{Hull20} combined with leakage may induce significant deviations. 

Given the results from this comparative analysis, we conclude that, for the purpose of the polarimetric analysis conducted in this paper, the {\it uv}-fitting method provides sufficiently accurate  flux values of Stokes\,$IQU$ in all cases, 
although for Sgr~A* we observe deviations in Stokes\,$QU$ 
when comparing the two flux extraction methods.

\begin{table*}
\caption{Comparison of the three flux-extraction methods for EHT targets.} \label{tab:methodcomp}
\centering
\begin{tabular}{cccccccc}
\hline\hline 
method$^{a}$ & I$^{b}$ & Q$^{b}$ & U$^{b}$ & V$^{b}$ & LP$^{b}$ & EVPA$^{c}$~[\dg] & RM$^{b}$ \\
\hline
{\sc 3x3} & 0.9995(0.0007)$^{d}$ & 0.989(0.008) & 0.998(0.004) & 1.00(0.04) & 0.993(0.006) & 0.0(0.1) & 1.06(0.08) \\
& 0\%$^{d}$ & 2\% & 1\% & 5\% & 10\% & 10\% & 0\% \\
\hline
{\sc intf} & 1.001(0.004) & 1.00(0.01) & 0.996(0.006) & 1.00(0.04) & 1.001(0.004) & 0.04(0.09) & 1.0(0.1)\\
& 0\% & 2\% & 17\% & 10\% & 5\% & 6\% & 0\% \\
\hline\hline
\end{tabular}
 \tablenotetext{a}{This table compares {\sc uvmultifit} ({\sc uvmf}) with: the sum of the nine central pixels ({\sc 3x3}) of the model image; and the integrated flux ({\sc intf}) from {\sc imfit}.}
 \tablenotetext{b}{The  values  reported in columns are ratios between the method referred in the first column and {\sc uvmf}, with the latter being in the denominator. }
 \tablenotetext{c}{The EVPA column shows the angle difference (in deg) between the methods.}
  \tablenotetext{d} {In each instance, three values are displayed: the median value of the distribution; the Median Absolute Deviation or MAD (the value in brackets);  the percentage of cases farther from the median than five times the MAD in quadrature with the measurement error (the value in the second row).}
\end{table*}

\section{Comparison of Stokes parameters with the AMAPOLA polarimetric Grid Survey} 
\label{app:amapola_comp}
For the purposes of absolute flux calibration, ALMA monitors the flux of bright sources (mainly blazars or QSOs) spread over the full range in right ascension (the Grid) by observing them together with solar system objects, the so-called Grid-Survey (GS), with a period of approximately 10\,days. These observations are  executed with the ACA, in Bands~3, 6, and 7.  Since full-polarization mode is adopted, it is possible to retrieve polarimetry information from the GS sources.
This is done with  AMAPOLA\footnote{\url{http://www.alma.cl/$\sim$skameno/AMAPOLA/}}, a set of CASA-friendly Python scripts used to reduce the full-Stokes polarimetry of GS observations with ACA.
Some of our targets are part of the  GS, including: 3C273 (J1229+0203); 3C279 (J1256-0547); 4C 01.28 (J1058+0133); 4C 09.57 (J1751+0939); J0510+1800; OJ287 (J0854+2006); J0006-0623; J1733-1304; and J1924-2914. 
Although the reported values by AMAPOLA are used for observation planning and the two arrays cover different $uv$-ranges, it is still useful to make a comparison with this database to assess any systematics or clear inconsistent variability within a week's time-frame.

\begin{figure*}[ht!]
\centering
\includegraphics[width=4cm]{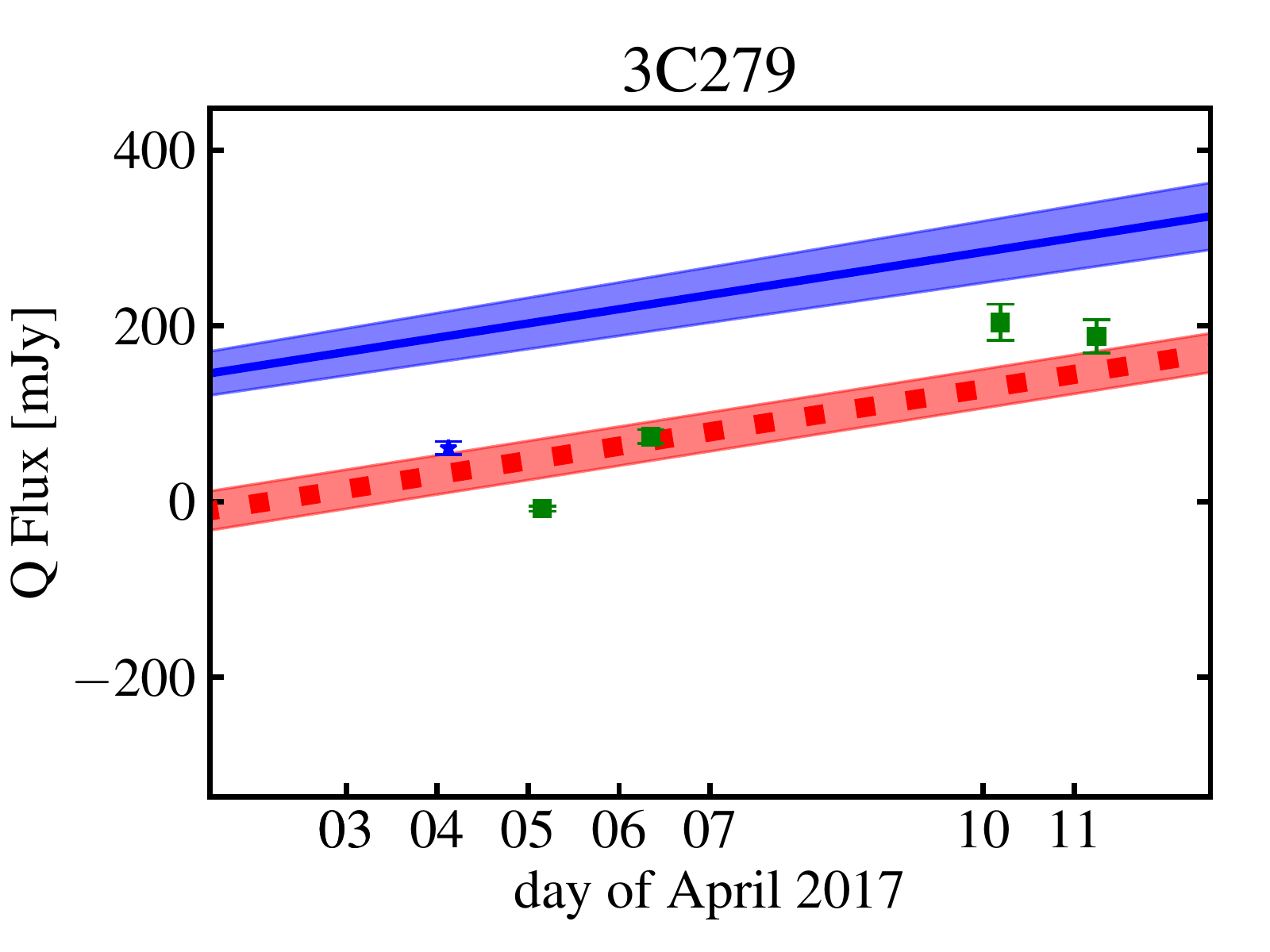} \hspace{-0.3cm}
\includegraphics[width=4cm]{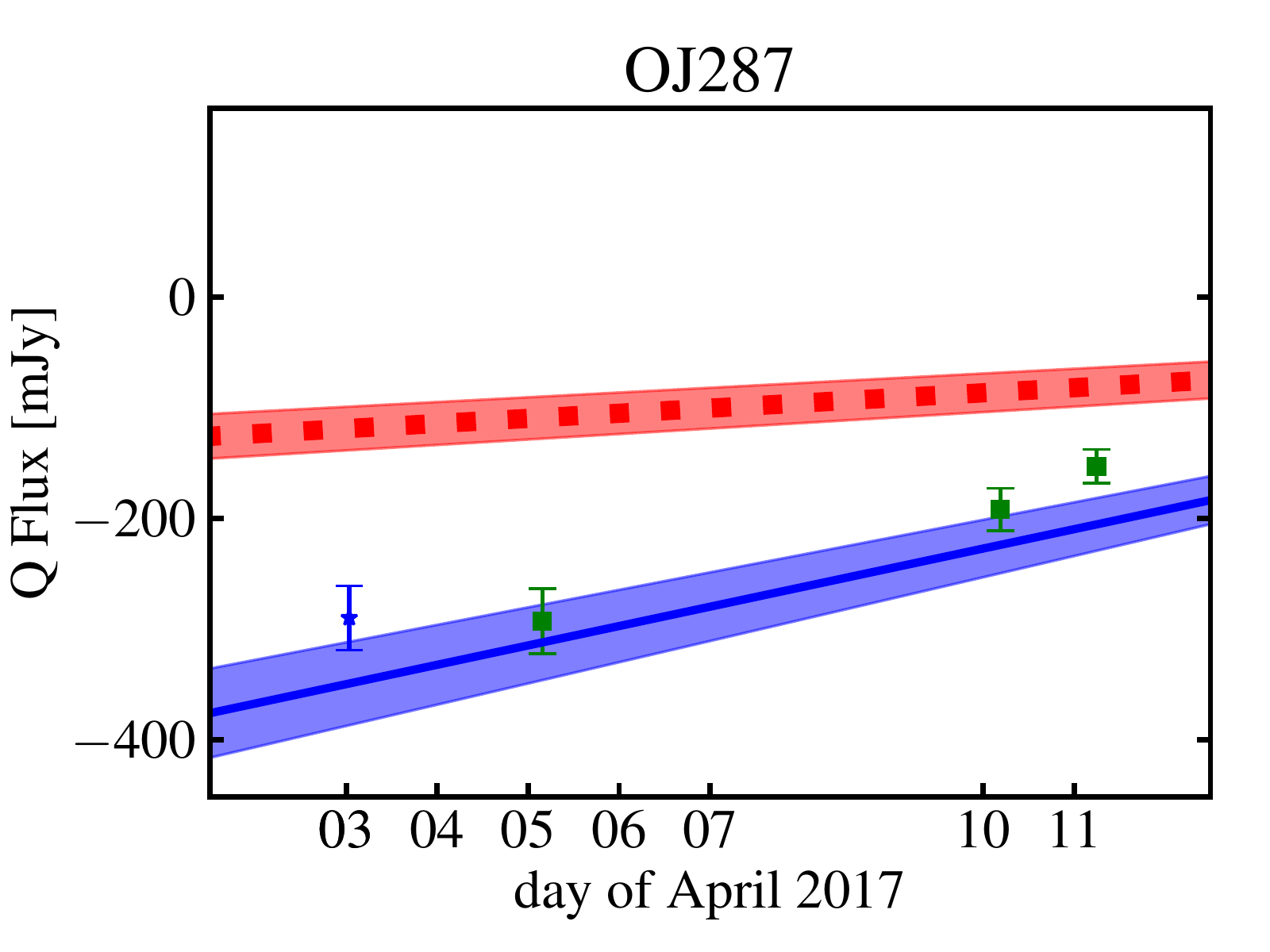}  \hspace{-0.3cm}
\includegraphics[width=4cm]{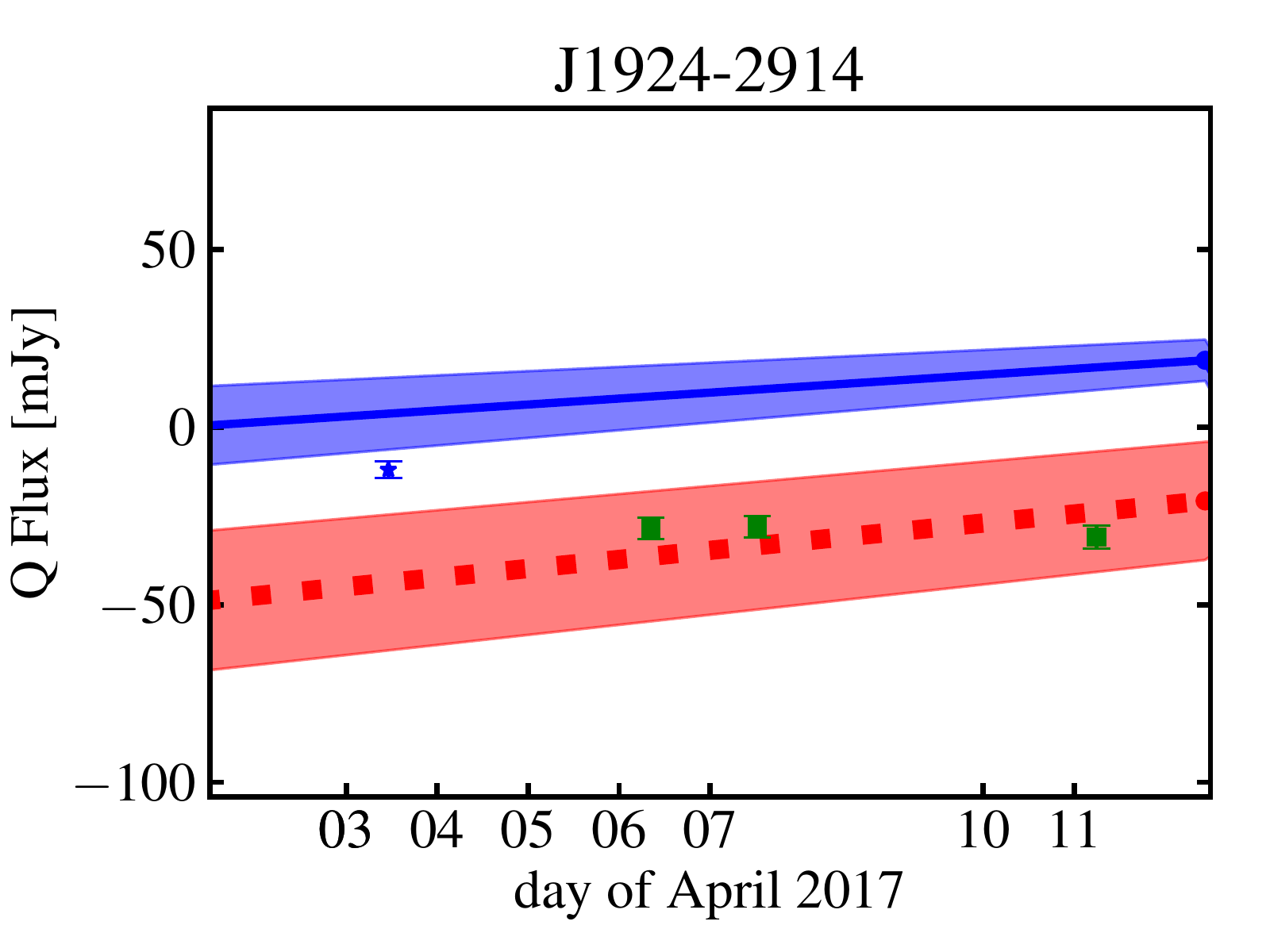}  \hspace{-0.3cm}
\includegraphics[width=4cm]{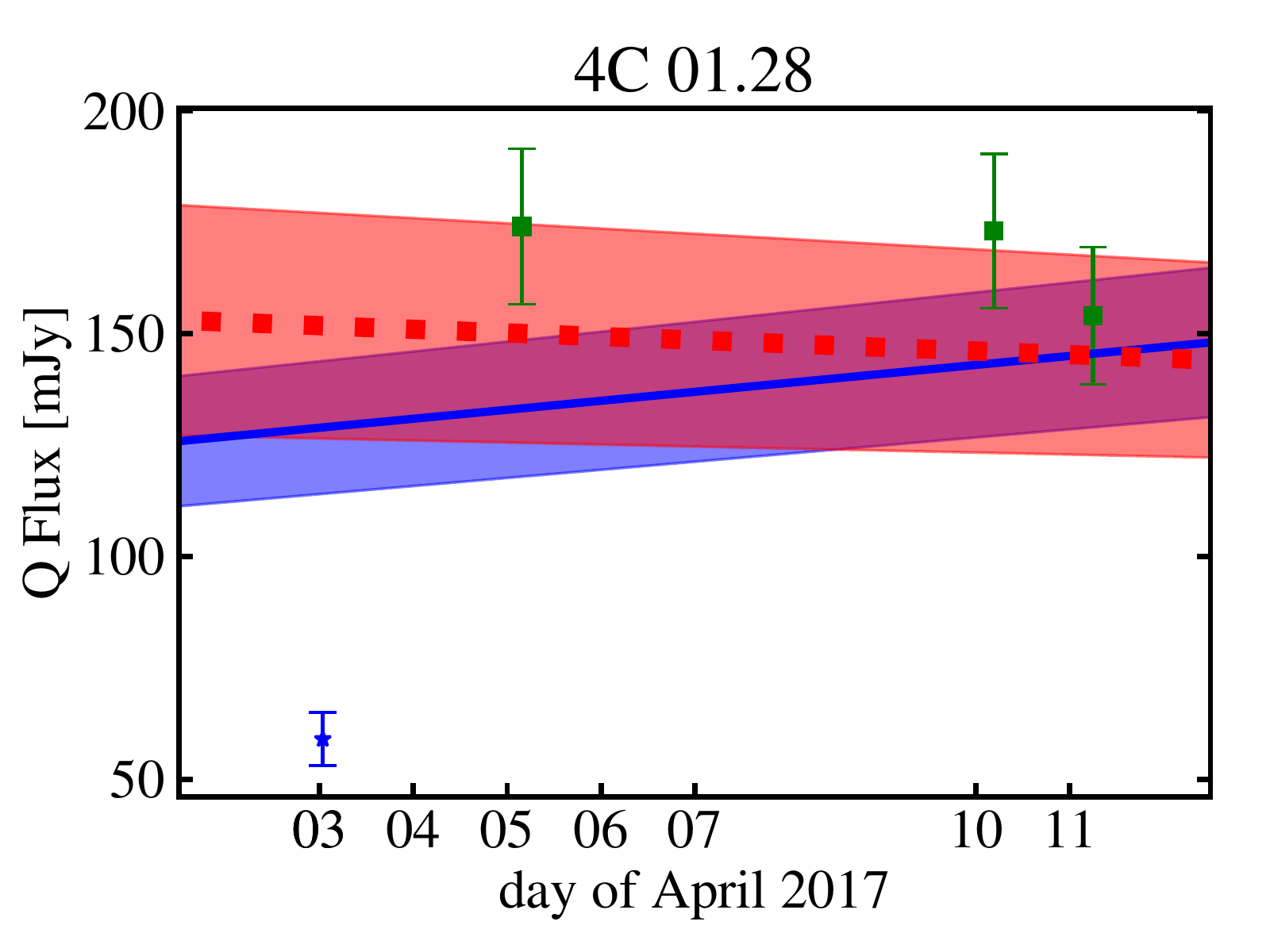}  
\includegraphics[width=4cm]{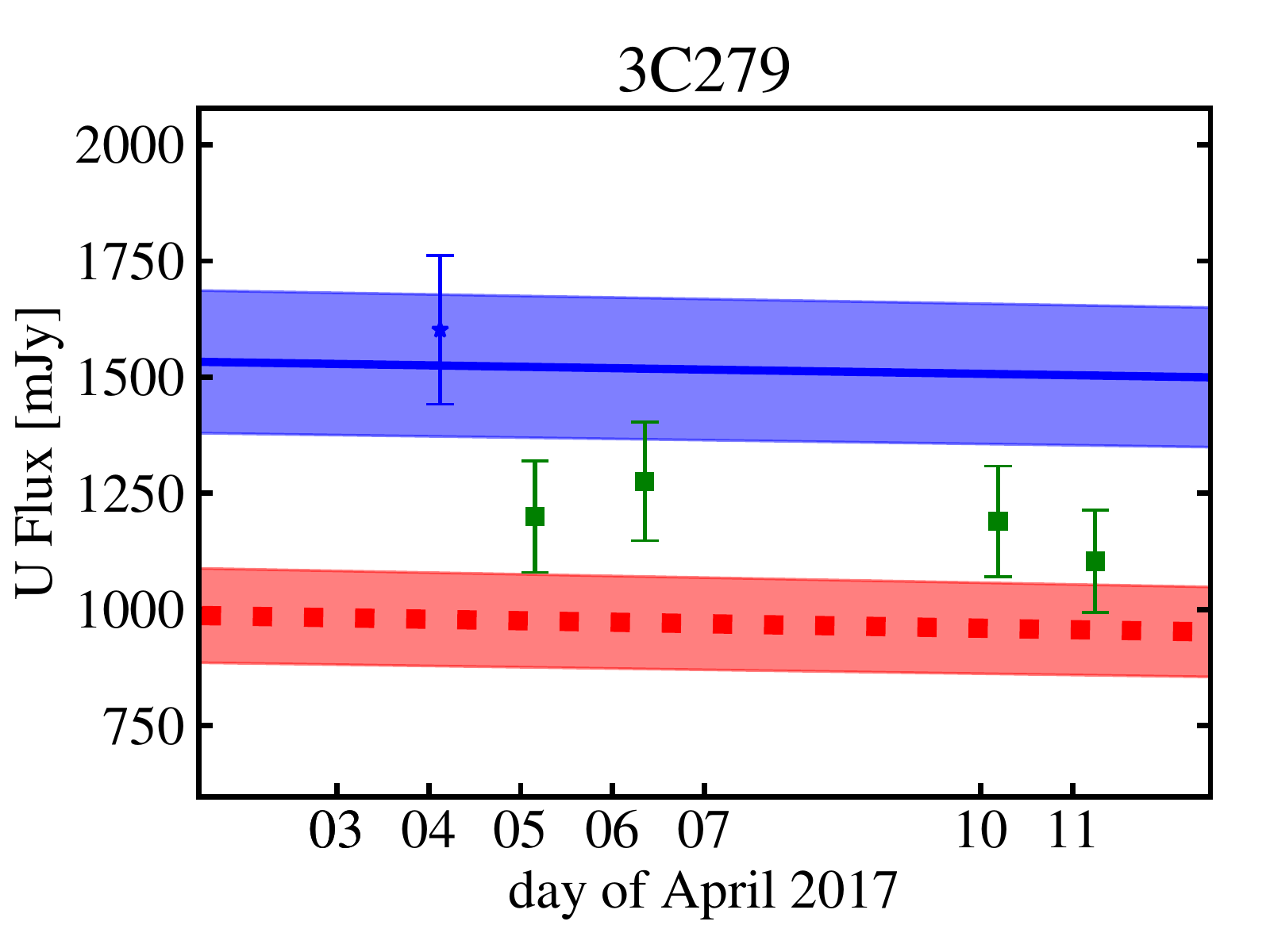} \hspace{-0.3cm}
\includegraphics[width=4cm]{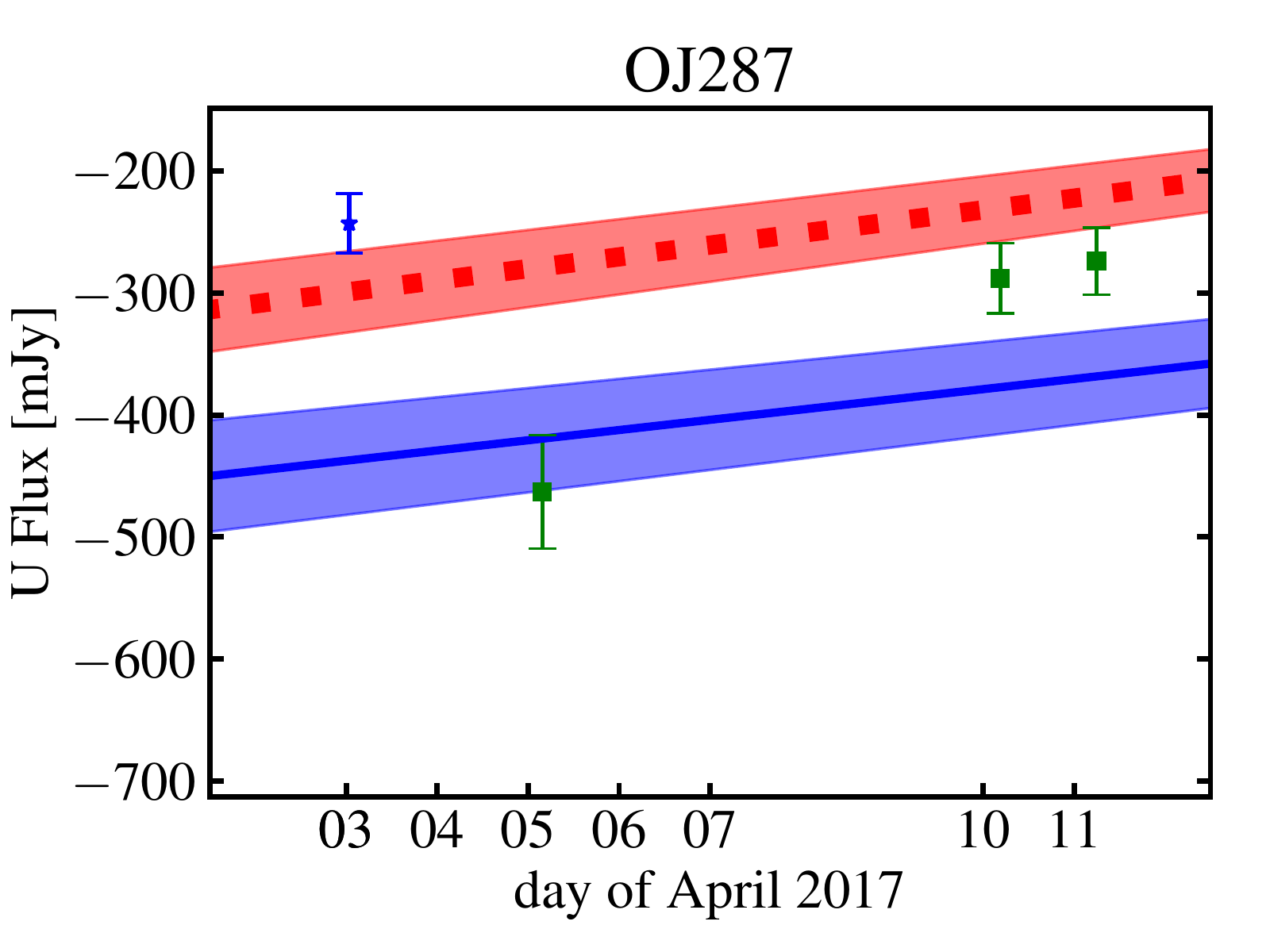}  \hspace{-0.3cm}
\includegraphics[width=4cm]{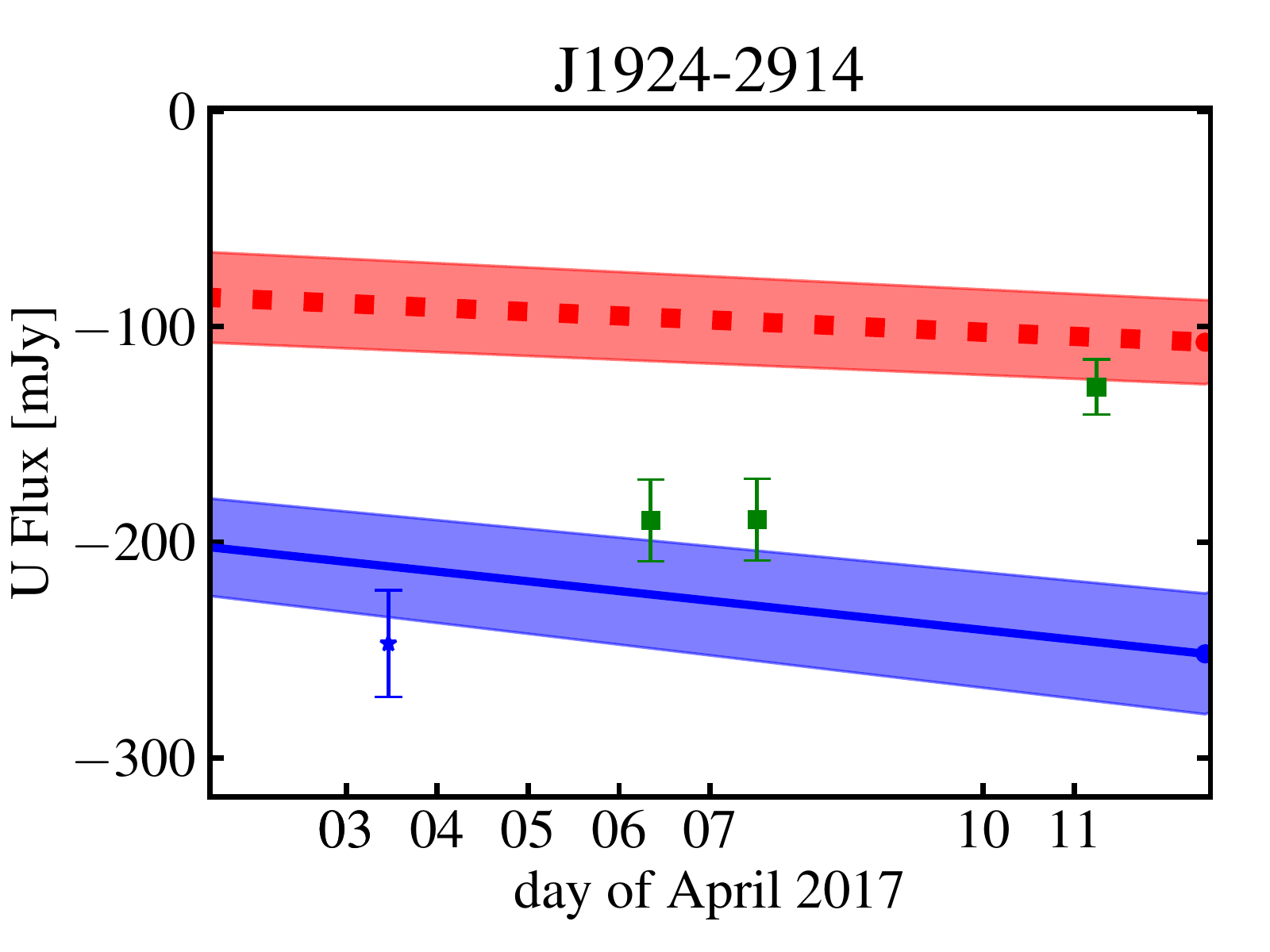}  \hspace{-0.3cm}
\includegraphics[width=4cm]{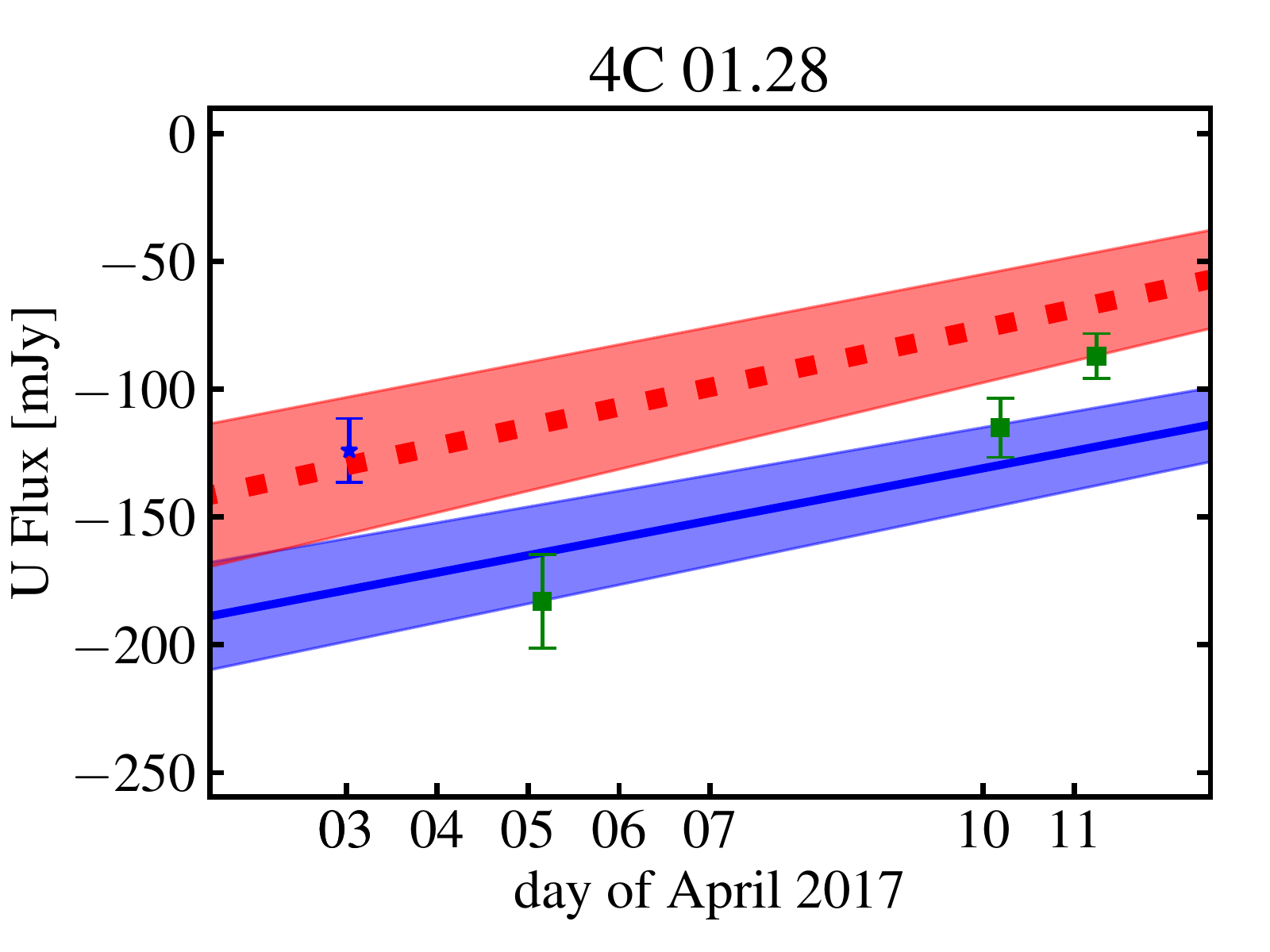}  
\includegraphics[width=4cm]{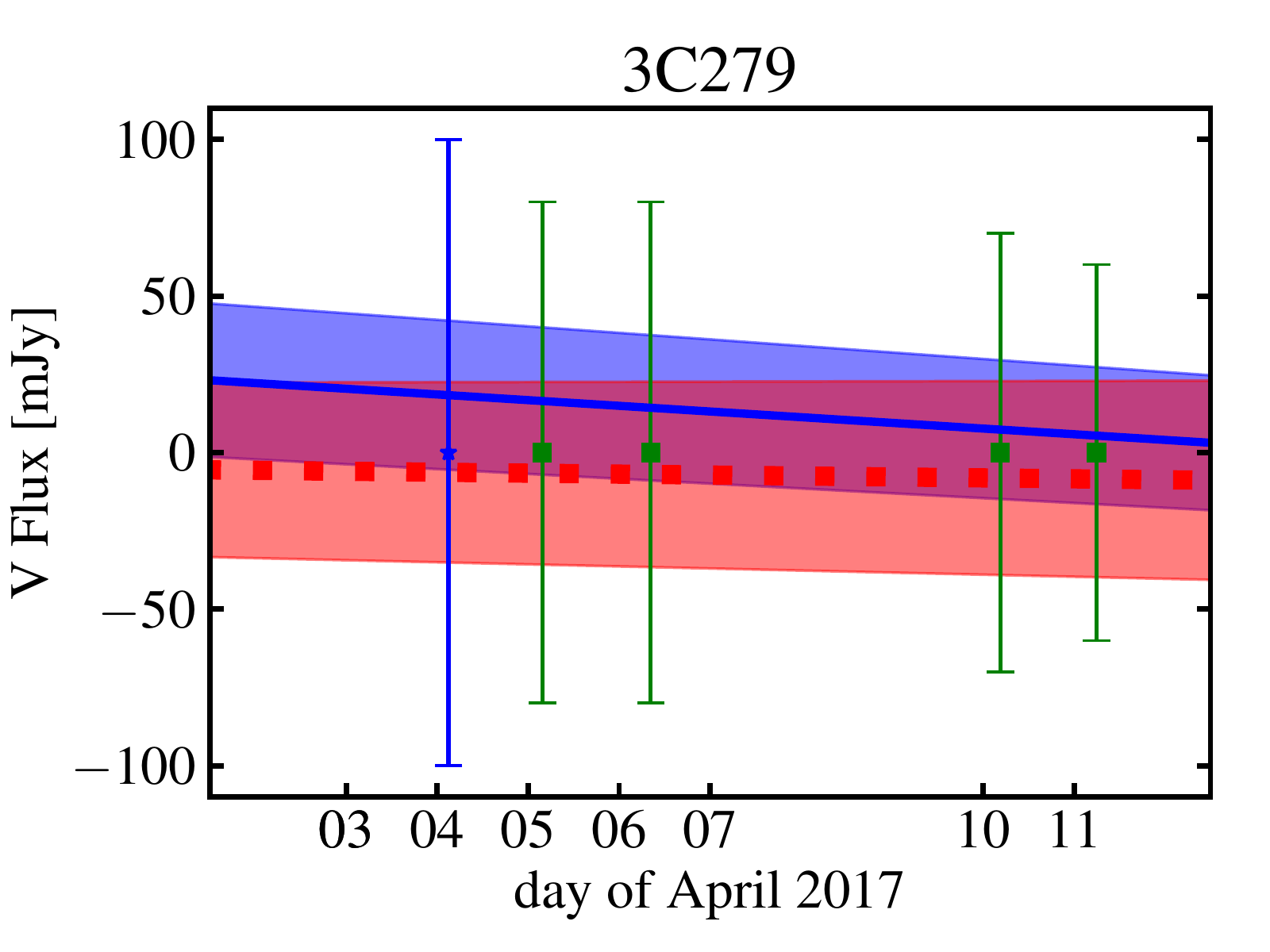} \hspace{-0.3cm}
\includegraphics[width=4cm]{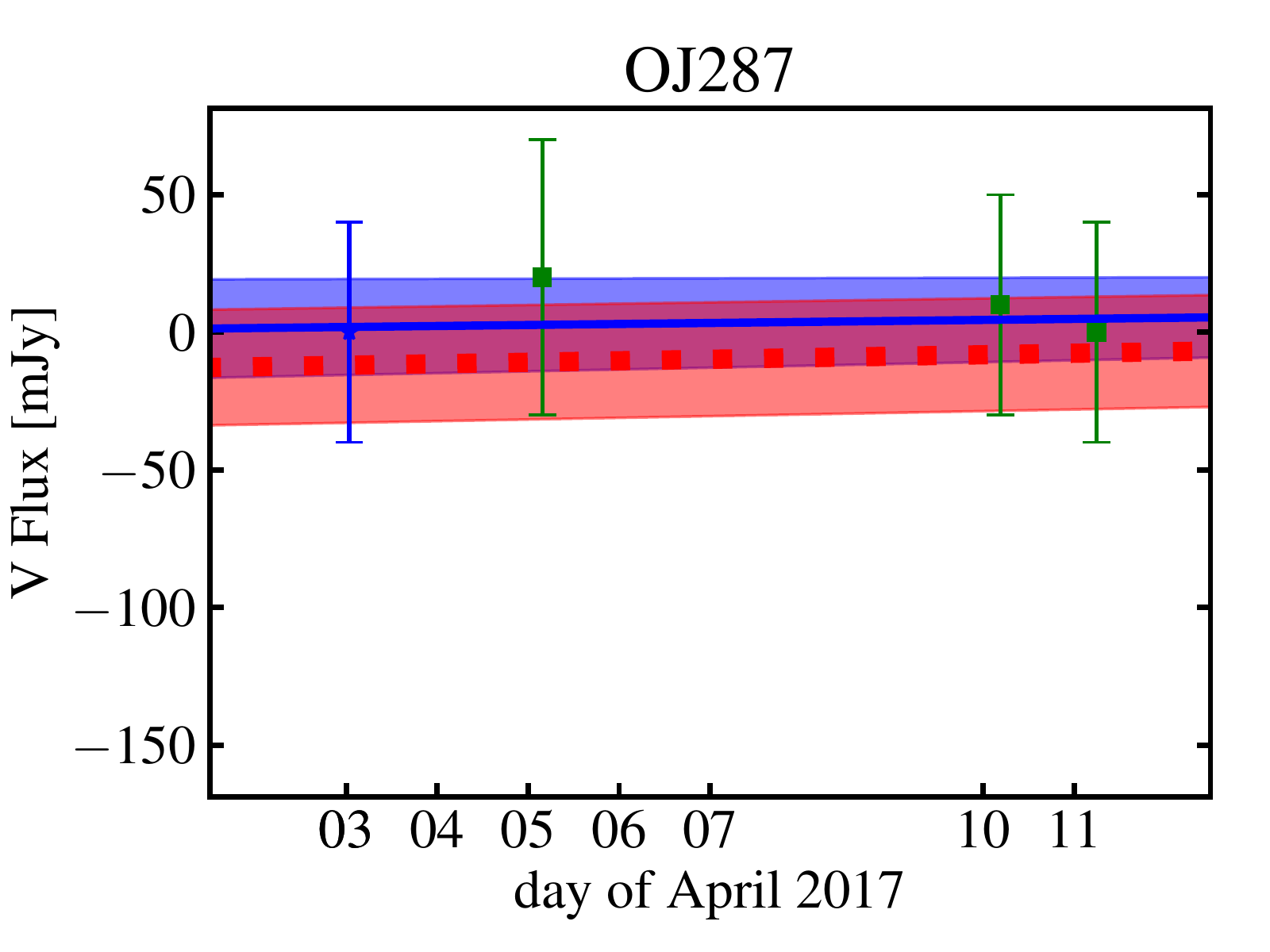}  \hspace{-0.3cm}
\includegraphics[width=4cm]{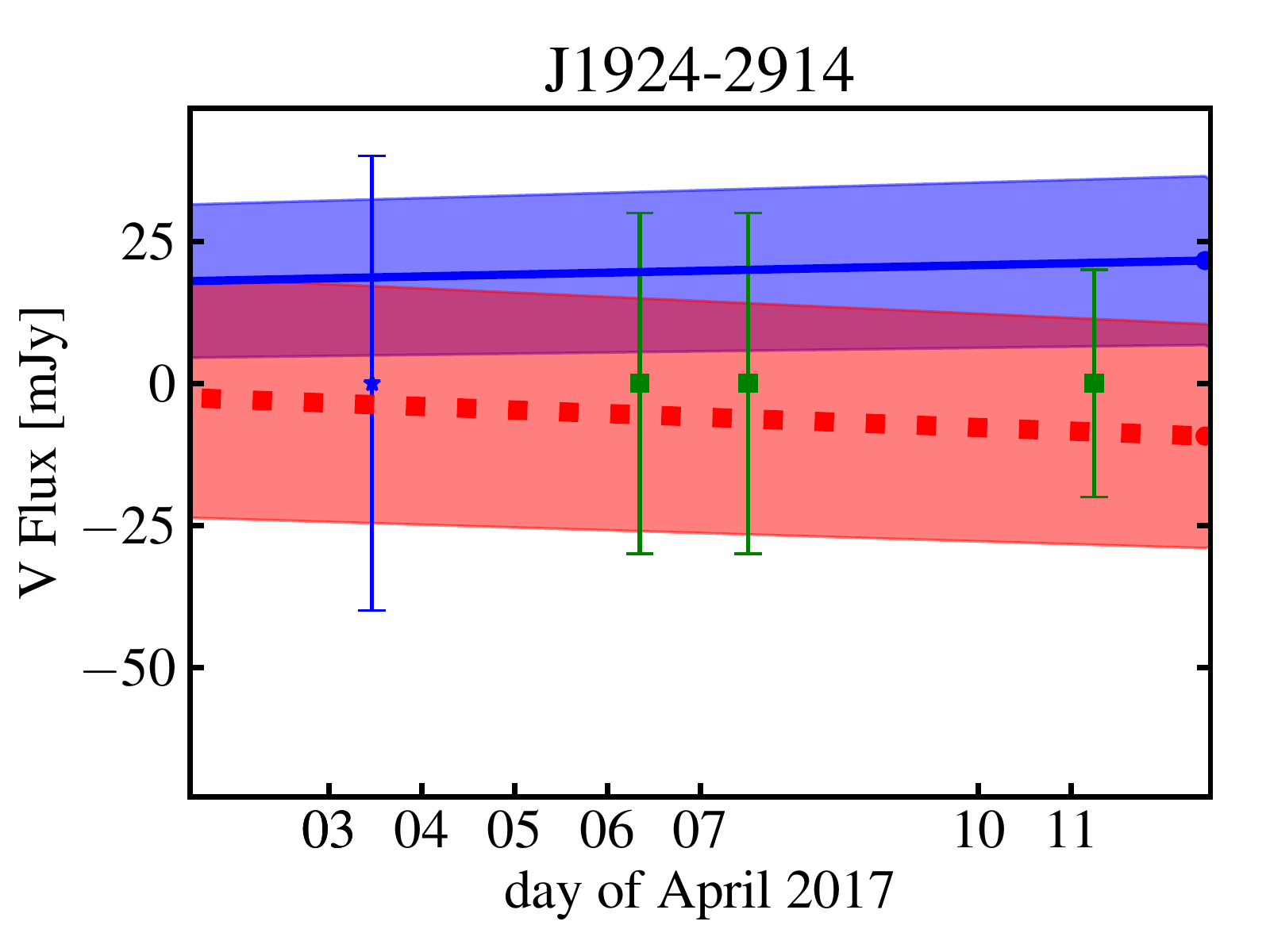}  \hspace{-0.3cm}
\includegraphics[width=4cm]{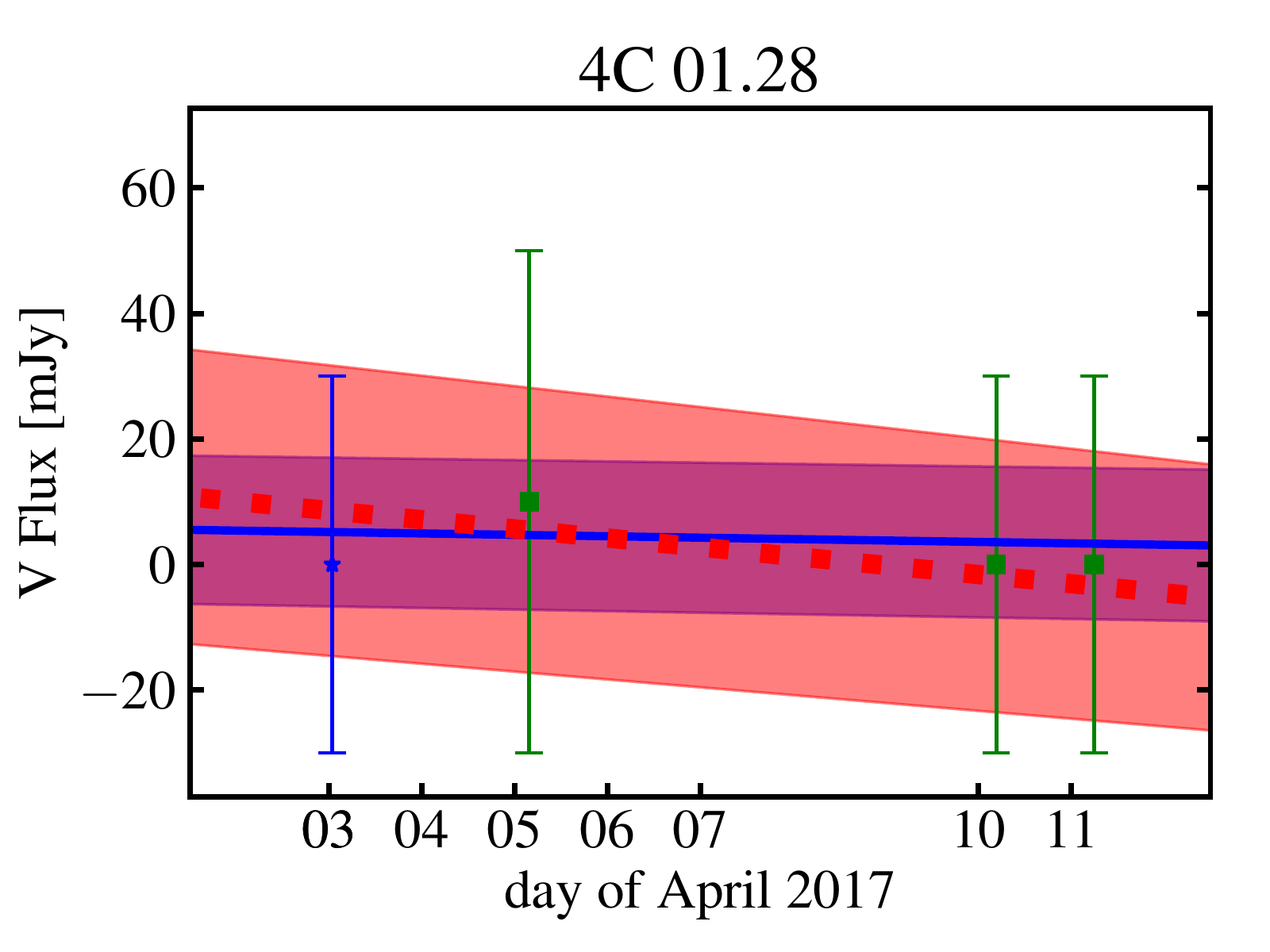}  
\includegraphics[width=4cm]{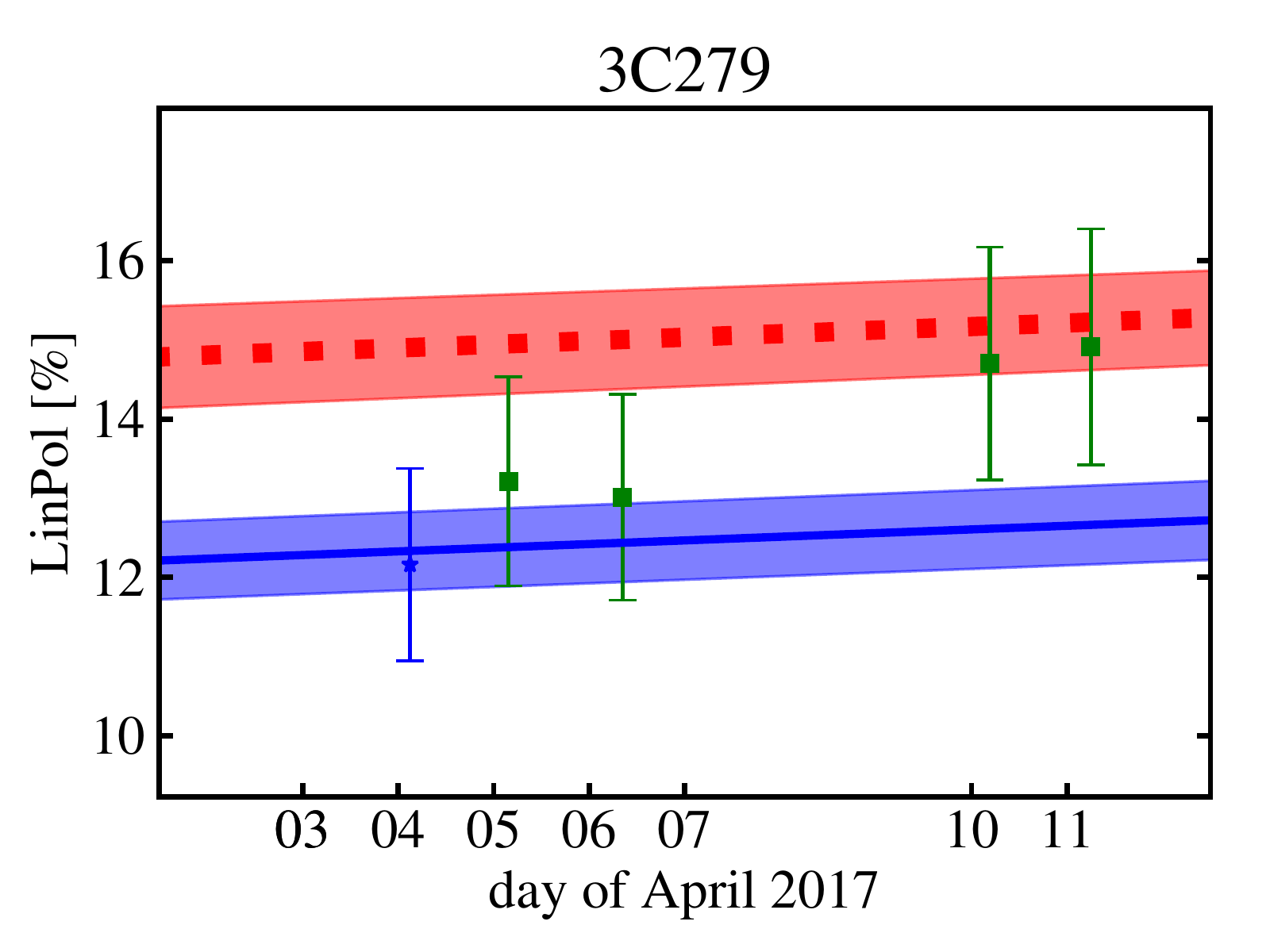} \hspace{-0.3cm}
\includegraphics[width=4cm]{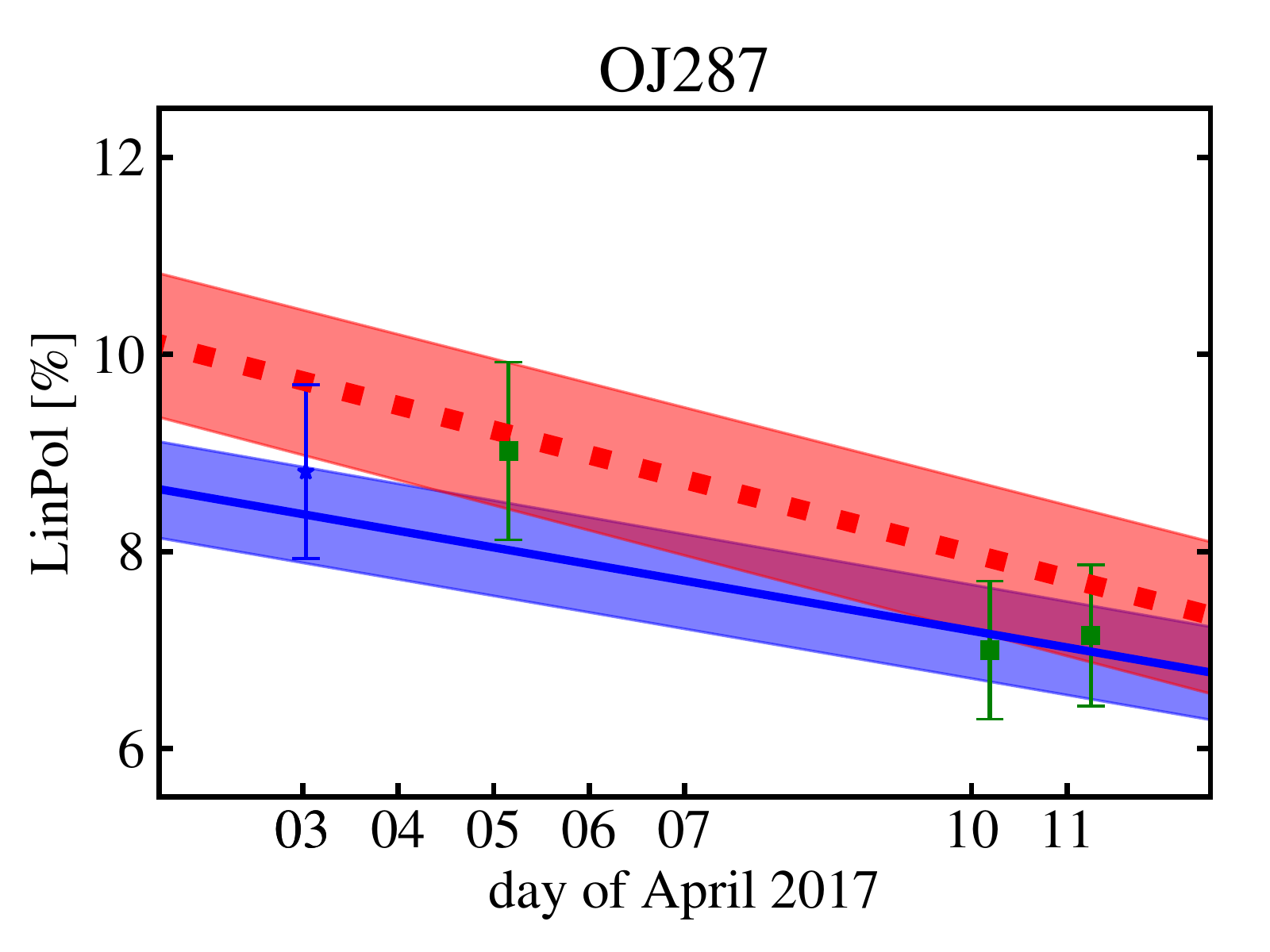}  \hspace{-0.3cm}
\includegraphics[width=4cm]{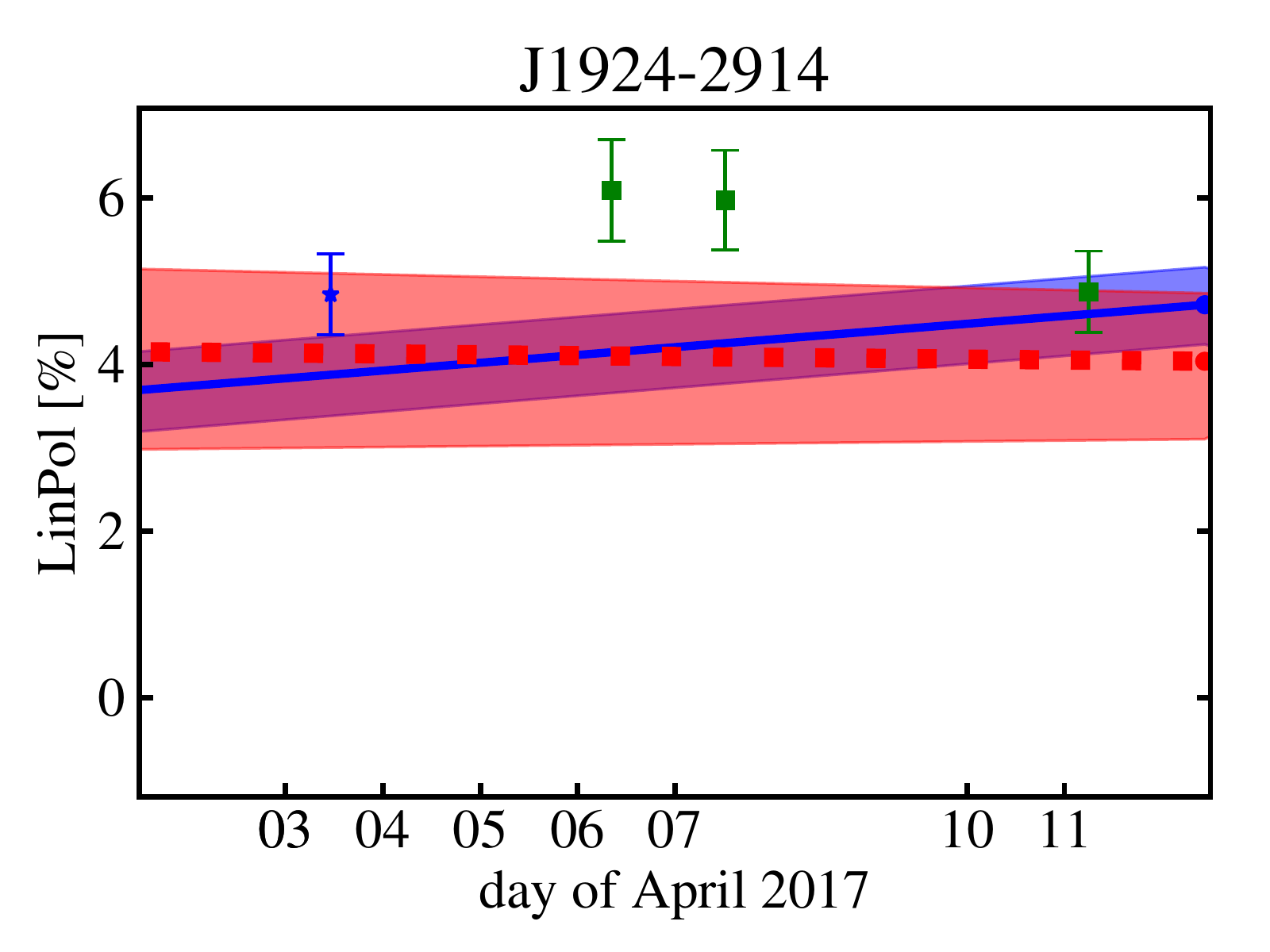}  \hspace{-0.3cm}
\includegraphics[width=4cm]{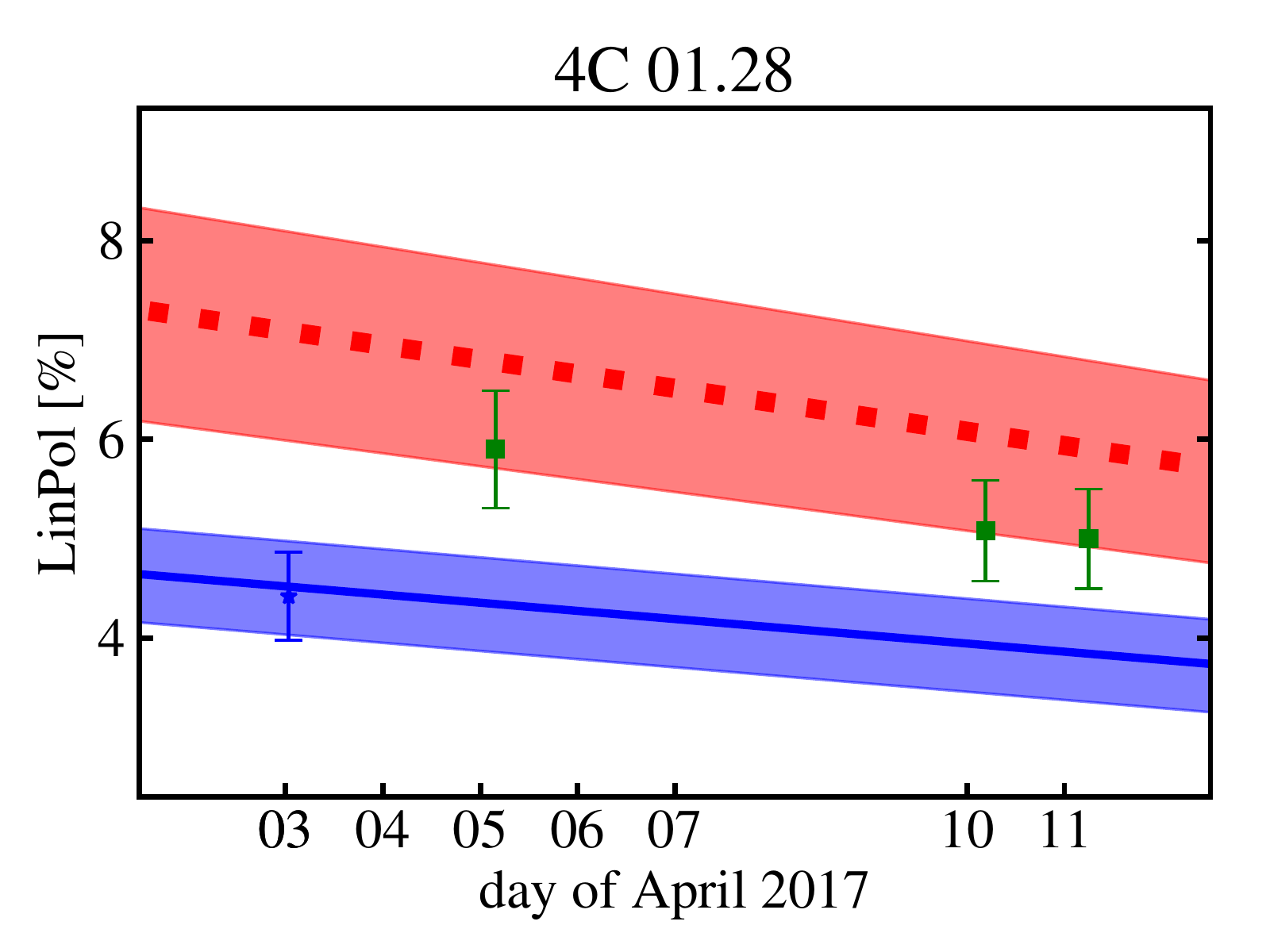}   
\includegraphics[width=4cm]{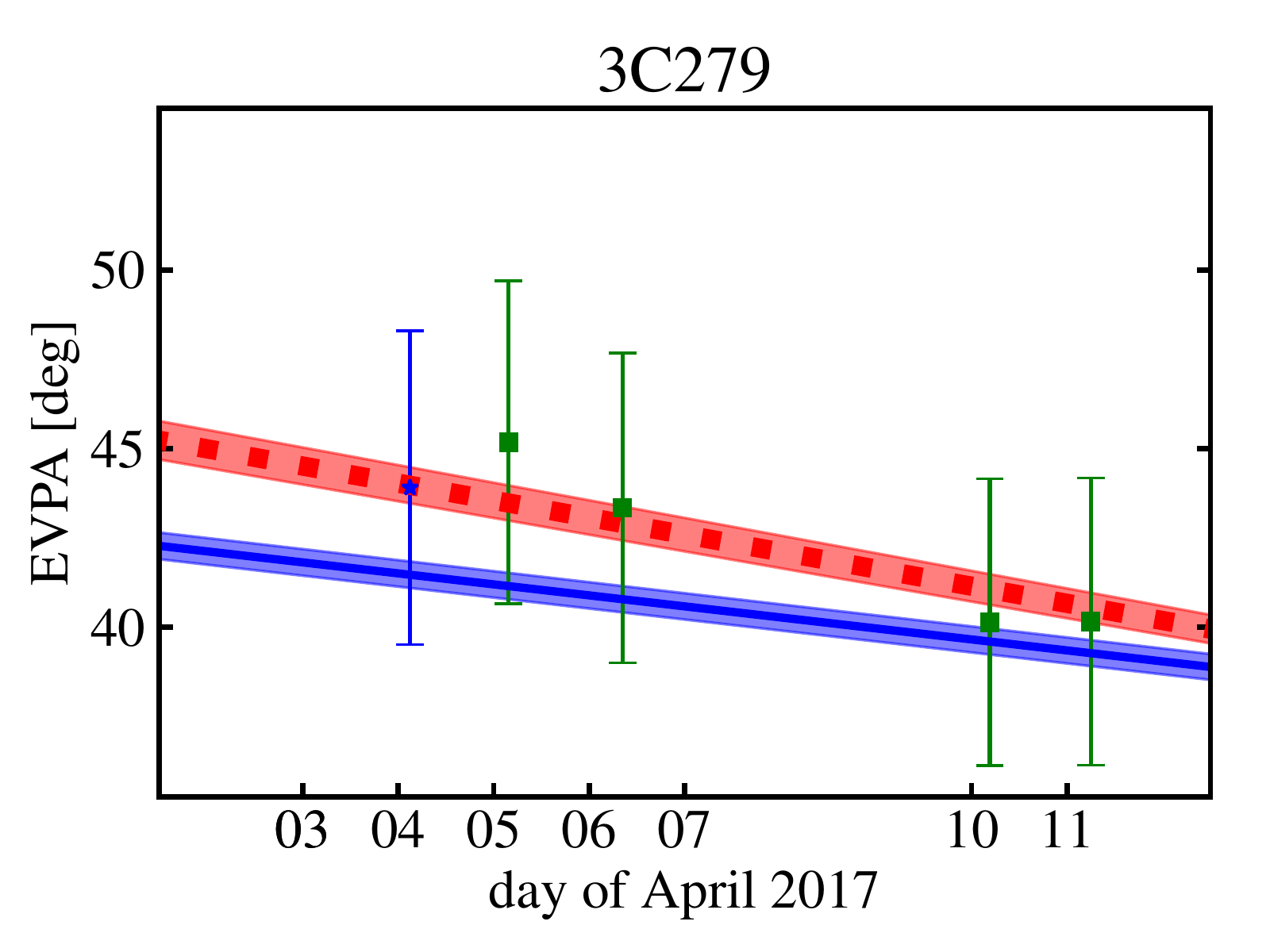} \hspace{-0.3cm}
\includegraphics[width=4cm]{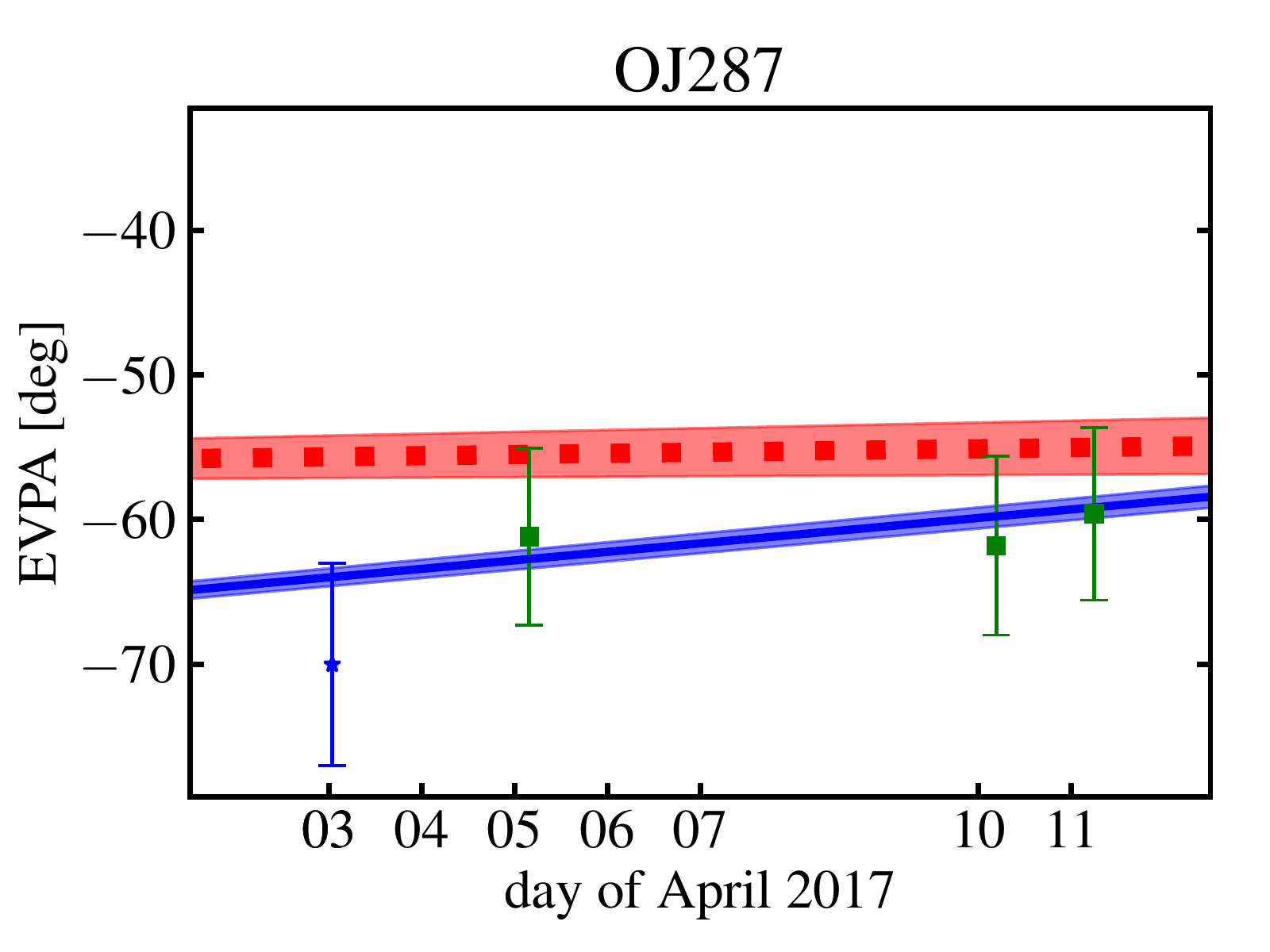}  \hspace{-0.3cm}
\includegraphics[width=4cm]{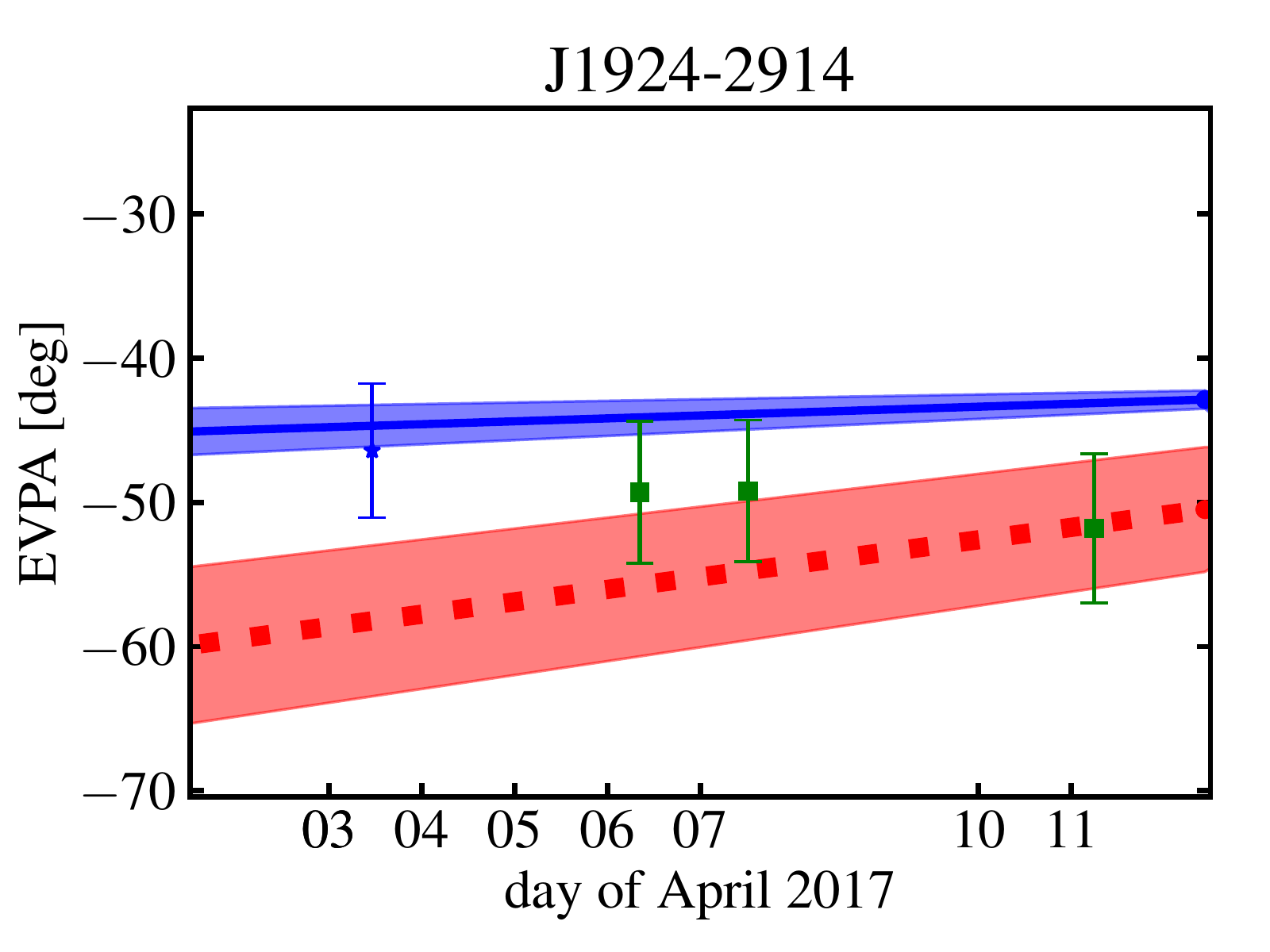}  \hspace{-0.3cm}
\includegraphics[width=4cm]{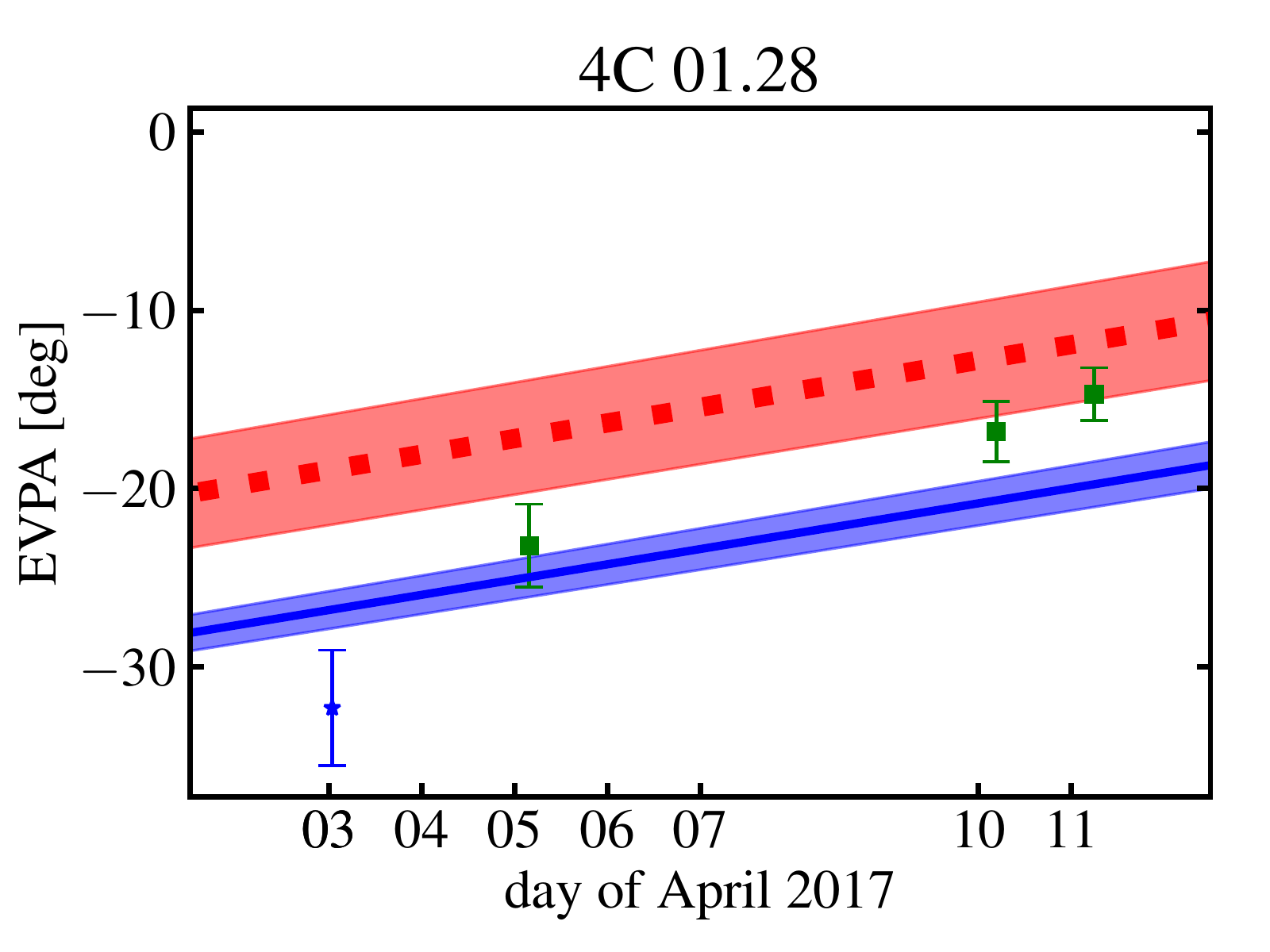}  

\caption{
Comparison with time between the polarimetric results obtained for all the sources observed in the ALMA-VLBI campaign and those retrieved from the AMAPOLA polarimetric analysis of Grid Survey data. Each row shows a parameter (from top to bottom: Stokes~Q, U, and V, LP, and EVPA) while each column corresponds to a source (from left to right: 3C279, OJ287, J1924-2914, and 4C~01.28; see Figure~\ref{fig:stokescomp_gs_2} for more).
Only sources with entries in the ALMA archive (close in time to the observations, i.e., between end of March and April 2017) are displayed. 
The measured flux values during the ALMA-VLBI observations are indicated as data points and corresponding errorbars (blue and green for Band~3 and 6 observations, respectively). 
The shaded regions indicate AMAPOLA's $\pm1\sigma$ uncertainty in Band\,3 (97.5\,GHz; blue shade) and Band\,7 (343.4\,GHz; red shade), respectively.
These are obtained from the ACA GS and their time evolution (lines) are obtained by interpolating between these measurements. 
}
\label{fig:stokescomp_gs_1}
\end{figure*}

\begin{figure*}[ht!]
\centering
\includegraphics[width=4cm]{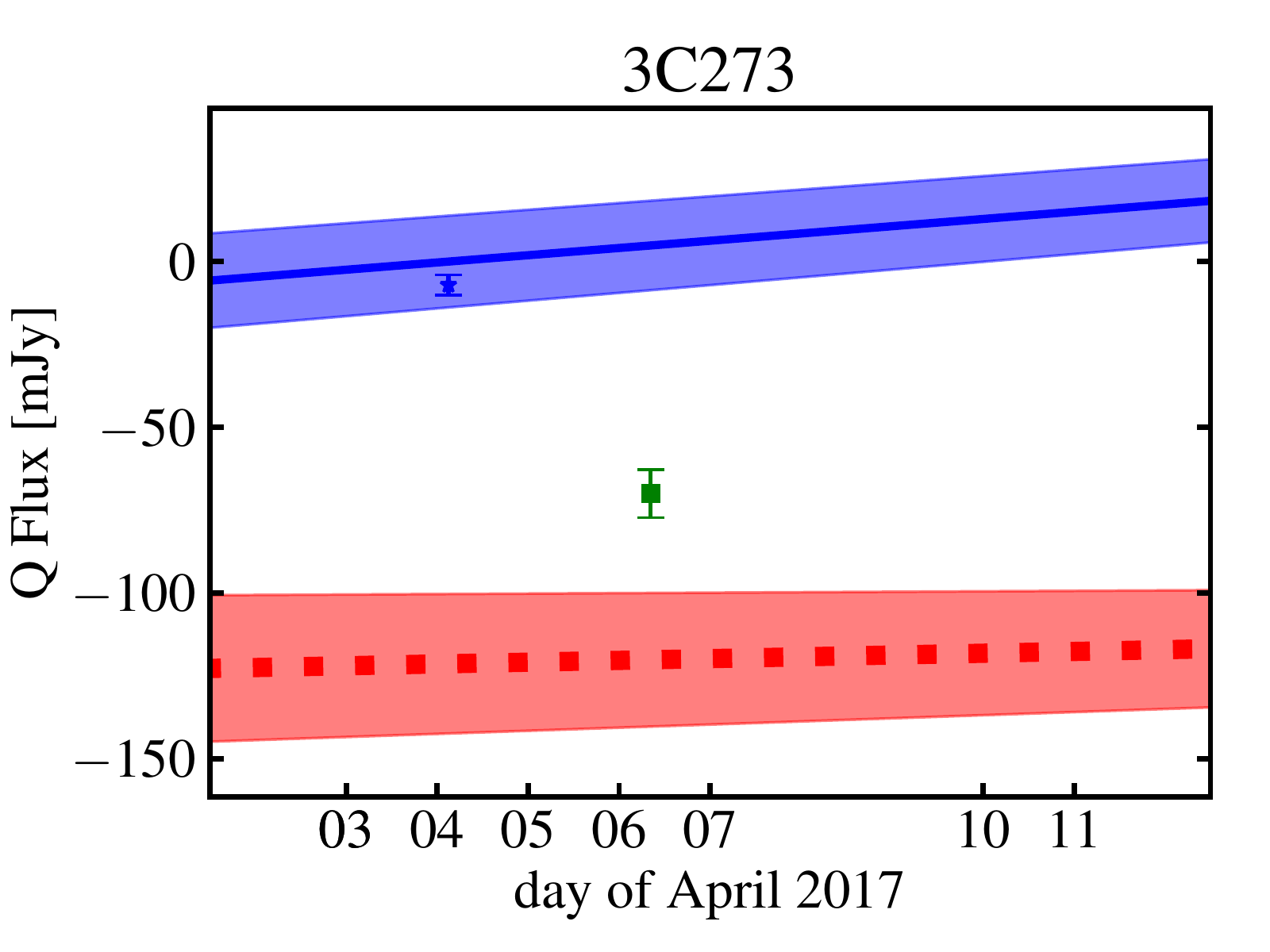} \hspace{-0.3cm}
\includegraphics[width=4cm]{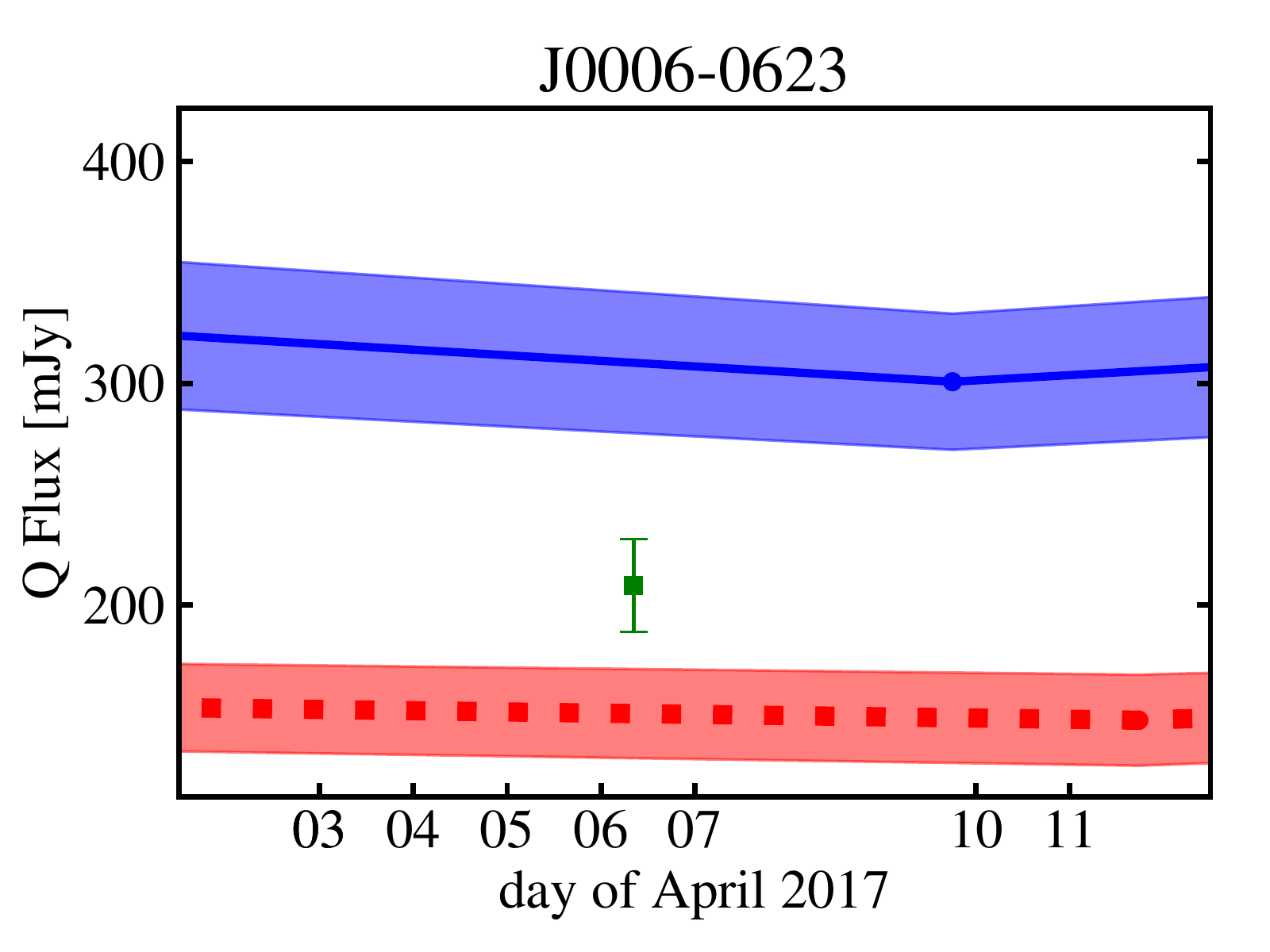}  \hspace{-0.3cm}
\includegraphics[width=4cm]{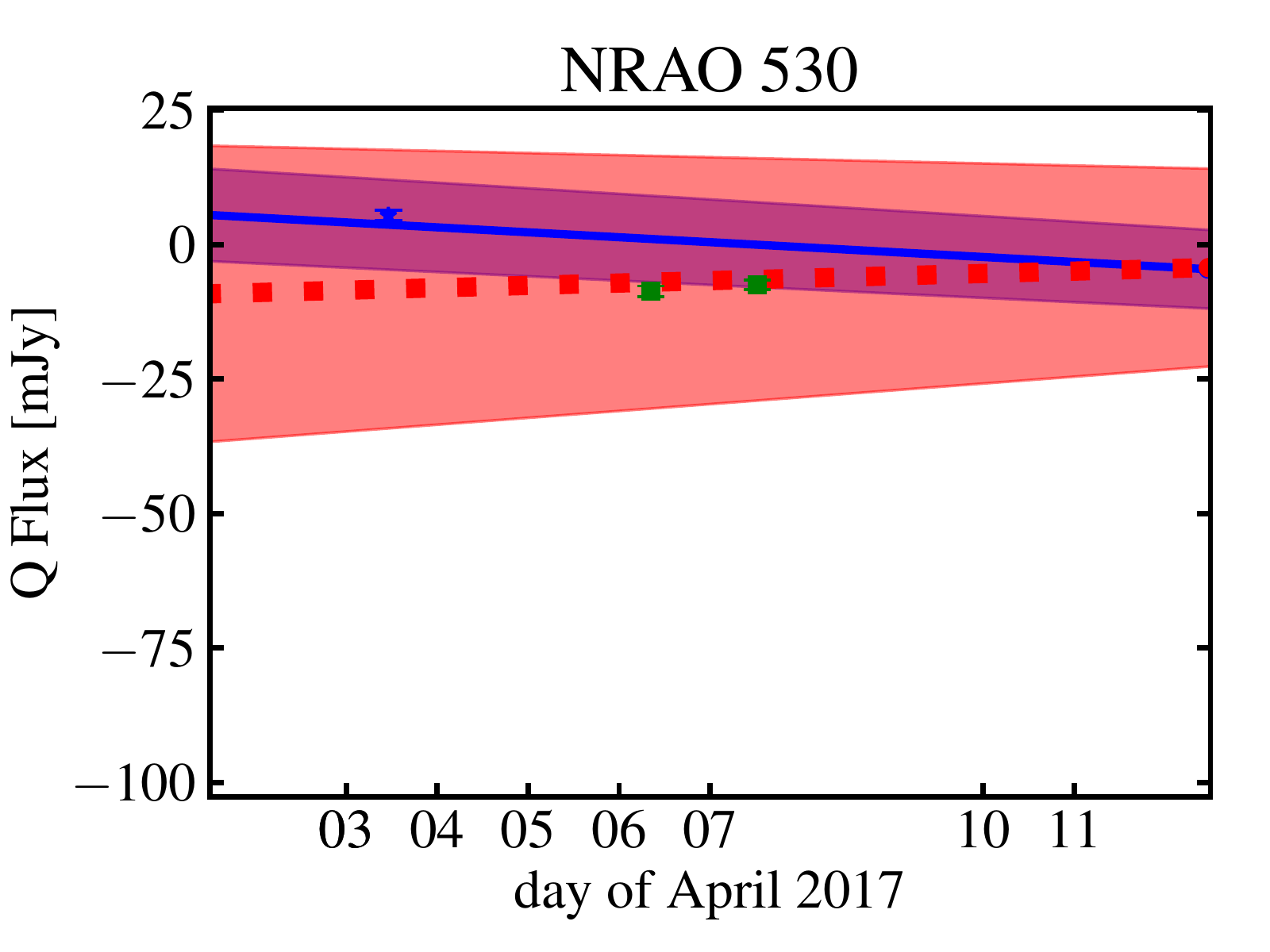}  \hspace{-0.3cm}
\includegraphics[width=4cm]{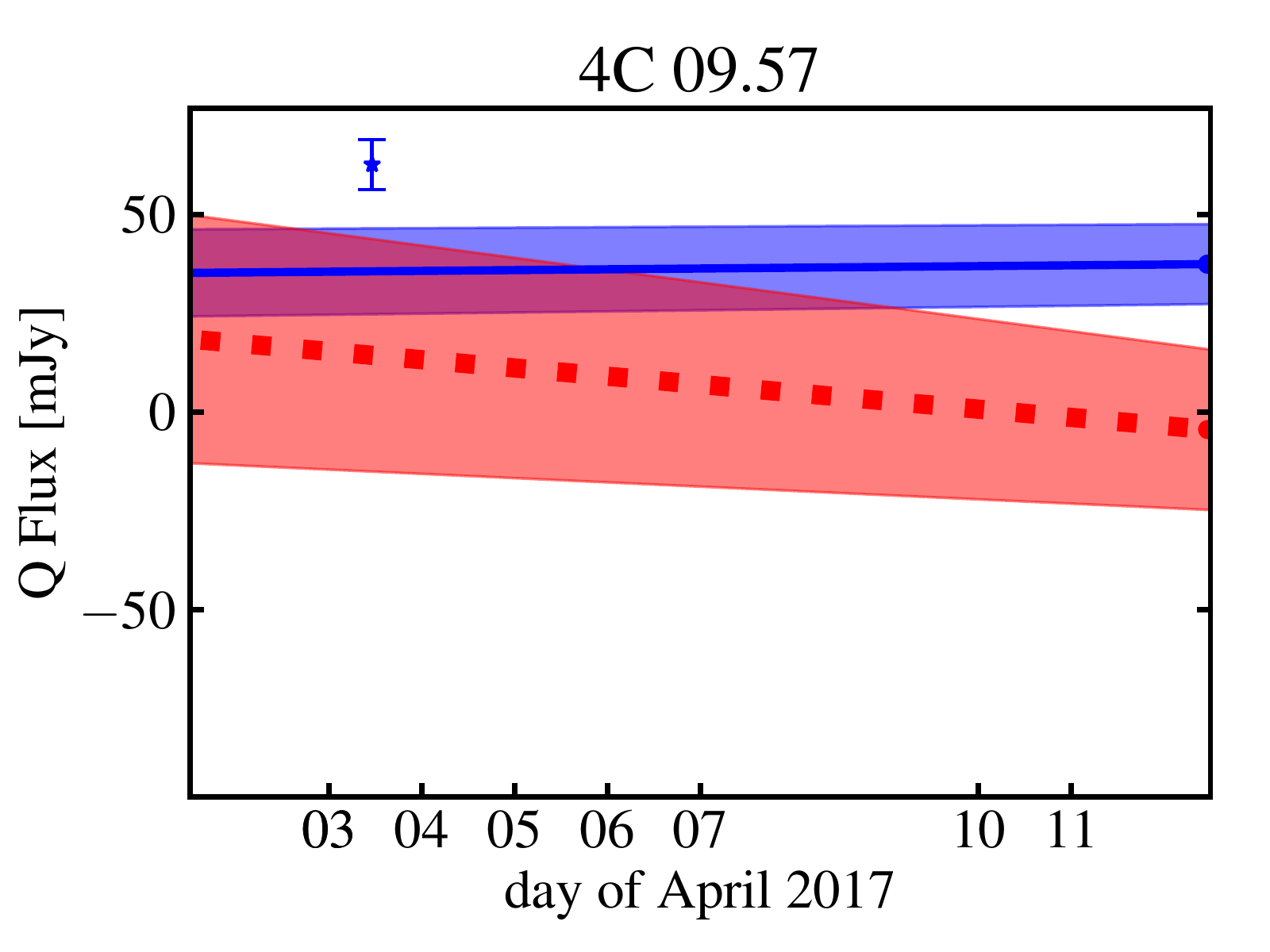}  
\includegraphics[width=4cm]{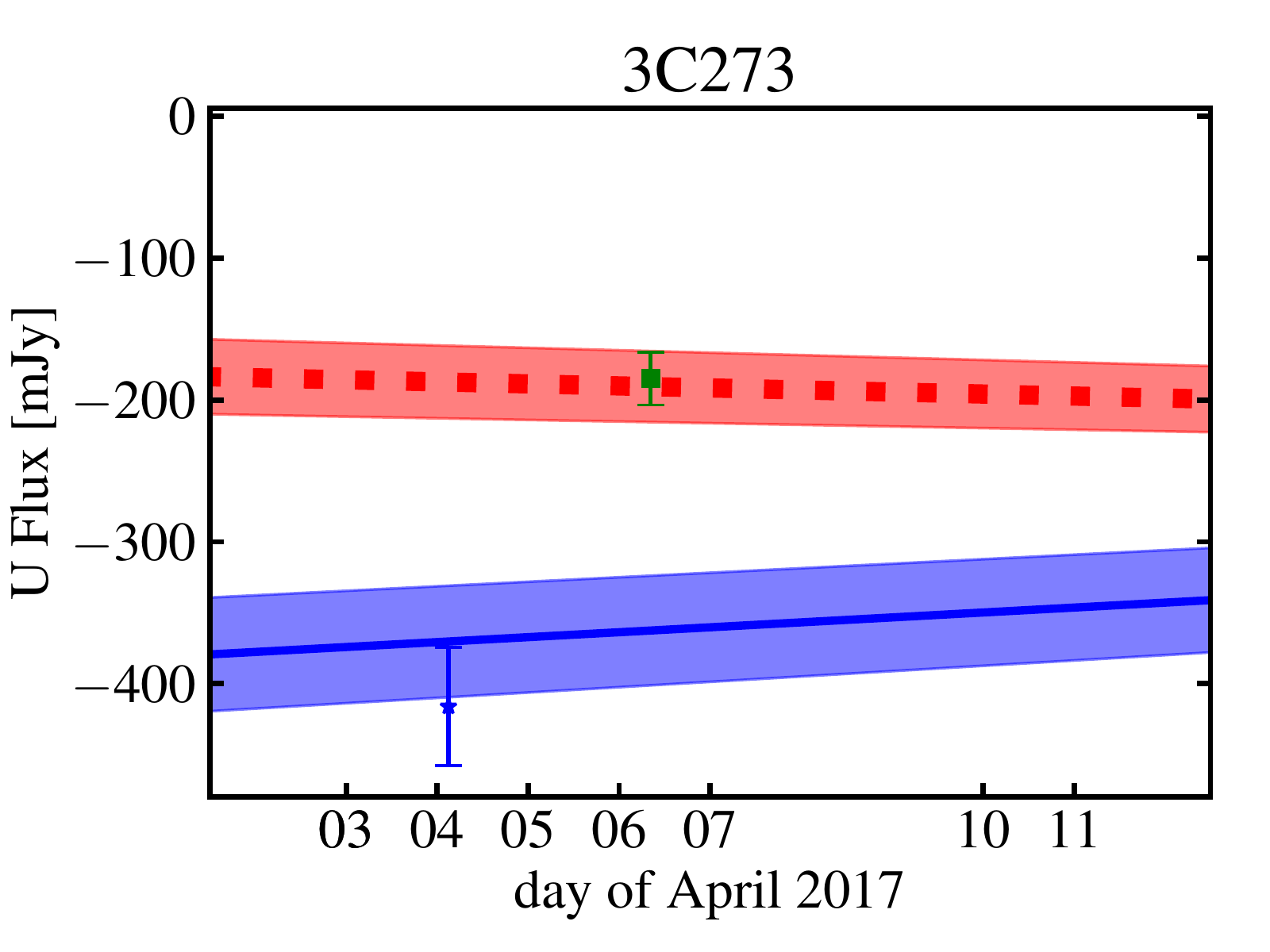} \hspace{-0.3cm}
\includegraphics[width=4cm]{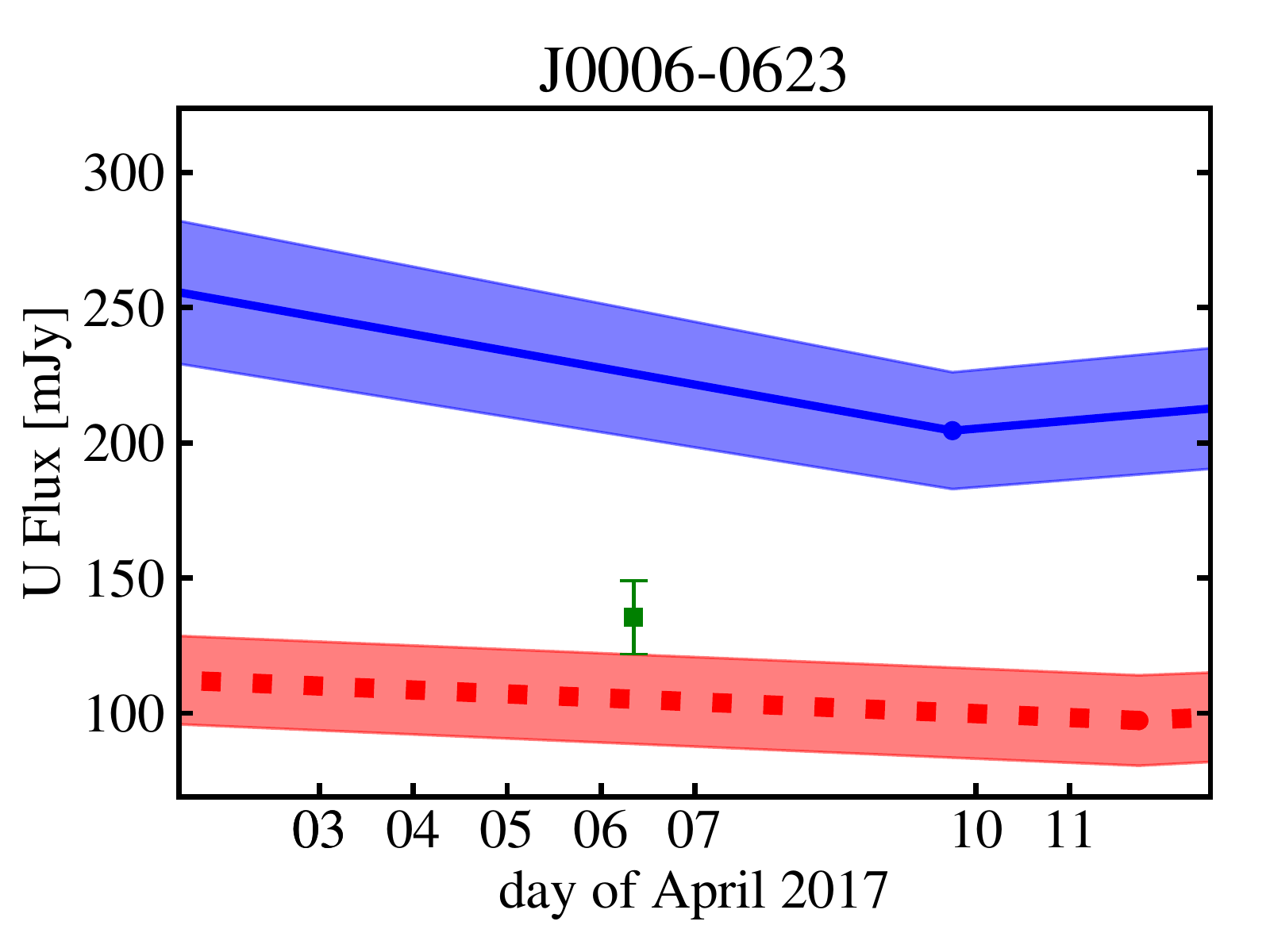}  \hspace{-0.3cm}
\includegraphics[width=4cm]{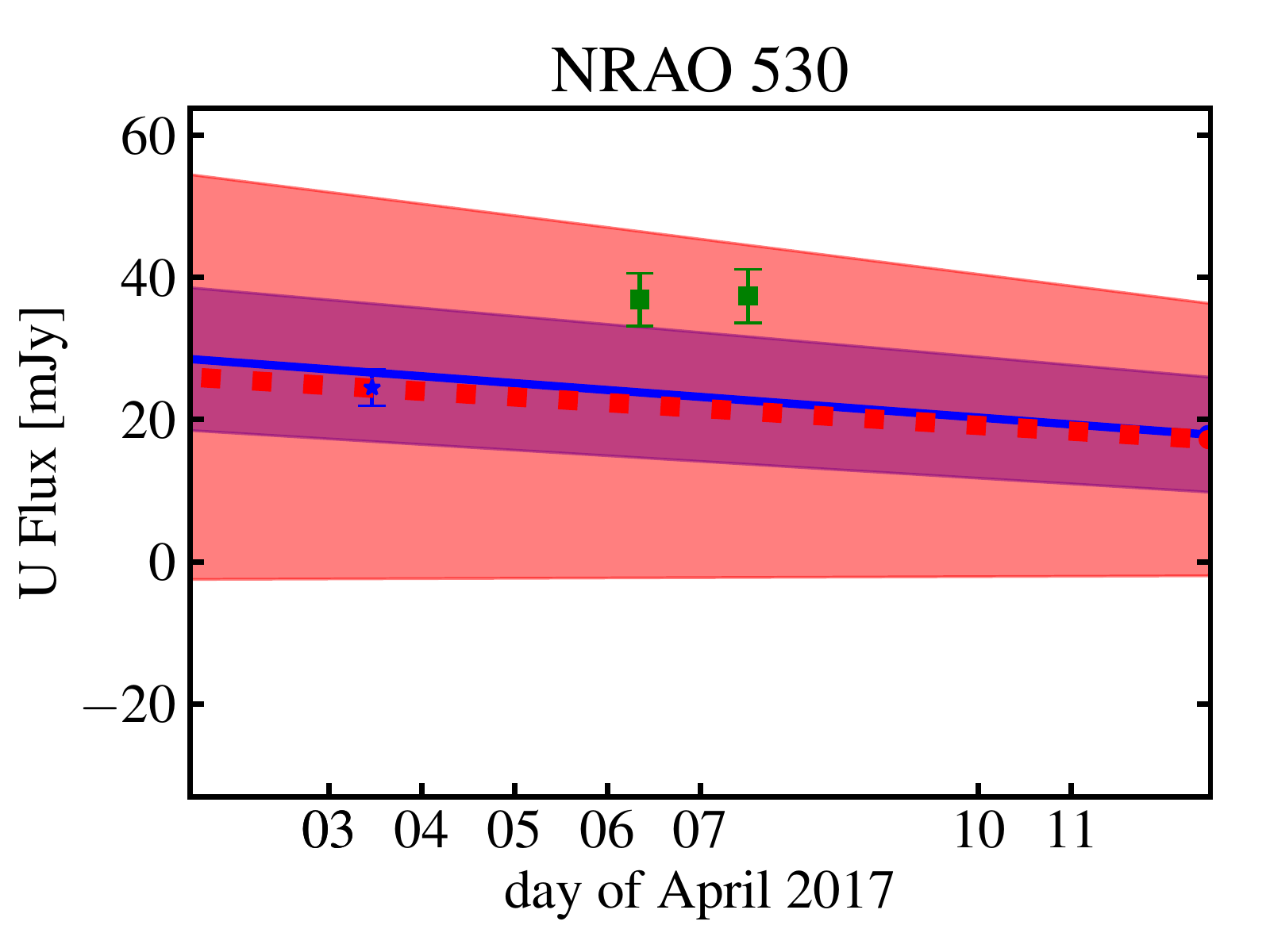}  \hspace{-0.3cm}
\includegraphics[width=4cm]{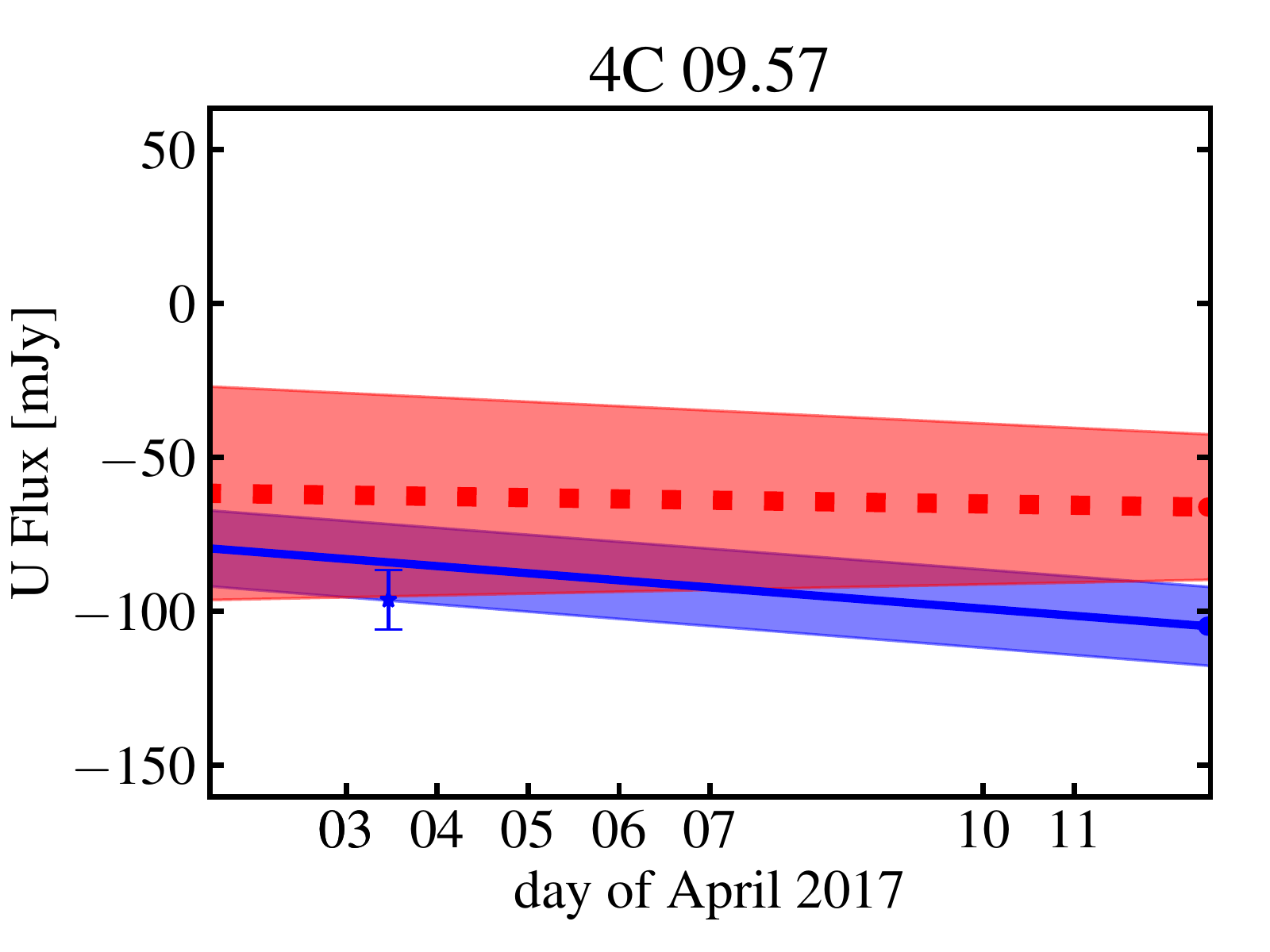}  
\includegraphics[width=4cm]{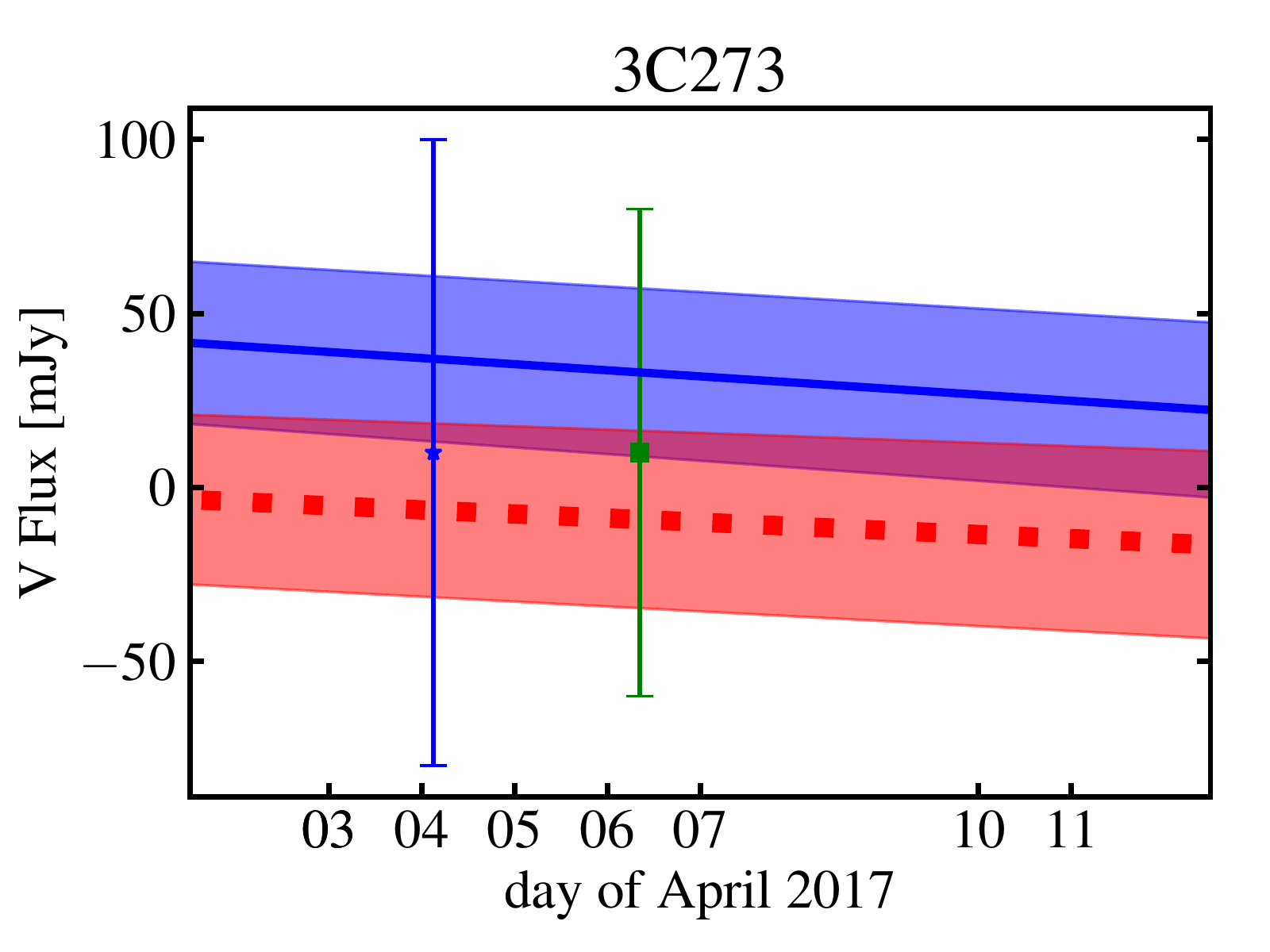} \hspace{-0.3cm}
\includegraphics[width=4cm]{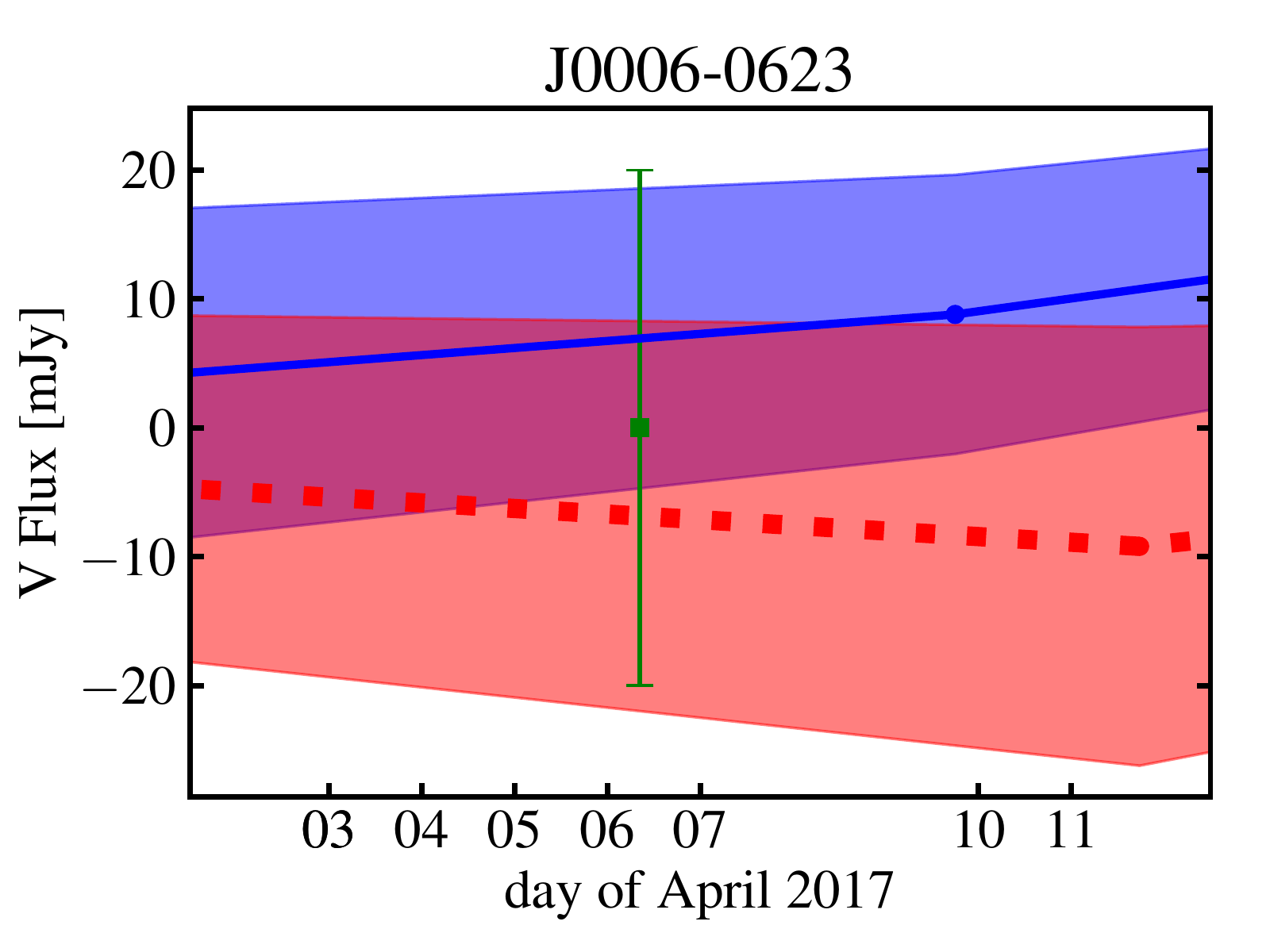}  \hspace{-0.3cm}
\includegraphics[width=4cm]{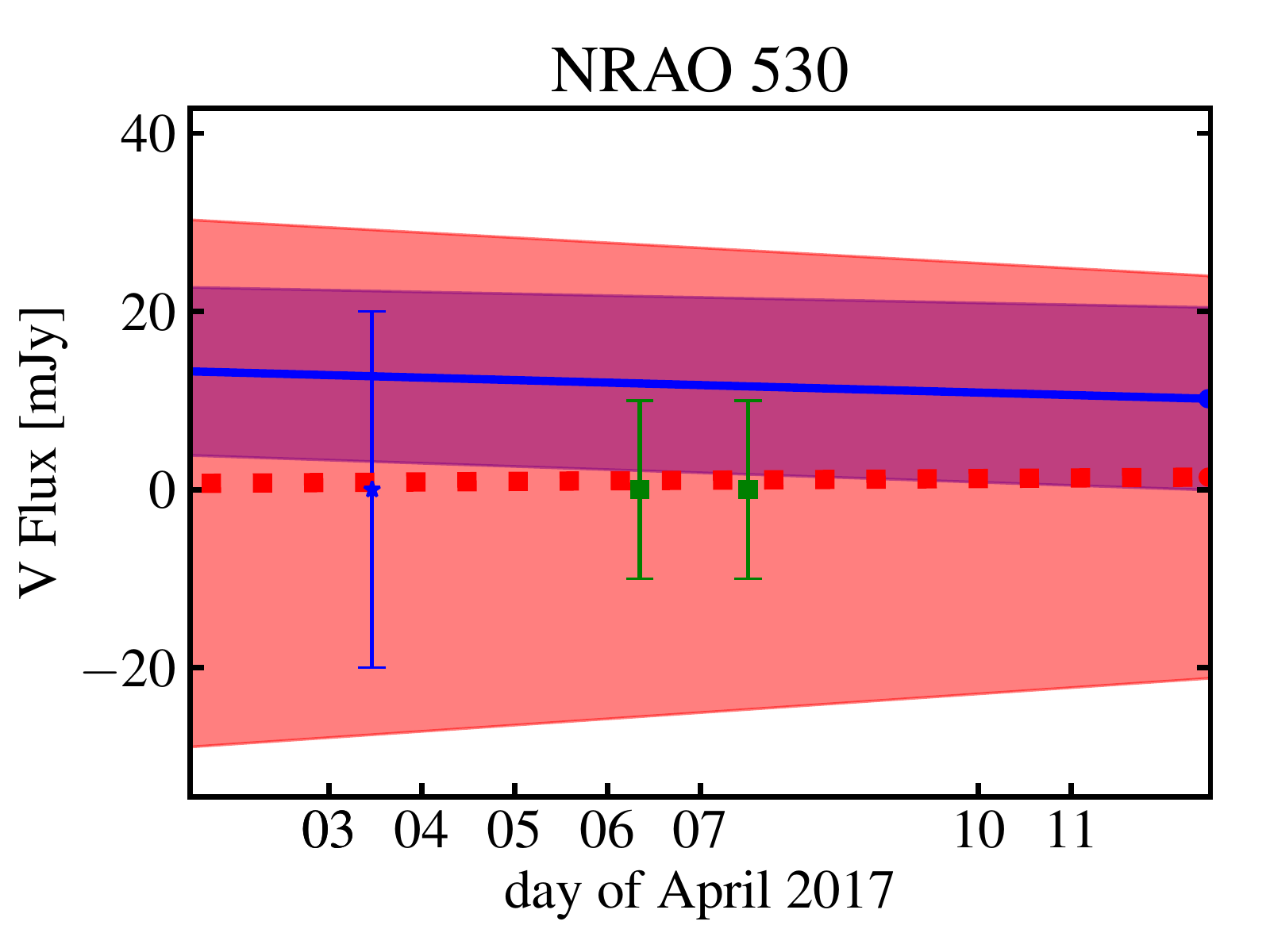}  \hspace{-0.3cm}
\includegraphics[width=4cm]{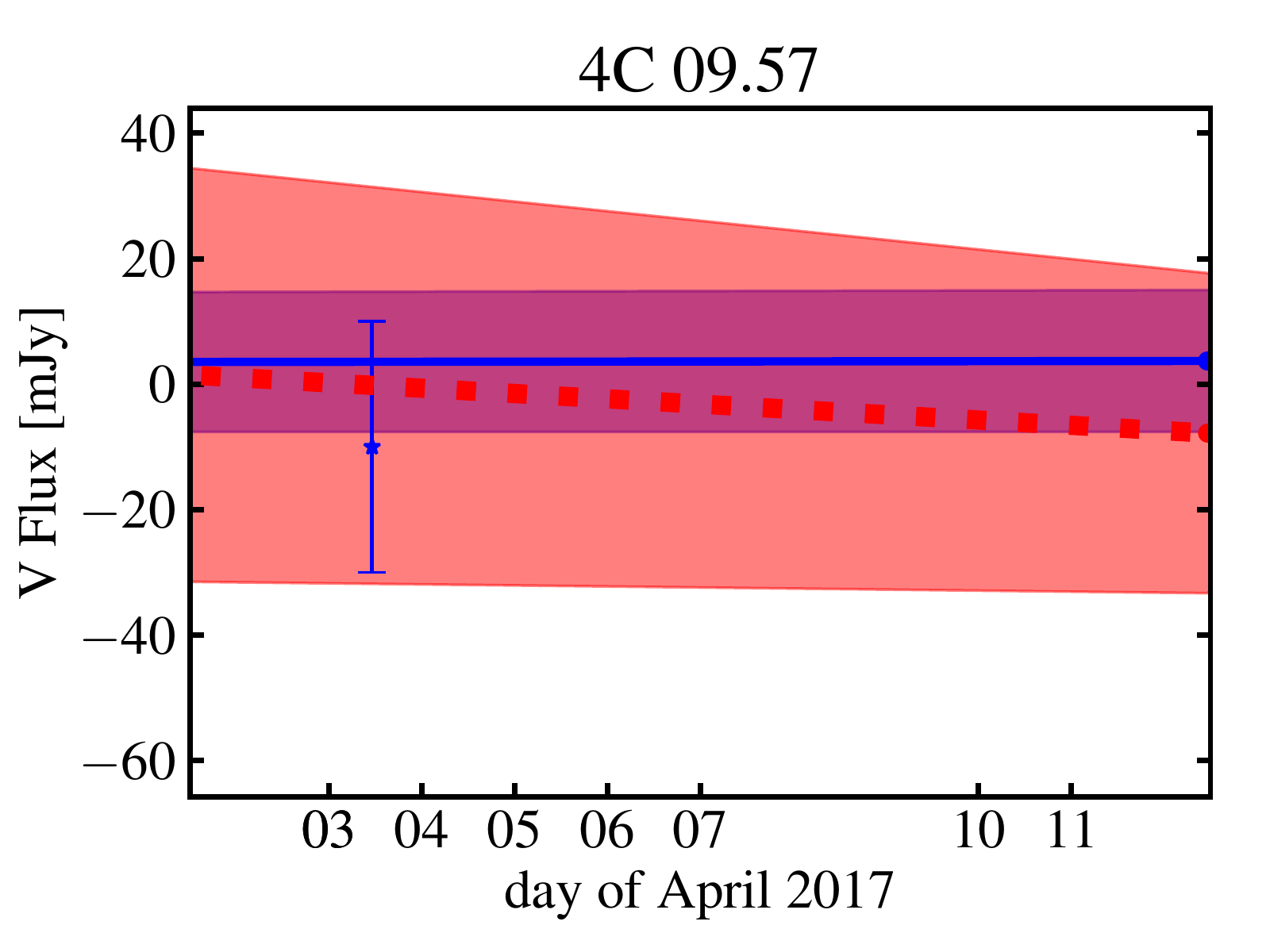}  
\includegraphics[width=4cm]{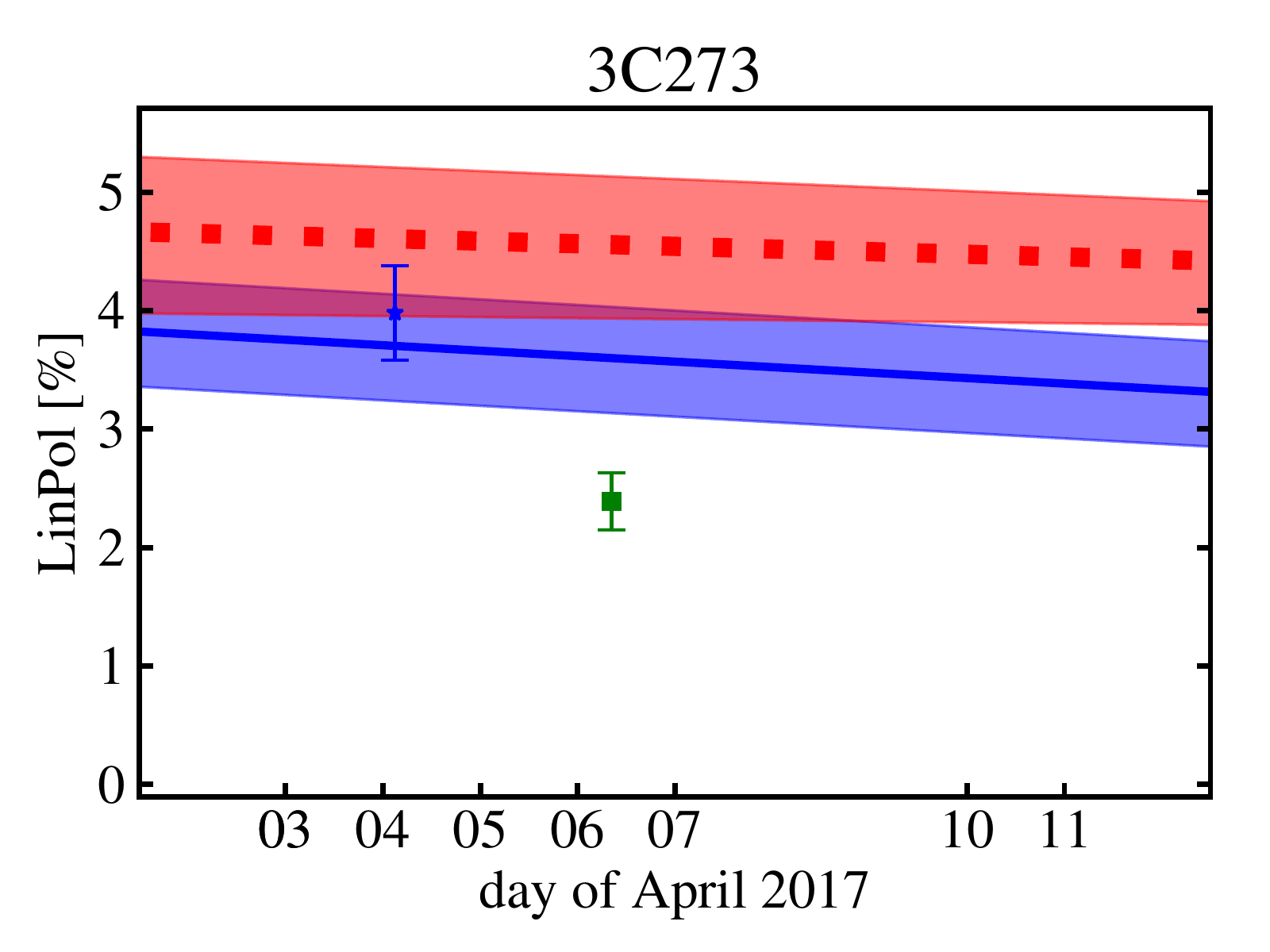} \hspace{-0.3cm}
\includegraphics[width=4cm]{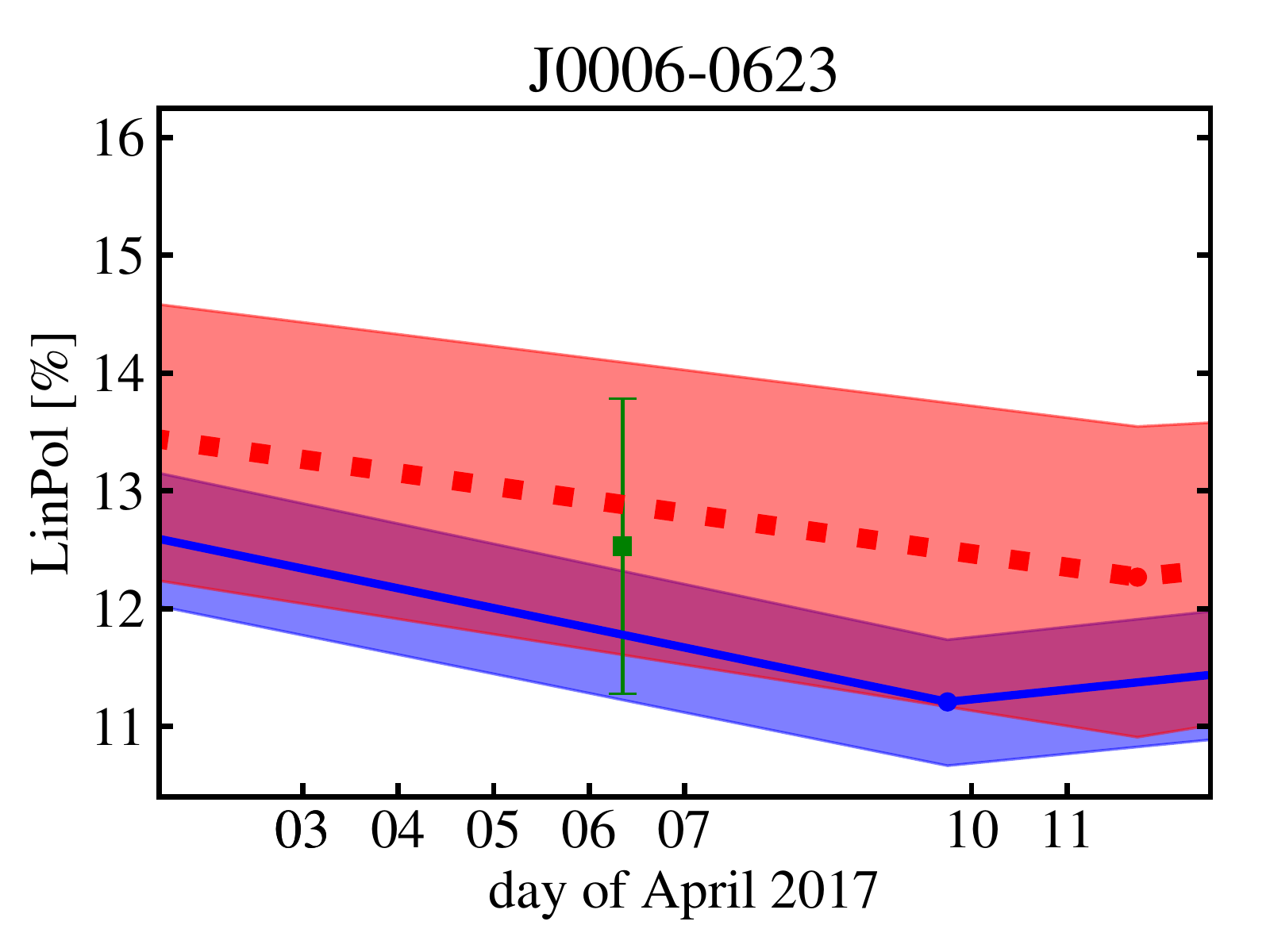}  \hspace{-0.3cm}
\includegraphics[width=4cm]{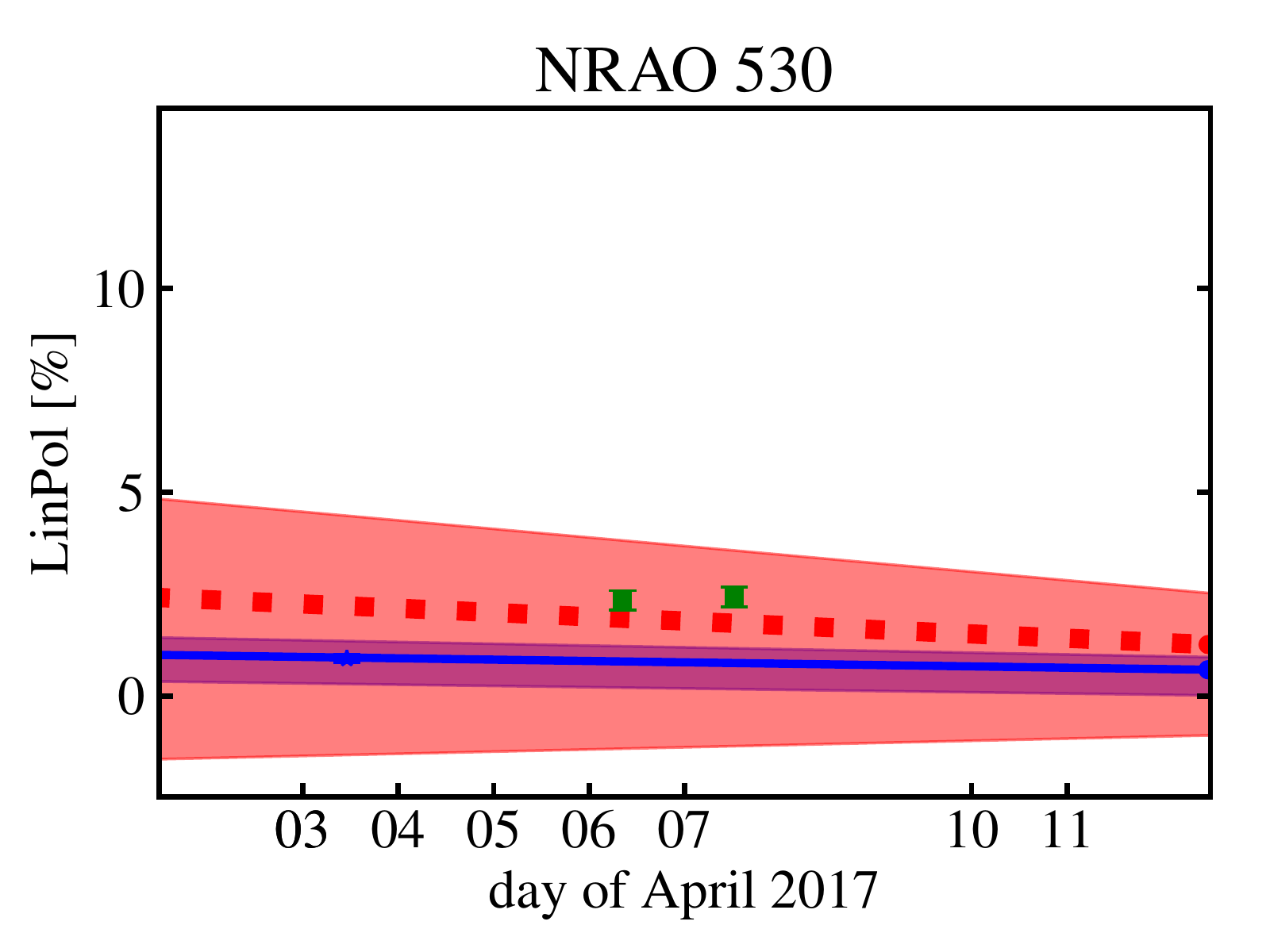}  \hspace{-0.3cm}
\includegraphics[width=4cm]{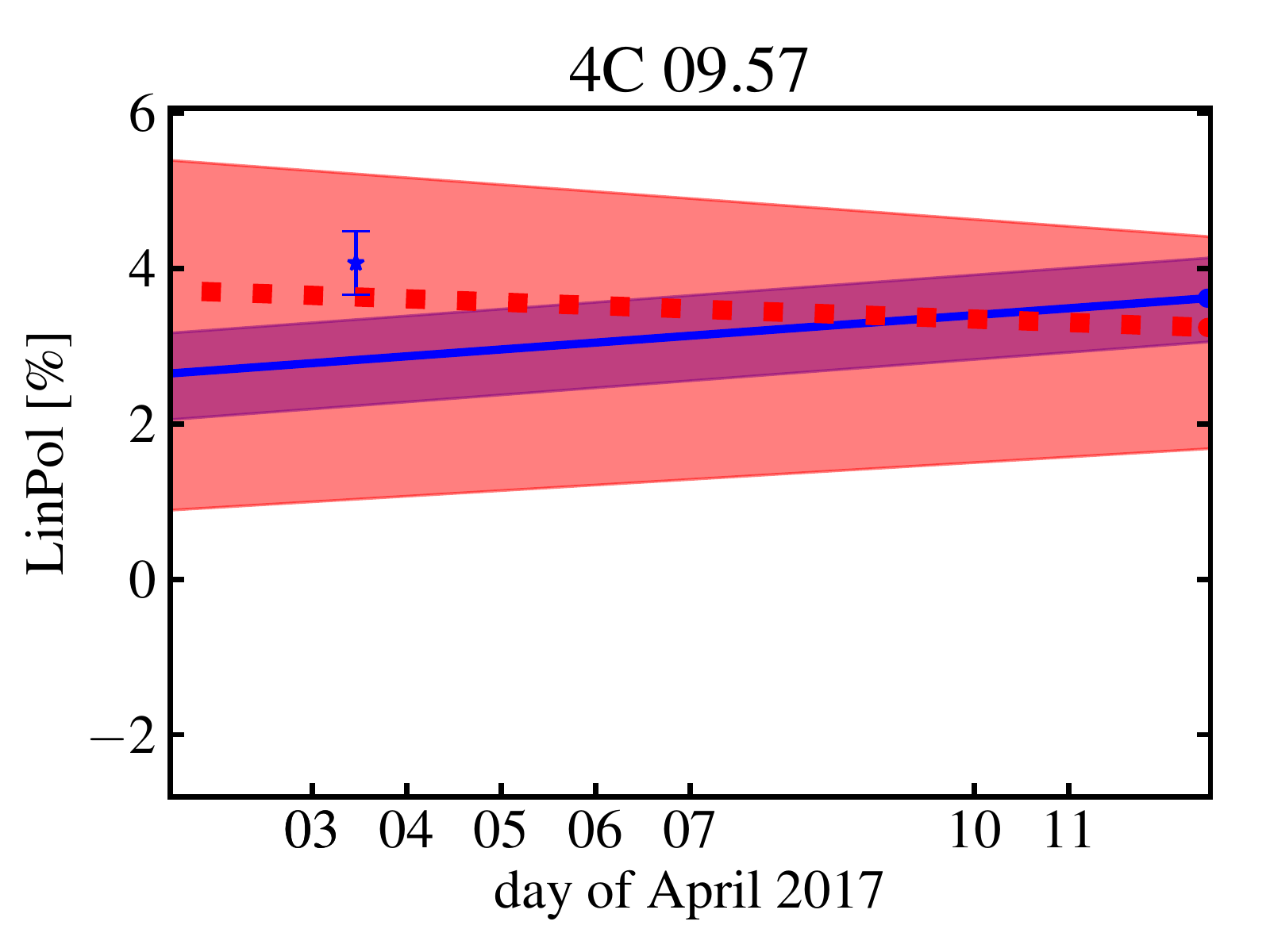}   
\includegraphics[width=4cm]{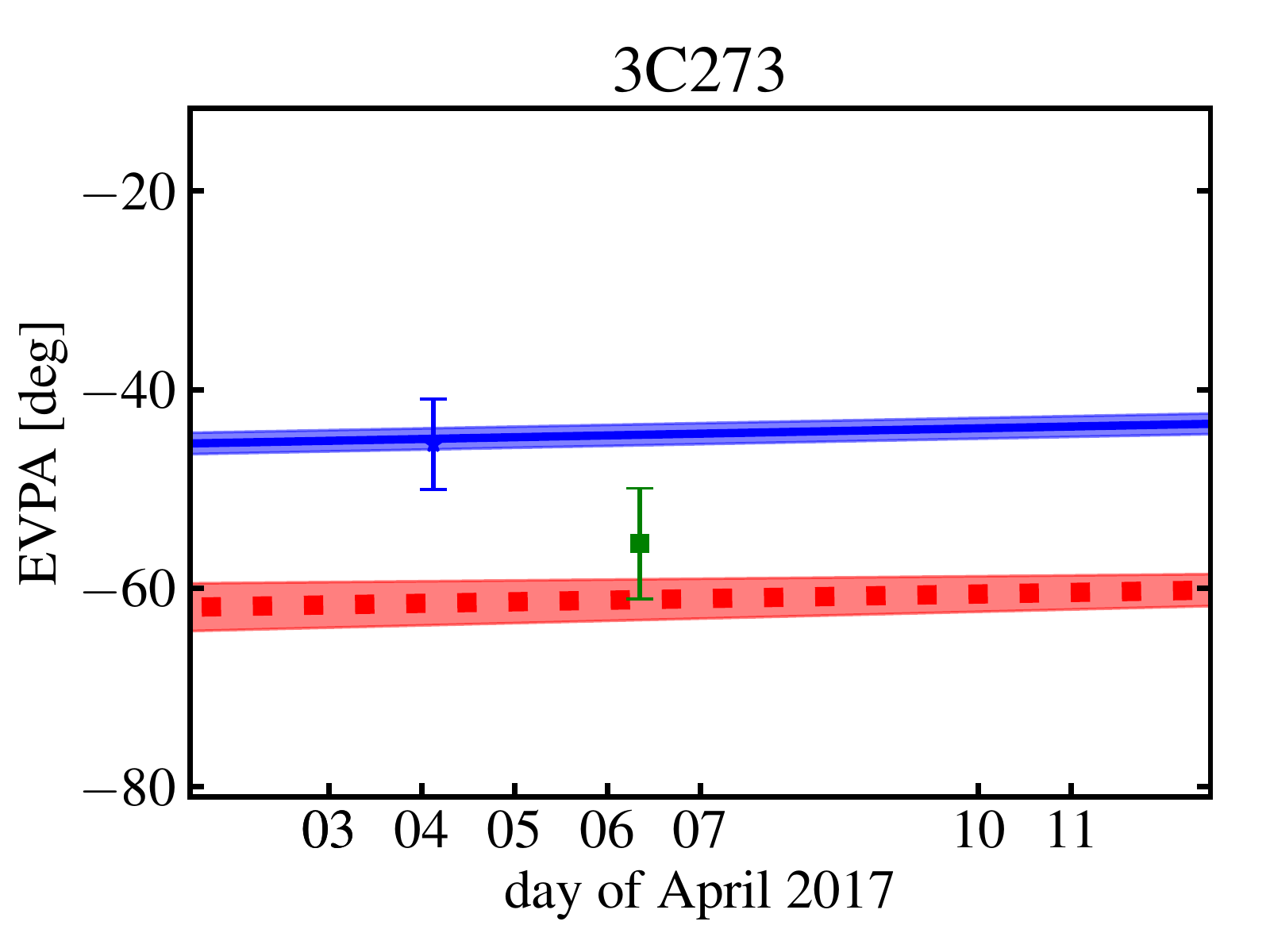} \hspace{-0.3cm}
\includegraphics[width=4cm]{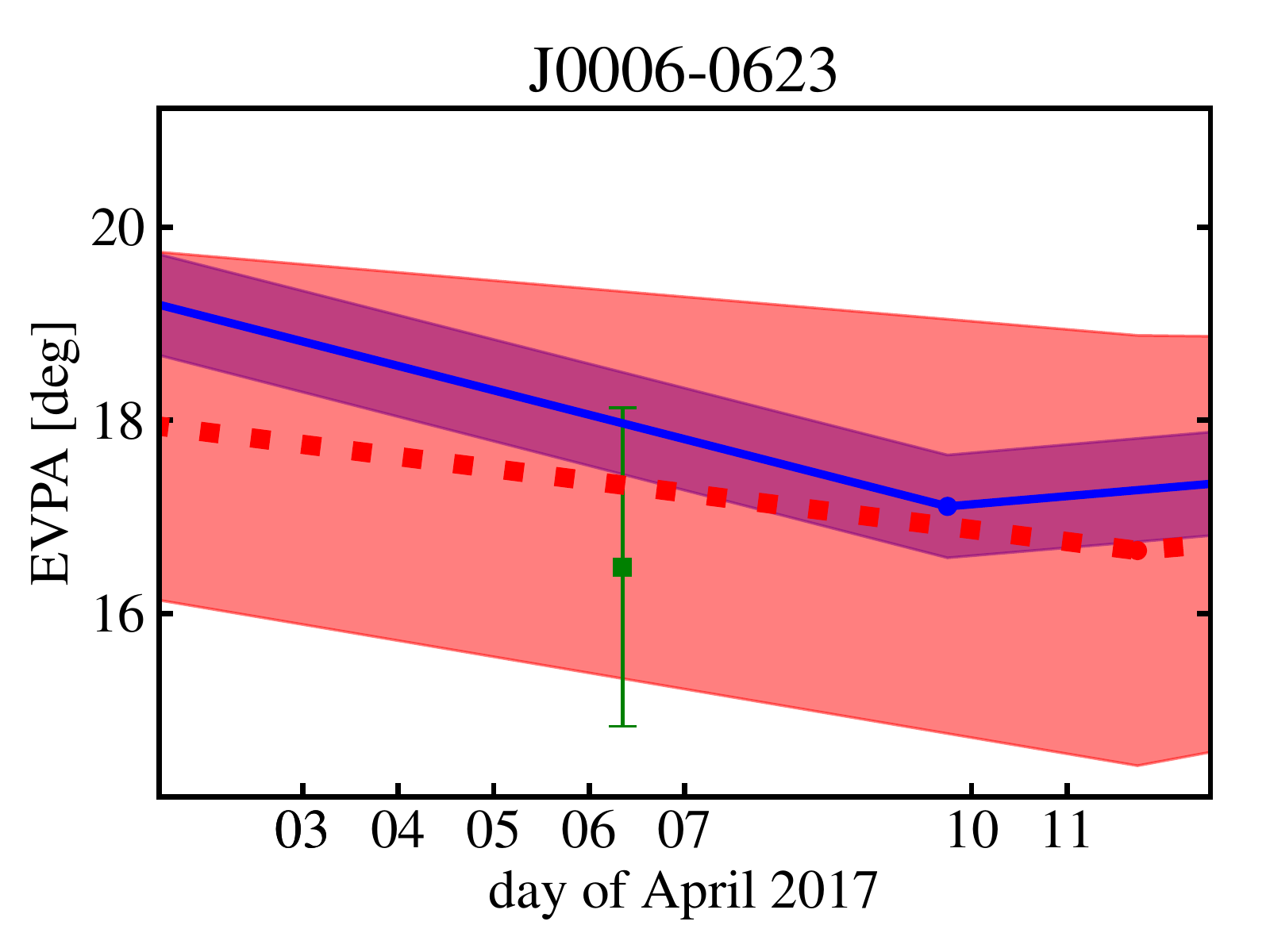}  \hspace{-0.3cm}
\includegraphics[width=4cm]{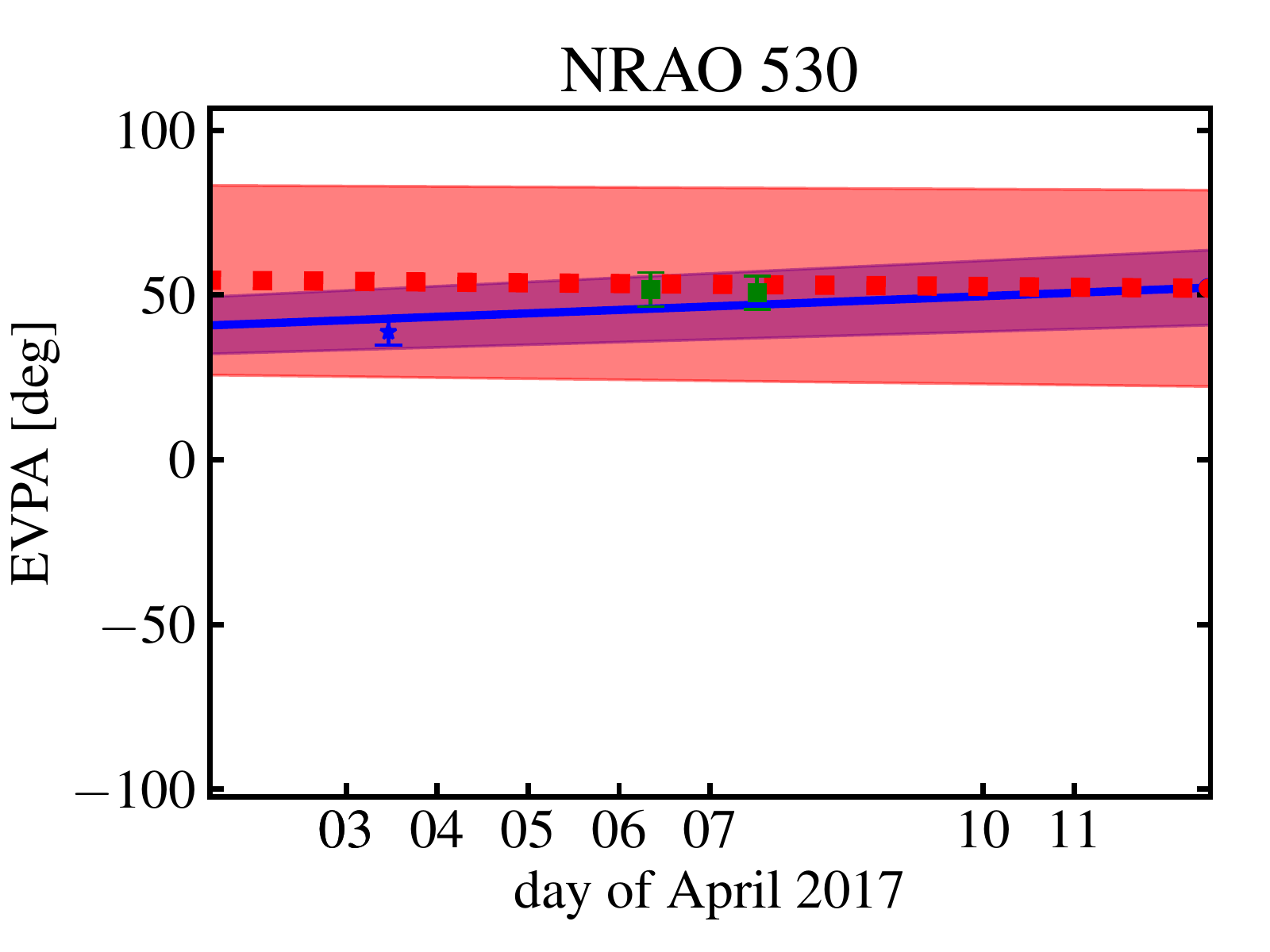}  \hspace{-0.3cm}
\includegraphics[width=4cm]{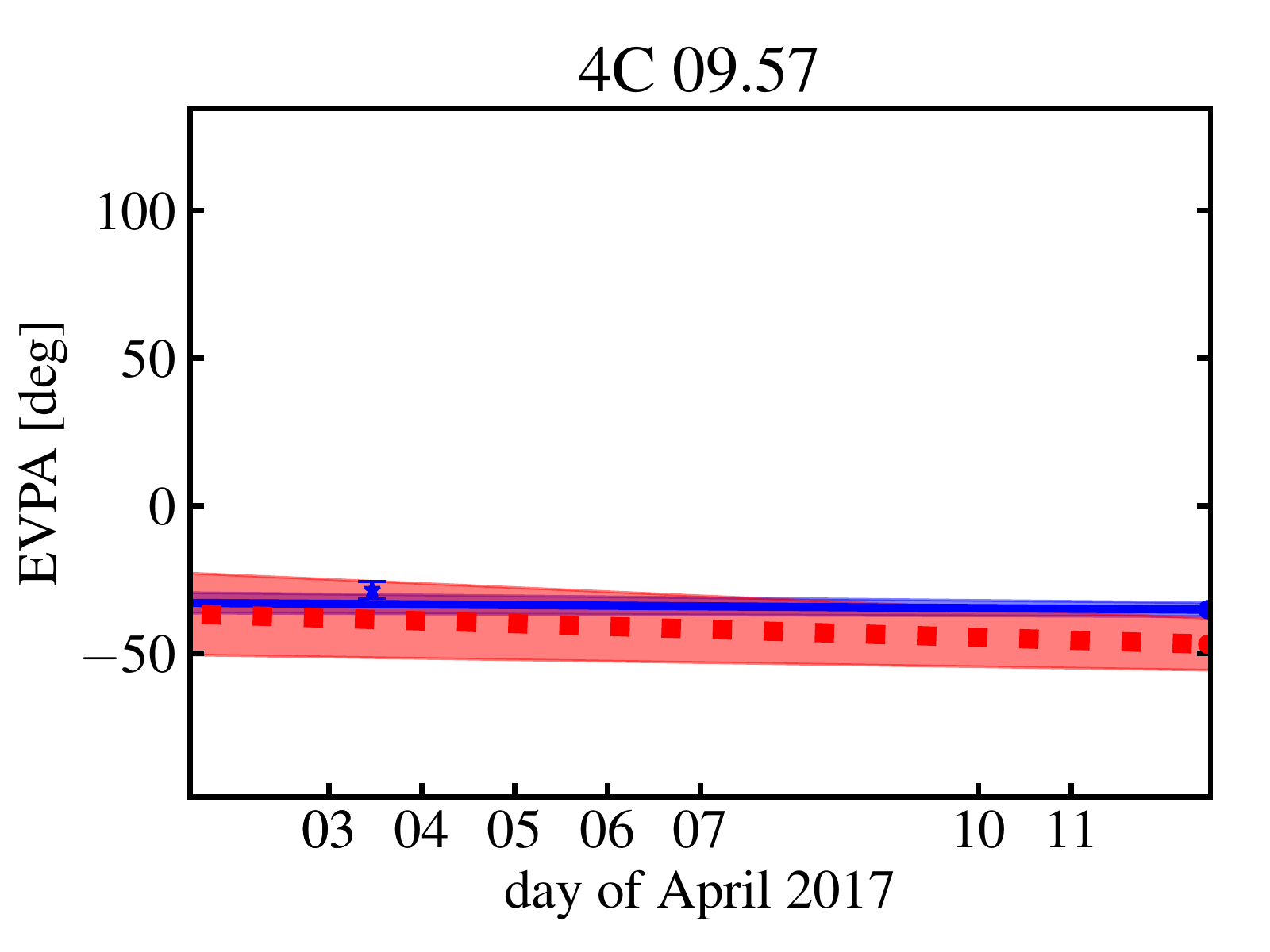} 

\caption{
Same as Fig.~\ref{fig:stokescomp_gs_1}, but for different sources (from left to right: 3C273, J0006-0623, NRAO 530, 4C~09.57). We note that source J0006-0623 was observed on Apr 7 but the calibrated data were of poor quality and were flagged before analysis (see \S~\ref{sect:datacal}). 
}
\label{fig:stokescomp_gs_2}
\end{figure*}

Figures~\ref{fig:stokescomp_gs_1} and \ref{fig:stokescomp_gs_2} show the observed polarimetry parameters for the 2017 VLBI sessions (data points and errorbars), specifically $Q$, $U$, and $V$ Stokes parameters, LP, and EVPA. The shaded $\pm1\sigma$ regions highlight the time variance of the same parameters as measured by AMAPOLA (an inflection in the trend means a time of GS observation). The colour-coding is such that blue refers to Band~3 measurements, green to Band~6, and red to Band~7 ones. 
These figures show that most Band~3 ALMA-VLBI measurements  fall within the blue regions, while most Band~6 measurements fall in-between the blue and red regions, indicating that our measurements are broadly consistent with the AMAPOLA trends.  Some potential conflicts can still be consistent with the inter-GS-cadence variability or differential time-variability between frequency bands (as also observed in some cases in the AMAPOLA monitoring). 
We conclude that, despite the different array specifications and data-reduction schemes, the ALMA-VLBI and AMAPOLA results are consistent between each other.

\section{Stokes parameters per ALMA frequency band (SPW)}
\label{app:stokes}

We report the polarimetric quantities (Stokes\,$IQU$, LP, EVPA)  per SPW in Tables~\ref{tab:GMVA_uvmf_spw} (GMVA sources) and~\ref{tab:EHT_uvmf_spw} (EHT sources). 
The Stokes parameters were fitted directly on the visibilities using \uvmultifit\  (see \S\ref{allstokes}).  Uncertainties are assessed with MC simulations, as the standard deviation of 1000 MC simulations for each Stokes parameter. 
The quoted uncertainties  include in quadrature the fitting error and 
the Stokes~$I$ leakage onto Stokes~$QU$ (0.03\% of $I$),  
as recommended by the ALMA observatory.
The LP uncertainty is dominated by such  systematic error, except for the weakest sources J0132-1654 and NGC~1052, for which the thermal noise starts to dominate. 

In order to correct for the LP bias in the low SNR regime,  
we estimate a {\it debiased} LP as $\sqrt{LP^2-\sigma_{LP^2}}$, where 
$LP =\sqrt{Q^2+U^2}/I$. \\
$\sigma_{LP}$ can be estimated using propagation of errors: 
\begin{equation*}
  \begin{aligned}
  \sigma_{LP}^2 I^2  = & 
{\frac{(Q \sigma_Q)^2 +  (U \sigma_U)^2 + 2 QU \sigma_{QU}}{Q^2+U^2}} + \\
& [(\frac{Q}{I})^2 + (\frac{U}{I})^2 ] \sigma^2_I - 2 \frac{Q}{I}  \sigma_{QI} - 2 \frac{U}{I}  \sigma_{UI}   
\end{aligned}
\end{equation*}
where $\sigma_{I}$, $\sigma_{Q}$, $\sigma_{U}$ are the uncertainties in $I$, $Q$ and $U$, respectively. 
Assuming  $\sigma_{QU} \sim \sigma_{QI} \sim \sigma_{UI} \sim  0$, then 
\begin{equation*}
  \begin{aligned}
  \sigma_{LP}  = \frac{1}{I} \sqrt{{\frac{(Q \sigma_Q)^2 +  (U \sigma_U)^2}{Q^2+U^2}}+ 
 [(\frac{Q}{I})^2 + (\frac{U}{I})^2 ] \sigma^2_I}.  
\end{aligned}
\end{equation*}

The LP values quoted in Tables~\ref{tab:GMVA_uvmf_RM}, \ref{tab:EHT_uvmf_RM}, \ref{tab:GMVA_uvmf_spw}, and \ref{tab:EHT_uvmf_spw} have this debiased correction applied. 
The latter does not affect most of the sources studied here, but it is especially important for the low-polarization sources such as NGC~1052 and Cen~A. 

The absolute flux-scale calibration systematic error  (5\% and 10\% of Stokes~$IQU$ fluxes in Band~3 and Band~6, respectively), is not included in  Tables~\ref{tab:GMVA_uvmf_spw} and \ref{tab:EHT_uvmf_spw}.  
The analysis on Stokes~$V$ is performed separately (see Appendix~\ref{app:stokesV} and Tables~\ref{tab:GMVA_uvmf_CP}, \ref{tab:EHT_uvmf_CP}). 

\begin{longtable*}{cccccc}
\caption{Polarization parameters of GMVA sources per frequency band (spw) and per day. } \label{tab:GMVA_uvmf_spw}\\
\hline\hline 
Frequency & I & Q & U & LP & EVPA\\
\ [GHz] & [mJy] & [mJy] & [mJy] & [\%] & [deg]\\
\hline\hline 
 &&&&& \\
 &&&&& \\
\hline
\multicolumn{6}{c}{Apr 2}\\
\hline
\multicolumn{6}{c}{OJ287}\\
86.3                &   6242.41$\pm$0.26    &    -409.6$\pm$1.9    &    -351.2$\pm$1.9     &    8.643$\pm$0.030 & -69.69$\pm$0.10 \\
88.3                &   6170.33$\pm$0.26    &    -410.2$\pm$1.9    &    -347.5$\pm$1.9     &    8.711$\pm$0.030 & -69.87$\pm$0.10 \\
98.3                &   5787.47$\pm$0.28    &    -398.9$\pm$1.7    &    -327.8$\pm$1.7     &    8.919$\pm$0.030 & -70.30$\pm$0.10 \\
100.3                &   5674.69$\pm$0.27    &    -392.4$\pm$1.7    &    -324.2$\pm$1.7     &    8.970$\pm$0.030 & -70.214$\pm$0.093 \\
\hline
\multicolumn{6}{c}{J0510+1800}\\
86.3                &   3259.30$\pm$0.57    &    -127.5$\pm$1.0    &    22.6$\pm$1.0     &    3.976$\pm$0.030 & 84.97$\pm$0.21 \\
88.3                &   3208.82$\pm$0.59    &    -126.2$\pm$1.0    &    31.9$\pm$1.0     &    4.054$\pm$0.030 & 82.91$\pm$0.20 \\
98.3                &   2996.03$\pm$0.62    &    -120.30$\pm$0.90    &    46.96$\pm$0.91     &    4.311$\pm$0.030 & 79.33$\pm$0.20 \\
100.3                &   2962.36$\pm$0.63    &    -121.52$\pm$0.90    &    43.17$\pm$0.90     &    4.353$\pm$0.030 & 80.22$\pm$0.21 \\
\hline
\multicolumn{6}{c}{4C 01.28}\\
86.3                &   5039.680$\pm$0.072    &    90.9$\pm$1.5    &    -195.6$\pm$1.5     &    4.281$\pm$0.030 & -32.54$\pm$0.20 \\
88.3                &   4970.067$\pm$0.055    &    93.2$\pm$1.5    &    -192.6$\pm$1.5     &    4.306$\pm$0.030 & -32.09$\pm$0.20 \\
98.3                &   4745.126$\pm$0.053    &    92.5$\pm$1.4    &    -194.1$\pm$1.4     &    4.533$\pm$0.030 & -32.26$\pm$0.20 \\
100.3                &   4665.997$\pm$0.054    &    92.1$\pm$1.4    &    -192.1$\pm$1.4     &    4.566$\pm$0.030 & -32.18$\pm$0.18 \\
 &&&&& \\
\hline
\multicolumn{6}{c}{Apr 3}\\
\hline
\multicolumn{6}{c}{Sgr A*}\\
86.3                &   2574.84$\pm$0.87    &    9.28$\pm$0.77    &    -3.64$\pm$0.77     &    0.388$\pm$0.030 & -10.7$\pm$2.2 \\
88.3                &   2480.44$\pm$0.82    &    12.51$\pm$0.74    &    -2.36$\pm$0.74     &    0.513$\pm$0.030 & -5.3$\pm$1.7 \\
98.3                &   2514.64$\pm$0.74    &    18.44$\pm$0.75    &    16.77$\pm$0.75     &    0.991$\pm$0.030 & 21.10$\pm$0.86 \\
100.3                &   2504.04$\pm$0.72    &    15.20$\pm$0.75    &    21.41$\pm$0.75     &    1.048$\pm$0.030 & 27.31$\pm$0.81 \\
\hline
\multicolumn{6}{c}{J1924-2914}\\
86.3                &   5273.93$\pm$0.18    &    -10.6$\pm$1.6    &    -245.0$\pm$1.6     &    4.650$\pm$0.030 & -46.25$\pm$0.19 \\
88.3                &   5244.90$\pm$0.18    &    -12.7$\pm$1.6    &    -248.9$\pm$1.6     &    4.752$\pm$0.030 & -46.46$\pm$0.18 \\
98.3                &   4984.67$\pm$0.19    &    -12.0$\pm$1.5    &    -248.2$\pm$1.5     &    4.986$\pm$0.030 & -46.39$\pm$0.17 \\
100.3                &   4920.17$\pm$0.21    &    -12.3$\pm$1.5    &    -244.6$\pm$1.5     &    4.978$\pm$0.030 & -46.44$\pm$0.18 \\
\hline
\multicolumn{6}{c}{NRAO 530}\\
86.3                &   2857.621$\pm$0.036    &    7.11$\pm$0.86    &    24.37$\pm$0.86     &    0.889$\pm$0.030 & 36.8$\pm$1.0 \\
88.3                &   2826.545$\pm$0.035    &    6.60$\pm$0.85    &    24.53$\pm$0.85     &    0.899$\pm$0.030 & 37.4$\pm$1.0 \\
98.3                &   2649.116$\pm$0.034    &    4.24$\pm$0.80    &    24.59$\pm$0.80     &    0.943$\pm$0.030 & 40.08$\pm$0.92 \\
100.3                &   2618.629$\pm$0.035    &    3.71$\pm$0.79    &    24.62$\pm$0.79     &    0.952$\pm$0.030 & 40.73$\pm$0.92 \\
\hline
\multicolumn{6}{c}{4C 09.57}\\
86.3                &   2920.272$\pm$0.085    &    65.23$\pm$0.88    &    -98.94$\pm$0.88     &    4.057$\pm$0.030 & -28.30$\pm$0.21 \\
88.3                &   2898.304$\pm$0.073    &    65.59$\pm$0.87    &    -96.53$\pm$0.88     &    4.028$\pm$0.030 & -27.91$\pm$0.22 \\
98.3                &   2805.452$\pm$0.071    &    61.49$\pm$0.84    &    -96.65$\pm$0.85     &    4.084$\pm$0.030 & -28.76$\pm$0.21 \\
100.3                &   2775.60$\pm$0.48    &    60.67$\pm$0.84    &    -96.55$\pm$0.84     &    4.107$\pm$0.030 & -28.93$\pm$0.22 \\
 &&&&& \\
\hline
\multicolumn{6}{c}{Apr 4}\\
\hline
\multicolumn{6}{c}{3C279}\\
86.3                &   13309.65$\pm$0.19    &    70.9$\pm$4.0    &    1612.7$\pm$4.0     &    12.129$\pm$0.030 & 43.746$\pm$0.071 \\
88.3                &   13168.27$\pm$0.19    &    67.6$\pm$4.0    &    1596.6$\pm$4.0     &    12.136$\pm$0.030 & 43.791$\pm$0.069 \\
98.3                &   12671.78$\pm$0.19    &    52.6$\pm$3.8    &    1541.8$\pm$3.8     &    12.175$\pm$0.030 & 44.021$\pm$0.070 \\
100.3                &   12575.88$\pm$0.19    &    49.9$\pm$3.8    &    1533.0$\pm$3.8     &    12.198$\pm$0.030 & 44.067$\pm$0.069 \\
\hline
\multicolumn{6}{c}{3C273}\\
86.3                &   10066.53$\pm$0.38    &    -13.8$\pm$3.0    &    -419.4$\pm$3.0     &    4.167$\pm$0.030 & -45.95$\pm$0.20 \\
88.3                &   10013.63$\pm$0.36    &    -12.3$\pm$3.0    &    -405.9$\pm$3.0     &    4.054$\pm$0.030 & -45.87$\pm$0.21 \\
98.3                &   9698.37$\pm$0.31    &    -3.3$\pm$2.9    &    -370.9$\pm$2.9     &    3.826$\pm$0.030 & -45.25$\pm$0.23 \\
100.3                &   9645.14$\pm$0.31    &    3.4$\pm$2.9    &    -375.2$\pm$2.9     &    3.891$\pm$0.030 & -44.74$\pm$0.23 \\
\hline\hline
\end{longtable*}
\begin{longtable*}{cccccc}
\caption{Polarization parameters of EHT sources per frequency band (spw) and per day. } \label{tab:EHT_uvmf_spw}\\
\hline\hline 
Frequency & I & Q & U & LP & EVPA\\
\ [GHz] & [mJy] & [mJy] & [mJy] & [\%] & [deg]\\
\hline\hline 
 &&&&& \\
 &&&&& \\
\hline
\multicolumn{6}{c}{Apr 5}\\
\hline
\multicolumn{6}{c}{4C 01.28}\\
 213.1                &   3572.28$\pm$0.11    &    144.5$\pm$1.1    &    -151.7$\pm$1.1   &     5.867$\pm$0.030 & -23.20$\pm$0.14 \\
 215.1                &   3617.41$\pm$0.11    &    145.8$\pm$1.1    &    -153.3$\pm$1.1   &     5.848$\pm$0.030 & -23.23$\pm$0.15 \\
 227.1                &   3429.88$\pm$0.11    &    138.9$\pm$1.0    &    -146.7$\pm$1.0   &     5.890$\pm$0.030 & -23.28$\pm$0.14 \\
 229.1                &   3419.77$\pm$0.11    &    141.9$\pm$1.0    &    -147.0$\pm$1.0   &     5.976$\pm$0.030 & -23.01$\pm$0.14 \\
\hline
\multicolumn{6}{c}{OJ287}\\
 213.1                &   4455.60$\pm$0.15    &    -214.2$\pm$1.3    &    -342.2$\pm$1.3   &     9.060$\pm$0.030 & -61.027$\pm$0.092 \\
 215.1                &   4476.37$\pm$0.17    &    -214.8$\pm$1.3    &    -339.6$\pm$1.3   &     8.976$\pm$0.030 & -61.161$\pm$0.092 \\
 227.1                &   4220.58$\pm$0.15    &    -205.0$\pm$1.3    &    -314.9$\pm$1.3   &     8.902$\pm$0.030 & -61.53$\pm$0.10 \\
 229.1                &   4198.43$\pm$0.15    &    -203.6$\pm$1.3    &    -325.3$\pm$1.3   &     9.139$\pm$0.030 & -61.022$\pm$0.093 \\
\hline
\multicolumn{6}{c}{M87}\\
 213.1                &   1336.29$\pm$0.13    &    30.07$\pm$0.40    &    -7.70$\pm$0.40   &     2.323$\pm$0.030 & -7.17$\pm$0.36 \\
 215.1                &   1325.68$\pm$0.13    &    29.94$\pm$0.40    &    -8.10$\pm$0.40   &     2.338$\pm$0.030 & -7.57$\pm$0.38 \\
 227.1                &   1236.46$\pm$0.11    &    29.62$\pm$0.37    &    -8.80$\pm$0.37   &     2.500$\pm$0.030 & -8.29$\pm$0.35 \\
 229.1                &   1227.58$\pm$0.11    &    29.81$\pm$0.37    &    -8.74$\pm$0.37   &     2.530$\pm$0.030 & -8.15$\pm$0.35 \\
\hline
\multicolumn{6}{c}{3C279}\\
 213.1                &   9202.50$\pm$0.11    &    -8.0$\pm$2.8    &    1214.2$\pm$2.8   &     13.195$\pm$0.030 & 45.187$\pm$0.066 \\
 215.1                &   9144.74$\pm$0.11    &    -7.2$\pm$2.7    &    1206.7$\pm$2.7   &     13.195$\pm$0.030 & 45.168$\pm$0.065 \\
 227.1                &   8845.31$\pm$0.12    &    -7.9$\pm$2.7    &    1169.0$\pm$2.7   &     13.216$\pm$0.030 & 45.193$\pm$0.063 \\
 229.1                &   8774.06$\pm$0.12    &    -6.9$\pm$2.6    &    1161.8$\pm$2.6   &     13.241$\pm$0.030 & 45.172$\pm$0.066 \\
 &&&&& \\
\hline
\multicolumn{6}{c}{Apr 6}\\
\hline
\multicolumn{6}{c}{Sgr A*}\\
 213.1                &   2631.24$\pm$0.32    &    -130.65$\pm$0.79    &    -113.68$\pm$0.79   &     6.581$\pm$0.030 & -69.49$\pm$0.13 \\
 215.1                &   2629.81$\pm$0.33    &    -128.22$\pm$0.79    &    -118.33$\pm$0.80   &     6.636$\pm$0.030 & -68.65$\pm$0.13 \\
 227.1                &     2533.72$\pm$0.27*    &    -105.98$\pm$0.76    &    -144.59$\pm$0.76   &     7.076$\pm$0.030 & -63.12$\pm$0.13 \\
 229.1                &   2625.81$\pm$0.28    &    -106.00$\pm$0.79    &    -156.22$\pm$0.79   &     7.188$\pm$0.030 & -62.07$\pm$0.13 \\
\hline
\multicolumn{6}{c}{J1924-2914}\\
 213.1                &   3342.261$\pm$0.089    &    -28.0$\pm$1.0    &    -202.0$\pm$1.0   &     6.101$\pm$0.030 & -48.95$\pm$0.14 \\
 215.1                &   3313.239$\pm$0.093    &    -28.5$\pm$1.0    &    -199.2$\pm$1.0   &     6.072$\pm$0.030 & -49.07$\pm$0.14 \\
 227.1                &   3179.165$\pm$0.088    &    -30.5$\pm$1.0    &    -190.4$\pm$1.0   &     6.064$\pm$0.030 & -49.55$\pm$0.14 \\
 229.1                &   3156.545$\pm$0.087    &    -30.65$\pm$0.95    &    -191.3$\pm$1.0   &     6.138$\pm$0.030 & -49.55$\pm$0.14 \\
\hline
\multicolumn{6}{c}{J0132-1654}\\
 213.1                &   426.29$\pm$0.15    &    7.28$\pm$0.19    &    4.11$\pm$0.19   &     1.963$\pm$0.044 & 14.68$\pm$0.64 \\
 215.1                &   420.00$\pm$0.16    &    7.19$\pm$0.20    &    4.32$\pm$0.20   &     1.999$\pm$0.047 & 15.48$\pm$0.65 \\
 227.1                &   409.56$\pm$0.16    &    7.16$\pm$0.20    &    4.59$\pm$0.20   &     2.076$\pm$0.049 & 16.34$\pm$0.66 \\
 229.1                &   404.98$\pm$0.16    &    6.62$\pm$0.19    &    4.05$\pm$0.19   &     1.919$\pm$0.048 & 15.68$\pm$0.74 \\
\hline
\multicolumn{6}{c}{NGC 1052}\\
 213.1                &   437.43$\pm$0.13    &    0.36$\pm$0.14    &    0.17$\pm$0.14   &     0.090$\pm$0.033 & 13$\pm$11 \\
 215.1                &   438.85$\pm$0.11    &    0.56$\pm$0.15    &    0.39$\pm$0.15   &     0.154$\pm$0.033 & 17.0$\pm$6.0 \\
 227.1                &   414.12$\pm$0.10    &    0.47$\pm$0.14    &    0.46$\pm$0.14   &     0.159$\pm$0.033 & 21.9$\pm$6.3 \\
 229.1                &   415.236$\pm$0.062    &    0.05$\pm$0.14    &    0.30$\pm$0.14   &     0.074$\pm$0.033 & 39$\pm$18 \\
\hline
\multicolumn{6}{c}{M87}\\
 213.1                &   1361.85$\pm$0.16    &    27.72$\pm$0.41    &    -6.24$\pm$0.41   &     2.088$\pm$0.030 & -6.34$\pm$0.43 \\
 215.1                &   1349.84$\pm$0.15    &    27.65$\pm$0.41    &    -6.70$\pm$0.41   &     2.108$\pm$0.030 & -6.83$\pm$0.41 \\
 227.1                &   1269.57$\pm$0.13    &    26.95$\pm$0.38    &    -8.34$\pm$0.38   &     2.224$\pm$0.030 & -8.61$\pm$0.38 \\
 229.1                &   1257.40$\pm$0.13    &    26.70$\pm$0.38    &    -8.28$\pm$0.38   &     2.224$\pm$0.030 & -8.61$\pm$0.39 \\
\hline
\multicolumn{6}{c}{J0006-0623}\\
 213.1                &   2044.52$\pm$0.32    &    213.57$\pm$0.63    &    139.74$\pm$0.65   &     12.482$\pm$0.031 & 16.598$\pm$0.071 \\
 215.1                &   2023.69$\pm$0.33    &    212.00$\pm$0.63    &    137.19$\pm$0.64   &     12.480$\pm$0.031 & 16.452$\pm$0.072 \\
 227.1                &   1944.0$\pm$1.1    &    205.43$\pm$0.62    &    130.51$\pm$0.62   &     12.520$\pm$0.033 & 16.215$\pm$0.073 \\
 229.1                &   1931.6$\pm$1.1    &    204.15$\pm$0.61    &    134.03$\pm$0.62   &     12.645$\pm$0.033 & 16.646$\pm$0.074 \\
\hline
\multicolumn{6}{c}{3C279}\\
 213.1                &   9571.40$\pm$0.12    &    72.2$\pm$2.9    &    1240.7$\pm$2.9   &     12.984$\pm$0.030 & 43.337$\pm$0.068 \\
 215.1                &   9517.18$\pm$0.13    &    72.4$\pm$2.9    &    1234.1$\pm$2.9   &     12.990$\pm$0.030 & 43.324$\pm$0.067 \\
 227.1                &   9180.18$\pm$0.12    &    69.2$\pm$2.8    &    1193.4$\pm$2.8   &     13.022$\pm$0.030 & 43.345$\pm$0.066 \\
 229.1                &   9168.61$\pm$0.11    &    69.3$\pm$2.8    &    1193.1$\pm$2.8   &     13.035$\pm$0.030 & 43.338$\pm$0.066 \\
\hline
\multicolumn{6}{c}{NRAO 530}\\
 213.1                &   1655.066$\pm$0.071    &    -8.41$\pm$0.50    &    36.58$\pm$0.50   &     2.267$\pm$0.030 & 51.46$\pm$0.38 \\
 215.1                &   1669.741$\pm$0.079    &    -8.95$\pm$0.51    &    37.51$\pm$0.51   &     2.309$\pm$0.030 & 51.70$\pm$0.38 \\
 227.1                &   1567.831$\pm$0.072    &    -8.86$\pm$0.48    &    36.81$\pm$0.48   &     2.414$\pm$0.030 & 51.76$\pm$0.37 \\
 229.1                &   1554.353$\pm$0.070    &    -8.41$\pm$0.47    &    36.72$\pm$0.47   &     2.424$\pm$0.030 & 51.44$\pm$0.36 \\
\hline
\multicolumn{6}{c}{3C273}\\
 213.1                &   7744.46$\pm$0.11    &    -58.5$\pm$2.3    &    -191.7$\pm$2.3   &     2.587$\pm$0.030 & -53.49$\pm$0.33 \\
 215.1                &   7707.72$\pm$0.11    &    -61.7$\pm$2.3    &    -186.8$\pm$2.3   &     2.553$\pm$0.030 & -54.15$\pm$0.34 \\
 227.1                &   7421.16$\pm$0.10    &    -68.4$\pm$2.2    &    -153.4$\pm$2.2   &     2.262$\pm$0.030 & -57.01$\pm$0.38 \\
 229.1                &   7357.33$\pm$0.10    &    -66.8$\pm$2.2    &    -145.2$\pm$2.2   &     2.170$\pm$0.030 & -57.36$\pm$0.39 \\
 &&&&& \\
\hline
\multicolumn{6}{c}{Apr 7}\\
\hline
\multicolumn{6}{c}{J1924-2914}\\
 213.1                &   3243.110$\pm$0.083    &    -26.8$\pm$1.0    &    -191.3$\pm$1.0   &     5.957$\pm$0.030 & -48.99$\pm$0.15 \\
 215.1                &   3219.06$\pm$0.10    &    -27.3$\pm$1.0    &    -190.0$\pm$1.0   &     5.963$\pm$0.030 & -49.09$\pm$0.15 \\
 227.1                &   3073.91$\pm$0.11    &    -27.90$\pm$0.92    &    -181.62$\pm$0.93   &     5.978$\pm$0.030 & -49.37$\pm$0.14 \\
 229.1                &   3048.69$\pm$0.12    &    -28.11$\pm$0.92    &    -180.17$\pm$0.92   &     5.982$\pm$0.030 & -49.44$\pm$0.15 \\
\hline
\multicolumn{6}{c}{J0132-1654}\\
 213.1                &   418.13$\pm$0.18    &    6.66$\pm$0.21    &    5.00$\pm$0.21   &     1.990$\pm$0.051 & 18.43$\pm$0.73 \\
 215.1                &   417.76$\pm$0.19    &    6.99$\pm$0.23    &    4.97$\pm$0.23   &     2.055$\pm$0.055 & 17.75$\pm$0.75 \\
 227.1                &   402.18$\pm$0.20    &    6.82$\pm$0.22    &    4.37$\pm$0.23   &     2.014$\pm$0.056 & 16.29$\pm$0.80 \\
 229.1                &   394.71$\pm$0.21    &    6.12$\pm$0.23    &    4.77$\pm$0.23   &     1.965$\pm$0.058 & 18.93$\pm$0.84 \\
\hline
\multicolumn{6}{c}{NRAO 530}\\
 213.1                &   1620.07$\pm$0.10    &    -7.36$\pm$0.49    &    37.57$\pm$0.49   &     2.363$\pm$0.030 & 50.55$\pm$0.35 \\
 215.1                &   1604.64$\pm$0.11    &    -7.57$\pm$0.49    &    37.41$\pm$0.49   &     2.379$\pm$0.030 & 50.72$\pm$0.36 \\
 227.1                &   1536.162$\pm$0.068    &    -7.65$\pm$0.47    &    36.73$\pm$0.47   &     2.443$\pm$0.030 & 50.89$\pm$0.37 \\
 229.1                &   1526.938$\pm$0.067    &    -7.37$\pm$0.46    &    37.77$\pm$0.46   &     2.520$\pm$0.030 & 50.51$\pm$0.35 \\
\hline
\multicolumn{6}{c}{Sgr A*}\\
 213.1                &   2418.25$\pm$0.25    &    -124.16$\pm$0.73    &    -113.45$\pm$0.73   &     6.954$\pm$0.030 & -68.79$\pm$0.12 \\
 215.1                &   2404.74$\pm$0.27    &    -121.16$\pm$0.72    &    -117.97$\pm$0.72   &     7.032$\pm$0.030 & -67.88$\pm$0.12 \\
 227.1                &     2318.41$\pm$0.22*    &    -100.51$\pm$0.70    &    -140.51$\pm$0.70   &     7.451$\pm$0.030 & -62.79$\pm$0.11 \\
 229.1                &   2403.69$\pm$0.23    &    -100.83$\pm$0.72    &    -148.69$\pm$0.72   &     7.472$\pm$0.030 & -62.07$\pm$0.11 \\
\hline
\multicolumn{6}{c}{NGC 1052}\\
 213.1                &   398.01$\pm$0.20    &    -0.31$\pm$0.14    &    0.87$\pm$0.14   &     0.231$\pm$0.035 & 54.9$\pm$4.5 \\
 215.1                &   397.25$\pm$0.18    &    -0.10$\pm$0.15    &    0.70$\pm$0.15   &     0.178$\pm$0.037 & 49.6$\pm$6.0 \\
 227.1                &   369.99$\pm$0.25    &    0.21$\pm$0.14    &    0.12$\pm$0.14   &     0.069$\pm$0.038 & 13$\pm$20 \\
 229.1                &   358.94$\pm$0.35    &    -0.46$\pm$0.14    &    0.44$\pm$0.14   &     0.177$\pm$0.039 & 68.1$\pm$7.9 \\
 &&&&& \\
\hline
\multicolumn{6}{c}{Apr 10}\\
\hline
\multicolumn{6}{c}{4C 01.28}\\
 213.1                &   3668.38$\pm$0.16    &    154.6$\pm$1.1    &    -102.5$\pm$1.1   &     5.055$\pm$0.030 & -16.77$\pm$0.16 \\
 215.1                &   3677.75$\pm$0.17    &    153.4$\pm$1.1    &    -102.8$\pm$1.1   &     5.020$\pm$0.030 & -16.91$\pm$0.17 \\
 227.1                &   3514.62$\pm$0.14    &    147.8$\pm$1.1    &    -99.7$\pm$1.1   &     5.074$\pm$0.030 & -17.00$\pm$0.18 \\
 229.1                &   3512.42$\pm$0.13    &    152.2$\pm$1.1    &    -99.6$\pm$1.1   &     5.179$\pm$0.030 & -16.60$\pm$0.17 \\
\hline
\multicolumn{6}{c}{OJ287}\\
 213.1                &   4319.36$\pm$0.10    &    -167.8$\pm$1.3    &    -253.8$\pm$1.3   &     7.044$\pm$0.030 & -61.73$\pm$0.12 \\
 215.1                &   4333.61$\pm$0.10    &    -166.9$\pm$1.3    &    -251.3$\pm$1.3   &     6.962$\pm$0.030 & -61.79$\pm$0.12 \\
 227.1                &   4119.79$\pm$0.10    &    -158.1$\pm$1.2    &    -235.9$\pm$1.2   &     6.894$\pm$0.030 & -61.91$\pm$0.12 \\
 229.1                &   4105.17$\pm$0.10    &    -162.0$\pm$1.2    &    -243.2$\pm$1.2   &     7.119$\pm$0.030 & -61.83$\pm$0.12 \\
\hline
\multicolumn{6}{c}{Cen A}\\
 213.1                &   5710.166$\pm$0.085    &    1.8$\pm$1.7    &    0.5$\pm$1.7   &     0.040$\pm$0.030 & 6$\pm$30 \\
 215.1                &   5677.690$\pm$0.090    &    3.1$\pm$1.7    &    2.3$\pm$1.7   &     0.070$\pm$0.030 & 18$\pm$15 \\
 227.1                &   5621.837$\pm$0.094    &    3.2$\pm$1.7    &    4.7$\pm$1.7   &     0.101$\pm$0.030 & 27.9$\pm$9.1 \\
 229.1                &   5628.36$\pm$0.10    &    0.2$\pm$1.7    &    3.9$\pm$1.7   &     0.071$\pm$0.030 & 43$\pm$14 \\
\hline
\multicolumn{6}{c}{M87}\\
 213.1                &   1382.93$\pm$0.26    &    37.15$\pm$0.42    &    -0.24$\pm$0.42   &     2.688$\pm$0.030 & -0.19$\pm$0.32 \\
 215.1                &   1371.95$\pm$0.25    &    37.12$\pm$0.42    &    -0.08$\pm$0.42   &     2.706$\pm$0.030 & -0.08$\pm$0.31 \\
 227.1                &   1285.70$\pm$0.23    &    35.68$\pm$0.39    &    0.47$\pm$0.39   &     2.777$\pm$0.030 & 0.37$\pm$0.31 \\
 229.1                &   1271.90$\pm$0.23    &    35.24$\pm$0.39    &    0.05$\pm$0.39   &     2.770$\pm$0.030 & 0.03$\pm$0.31 \\
\hline
\multicolumn{6}{c}{3C279}\\
 213.1                &   8750.09$\pm$0.11    &    216.5$\pm$2.6    &    1264.2$\pm$2.6   &     14.659$\pm$0.030 & 40.141$\pm$0.057 \\
 215.1                &   8699.92$\pm$0.11    &    215.5$\pm$2.6    &    1257.5$\pm$2.6   &     14.665$\pm$0.030 & 40.137$\pm$0.058 \\
 227.1                &   8415.64$\pm$0.11    &    209.4$\pm$2.5    &    1220.6$\pm$2.5   &     14.717$\pm$0.030 & 40.133$\pm$0.058 \\
 229.1                &   8373.72$\pm$0.11    &    208.6$\pm$2.5    &    1216.4$\pm$2.5   &     14.739$\pm$0.030 & 40.135$\pm$0.061 \\
 &&&&& \\
\hline
\multicolumn{6}{c}{Apr 11}\\
\hline
\multicolumn{6}{c}{Sgr A*}\\
 213.1                &   2388.59$\pm$0.31    &    -41.33$\pm$0.72    &    -169.55$\pm$0.72   &     7.307$\pm$0.030 & -51.85$\pm$0.12 \\
 215.1                &   2383.72$\pm$0.30    &    -37.26$\pm$0.72    &    -169.35$\pm$0.72   &     7.275$\pm$0.030 & -51.21$\pm$0.12 \\
 227.1                &     2266.69$\pm$0.26*    &    -13.97$\pm$0.68    &    -170.91$\pm$0.68   &     7.565$\pm$0.030 & -47.34$\pm$0.11 \\
 229.1                &   2362.91$\pm$0.27    &    -12.14$\pm$0.71    &    -181.89$\pm$0.71   &     7.714$\pm$0.030 & -46.91$\pm$0.12 \\
\hline
\multicolumn{6}{c}{4C 01.28}\\
 213.1                &   3643.15$\pm$0.20    &    158.8$\pm$1.1    &    -87.2$\pm$1.1   &     4.972$\pm$0.030 & -14.38$\pm$0.18 \\
 215.1                &   3632.17$\pm$0.20    &    157.7$\pm$1.1    &    -88.5$\pm$1.1   &     4.980$\pm$0.030 & -14.65$\pm$0.17 \\
 227.1                &   3503.45$\pm$0.21    &    152.0$\pm$1.1    &    -88.4$\pm$1.1   &     5.018$\pm$0.030 & -15.10$\pm$0.17 \\
 229.1                &   3485.19$\pm$0.22    &    152.9$\pm$1.1    &    -87.0$\pm$1.1   &     5.047$\pm$0.030 & -14.82$\pm$0.18 \\
\hline
\multicolumn{6}{c}{OJ287}\\
 213.1                &   4366.345$\pm$0.094    &    -148.7$\pm$1.3    &    -274.2$\pm$1.3   &     7.143$\pm$0.030 & -59.23$\pm$0.12 \\
 215.1                &   4356.28$\pm$0.10    &    -151.4$\pm$1.3    &    -271.9$\pm$1.3   &     7.143$\pm$0.030 & -59.56$\pm$0.12 \\
 227.1                &   4170.47$\pm$0.36    &    -149.1$\pm$1.3    &    -258.1$\pm$1.3   &     7.146$\pm$0.030 & -60.01$\pm$0.12 \\
 229.1                &   4152.92$\pm$0.11    &    -145.7$\pm$1.2    &    -259.6$\pm$1.2   &     7.168$\pm$0.030 & -59.65$\pm$0.12 \\
\hline
\multicolumn{6}{c}{J1924-2914}\\
 213.1                &   3299.39$\pm$0.19    &    -35.7$\pm$1.0    &    -156.7$\pm$1.0   &     4.870$\pm$0.030 & -51.42$\pm$0.18 \\
 215.1                &   3289.89$\pm$0.19    &    -36.8$\pm$1.0    &    -155.1$\pm$1.0   &     4.845$\pm$0.030 & -51.68$\pm$0.18 \\
 227.1                &   3155.4$\pm$1.0    &    -37.30$\pm$0.95    &    -147.9$\pm$1.0   &     4.834$\pm$0.030 & -52.08$\pm$0.18 \\
 229.1                &   3146.13$\pm$0.17    &    -38.04$\pm$0.95    &    -149.99$\pm$0.95   &     4.918$\pm$0.030 & -52.12$\pm$0.17 \\
\hline
\multicolumn{6}{c}{M87}\\
 213.1                &   1393.36$\pm$0.11    &    35.91$\pm$0.42    &    -1.16$\pm$0.42   &     2.579$\pm$0.030 & -0.94$\pm$0.33 \\
 215.1                &   1380.69$\pm$0.11    &    36.33$\pm$0.42    &    -1.06$\pm$0.42   &     2.634$\pm$0.030 & -0.85$\pm$0.33 \\
 227.1                &   1290.61$\pm$0.10    &    36.30$\pm$0.39    &    -0.68$\pm$0.39   &     2.813$\pm$0.030 & -0.54$\pm$0.31 \\
 229.1                &   1278.47$\pm$0.10    &    36.23$\pm$0.39    &    -0.31$\pm$0.39   &     2.833$\pm$0.030 & -0.25$\pm$0.30 \\
\hline
\multicolumn{6}{c}{3C279}\\
 213.1                &   8365.96$\pm$0.23    &    208.5$\pm$2.5    &    1223.0$\pm$2.5   &     14.830$\pm$0.030 & 40.163$\pm$0.059 \\
 215.1                &   8317.40$\pm$0.23    &    207.2$\pm$2.5    &    1217.3$\pm$2.5   &     14.846$\pm$0.030 & 40.168$\pm$0.058 \\
 227.1                &   8000.38$\pm$0.25    &    201.7$\pm$2.4    &    1180.8$\pm$2.4   &     14.974$\pm$0.030 & 40.154$\pm$0.058 \\
 229.1                &   7972.54$\pm$0.25    &    200.5$\pm$2.4    &    1178.2$\pm$2.4   &     14.991$\pm$0.030 & 40.171$\pm$0.058 \\
\hline\hline
\multicolumn{6}{l}{*The flux of Sgr A* at 227 GHz (spw=2) is systematically $\sim$5\% lower than at 229 GHz (spw=3),}\\
\multicolumn{6}{l}{owing to the presence of spectral absorption lines (see Appendix~\ref{appendix:sgr_spw2}).}\\
\multicolumn{6}{l}{Given its flat spectral index, $F_{spw=2}=F_{spw=3}$ should be assumed for Sgr A*.} \\
\end{longtable*} 

\section{Faraday RM plots} 
\label{app:RMplots}

\begin{figure*}
\centering
\hspace{-0.35cm}
\includegraphics[width=0.25\textwidth]{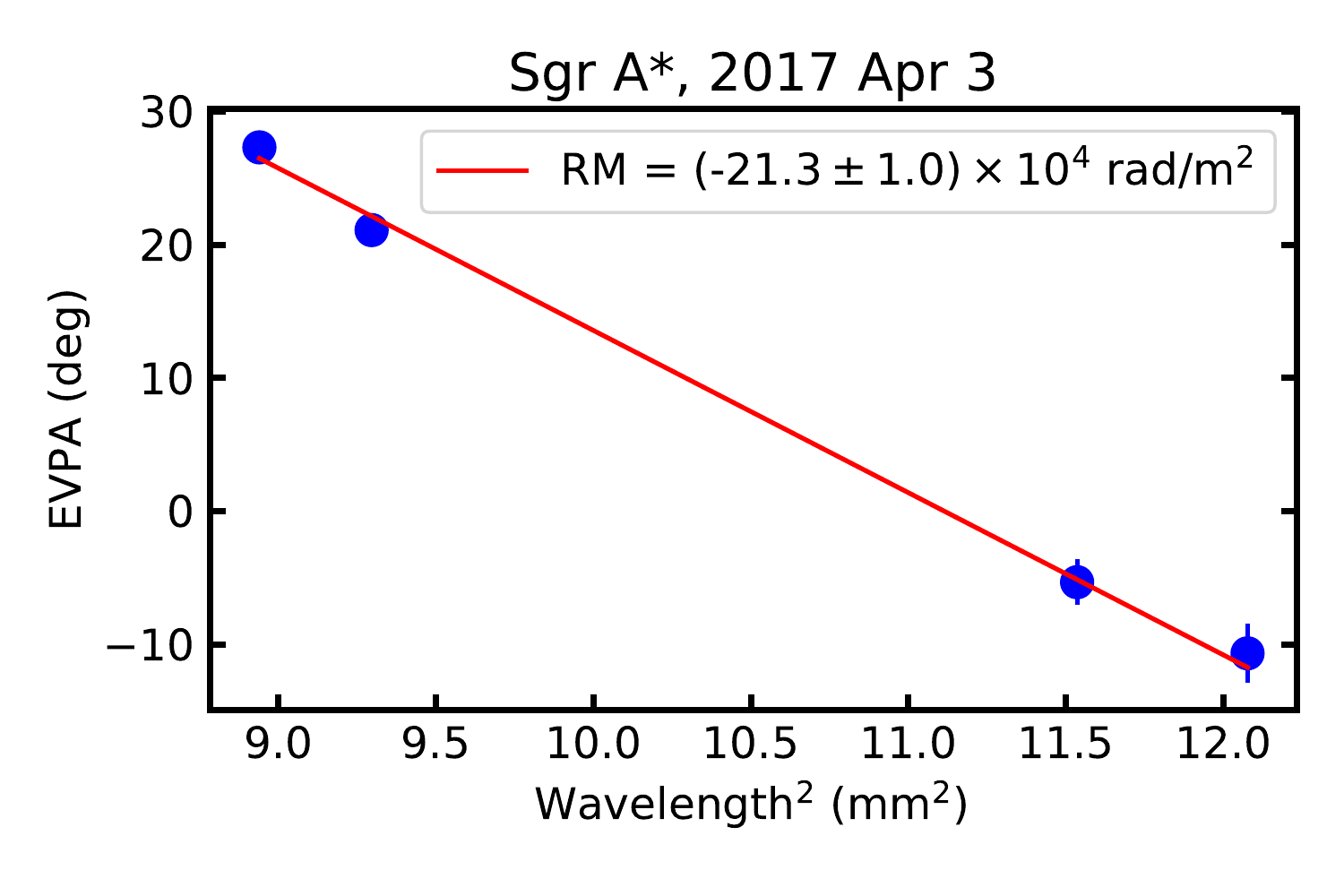}
\hspace{-0.35cm}
\includegraphics[width=0.25\textwidth]{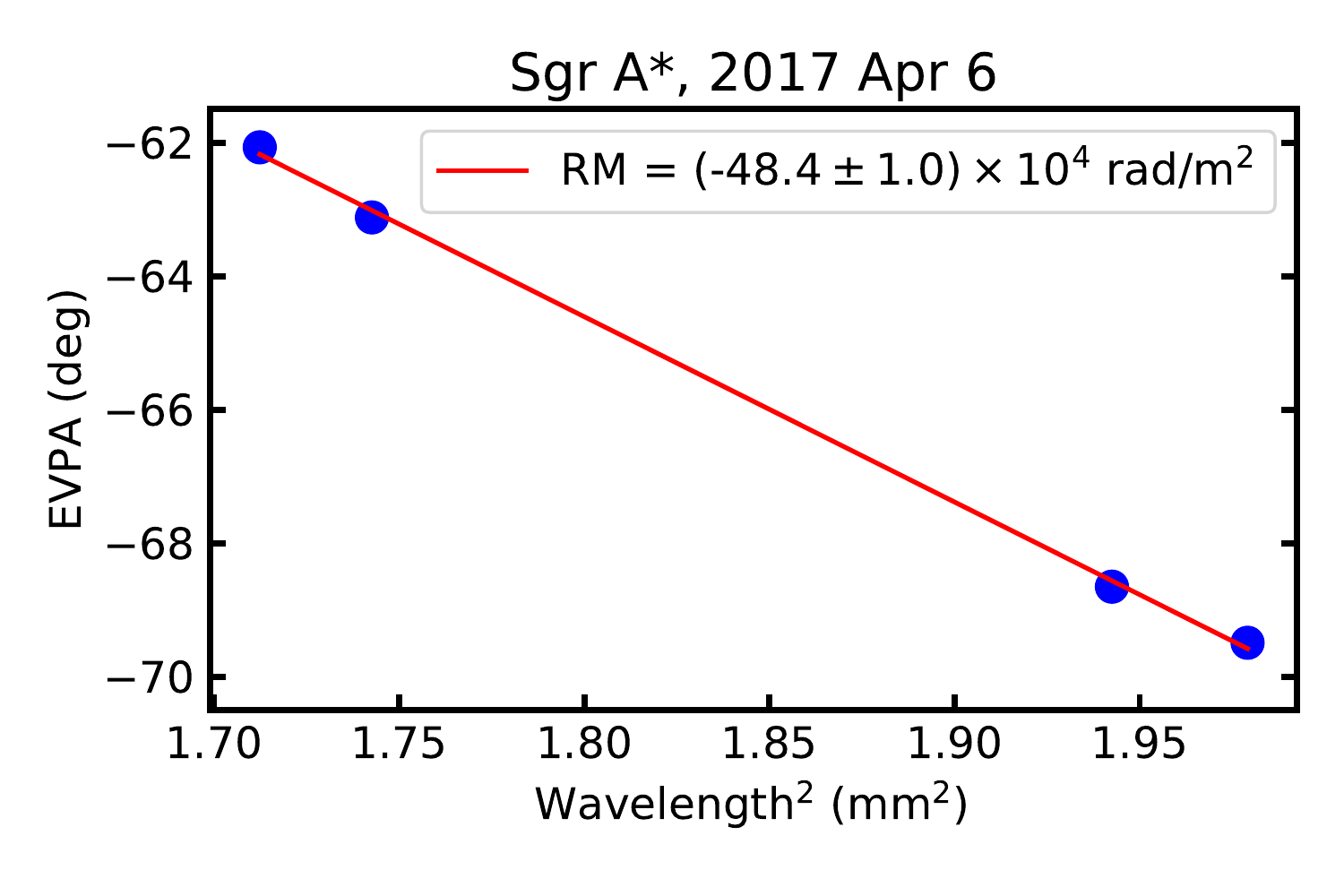} \hspace{-0.35cm}
\includegraphics[width=0.25\textwidth]{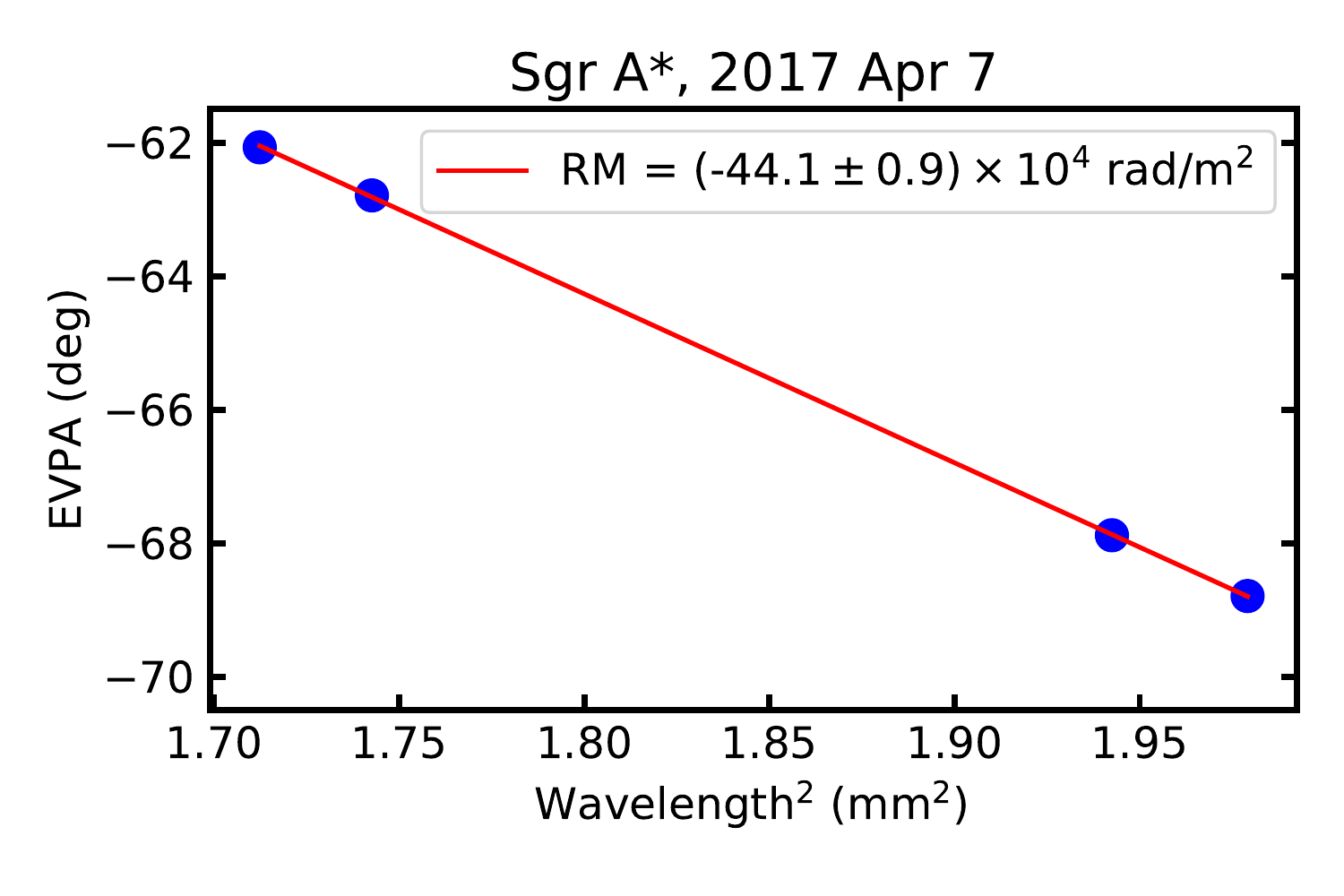} \hspace{-0.35cm}
\includegraphics[width=0.25\textwidth]{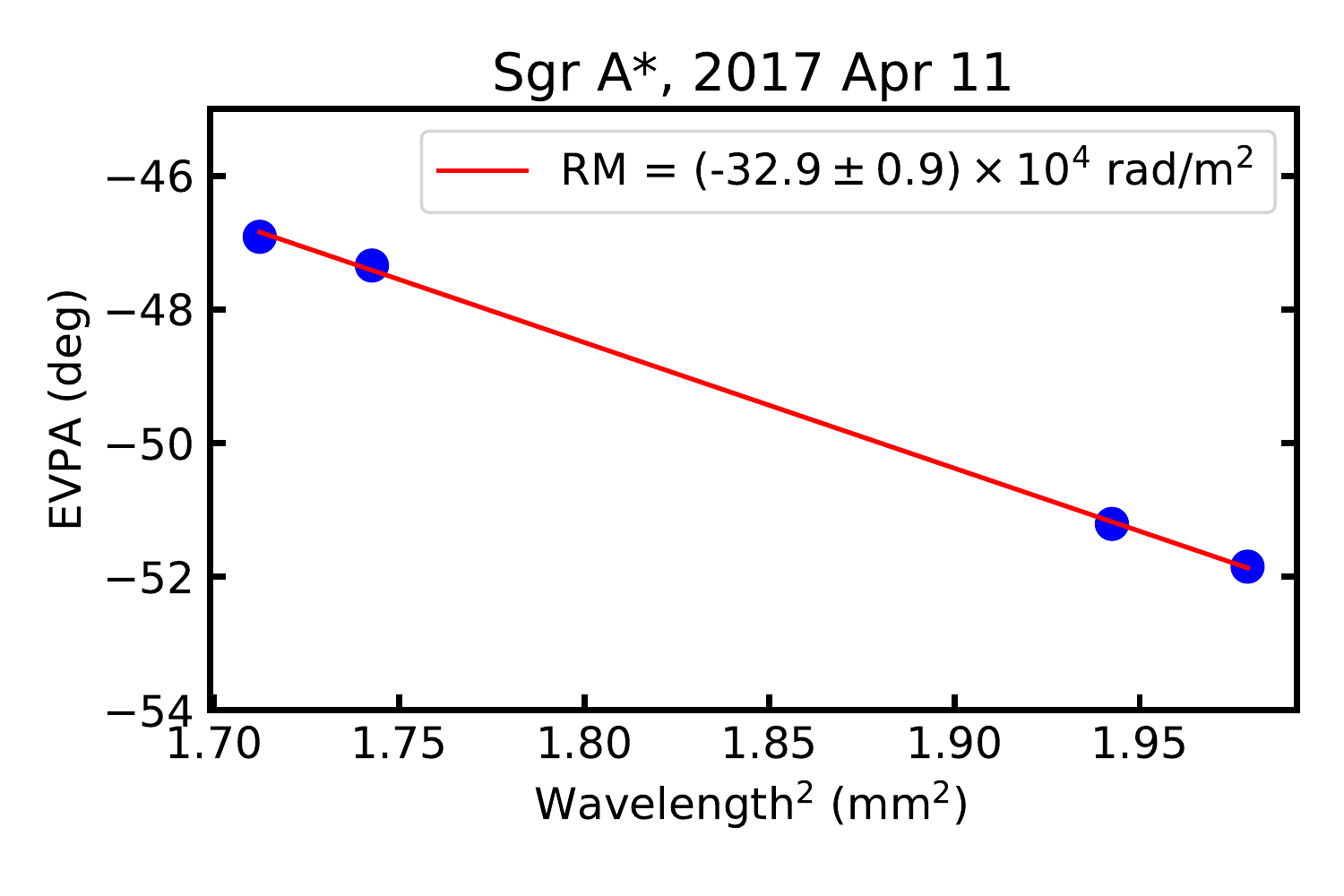} 
\hspace{-0.35cm}
\includegraphics[width=0.25\textwidth]{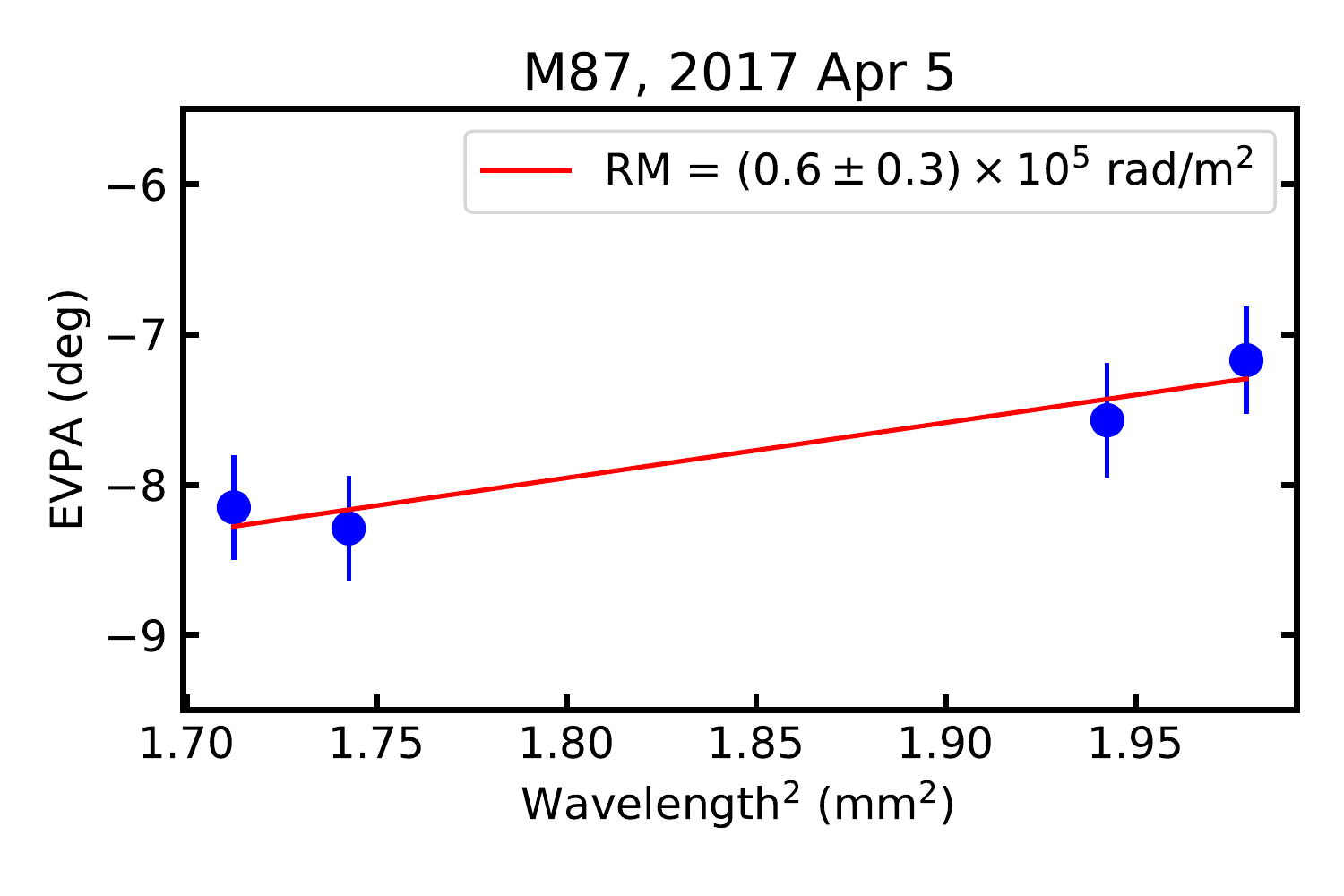} \hspace{-0.35cm}
\includegraphics[width=0.25\textwidth]{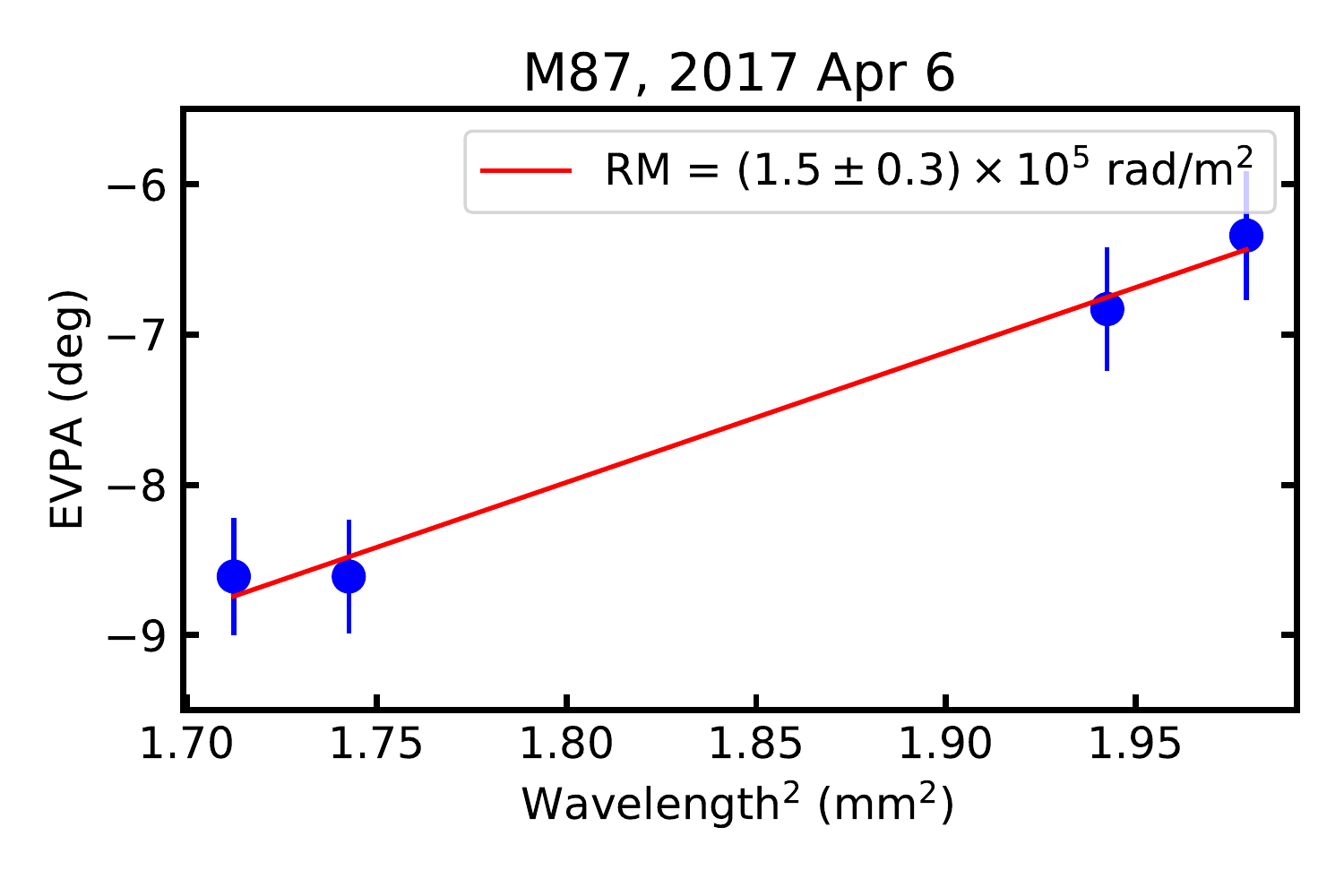} \hspace{-0.35cm}
\includegraphics[width=0.25\textwidth]{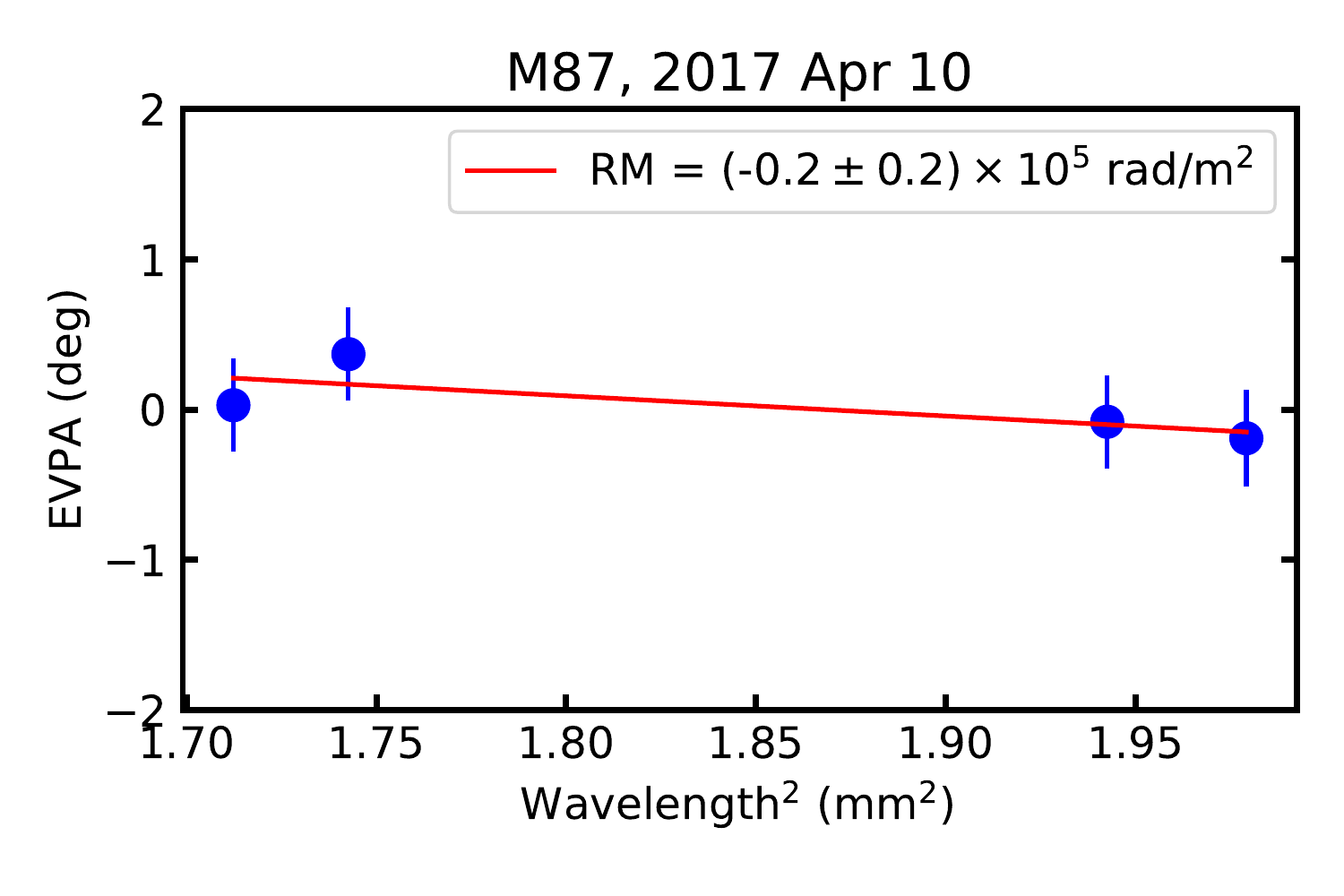} \hspace{-0.35cm}
\includegraphics[width=0.25\textwidth]{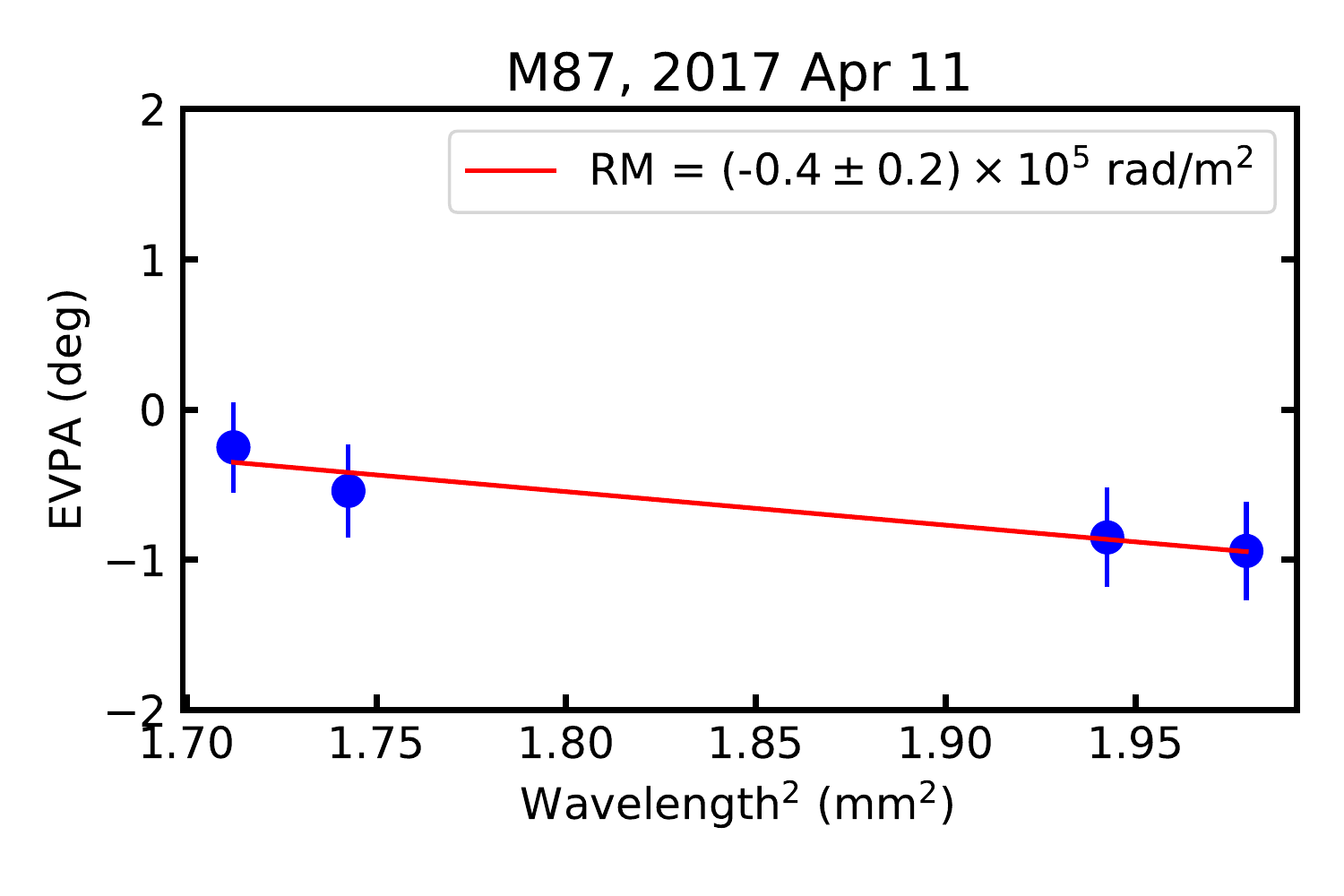} \hspace{-0.35cm}
\includegraphics[width=0.25\textwidth]{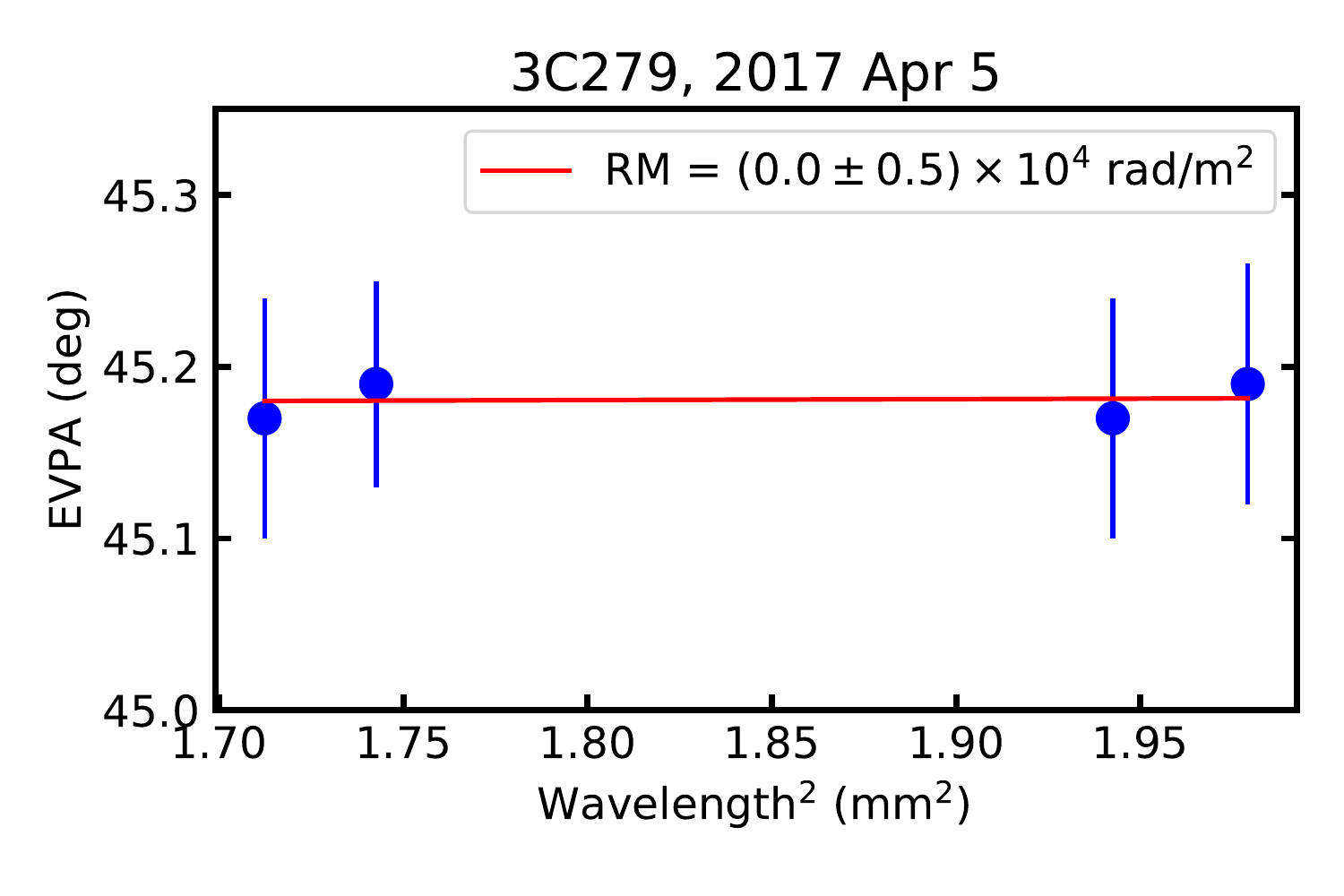} \hspace{-0.35cm}
\includegraphics[width=0.25\textwidth]{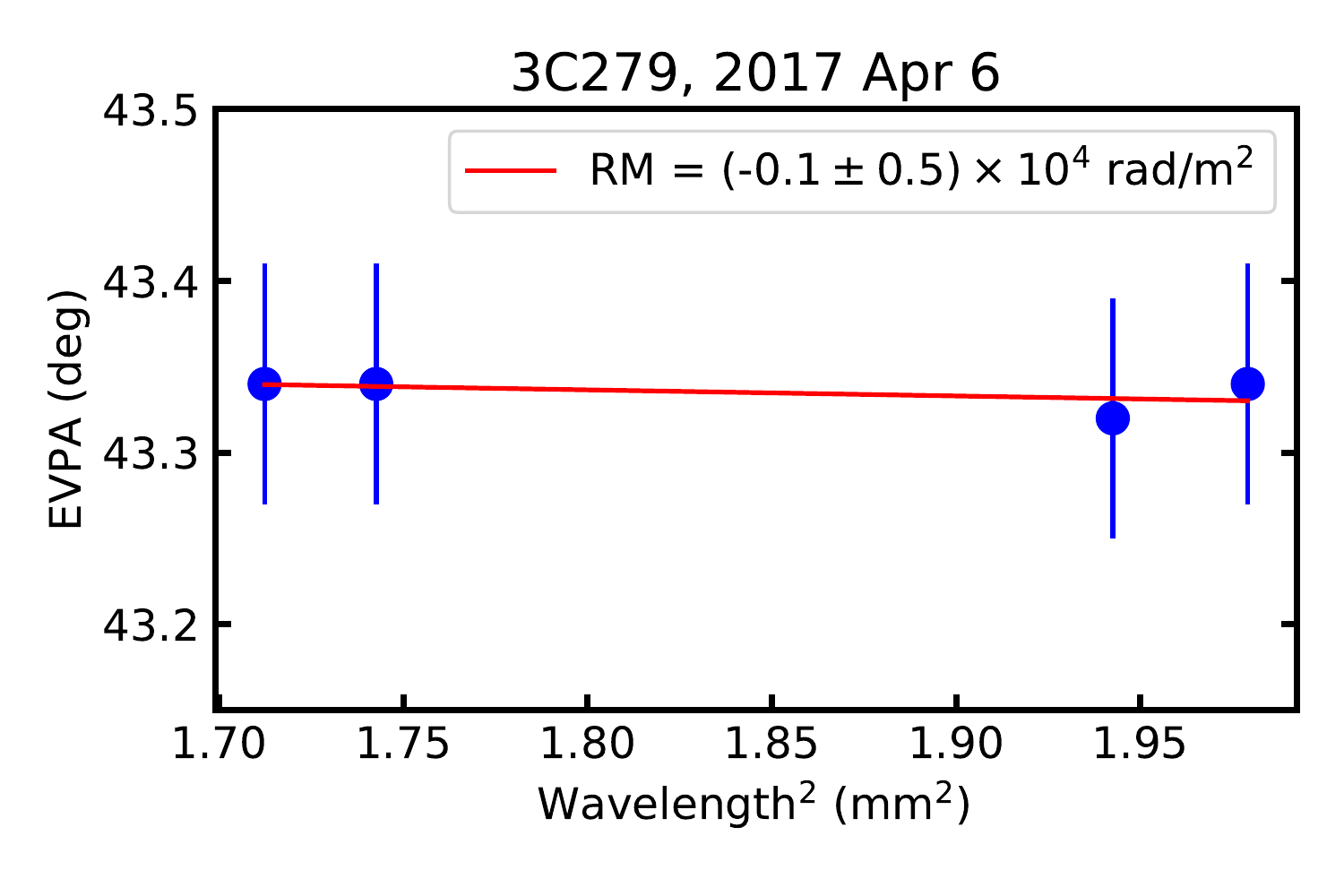} \hspace{-0.35cm}
\includegraphics[width=0.25\textwidth]{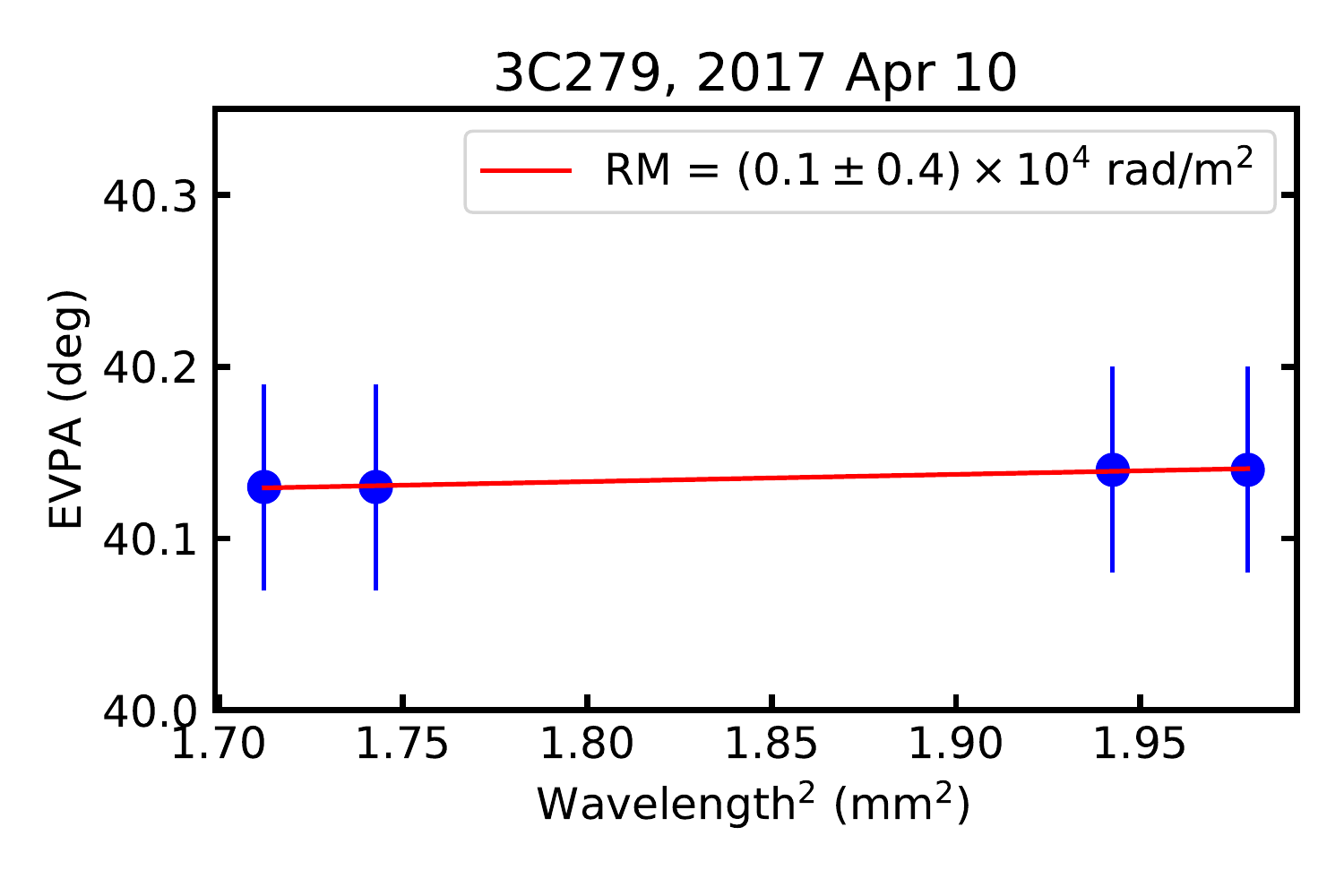} \hspace{-0.35cm}
\includegraphics[width=0.25\textwidth]{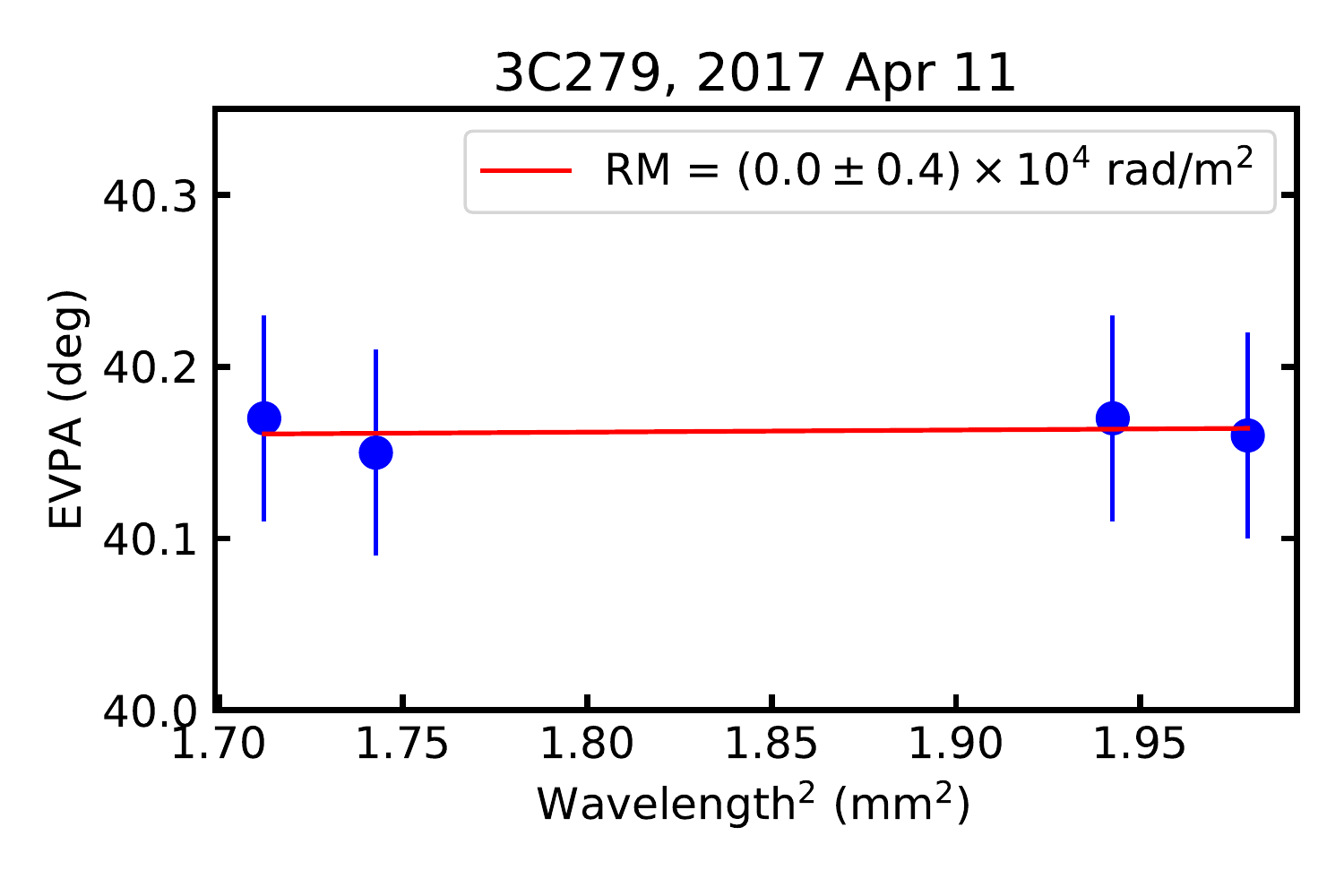} \hspace{-0.35cm}
\caption{
Faraday RM of Sgr A* (upper panels), M87 (middle panels), and 3C~279 (lower panels). 
The EVPA as a function of wavelength-squared is presented with $1\sigma$ error bars for each SPW and for each day. 
The EVPAs are measured in the 1.3~mm band, except for the  upper left panel, presenting the EVPAs measured at 3~mm in Sgr~A*. 
The error bars are derived 
 adding in quadrature to the thermal uncertainty  of the Q and U maps ($1\sigma$ image RMS)  a systematic error of 0.03\% of Stokes I (error bars for Sgr A* are smaller than the displayed points). 
 The  line is a linear fit to the data giving the RM reported inside the box. 
Plots in each row span the same vertical axis range in degrees  in order to highlight differences in slope (the upper left panel is an exception). 
It is remarkable that Sgr~A* showcases a very consistent slope across days, while in M87 the slope appears to change sign.  
Note also the different EVPA values between Apr 6/7 and Apr 11 in Sgr~A* and 
 between Apr 5/6 and Apr 10/11 in M87 and 3C279.
 The measured RM in 3C~279 (used as polarization calibrator on Apr 5, 6, 10, and 11) is consistent with zero across all days, demonstrating the stability of the polarization measurements on M87 and on Sgr~A* on the same days.
}
\label{fig:RM1}
\end{figure*}

\begin{figure*}[ht!]
\centering
\hspace{-0.35cm}
\includegraphics[width=0.33\textwidth]{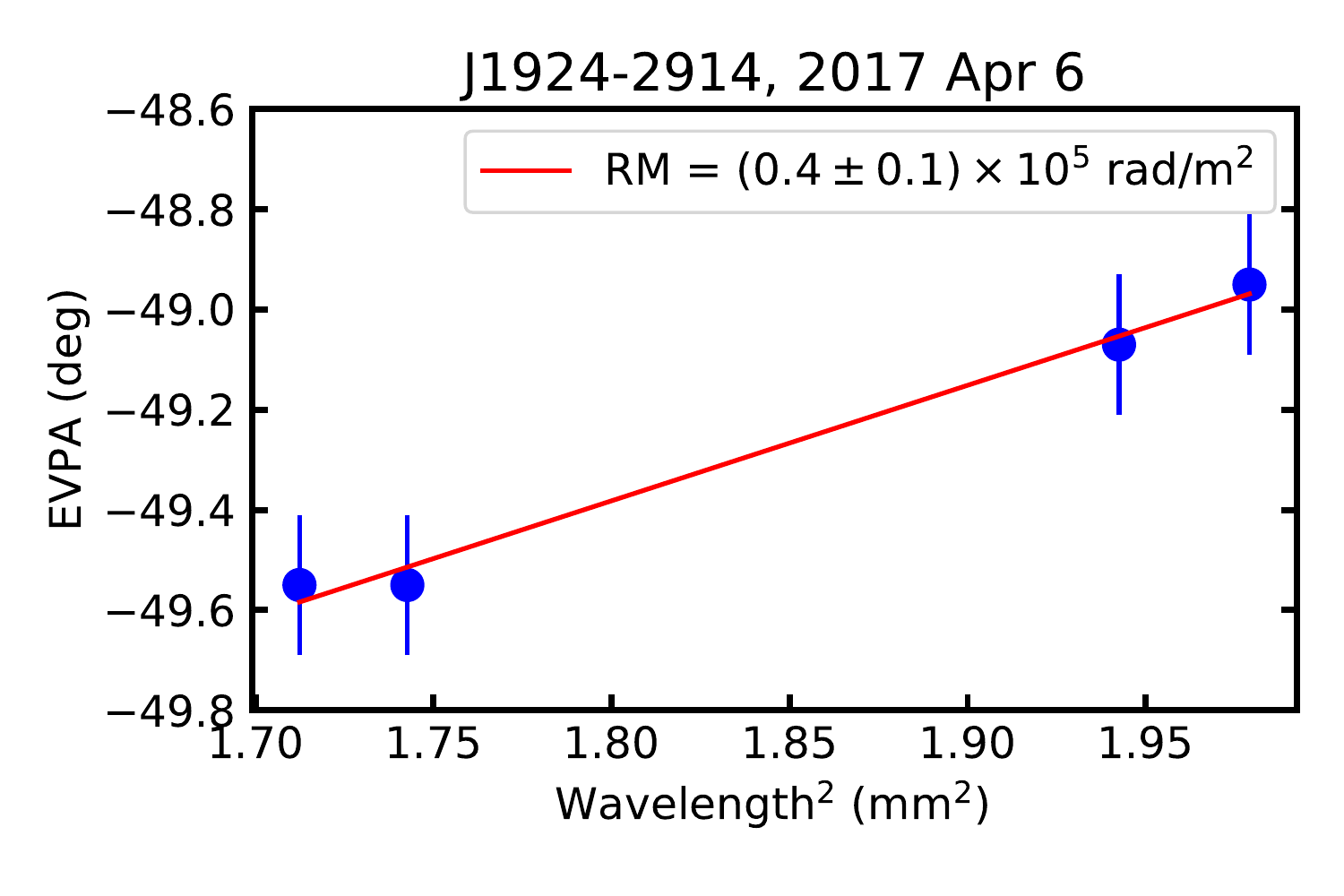} \hspace{-0.35cm}
\includegraphics[width=0.33\textwidth]{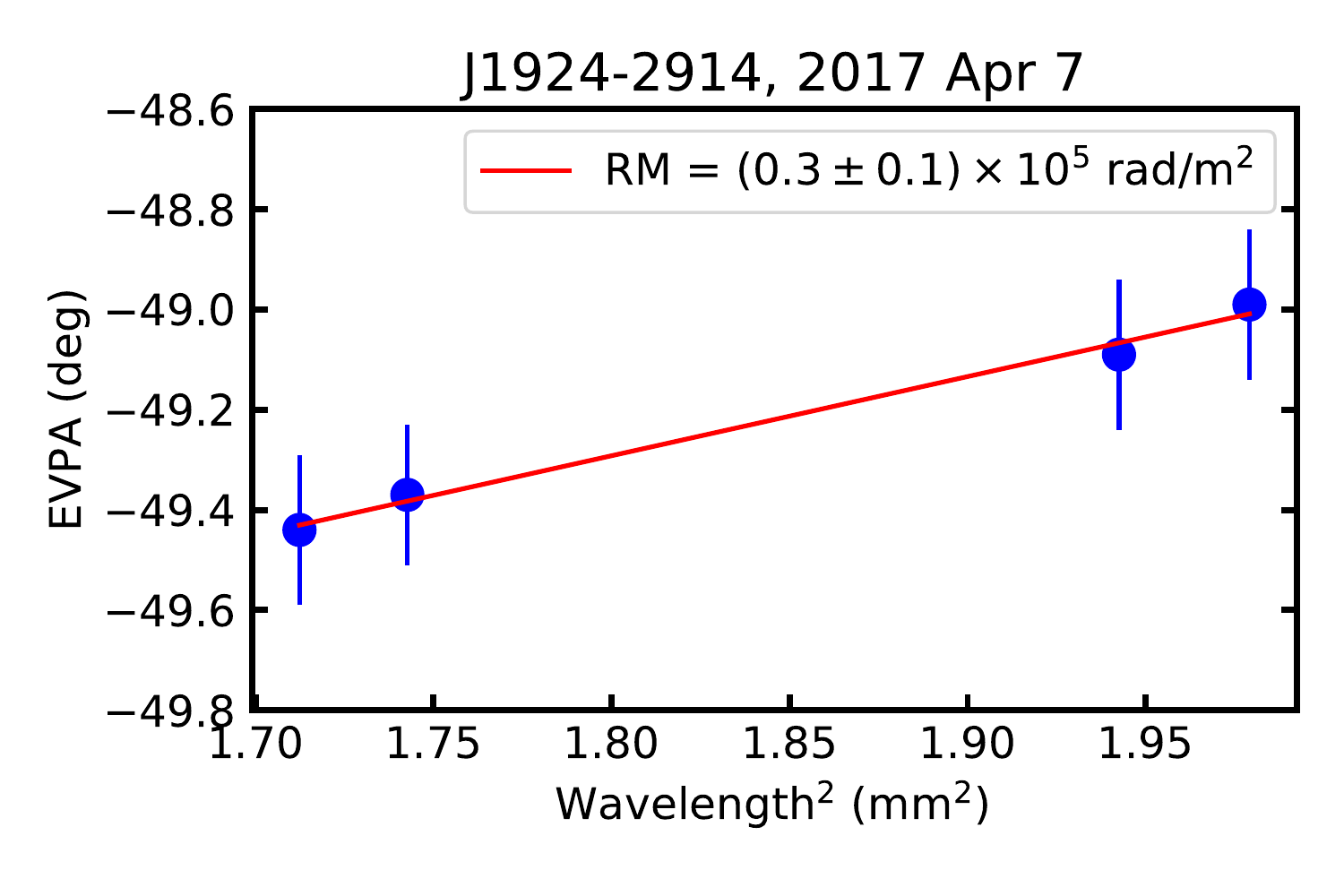} \hspace{-0.35cm}
\includegraphics[width=0.33\textwidth]{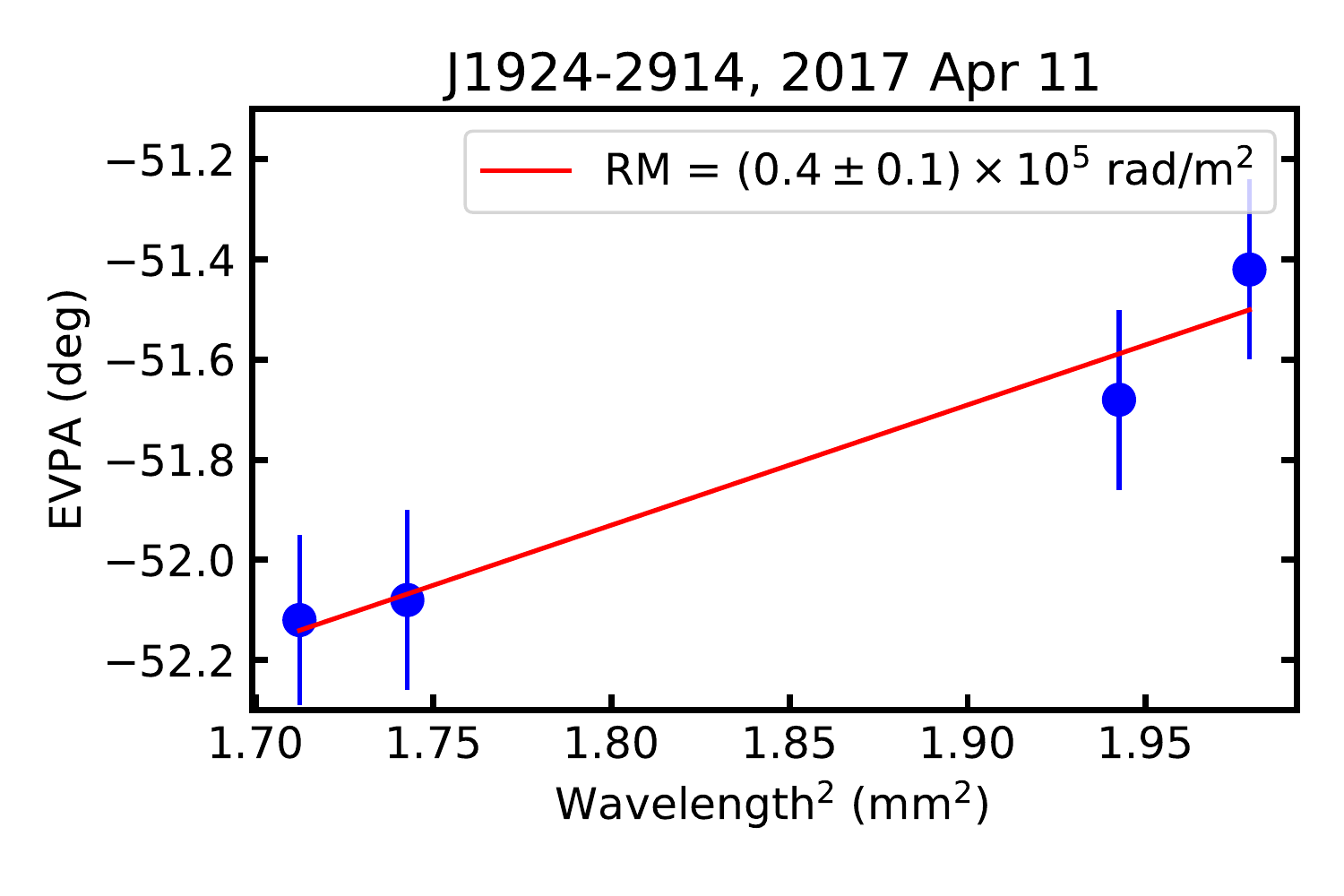} \hspace{-0.35cm}
\includegraphics[width=0.33\textwidth]{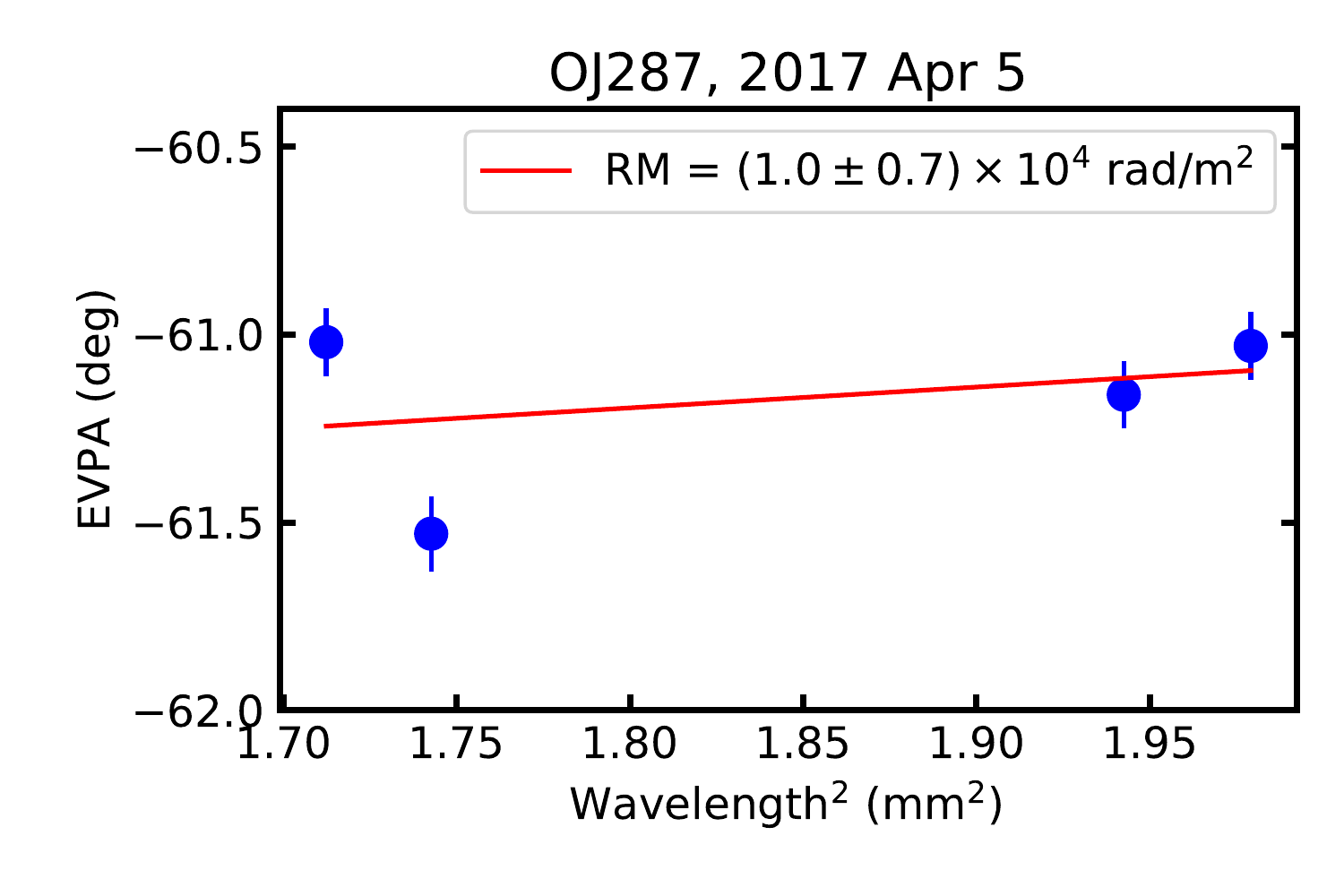} \hspace{-0.35cm}
\includegraphics[width=0.33\textwidth]{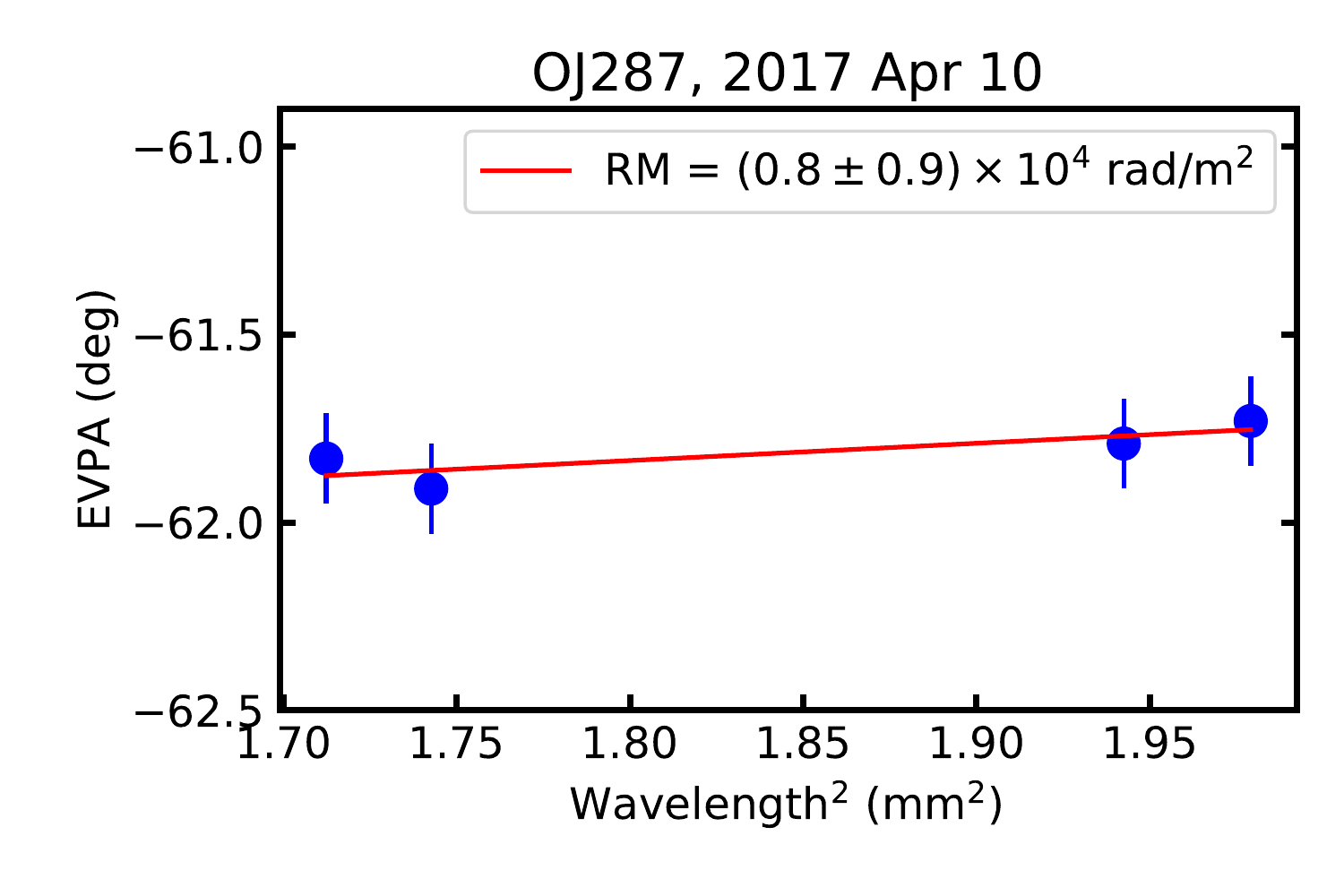} \hspace{-0.35cm}
\includegraphics[width=0.33\textwidth]{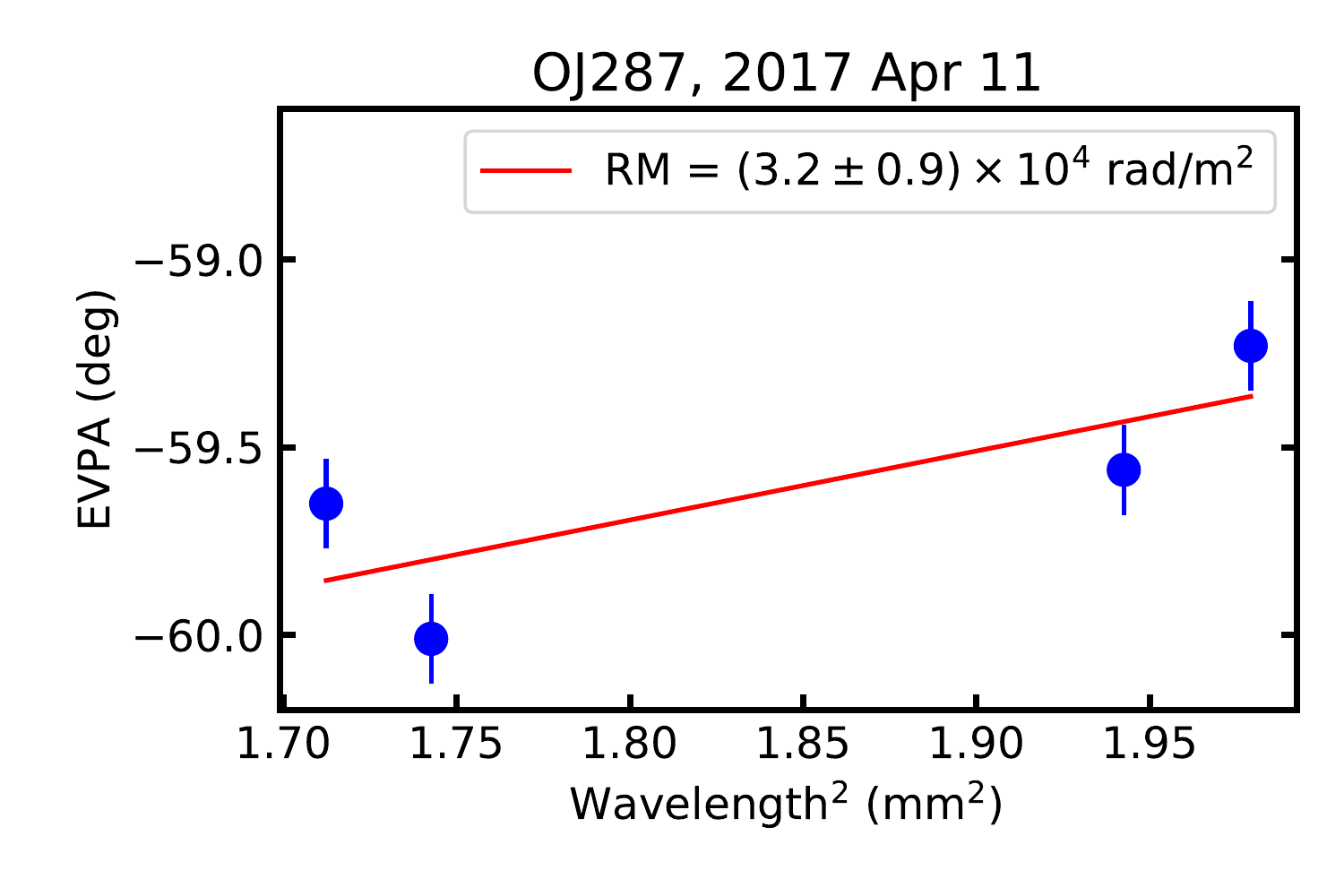} \hspace{-0.35cm}
\includegraphics[width=0.33\textwidth]{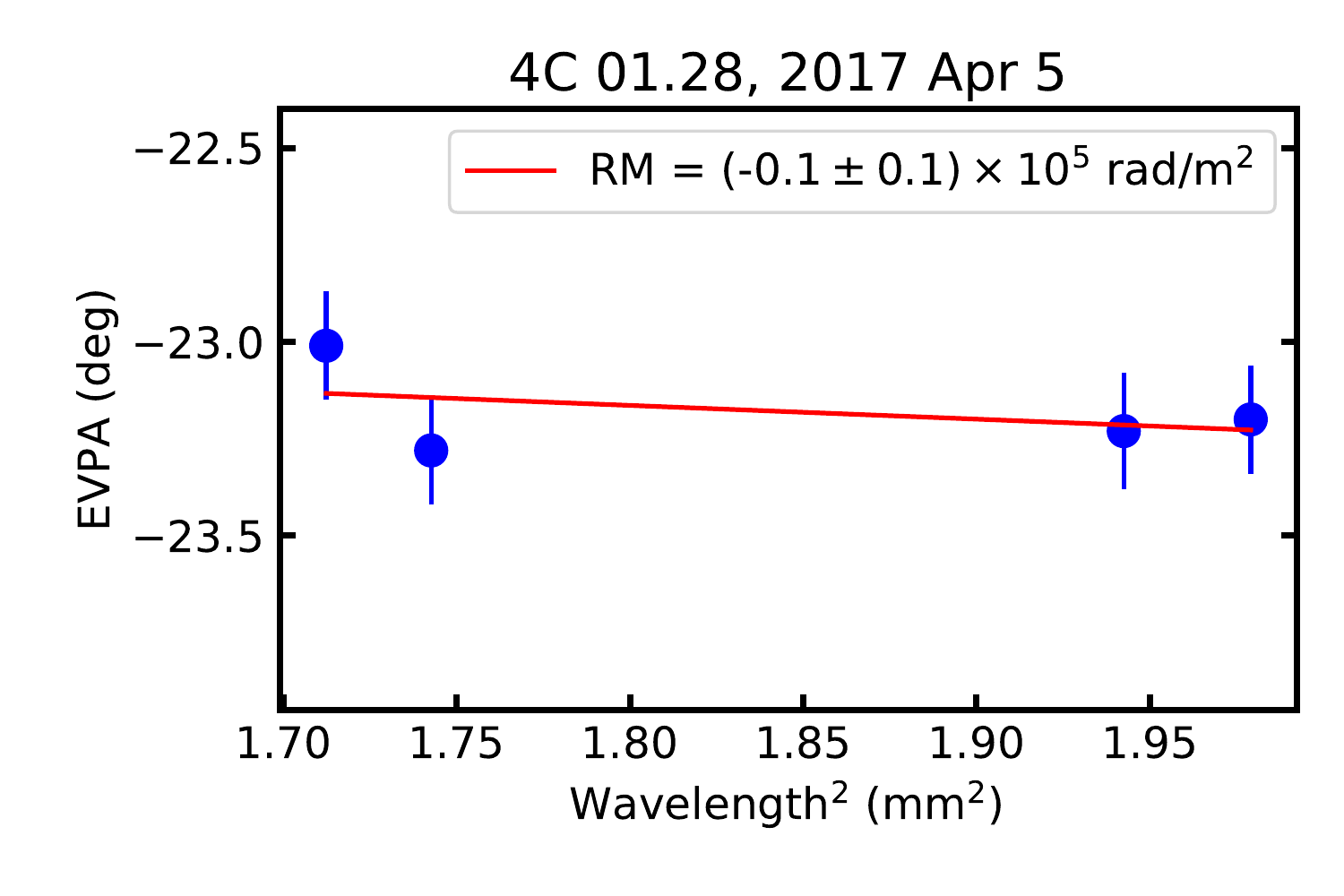} \hspace{-0.35cm}
\includegraphics[width=0.33\textwidth]{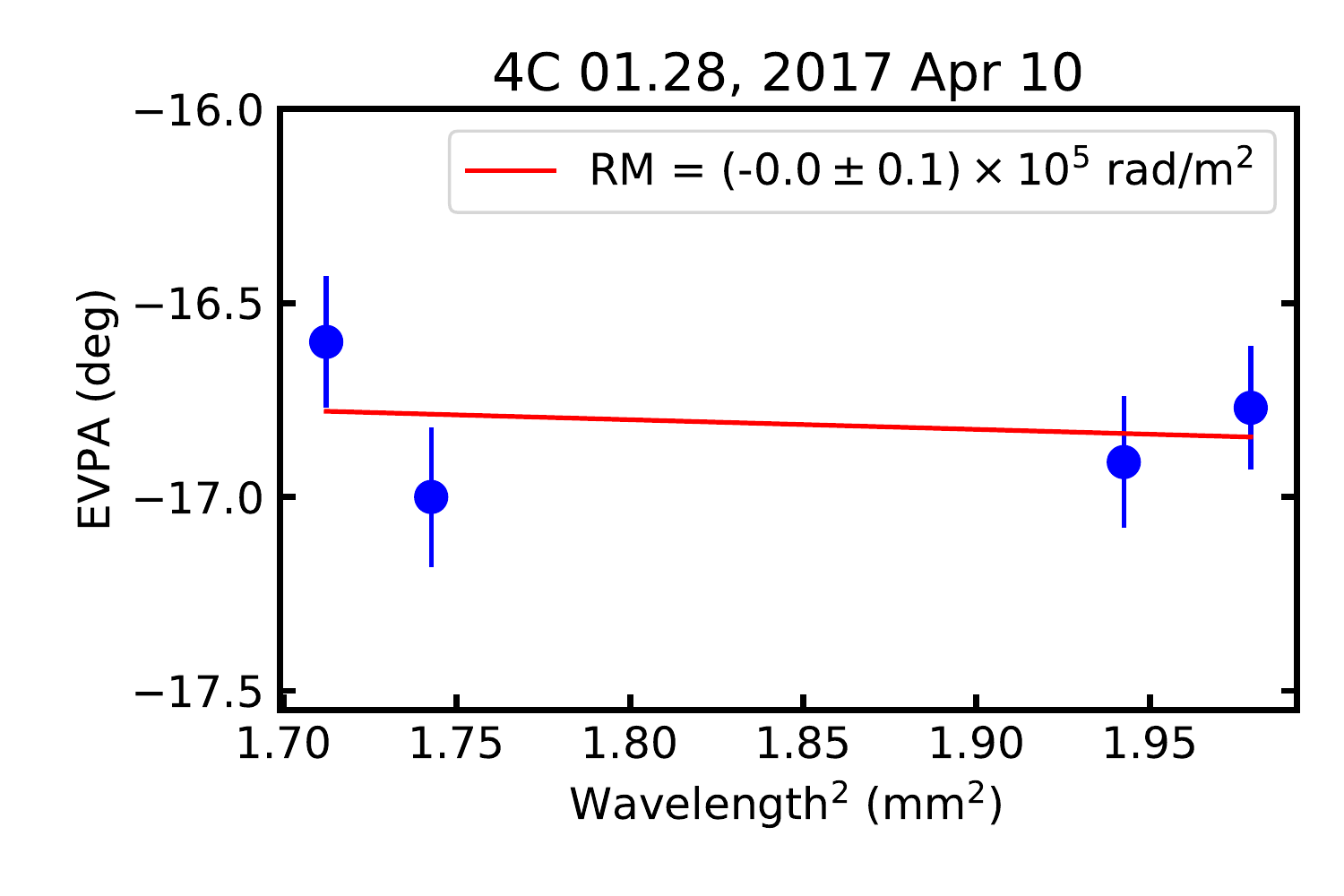} \hspace{-0.35cm}
\includegraphics[width=0.33\textwidth]{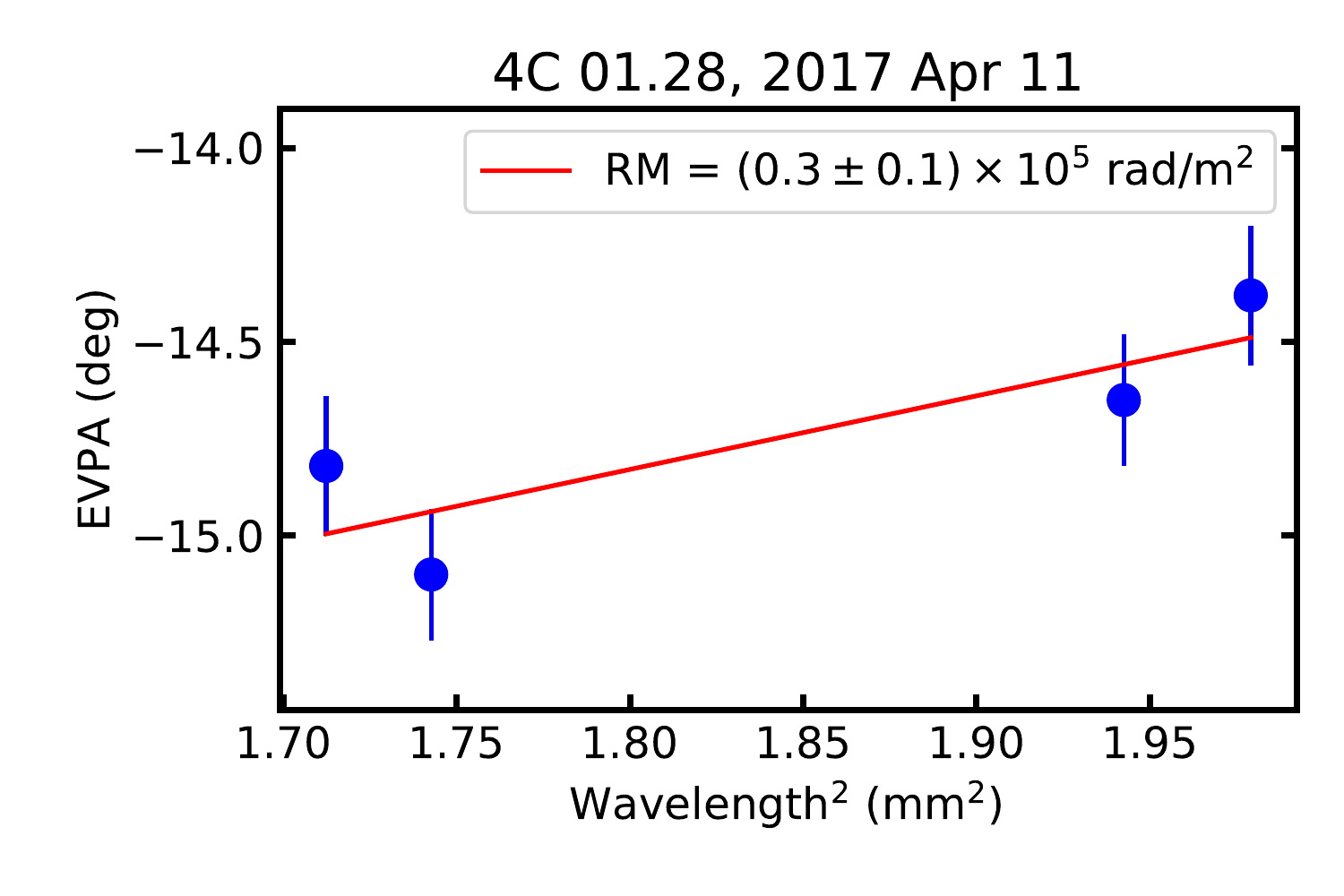} \hspace{-0.35cm}
\includegraphics[width=0.33\textwidth]{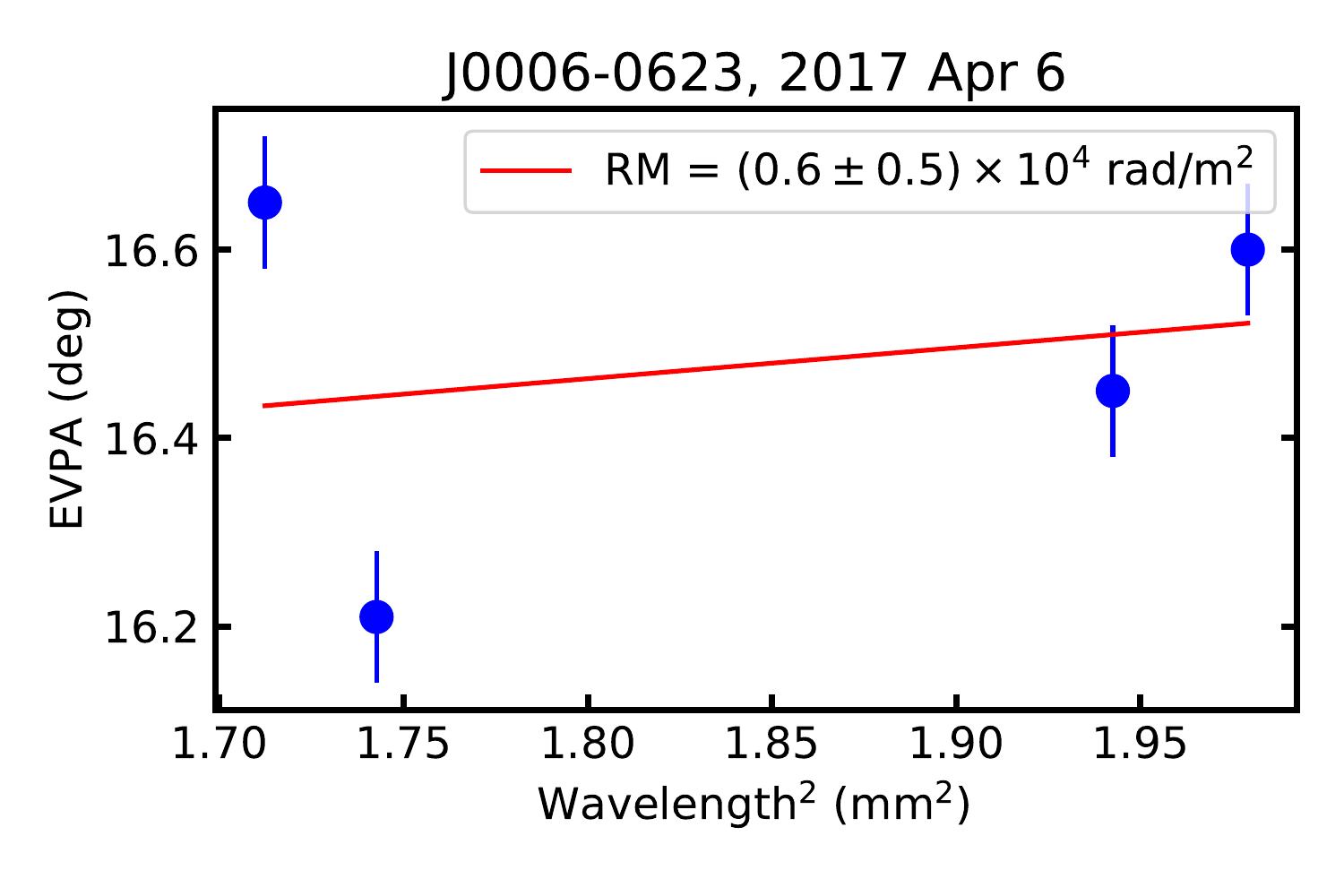} \hspace{-0.35cm}
\includegraphics[width=0.33\textwidth]{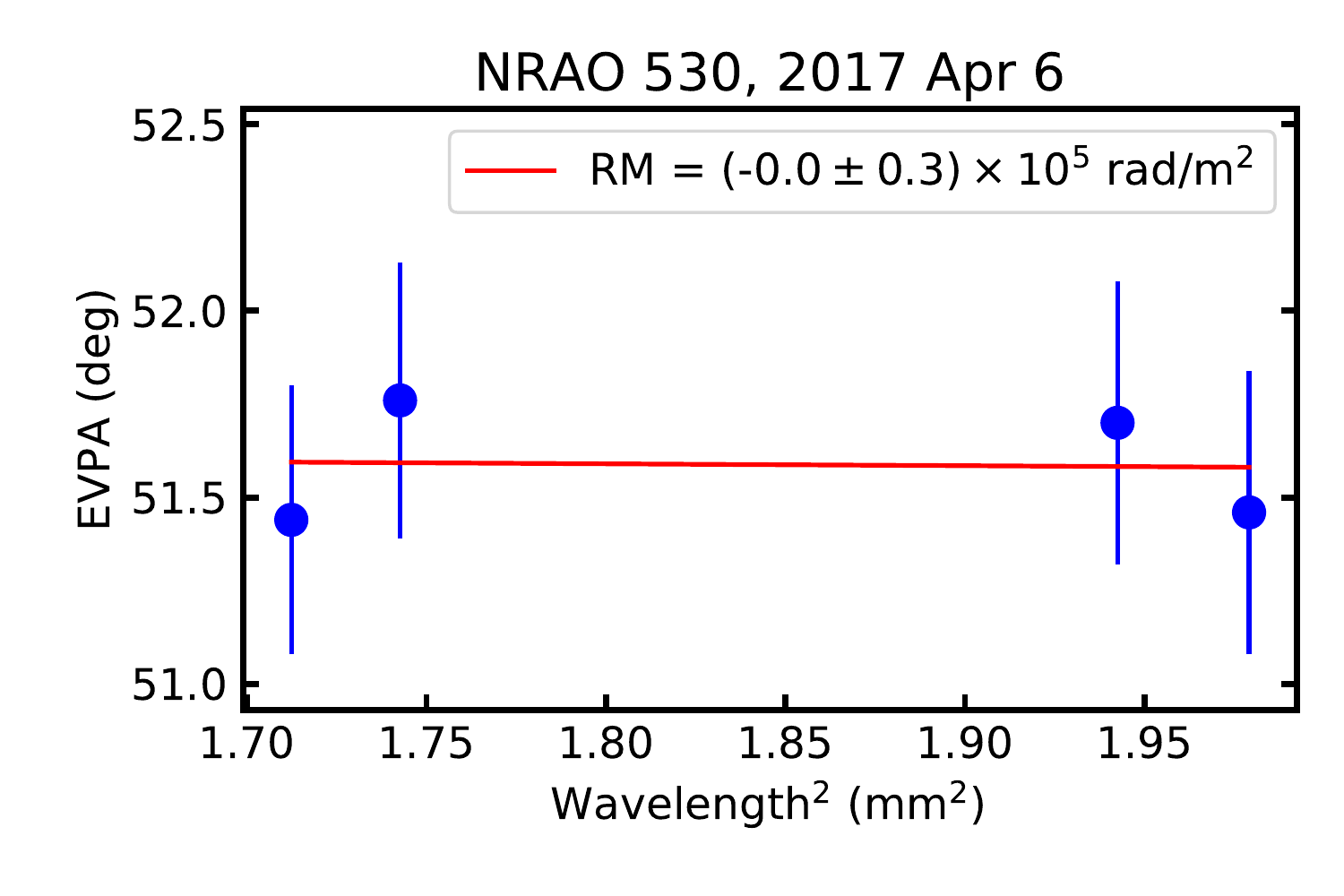} \hspace{-0.35cm}
\includegraphics[width=0.33\textwidth]{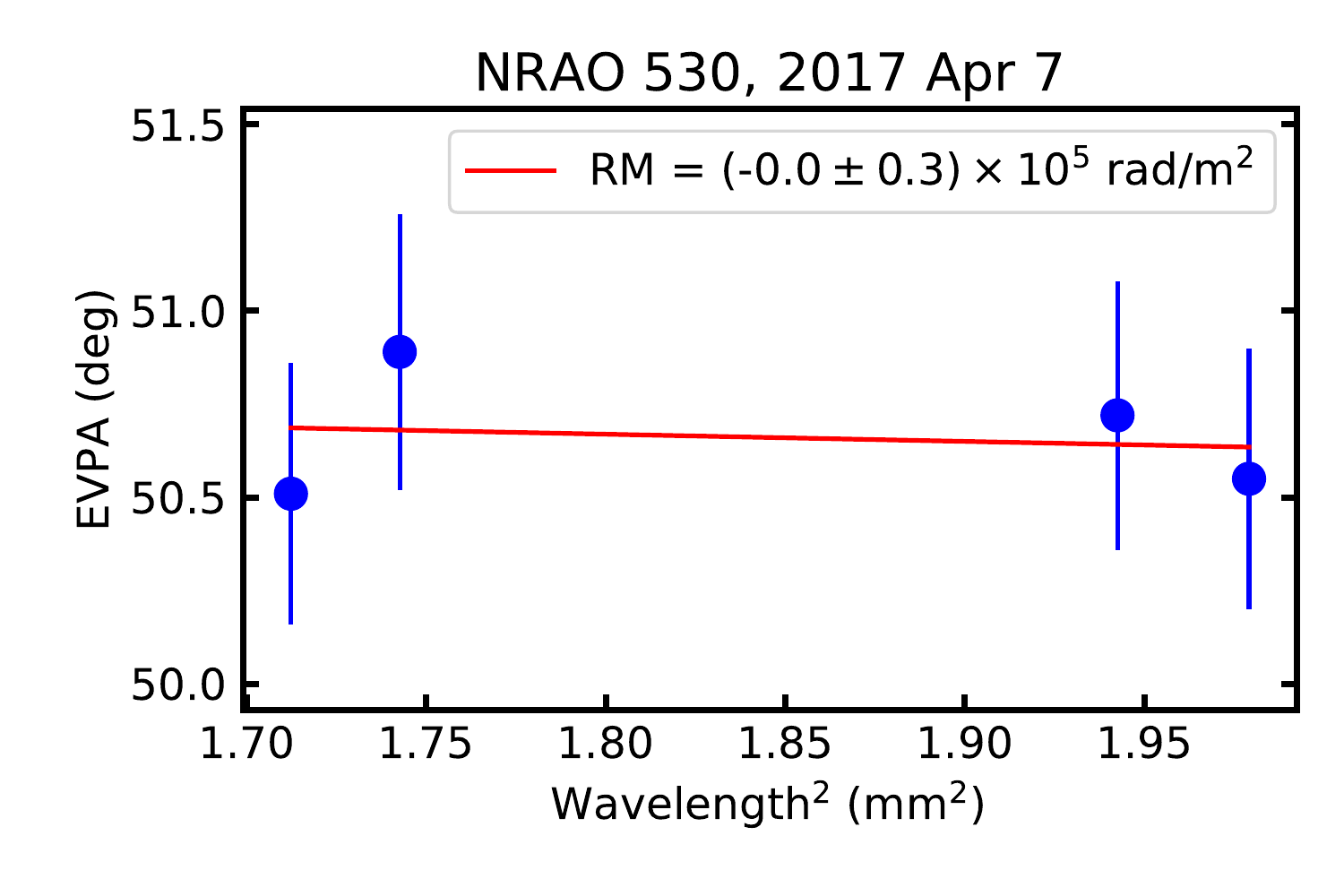} \hspace{-0.35cm}
\includegraphics[width=0.33\textwidth]{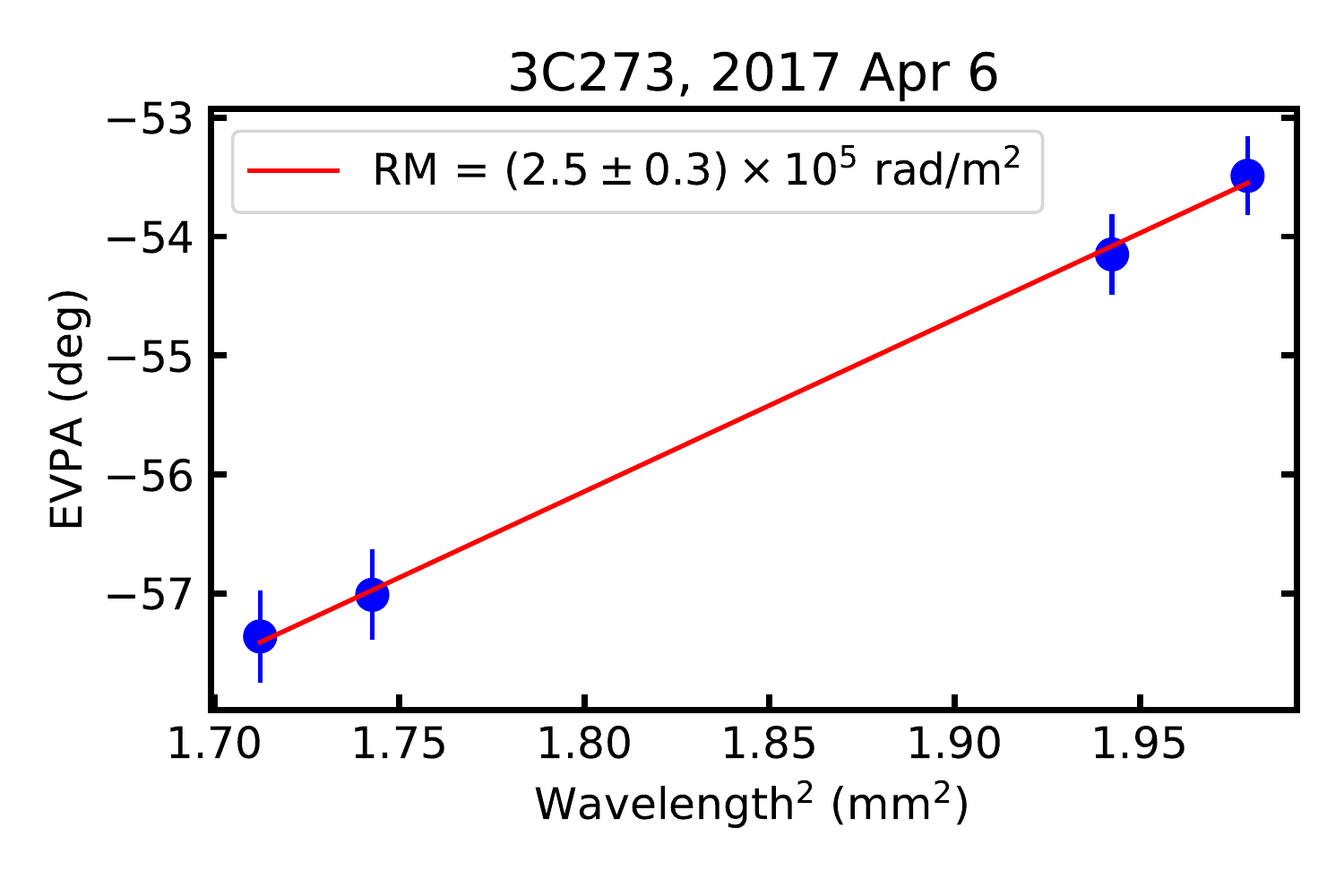} \hspace{-0.35cm}
\includegraphics[width=0.33\textwidth]{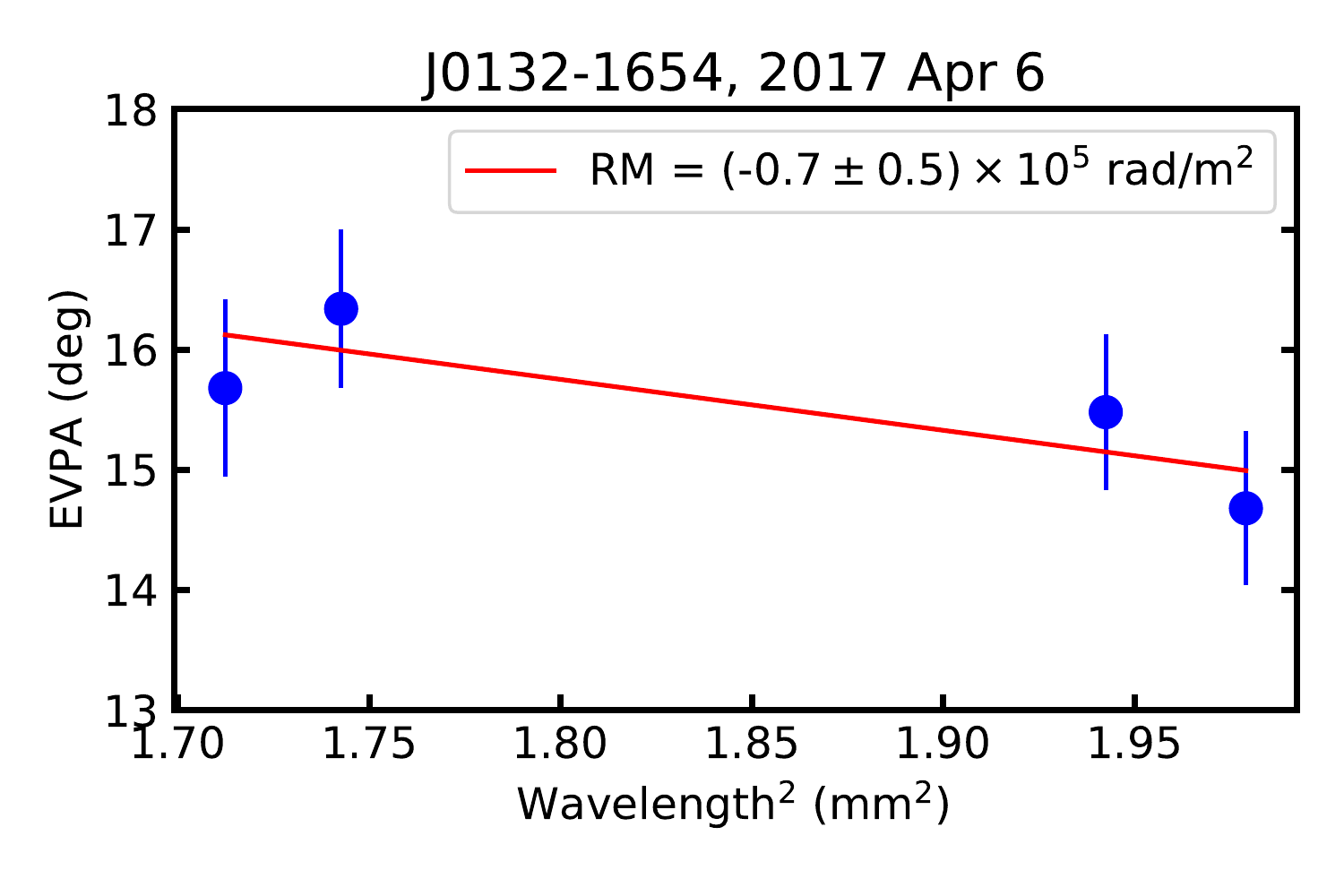} \hspace{-0.35cm}
\includegraphics[width=0.33\textwidth]{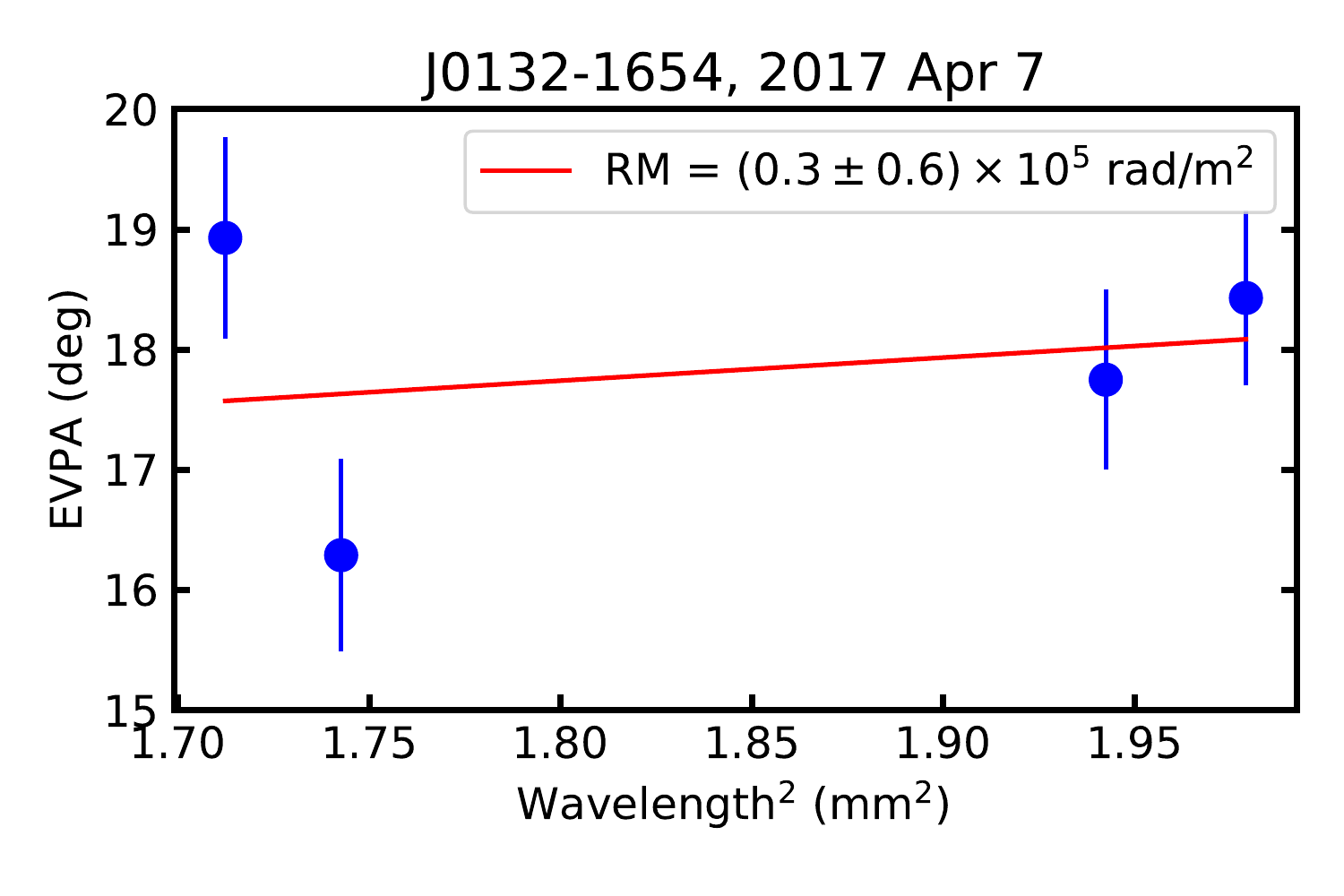}\hspace{-0.35cm} 
\caption{
RM fits for remaining 2017 EHT targets with EVPA measurements at 1.3~mm (see Table~\ref{tab:EHT_uvmf_RM}).
Plots in each row span the same vertical axis range for the same source in order to highlight differences in slope. 
}
\label{fig:RM3}
\end{figure*}

\begin{figure*}[ht!]
\centering
\includegraphics[width=0.25\textwidth]{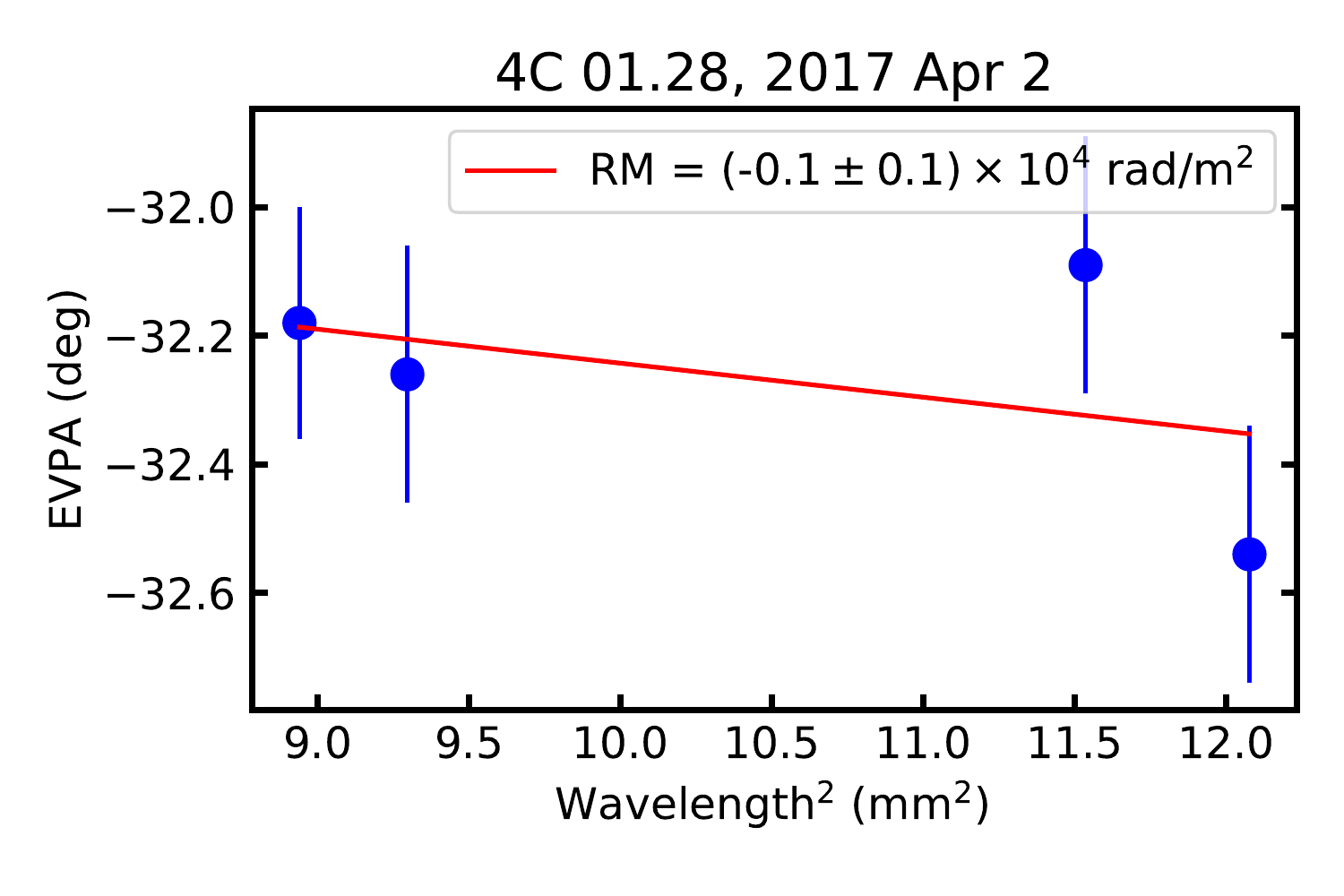} \hspace{-0.35cm}
\includegraphics[width=0.25\textwidth]{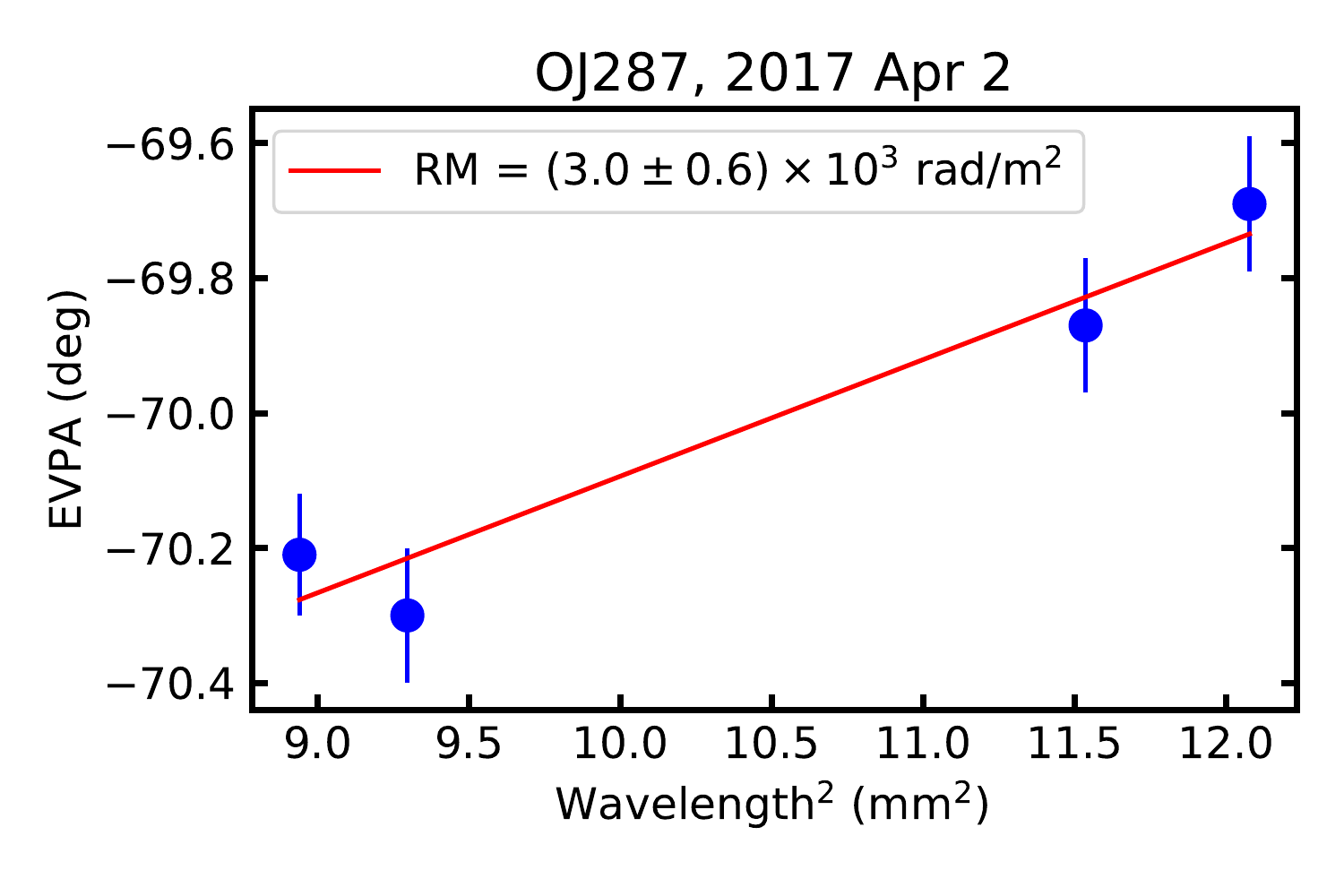} \hspace{-.35cm}
\includegraphics[width=0.25\textwidth]{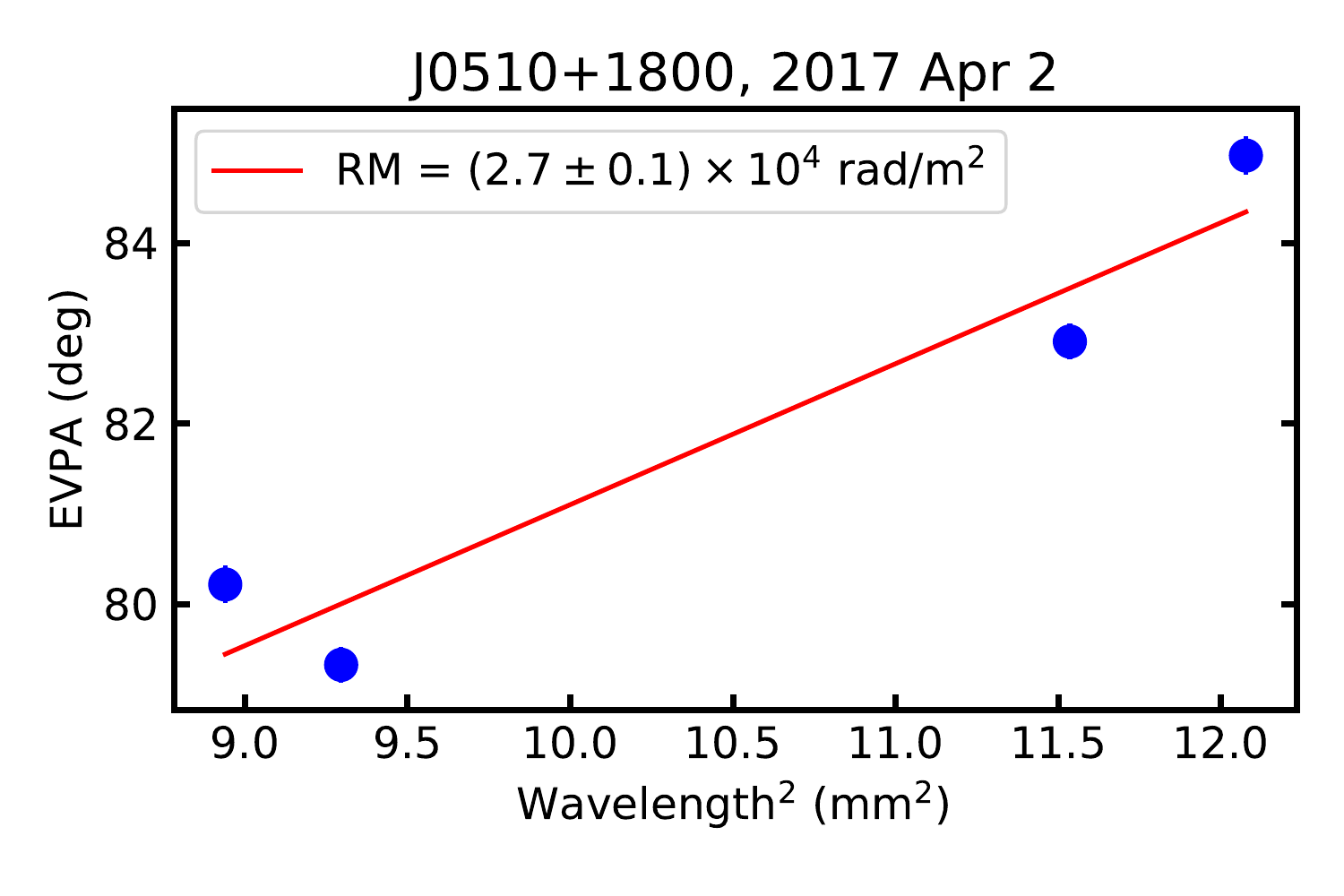} \hspace{-.35cm}
\includegraphics[width=0.25\textwidth]{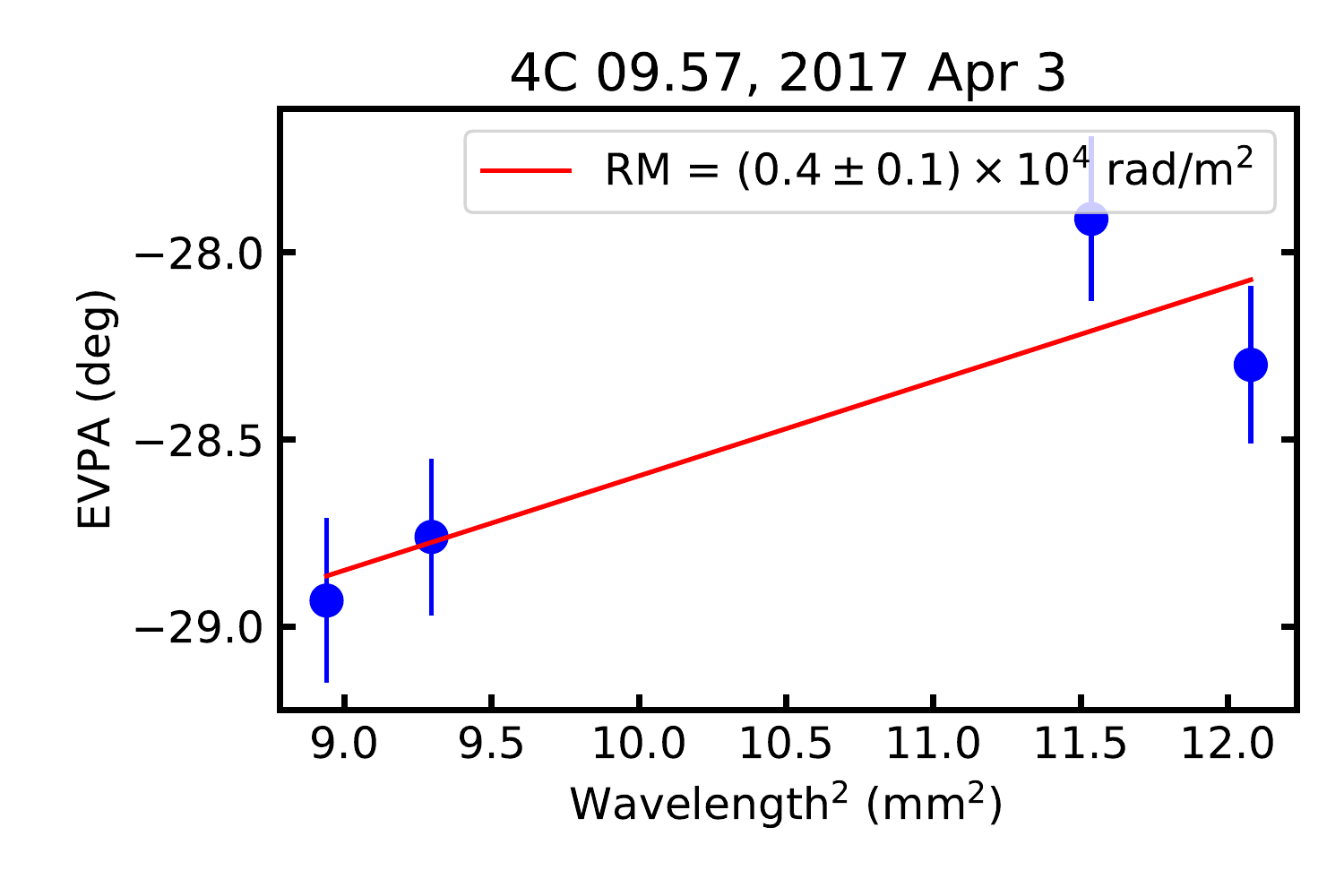} \hspace{-0.35cm}
\includegraphics[width=0.25\textwidth]{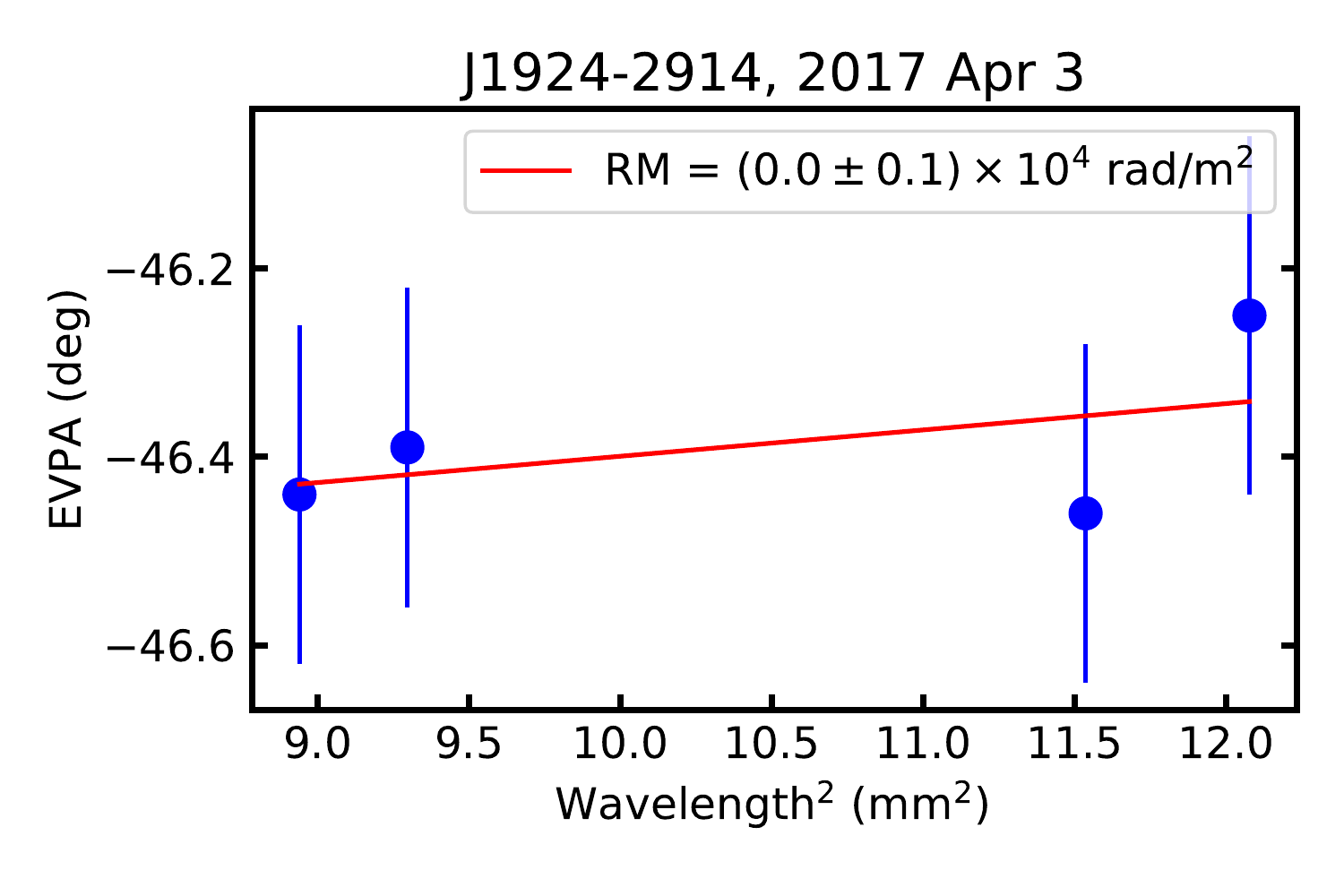} \hspace{-0.35cm}
\includegraphics[width=0.25\textwidth]{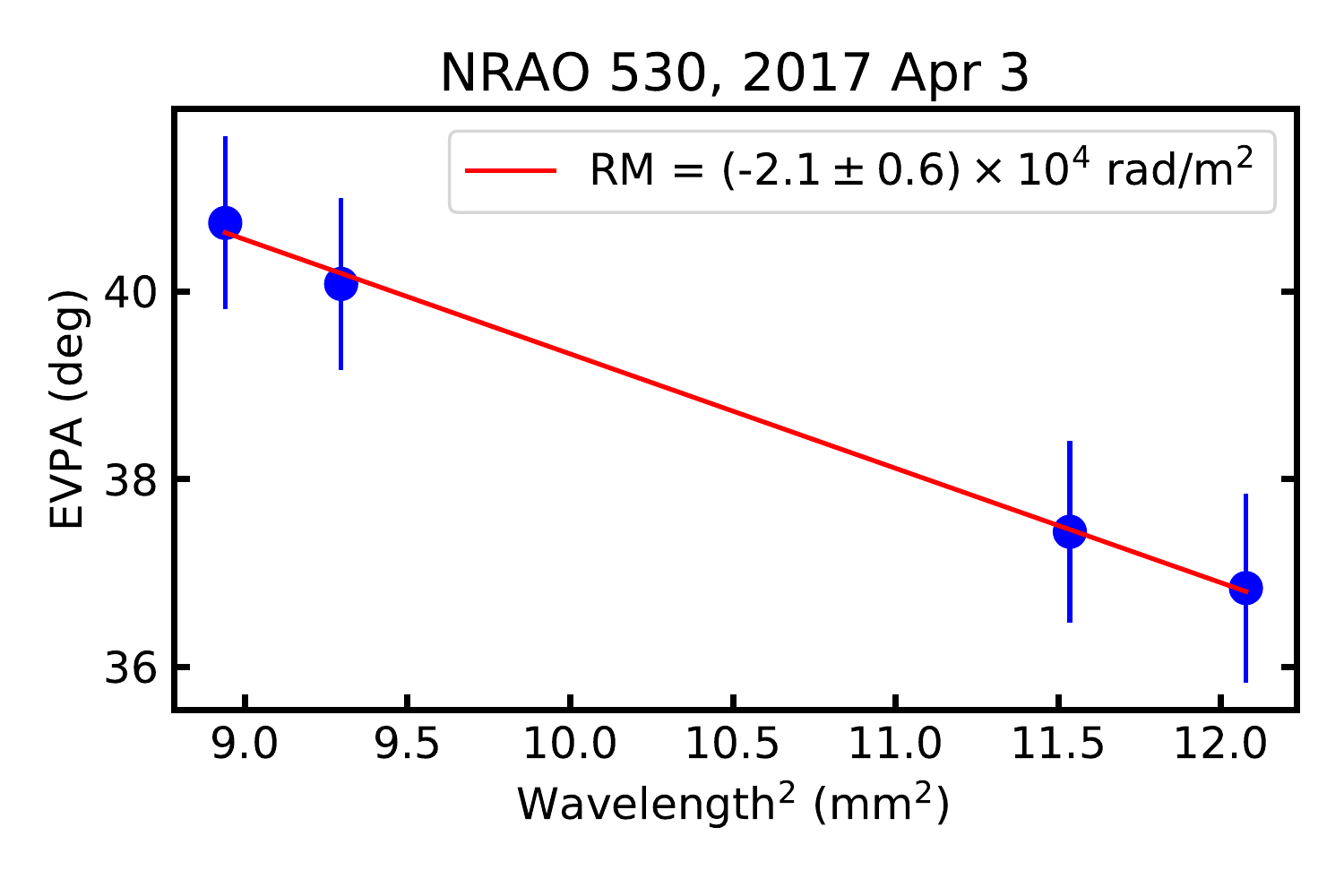} \hspace{-0.35cm}
\includegraphics[width=0.25\textwidth]{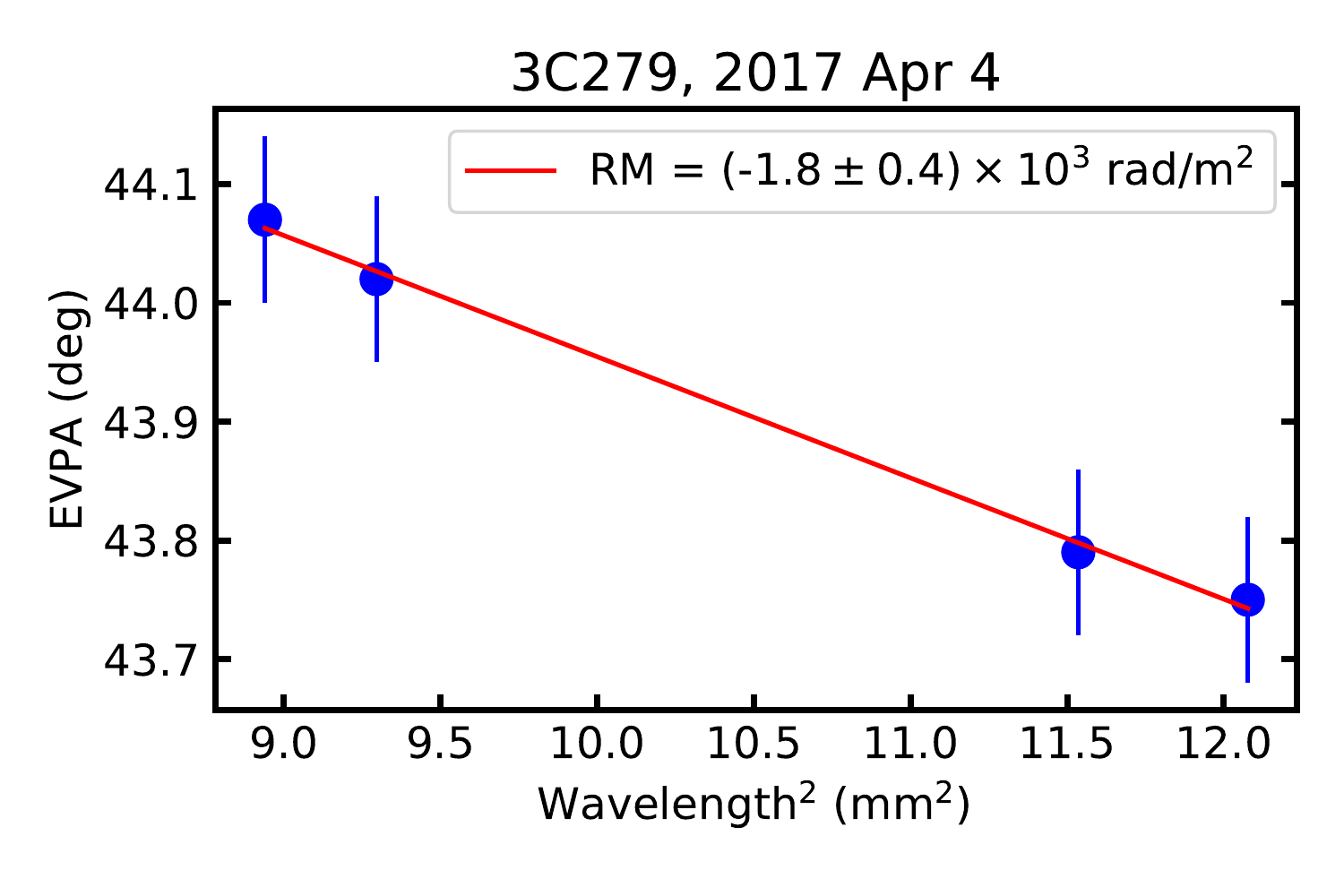} \hspace{-0.35cm}
\includegraphics[width=0.25\textwidth]{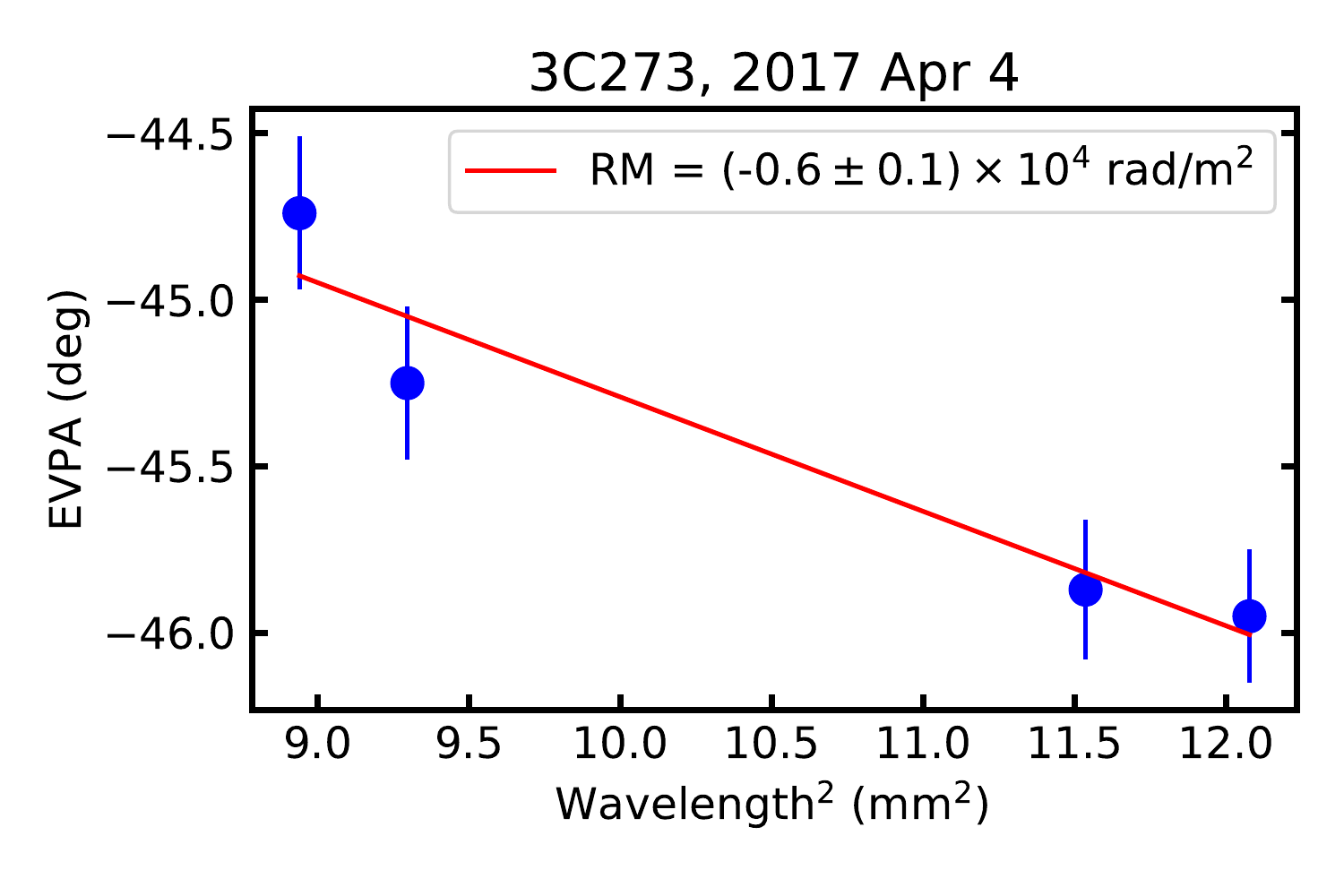} \hspace{-0.35cm}
\caption{
RM fits for the 2017 GMVA targets with EVPA measurements at 3~mm (see Table~\ref{tab:GMVA_uvmf_RM}). 
}
\label{fig:RM_3mm}
\end{figure*}

\begin{figure*}
\centering
\includegraphics[width=0.33\textwidth]{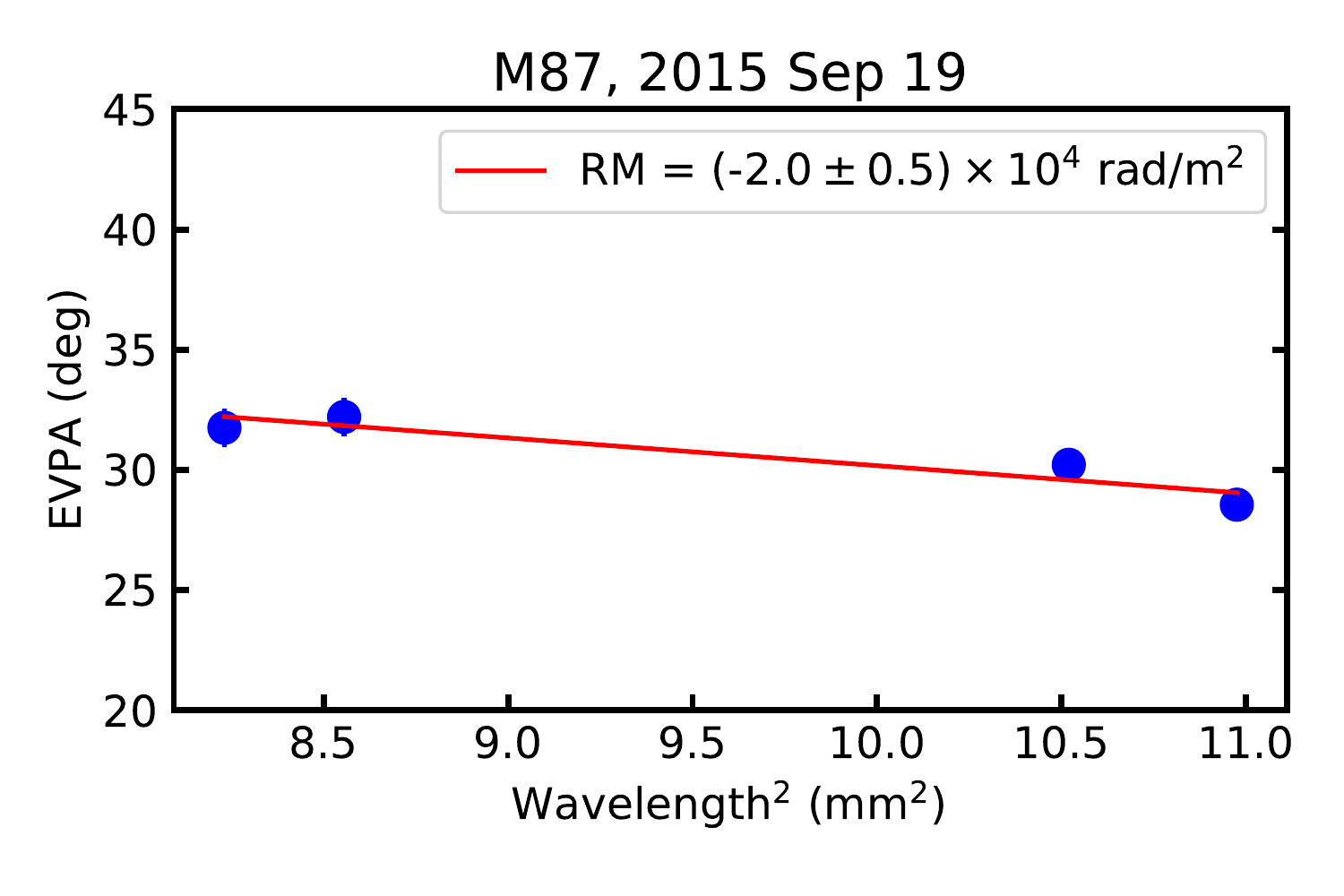} \hspace{-0.35cm}
\includegraphics[width=0.33\textwidth]{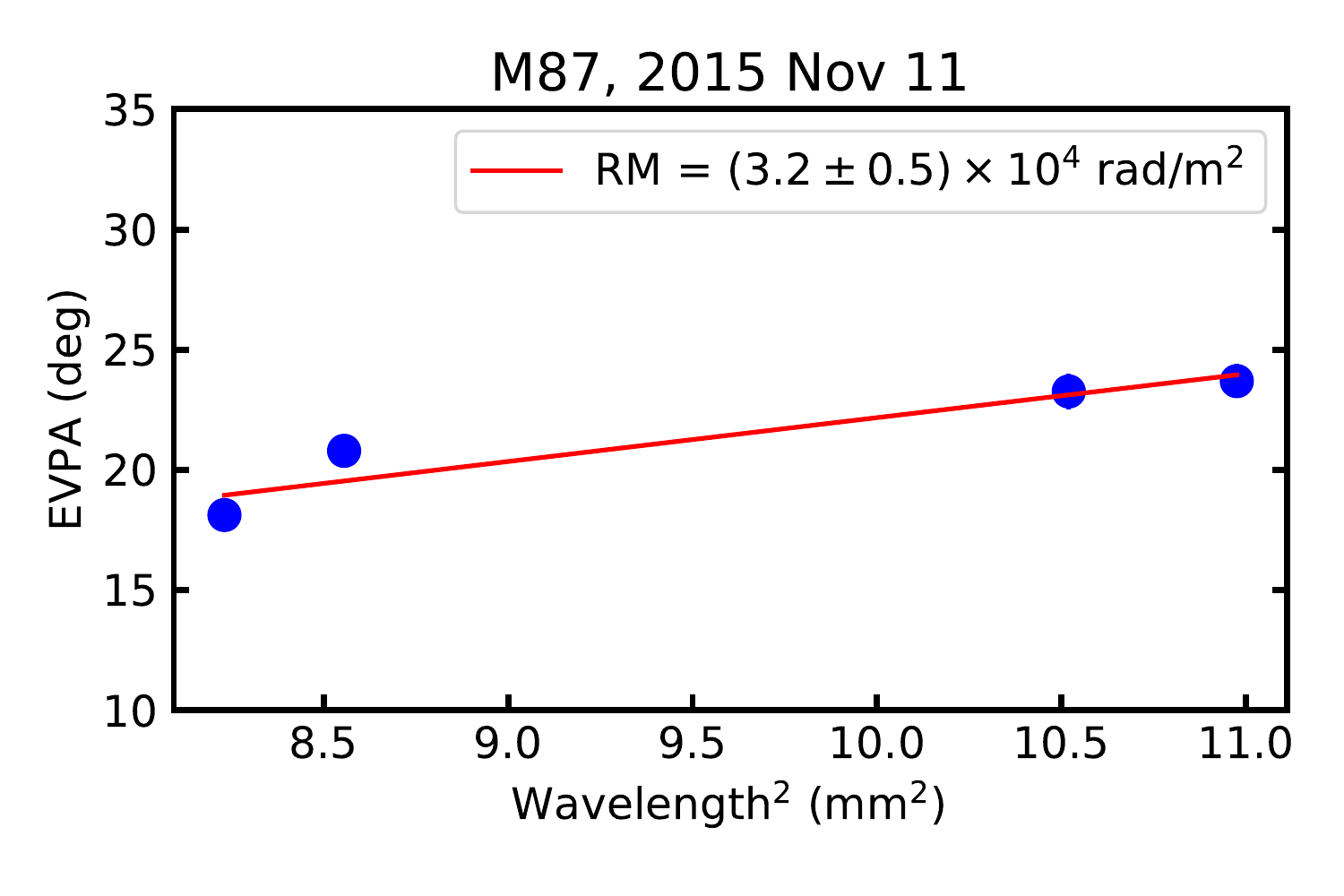} \hspace{-0.35cm}
\includegraphics[width=0.33\textwidth]{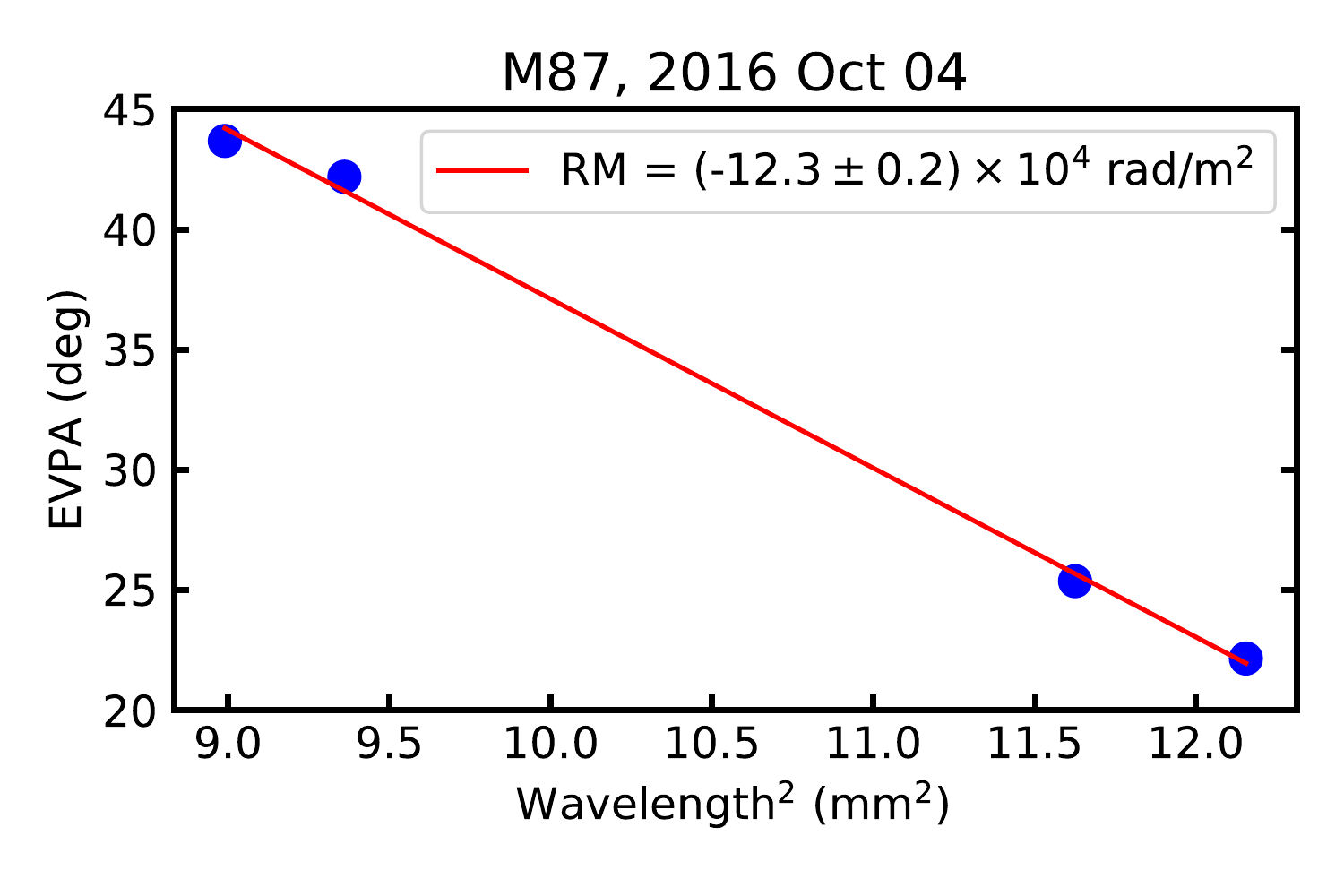} \includegraphics[width=0.33\textwidth]{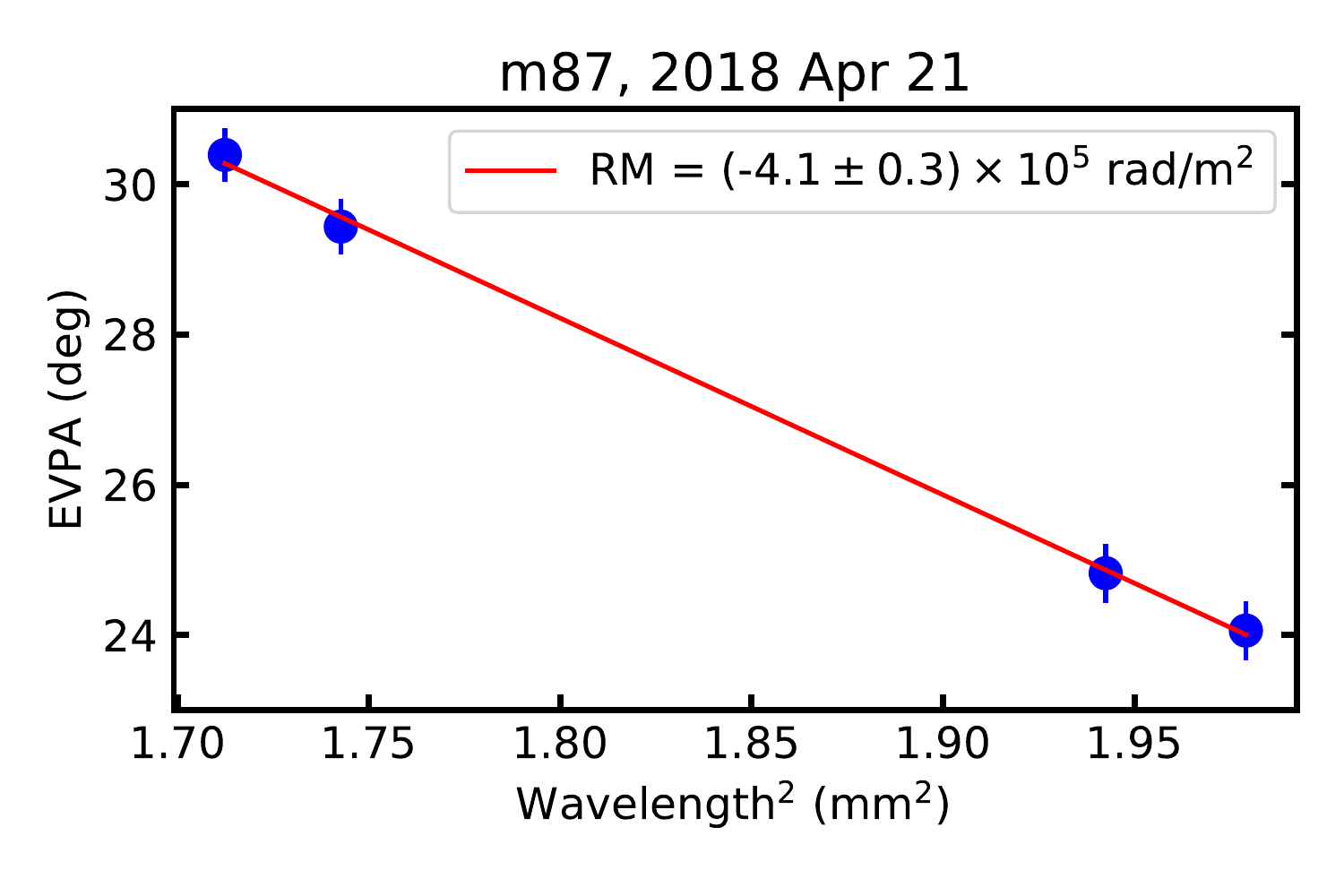} \hspace{-0.35cm}
\includegraphics[width=0.33\textwidth]{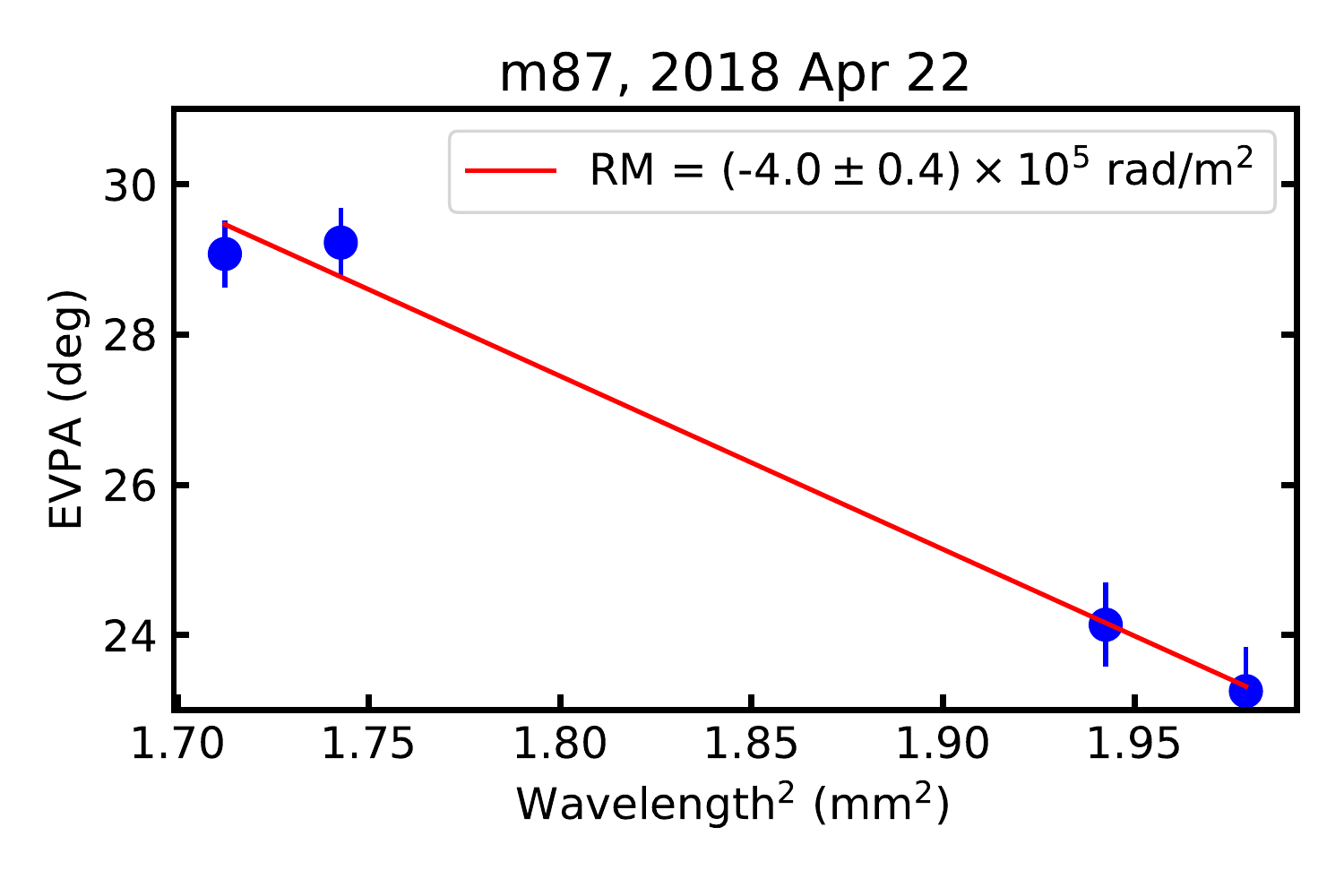} \hspace{-0.35cm}
\includegraphics[width=0.33\textwidth]{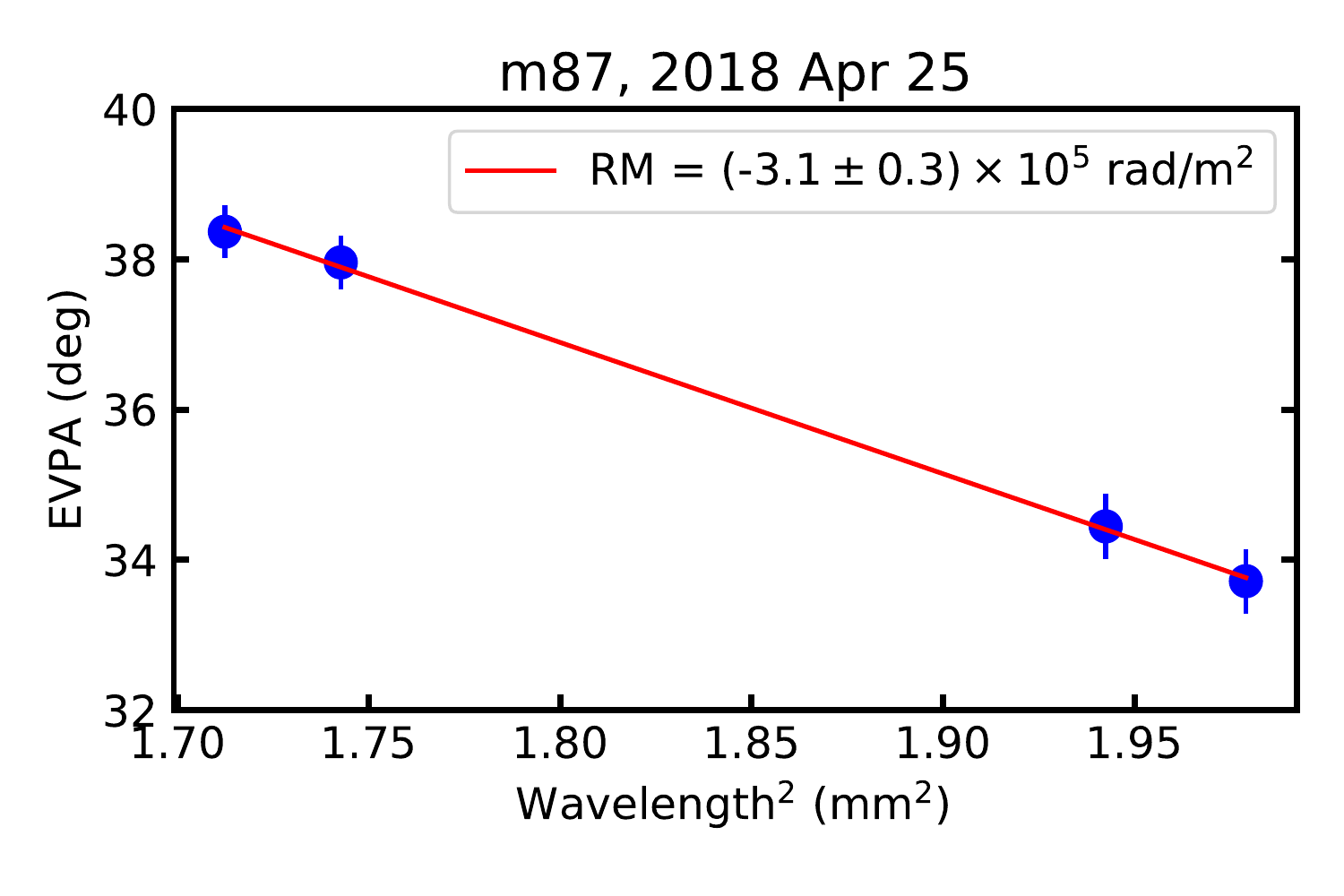} \hspace{-0.35cm}
\caption{
RM of M87 from observations carried out in Sep \& Nov 2015 and Oct 2016 at 3~mm (top row) and in Apr  2018 at 1.3~mm (bottom row). 
Plots in both rows span the same vertical axis range (25\dg\ and 8\dg\ respectively) in order to highlight differences in slope (but note the different EVPA values in different panels).
}
\label{fig:RM_M87_3+1mm}
\end{figure*}

We display the EVPA data  as a function of $\lambda^2$  and their RM fitted models 
 in Figures~\ref{fig:RM1}, \ref{fig:RM3},  \ref{fig:RM_3mm}, and \ref{fig:RM_M87_3+1mm}; EVPAs are measured  in each of the four ALMA SPWs. 
Let us first focus on Figure~\ref{fig:RM1}, showing the
Faraday RM of Sgr A* (upper panels), M87 (middle panels), and 3C~279 (lower panels). 
The EVPAs measured in Sgr~A* reveal  a remarkably precise $\lambda^2$ dependence both at $\lambda$3\,mm (left panel) and $\lambda$1.3\,mm  (second to fourth panels). 
It is also remarkable that Sgr~A* showcases a very consistent slope across all days, while in M87 the slope appears to change sign from Apr 5/6 to Apr 10/11. 
3C~279 was used as polarization calibrator on Apr 5, 6, 10, and 11, and its measured RM (consistent with zero across all days) demonstrates the stability of the polarization measurements on M87 and on Sgr~A* on the same days. 
On Apr 7, the polarization calibrator was J1924-2914, which also shows consistent RM across days  (see Figure~\ref{fig:RM3}, upper panels). 
Collectively, they demonstrate the stability of the polarization measurements in Apr 2017 on Sgr~A*, M87, and, by extension, on the remaining AGN observed at $\lambda$1.3\,mm (displayed in Figure~\ref{fig:RM3}) and at $\lambda$3\,mm (displayed in Figure~\ref{fig:RM_3mm}).

We notice that, for some of the targets, the RM fits are almost perfectly consistent with the measured EVPAs; in particular, the plots for  Sgr A*, 3C~279  (Fig.~\ref{fig:RM1}), 3C~273, J1924-2914 (Fig.~\ref{fig:RM3}),  and M87 (Fig.~\ref{fig:RM_M87_3+1mm}), look very unlikely given the  error bars.
In fact, a standard $\chi^2$ analysis for these sources would give values close to 0 
while for 4 points/2-parameter fit, we would expect  $\chi^2\sim$2.
 This apparently unexpected behaviour  can be explained by the fact that the EVPA uncertainties displayed in the RM plots are not just the thermal errors (which would naturally introduce scatter in the measurements)  but also include a systematic error (0.03\% of $I$ into $QU$ errors), which in fact  dominates the total error budget (especially for the strongest sources with high Stokes $I$). 
We assessed that once  such  systematic error is removed,
the EVPA data points are no longer consistent with the line to within their (thermal-noise-only) uncertainties. The fact that the thermal-only error bars are too small, but the error bars in the plots of this paper that include the systematic 0.03\% Stokes\,$I$ leakage into the $QU$ error budget are too large, suggests that there is a real systematic error in the EVPA measurements but that it is smaller than the ALMA standard value. The fact that we are being conservative in our error estimates should ensure that we are not over-interpreting our measurements.

Finally, we notice that, for some of the targets and/or on specific days, there are $>1\sigma$ deviations between the observed EVPA and the RM fit-predicted values (e.g., OJ287,  4C 01.28, J0006-0623, J0510+1800; see selected panels in  Figs.~\ref{fig:RM3} and \ref{fig:RM_3mm}). 
This may suggest either the presence of an additional systematic error not accounted for in our error budget or that the assumption of the EVPA $\lambda^2$-dependence is not valid in some cases. 
In fact, in \S~\ref{subsec:depol},\S~\ref{sect:LP_1mmvs3mm}, and \S~\ref{sect:m87_disc} we discuss the possibility that some of the observed Faraday rotation may be partly internal,  which would imply a more complex RM model than assumed in our analysis. 
Without a wider wavelength coverage, we cannot distinguish between the two cases.  Therefore, we will not  consider any additional systematic error in the RM analysis presented in this paper.  

\color{black}


\section{Circular Polarization and Assessment of the ALMA Polarimetry}
  \label{app:stokesV}

A reliable detection of Stokes V in interferometric observations with linear-polarization feeds (i.e., the case of ALMA) strongly depends on the correct estimate of two instrumental quantities. On the one hand, the relative phase, $\Delta$, between the X and Y polarizers (i.e., the cross-polarization phase) at the reference antenna \citep[i.e., the quantity stored in the {\tt XY0.APP} table, see][]{QA2Paper}. On the other hand, the imaginary parts of the Dterms that describe the polarization leakage of the individual ALMA elements \citep[i.e., the quantities stored either in the {\tt Df0.APP} or in the {\tt Df0.ALMA} table, depending on the calibration strategy, as described in][]{QA2Paper}. 
Following the standard ALMA calibration procedure, $\Delta$ is estimated by assuming a negligible Stokes V in the polarization calibrator. With this assumption, and neglecting also the effects from polarization leakage, the cross-polarization correlations ($XY^*$ and $YX^*$) between two ALMA antennas observing a polarized point source are

\begin{equation}
XY^* = e^{-j\Delta} p \sin{\left[2(\phi - \psi)\right]} ~~ \mathrm{and} ~~ YX^* = e^{j\Delta} p \sin{\left[2(\phi - \psi)\right]},
\label{CrossPolEq}
\end{equation}

\noindent where the data are assumed to be already corrected for phase and amplitude gains, $\psi$ and $\phi$ are the feed angle of the antennas\footnote{The parallactic angle plus the rotation of the receiver cartridge with respect to the antenna mount.} and the EVPA of the observed source (respectively), and $p$ is the calibrator's linearly-polarized flux density ($p= LP \times I$).

The only complex quantity that appears in Eqs. \ref{CrossPolEq} is given by the factor $e^{j\Delta}$. Hence, the $XY^*$ (and $YX^*$) visibilities may change their amplitudes as a function of time (via the changes in $\psi$), but their phases will remain constant and equal to $\Delta$. The CASA-based calibration algorithm for ALMA (provided in the \texttt{polcal} task) takes advantage of this fact, and estimates the value of $\Delta$ from the fit of $XY^*$ and/or $YX^*$ (over different values of $\psi$) to a model with a constant phase. We call the phase estimated in this way as $\Delta_{\mathrm{QA2}}$.

\subsection{Effects of an inaccurate ALMA polarization calibration}

\subsubsection{Polarization calibrator}

If the true value of the cross-polarization phase, $\Delta$, is offset from $\Delta_{\mathrm{QA2}}$ by an unknown quantity, $\beta$ (i.e., $\Delta = \beta + \Delta_{\mathrm{QA2}}$), this offset will introduce a leakage-like effect into the polconverted VLBI visibilities, which will be described by the Dterms given in Eq. 13 of \cite{QA2Paper}. If $\beta$ is very small, that equation predicts an ALMA leakage for VLBI which has two remarkable properties: 1) it is pure imaginary, and 2) it is the same for the R and L polarization hands. The value of this Dterm is directly related to $\beta$ via the equation: 

$$ D^{VLBI}_R = D^{VLBI}_L \sim j \beta. $$

Therefore, if the QA2 calibration procedure has introduced an offset $\beta$ into the cross-polarization phase, $\Delta_{\mathrm{QA2}}$, we predict a pure imaginary Dterm in the ALMA-VLBI visibilities. The actual Dterms estimated from the ALMA-VLBI observations during the EHT 2017 campaign are reported in \cite{EHTC2020_1}.

The phase offset $\beta$ may also introduce a spurious V signal into the data, which can be described by the equation \citep[e.g., ][]{Hovatta2019}

\begin{equation}
V^{tot} = V^{const}\cos{\beta} + p\,\sin{\left[2\left(\phi - \psi\right)\right]}\sin{\beta},
\label{SpuriousVEq}
\end{equation}

\noindent where $V^{tot}$ is the total V signal recovered from the (QA2-calibrated) data and $V^{const}$ is a constant V signal, independent of $\psi$. If the QA2 Dterm estimates for ALMA are correct, $V^{const}$ is equal to the true Stokes V of the source, (i.e., $V^{const} = V^{true}$). 
However, if the ALMA Dterms estimated in the QA2 are offset from their true values, there is another instrumental contribution to $XY^*$ and $YX^*$, which couples to  $V^{const}$ as following  

\begin{equation}
  \begin{aligned}
XY^* &= p\,\sin{\left[ 2 \left( \phi - \psi \right) \right]} + jV^{true} + \\
&\left( D_X^a + (D_Y^b)^* \right) I 
+ \mathcal{O}(p\, D ) + \mathcal{O}(V^{true}\, D ).
  \end{aligned}
\label{CrossPolEq3}
\end{equation}

In this equation, $D_X^k$ is the part of the Dterm of the $X$ polarizer of antenna $k$ that remains uncalibrated after the QA2 (and similarly for the $Y$ polarizer). Therefore, if the effects of the antenna Dterms are not fully removed from the data, the calibrator source will show a non-negligible $V^{true}$, while the CASA model assumes it to be null. 

\begin{figure}[t!]
\centering
\includegraphics[width=8cm]{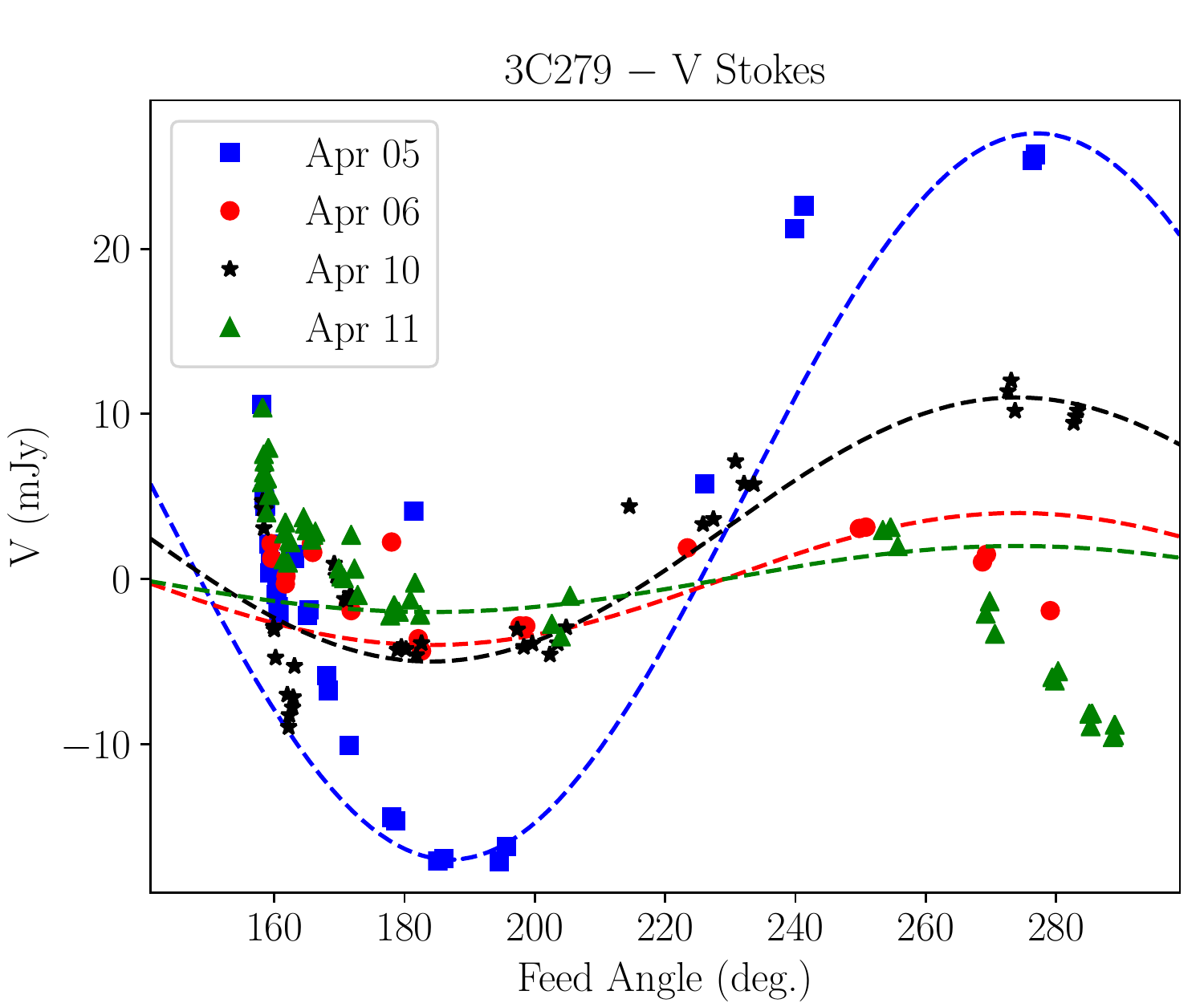}
\caption{Reconstructed Stokes V of 3C\,279 as a function of feed angle,  $\phi$, during the EHT campaign.
 Stokes V is computed as the real part of $(XY^* - YX^*)/j$. 
Dashed lines show the simplified model of Eq. \ref{SpuriousVEq}, based on cross-polarization phase offset, $\beta$, of the phased-ALMA reference antenna.}
\label{VFigure}
\end{figure}

In Fig. \ref{VFigure}, we show the $V^{tot}$ signal (computed as the average real part of $(XY^* - YX^*)/j$ among all ALMA antennas) of the polarization calibrator 3C\,279, as a function of parallactic angle, for the epochs where this source was observed. We have averaged the visibilities in time bins of 120 seconds and the data taken with antenna elevations below 30 degrees have been discarded. In the figure, we also show the simple model given by Eq. \ref{SpuriousVEq}, where $V^{const}$ and $\Delta$ are the only two free parameters used in the fit. 

The V signals from the polarization calibrator show a clear dependence with parallactic angle, which indicates that there are offsets, $\beta$ in the estimated cross-polarization phases, $\Delta_{\mathrm{QA2}}$. Using the values of $p$ and $\phi$ estimated for 3C\,279 on  different days, we fit an X-Y phase offset of $\beta = 0.2-0.5$~degrees in all tracks except April 5, where $\beta \sim 1-2$ degrees. We notice that these ranges assume no bias in the QA2 estimates of the ALMA Dterms.  

It is interesting to note that the data depart from the sinusoidal model of Eq. \ref{SpuriousVEq}, especially for the epochs on April 6 and 11, and for observations far from transit. These deviations may be related to other instrumental effects (e.g., the second-order leakage contributions, like $\mathcal{O}(p\,D)$ in Eq. \ref{CrossPolEq3}).

From the values of $\beta$ fitted with Eq. \ref{SpuriousVEq}, we can estimate a rough upper bound for the Stokes V of the calibrator, assuming a perfect calibration of the ALMA Dterms. 
For 3C\,279, our analysis yields V = [27, 4, 4, 2] mJy, corresponding to  CP = [0.30, 0.04, 0.07, 0.02] \%, and $\beta$ = [2.2, 0.67, 0.57, 0.35] degrees, on Apr 5, 6, 10, 11, respectively. For J1924-2914, our analysis yields V = 0.7 mJy, corresponding to  CP =  0.02\%, and $\beta$ = -0.91 degrees, on Apr 7 (3C279 was not observed on Apr 7).

The estimated values of $V^{const}$ obtained from Eq. \ref{SpuriousVEq} are all of the order of 1\,mJy, at most. These values are small compared to the $V^{true}$ coming from the $\beta$ estimates. This is especially true for the epoch of April 5. As discussed at the beginning of this section, a low $V^{const}$ can be explained by the compensating effect of (small) biases in the QA2 estimates of the ALMA Dterms (Eq. \ref{CrossPolEq3}). Such systematics would force $V^{const}$ to be close to zero, regardless of the value of $V^{true}$.

In summary, two clear conclusions can be drawn from Fig. \ref{VFigure}. On the one hand, there is an offset, $\beta$, in the X-Y cross-polarization phase of the reference antenna. On the other hand, there may be other systematics in the estimates of the antenna Dterms that may compensate the imprint of a true circular polarization of the calibrator (since CASA always assumes a null Stokes V in the calibrator).

\subsubsection{Other sources}

The Dterms and cross-polarization phases discussed in the previous subsection are applied to all sources in the data. Hence, the V Stokes from the polarization calibrator, which may have introduced a cross-polarization phase offset, $\beta$, and other biases to the Dterm estimates, will be systematically put back (after applying the polarization calibration) into the Stokes V signals of the rest of sources. 

Thus, if we find different values of the fractional V Stokes among different sources, there has to be a contribution to their $V^{const}$ values (Eqs. \ref{SpuriousVEq} and \ref{CrossPolEq3}) that is independent of that introduced by the polarization calibration. In the frame of our modelling of the instrumental polarization, such a contribution would likely be related to $V^{true}$, i.e., a true circular polarization associated to the sources.

\begin{figure*}
\centering
\includegraphics[width=8cm]{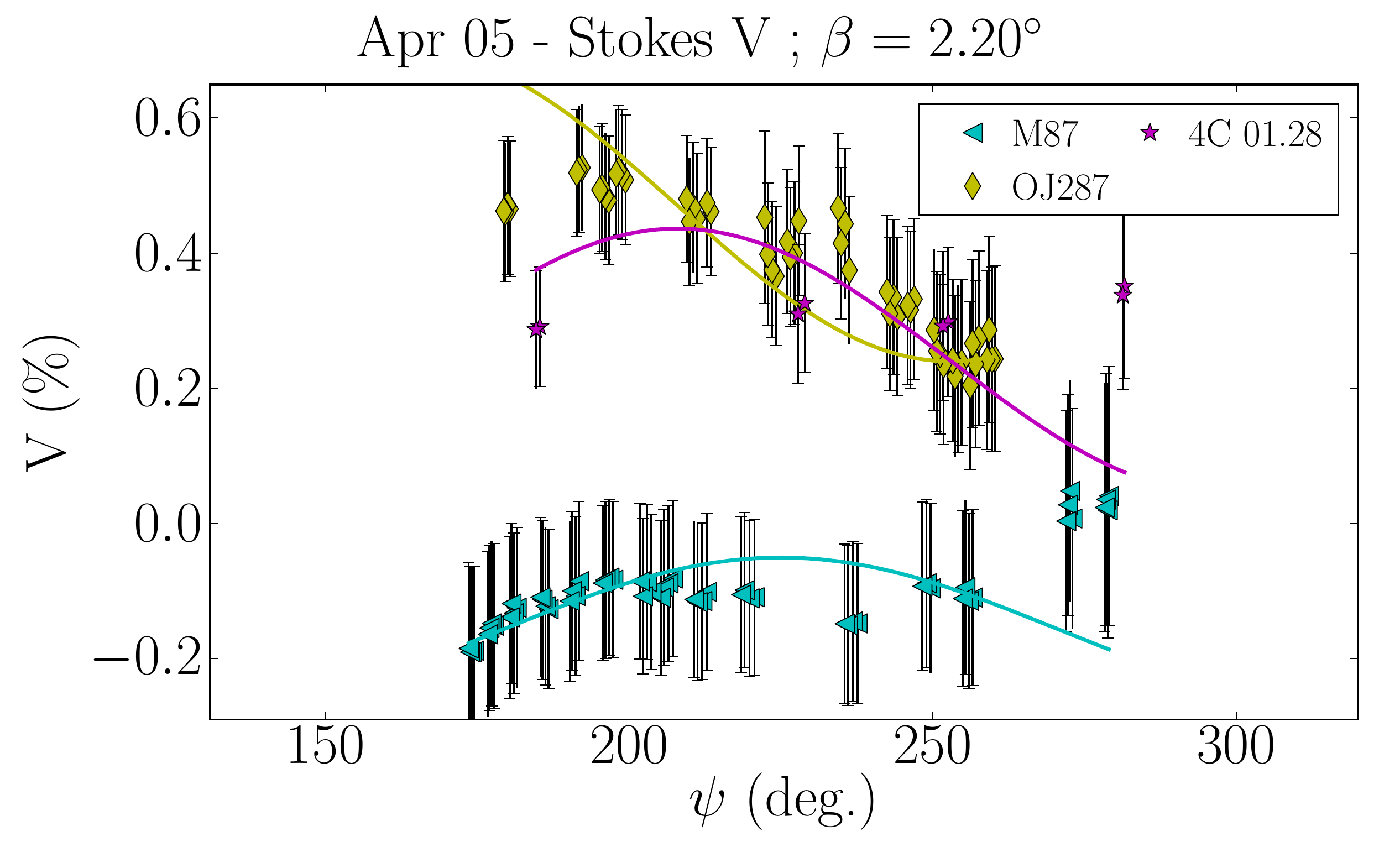}
\includegraphics[width=8cm]{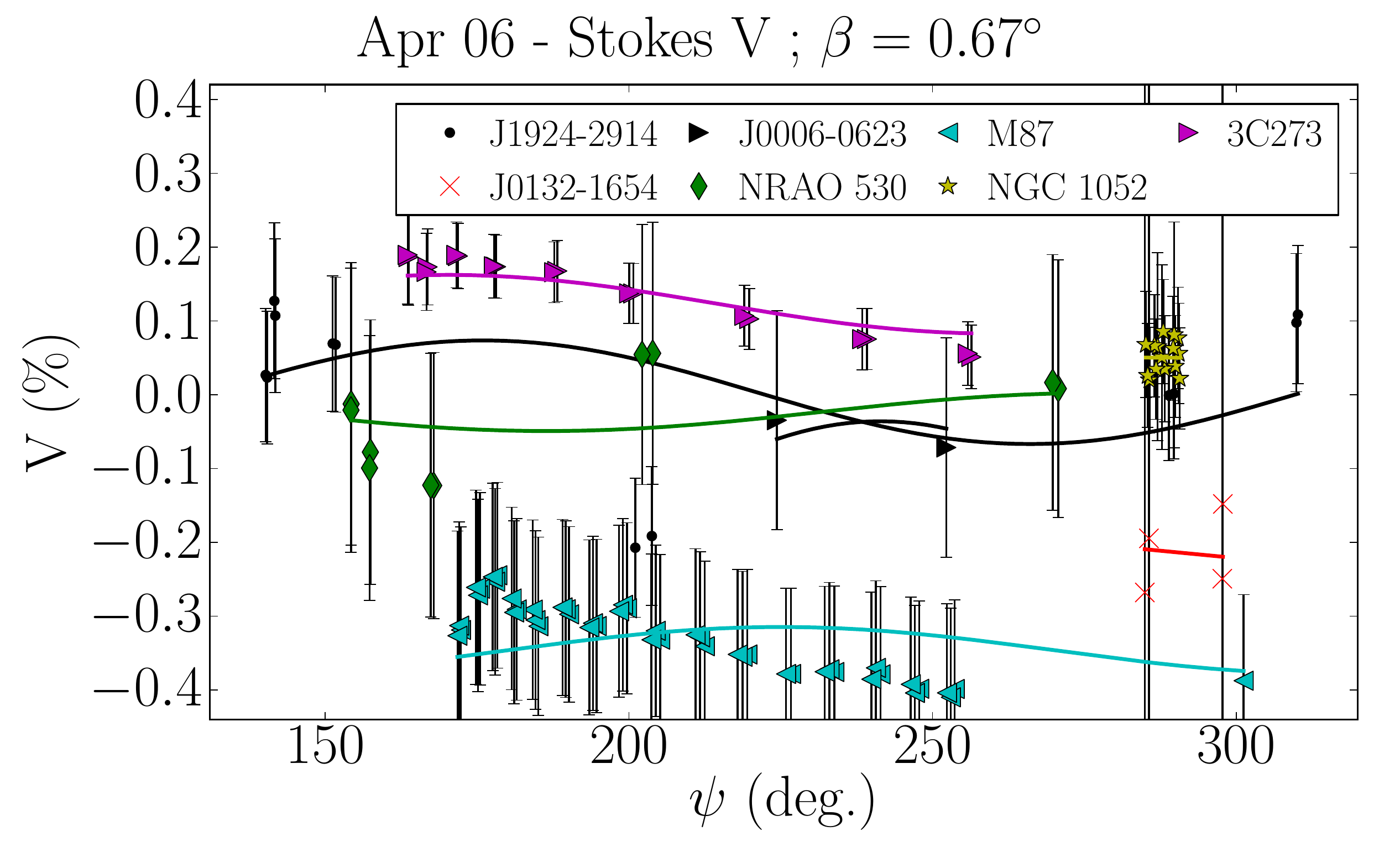}
\includegraphics[width=8cm]{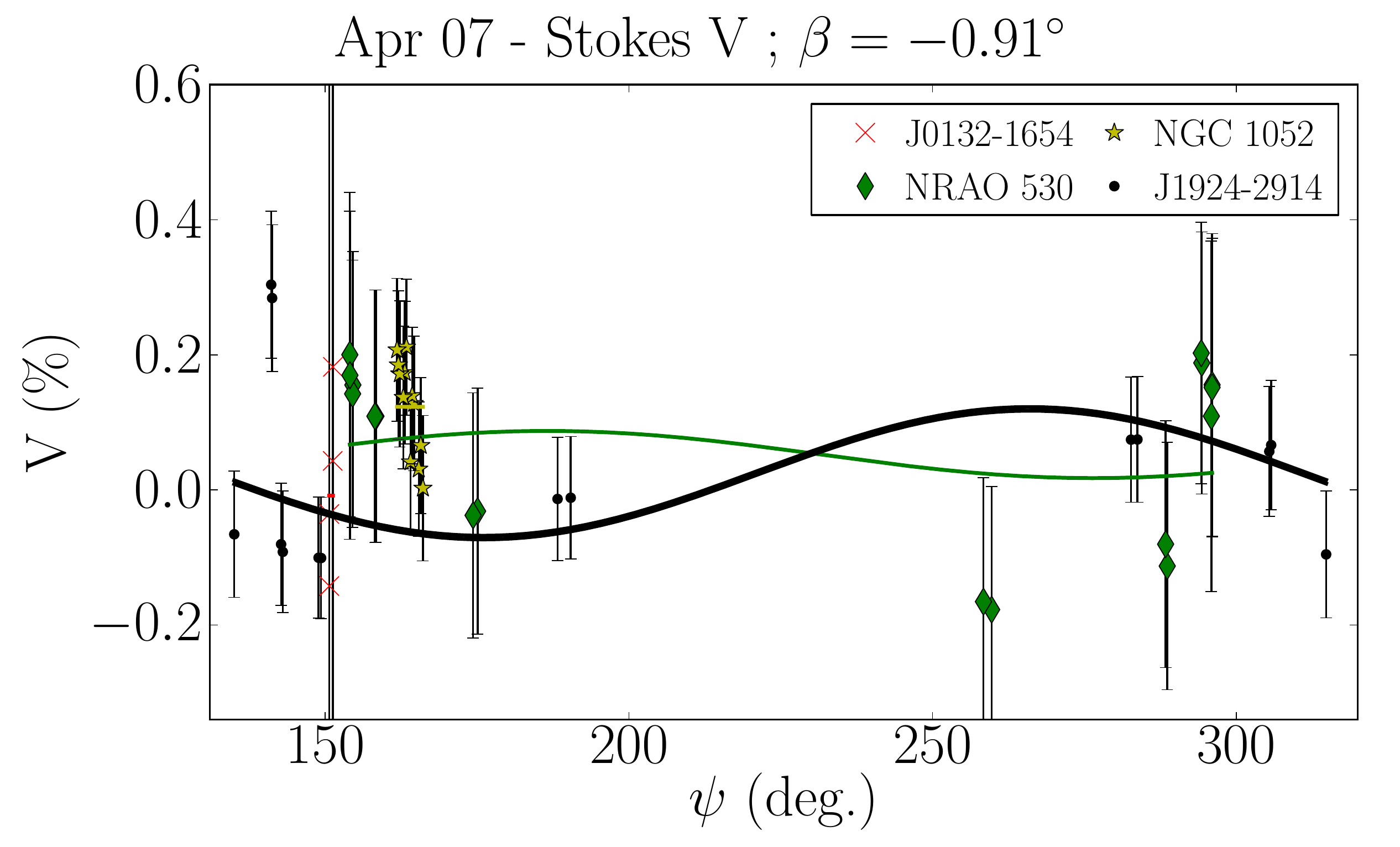}
\includegraphics[width=8cm]{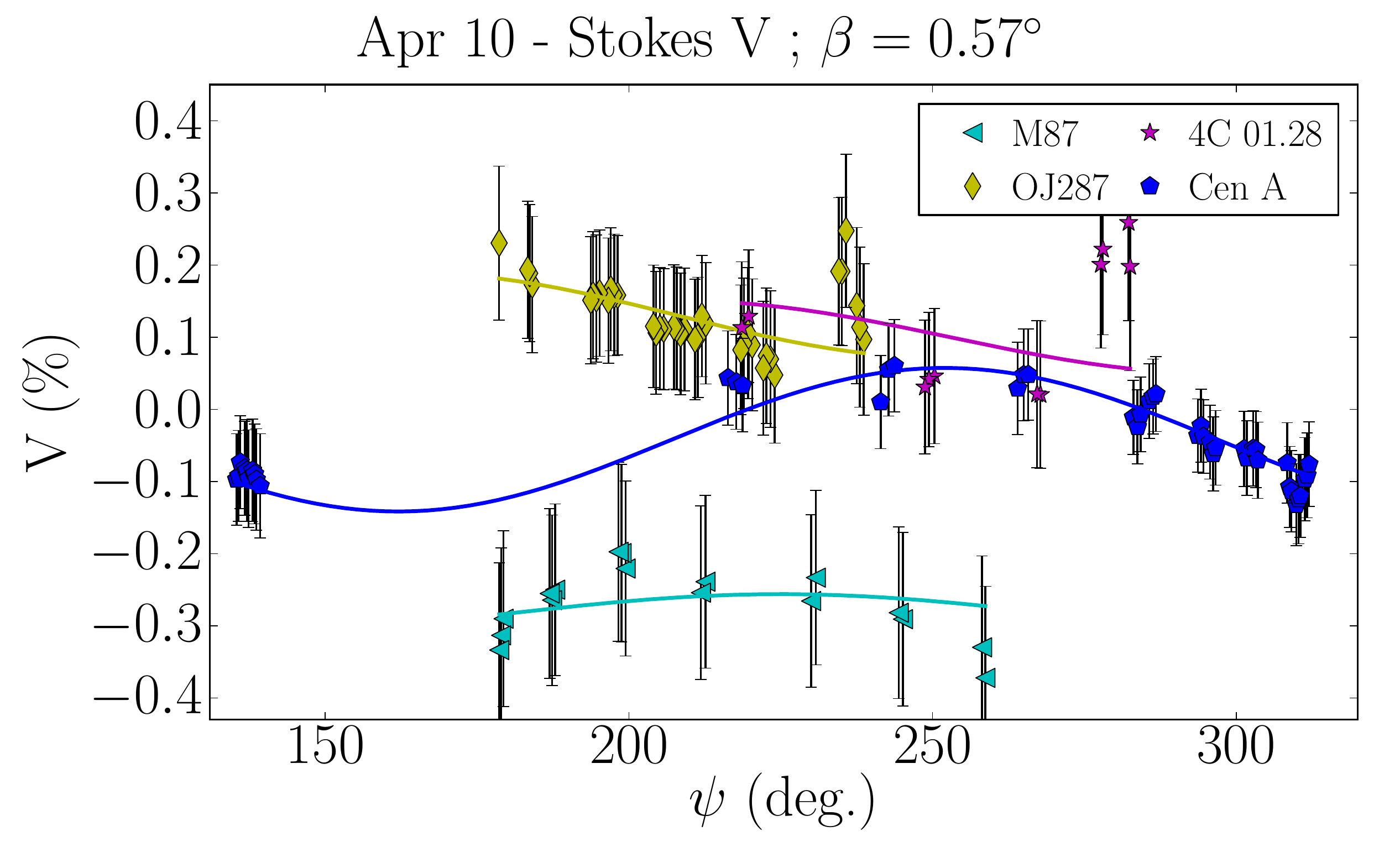}
\includegraphics[width=8cm]{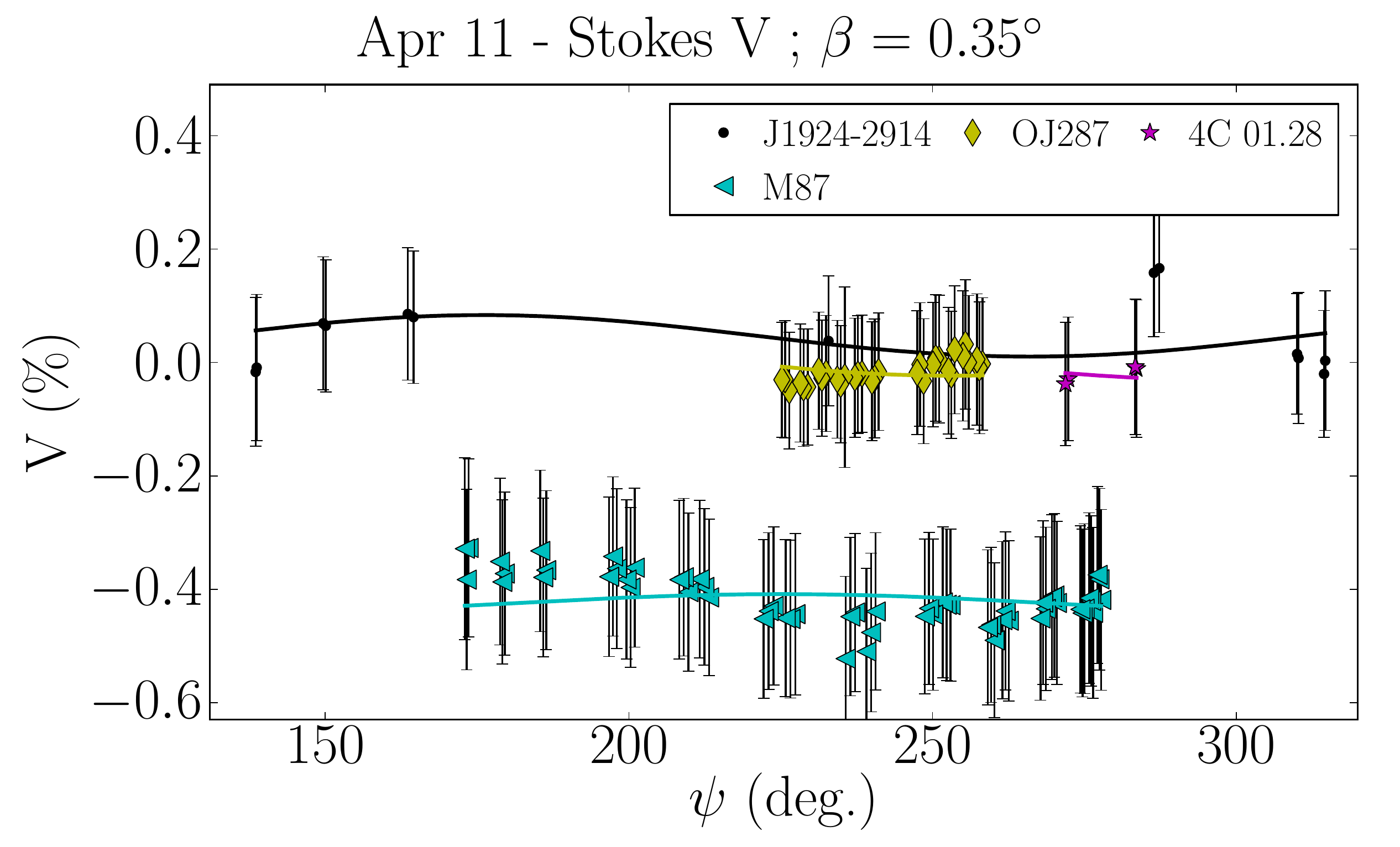}
\caption{Reconstructed fractional Stokes V of all sources (but 3C\,279), as a function of feed angle, during the EHT campaign. 
 Stokes V is computed as the real part of $(XY^* - YX^*)/j$. 
Continuum lines show a simplified model based on cross-polarization phase offsets at ALMA, which have been fixed to the $\beta$ values shown on top of each figure. Note that Sgr~A* is not plotted  because it displays significant intrinsic variability on the timescales (of hours) plotted here  \citep[e.g.,][]{Bower2018}.}
\label{AllVStokes}
\end{figure*}

In Fig. \ref{AllVStokes}, we show the fractional Stokes V values for all the sources, with the exception of 3C\,279, as a function of feed angle. As with Fig. \ref{VFigure}, a time binning of 120 seconds has been applied and data with elevations lower than 30\,degrees have not been used. We also show the model fitted with Eq. \ref{SpuriousVEq}, where $\beta$ is fixed to the value derived from the 3C\,279 data (so the only free parameter is $V^{const}$). The only exception to this modelling is for the epoch on April 7, where J1924-291 was used instead as the polarization calibrator.

Some sources show a clear dependence of $V^{tot}$ with feed angle, being 3C\,273 (on April 6) and OJ287 (on April 5) remarkable examples. However, the model prediction for such a variability, based on the cross-polarization phase offsets, $\beta$, estimated from 3C\,279, cannot reproduce all the data. A possible explanation for this discrepancy could be, for instance, a small (within 1 degree) variation of $\Delta$ with pointing direction (i.e., antenna elevation and azimuth) and/or an effect related to residual Dterms (see Eq. \ref{CrossPolEq3}). A deep analysis of these possibilities is out of the scope of this paper and should indeed be carried out at the Observatory level.

In any case, the fractional circular polarizations shown in Fig. \ref{AllVStokes} are very different among sources, which is a good indicative that these are not dominated by the Dterm systematics (at least, to a first-order approximation). Some sources do show the sinusoidal dependence with feed angle, whereas others (like M87) are dominated by $V^{const}$. Since we do not know the exact effects related to residual Dterms, it is not possible to derive $V^{true}$ from the estimated $V^{const}$. 
A robust conclusion, however, is that there {\em is} circular polarization detected in most of the sources (with amplitudes of the order of a few 0.1\%) and that, to our understanding, the instrumental effects, alone, cannot explain the results for all sources in a self-consistent way.
\begin{table*}
\caption{Frequency-averaged circular polarization fraction of GMVA targets (at a representative frequency of 93 GHz). }
\centering  
\begin{tabular}{ccccccc}
\hline\hline 
Source & Day & I &  V$_{meas}$ & V$_{true}$ & CP & LP \\
  &  [2017]  & [Jy]& [mJy]  & [mJy]  & [\%]  & [\%] \\
\hline\hline 
               OJ287             &  Apr 2               &  5.97$\pm$0.30    &      -4$\pm$36               &      -3        &      -0.05$\pm$0.60         &      8.811$\pm$0.030     \\
          J0510+1800             &  Apr 2               &  3.11$\pm$0.16    &      -4$\pm$19               &      -4        &      -0.14$\pm$0.60         &      4.173$\pm$0.031     \\
            4C 01.28$^a$             &  Apr 2               &  4.86$\pm$0.24    &      0$\pm$29              &      1.0         &      0.02$\pm$0.60         &      4.420$\pm$0.030     \\
              Sgr A*             &  Apr 3               &  2.52$\pm$0.13    &      -0$\pm$15               &      -0.5        &      -0.02$\pm$0.60         &      0.734$\pm$0.030     \\
          J1924-2914             &  Apr 3               &  5.11$\pm$0.26    &      0$\pm$31               &      5        &      0.09$\pm$0.60         &      4.841$\pm$0.031     \\
            NRAO 530$^a$             &  Apr 3               &  2.74$\pm$0.14    &      0$\pm$16              &      0.0         &      0.00$\pm$0.60         &      0.919$\pm$0.031     \\
            4C 09.57             &  Apr 3               &  2.85$\pm$0.14    &      -9$\pm$17               &      -10        &      -0.34$\pm$0.60         &      4.069$\pm$0.030     \\
               3C279$^a$             &  Apr 4               &  12.93$\pm$0.65    &      -0$\pm$78              &      -10         &      -0.1$\pm$1.2         &      12.159$\pm$0.030     \\
               3C273             &  Apr 4               &  9.86$\pm$0.49    &      11$\pm$59               &      14        &      0.14$\pm$0.60         &      3.984$\pm$0.029     \\
\hline\hline
\multicolumn{7}{l}{$^a$The polarization calibrator is assumed to have Stokes V=0 }\\
\multicolumn{7}{l}{for polarization calibration purposes \citep[see][]{QA2Paper}.}\\
\label{tab:GMVA_uvmf_CP}
\end{tabular}
\end{table*} 
\begin{longtable*}{ccccccc}
\caption{Frequency-averaged circular polarization fraction of EHT targets (at a representative frequency of 221 GHz). }\\
\hline 
Source & Day & I &  V$_{meas}$ & V$_{true}$ & CP & LP \\
  &  [2017]  & [Jy]& [mJy]  & [mJy]  & [\%]  & [\%]   \\
\hline\hline 
 &&&&& \\
               3C279$^a$         &  Apr 5               &  8.99$\pm$0.90    &      0$\pm$54               &      27        &      0.30$\pm$0.60         &      13.210$\pm$0.030     \\
               3C279$^a$         &  Apr 6               &  9.36$\pm$0.94    &      0$\pm$56               &      4        &      0.04$\pm$0.60         &      13.010$\pm$0.030     \\
               3C279$^a$         &  Apr 10               &  8.56$\pm$0.86    &      0$\pm$51               &      6        &      0.07$\pm$0.60         &      14.690$\pm$0.030     \\
               3C279$^a$         &  Apr 11               &  8.16$\pm$0.82    &      0$\pm$49               &      2        &      0.02$\pm$0.60         &      14.910$\pm$0.030     \\
 &&&&& \\
\hline 
 &&&&& \\
                 M87             &  Apr 5               &  1.28$\pm$0.13    &      -1.5$\pm$7.7               &      -2        &      -0.15$\pm$0.60         &      2.420$\pm$0.030     \\
                 M87             &  Apr 6               &  1.31$\pm$0.13    &      -4.4$\pm$7.9               &      -5        &      -0.34$\pm$0.60         &      2.160$\pm$0.030     \\
                 M87             &  Apr 10               &  1.33$\pm$0.13    &      -3.5$\pm$8.0               &      -4        &      -0.28$\pm$0.60         &      2.730$\pm$0.030     \\
                 M87             &  Apr 11               &  1.34$\pm$0.13    &      -5.4$\pm$8.0               &      -6        &      -0.41$\pm$0.60         &      2.710$\pm$0.030     \\
 &&&&& \\
\hline 
 &&&&& \\
              Sgr A*             &  Apr 6               &  2.63$\pm$0.26    &      -40$\pm$16               &      -40        &      -1.51$\pm$0.61         &      6.870$\pm$0.030     \\
              Sgr A*             &  Apr 7               &  2.41$\pm$0.24    &      -27$\pm$15               &      -27        &      -1.14$\pm$0.61         &      7.230$\pm$0.030     \\
              Sgr A*             &  Apr 11               &  2.38$\pm$0.24    &      -24$\pm$14               &      -24        &      -1.01$\pm$0.60         &      7.470$\pm$0.030     \\
 &&&&& \\
\hline 
 &&&&& \\
          J1924-2914             &  Apr 6               &  3.25$\pm$0.32    &      0$\pm$19               &      0.1        &      0.00$\pm$0.60         &      6.090$\pm$0.030     \\
          J1924-2914$^a$              &  Apr 7               &  3.15$\pm$0.31    &      0$\pm$19               &      0.7        &      0.02$\pm$0.60         &      5.970$\pm$0.030     \\
          J1924-2914             &  Apr 11               &  3.22$\pm$0.32    &      2$\pm$19               &      2        &      0.05$\pm$0.60         &      4.870$\pm$0.030     \\
 &&&&& \\
\hline 
 &&&&& \\
               OJ287             &  Apr 5               &  4.34$\pm$0.43    &      17$\pm$26               &      20        &      0.46$\pm$0.60         &      9.020$\pm$0.030     \\
               OJ287             &  Apr 10               &  4.22$\pm$0.42    &      5$\pm$25               &      5        &      0.12$\pm$0.60         &      7.000$\pm$0.030     \\
               OJ287             &  Apr 11               &  4.26$\pm$0.43    &      -1$\pm$26               &      0.6        &      0.01$\pm$0.60         &      7.150$\pm$0.030     \\
 &&&&& \\
\hline 
 &&&&& \\
            4C 01.28             &  Apr 5               &  3.51$\pm$0.35    &      10$\pm$21               &      9        &      0.25$\pm$0.60         &      5.890$\pm$0.030     \\
            4C 01.28             &  Apr 10               &  3.59$\pm$0.36    &      3$\pm$22               &      3        &      0.09$\pm$0.60         &      5.080$\pm$0.030     \\
            4C 01.28             &  Apr 11               &  3.57$\pm$0.36    &      -1$\pm$21               &      0.2        &      0.01$\pm$0.60         &      5.000$\pm$0.030     \\
 &&&&& \\
\hline 
 &&&&& \\
            NRAO 530             &  Apr 6               &  1.61$\pm$0.16    &      -0$\pm$10               &      -0.4        &      -0.02$\pm$0.60         &      2.350$\pm$0.030     \\
            NRAO 530             &  Apr 7               &  1.57$\pm$0.16    &      0.5$\pm$9.4               &      0.8        &      0.05$\pm$0.60         &      2.430$\pm$0.030     \\
 &&&&& \\
\hline 
 &&&&& \\
          J0132-1654             &  Apr 6               &  0.420$\pm$0.040    &      -0.7$\pm$2.5               &      -0.8        &      -0.19$\pm$0.60         &      1.990$\pm$0.050     \\
          J0132-1654             &  Apr 7               &  0.410$\pm$0.040    &      -0.1$\pm$2.5               &      -0.2        &      -0.04$\pm$0.60         &      2.000$\pm$0.050     \\
 &&&&& \\
\hline 
 &&&&& \\
            NGC 1052             &  Apr 6               &  0.430$\pm$0.040    &      0.2$\pm$2.6               &      0.2        &      0.05$\pm$0.60         &      0.110$\pm$0.030     \\
            NGC 1052             &  Apr 7               &  0.380$\pm$0.040    &      0.4$\pm$2.3               &      0.3        &      0.08$\pm$0.60         &      0.150$\pm$0.040     \\
 &&&&& \\
\hline 
 &&&&& \\
               Cen A             &  Apr 10               &  5.66$\pm$0.57    &      -2$\pm$34               &      -2        &      -0.04$\pm$0.60         &      0.060$\pm$0.030     \\
 &&&&& \\
\hline 
 &&&&& \\
               3C273             &  Apr 6               &  7.56$\pm$0.76    &      10$\pm$45               &      9        &      0.12$\pm$0.60         &      2.390$\pm$0.030     \\
 &&&&& \\
\hline 
 &&&&& \\
          J0006-0623             &  Apr 6               &  1.99$\pm$0.20    &      -1$\pm$12               &      -3        &      -0.16$\pm$0.60         &      12.530$\pm$0.030     \\
 &&&&& \\
\hline 
 &&&&& \\
\hline\hline
\multicolumn{7}{l}{$^a$The polarization calibrator is assumed to have Stokes V=0 }\\
\multicolumn{7}{l}{for polarization calibration purposes \citep[see][]{QA2Paper}.}\\
\label{tab:EHT_uvmf_CP}
\end{longtable*} 

\section{Spectral Indices of total intensity} 
\label{app:alpha}

We compute the total intensity spectral index for all the sources observed in the VLBI campaign at 3~mm and 1.3 mm.
For each source, the spectral index $\alpha$, defined as $I(\nu) \propto \nu^{\alpha}$, is derived "in-band", performing a weighted least-squares fit across the four flux-density values estimated with {\sc uvmultifit}  in each SPW, i.e. at frequencies of 213, 215, 227, 229 GHz in the 1.3~mm band, and 86, 88, 98, and 100 GHz in the 3~mm band, respectively. 
For Sgr A*, the spectral index of the compact core is around 0 both at 3~mm ($\alpha = 0.01\pm 0.1$) and 1.3~mm ($\alpha=[-0.03,-0.15]\pm 0.06$).  
For M87, the spectral index of the compact core at 1.3~mm is negative ($\alpha$=[--1.2,--1.1]). 
Cycle 0 ALMA observations at 3~mm  at a comparable angular resolution ($2.6'' \times 1.4''$) yields a much flatter $\alpha$=[-0.2,-0.3] \citep{Doi2013}, consistent with VLA measurements at radio frequency bands (8.4--43 GHz).  
VLBA observations revealed a flat-spectrum ($\alpha$=0) compact core \citep[e.g.,][]{Kravchenko2020}.
The steeper spectrum measured in the 1.3~mm band suggests that a spectral break must occur between 3\,mm and 1.3\,mm.
Spectral steepening  is also observed in  other AGN sources in the sample, which vary in the range $\alpha$=[--0.7,--0.3] at 3~mm and $\alpha$=[--1.3,--0.6] at 1.3~mm (Cen~A being the only exception, with $\alpha$=--0.2).  
When put together, these results indicates that the spectrum of AGN cores becomes progressively more optically thin at mm wavelengths 
(see \S~\ref{sect:LP_1mmvs3mm}).

\subsection{Foreground absorption at 226.91\,GHz toward Sgr A*} 
\label{appendix:sgr_spw2}

\begin{figure}
\centering
\includegraphics[width=0.48\textwidth]{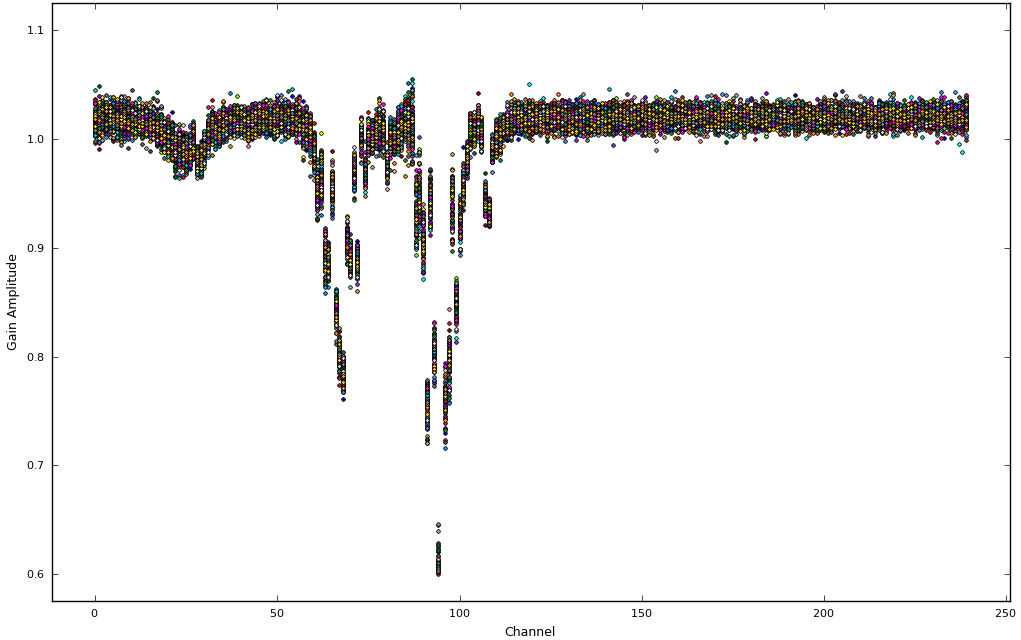} \hspace{-0.35cm}
\vspace{-.5cm}
\caption{
Bandpass for Sgr A* core (after removing instrumental bandpass with the calibrator NRAO~530) for 2017 Apr 07 observations. Only baselines longer than 60-m were used to remove virtually all of the extended arc-second emission.  For each channel, every antenna bandpass amplitude is plotted (represented by different colors). All three April observations show virtually the same `bandpass'.
}
\label{fig:alpha_1mm_sgra}
\end{figure}

Figure~\ref{fig:alpha_1mm_sgra} shows the spectrum of Sgr~A* in SPW=2 on 2017 Apr 07. The spectrum is virtually the same in the remainder days of observation in Apr 2017, suggesting that the absorption is probably associated with material that is  in front of the galactic center core. Cyanide Radical (CN) and its hyperfine structure at 226.3600\,GHz (N=2-1, J=3/2-3/2), 226.6595\,GHz (N=2-1, J=3/2-1/2), and 226.8748\,GHz (N=2-1, J=5/2-3/2)\footnote{NIST Recommended Rest Frequencies for Observed Interstellar Molecular Microwave Transitions; \\  \url{https://physics.nist.gov/cgi-bin/micro/table5/start.pl}} are identified as the  carriers of the absorption features. The lines predict a loss of integrated emission over the 1.8\,GHz bandwidth of about 2\%, in reasonable agreement with the decrements seen in SPW=2 in Fig.~\ref{fig:alpha_1mm_sgra}. 

\section{Two-component Polarization Model for M87}
  \label{app:twocomp}
  
Here we present the two-component polarization model for M87 summarized in Section~\ref{2polcomp} and shown schematically in Figure~\ref{fig:twocomp_cartoon}.  Details of the model itself, analysis, and parameter constraints can be found below.  Note that unlike the detailed image models presented in \citet{EHTC2020_2}, these models seek to reconstruct only the Stokes $I$, $Q$, and $U$ integrated over the EHT map and within the ALMA core.
 
The motivation of this modeling is to assess whether or not the significant interday variations seen in the RMs of M87 can be accommodated by a model in which the only variable element is the intrinsic polarization of the horizon-scale emission, holding all other properties of the Faraday rotation and a putative large-scale polarized component fixed.  In summary, we find that it is possible to do so, though it does require the RM of the horizon-scale component to be significantly larger than the RMs associated with the ALMA core in M87 reported in Table~\ref{tab:EHT_uvmf_RM}.

\subsection{Model Definitions}
\label{app:twocomp:models}
We consider two two-component models for the polarimetric properties of M87, differing in the location of the small-scale Faraday screen.  For both components, we construct the set of Stokes $I$, $Q$, $U$, which may depend on observation day and wavelength.  The directly compared quantities that comprise the model are the integrated Stokes parameters of the compact component, $I_{\rm com,day,\lambda}$, $Q_{\rm com,day,\lambda}$, and $U_{\rm com,day,\lambda}$, and of the combination of the compact and extended components, $I_{\rm tot,day,\lambda}=I_{\rm com,day,\lambda}+I_{\rm ext,\lambda}$, $Q_{\rm tot,day,\lambda}=Q_{\rm com,day,\lambda}+Q_{\rm ext,\lambda}$, and $U_{\rm tot,day,\lambda}=U_{\rm com,day,\lambda}+U_{\rm ext,\lambda}$.  
 A summary list of the model parameters for the models described in Sections~\ref{app:twocomp:models:ext}-\ref{app:twocomp:models:comint} is contained in Table~\ref{tab:twocomp_fits}.

\subsubsection{Extended component}
\label{app:twocomp:models:ext}
Both models contain a large-scale component defined by an intensity normalization $I_{0,\rm ext}$ at a reference wavelength $\lambda_0$, spectral index $\alpha_{0,\rm ext}$, polarization fraction $m_{\rm ext}$, and EVPA $\psi_{\rm ext}$ at $\lambda_0$.  This is further processed through an external Faraday screen with a rotation measure of ${\rm RM}_{\rm ext}$.  The contribution to the Stokes $I$, $Q$, and $U$ are, then
\begin{equation}
  \begin{aligned}
    I_{{\rm ext},\lambda} &= I_{0,\rm ext} \left(\lambda/\lambda_0\right)^{\alpha_{\rm ext}}\\
    Q_{{\rm ext},\lambda} &= m_{\rm ext} I_{{\rm ext},\lambda} \cos\left[2\psi+2{\rm RM}_{\rm ext}(\lambda^2-\lambda_0^2)\right]\\
    U_{{\rm ext},\lambda} &= m_{\rm ext} I_{{\rm ext},\lambda} \sin\left[2\psi+2{\rm RM}_{\rm ext}(\lambda^2-\lambda_0^2)\right]
  \end{aligned}
\end{equation}
This introduces five parameters: $I_{0,\rm ext}$, $\alpha_{\rm ext}$, ${\rm RM_{\rm ext}}$, $m_{\rm ext}$, and $\psi$.

\subsubsection{Compact component: external Faraday screen}
\label{app:twocomp:models:comext}
For each day on which M87 was observed by ALMA and the EHT (i.e., April 5, 6, 10 and 11) we specify a similar model for the compact component.  When the screen is assumed to be external, we adopt a similar model to the large-scale component with the exception that the rotation is due to both screens:
\begin{equation}
  \begin{aligned}
    I^{\rm Ex}_{{\rm com},{\rm day},\lambda} 
    &= 
    I_{0,\rm com} \left(\lambda/\lambda_0\right)^{\alpha_{\rm com}}\\
    Q^{\rm Ex}_{{\rm com},{\rm day},\lambda} 
    &= 
    m_{\rm com, day} I^{\rm Ex}_{{\rm com},{\rm day},\lambda} \\
    &\cos\left[2\psi_{\rm com, day}
    +2({\rm RM_{\rm com}}+{\rm RM}_{\rm ext})(\lambda^2-\lambda_0^2)\right]\\
    U^{\rm Ex}_{{\rm com},{\rm day},\lambda} 
    &= 
    m_{\rm com, day} I^{\rm Ex}_{{\rm com},{\rm day},\lambda} \\
    &\sin\left[2\psi_{\rm com, day}
    +2({\rm RM_{\rm com}}+{\rm RM}_{\rm ext}(\lambda^2-\lambda_0^2)\right].
  \end{aligned}
\end{equation}
This introduces an additional eleven parameters: $I_{0,\rm com}$, $\alpha_{\rm com}$, ${\rm RM_{\rm com}}$, and four each $m_{\rm com, day}$ and $\psi_{\rm com, day}$.
 The polarization fraction and EVPA of the compact component, $m_{\rm com,day}$ and $\psi_{\rm com,day}$, are distinct from all of the other parameters in the model in that they vary among observation days.  Therefore, where useful, we will distinguish these as ``dynamic'' parameters, with the remainder of the parameters being ``static'' in the limited sense that they do not vary across the observation campaign.

\subsubsection{Compact component: internal Faraday screen model}
\label{app:twocomp:models:comint}
Many simulations of M87 indicate the presence of large Faraday depths in the emission region \citep{BroderickMcKinney2010,Moscibrodzka2017,Ricarte2020}.  Therefore, we also consider a simple model for the compact component in which the emission and rotation are co-located, i.e., an internal Faraday screen.  In principle, this is inextricably linked to the detailed properties of the emission region.  Here we employ the gross simplification of a single-zone, or slab, model: the emission and Faraday rotation within the compact component occurs within a homogeneous region.  We begin with a summary of the polarimetric properties of such a slab.

For a plane-parallel source with physical depth $L$ and at some reference wavelength $\lambda_0$ a uniform emissivity $j$, polarization fraction $m_{\rm em}$ at emission, EVPA at emission $\psi_{\rm em}$, Faraday rotativity $R$, we have total intensity,
\begin{equation}
  I = \int_0^L j dz = j L,
\end{equation}
and Stokes $Q$ and $U$,
\begin{equation}
\begin{aligned} 
  Q+iU &= \int_0^L j m_{\rm em} e^{2 i [R x \lambda^2 + \psi_{\rm em}]} dx\\
  &= \frac{m_{\rm em} I}{2 RL\lambda^2}
  \left[\sin(2\psi_{\rm em} + 2RL\lambda^2) - \sin(2\psi_{\rm em})
  \right]\\
  &+
  i\frac{m_{\rm em} I}{2RL\lambda^2}
  \left[
    \cos(2\psi_{\rm em}) - \cos(2\psi_{\rm em} + 2RL\lambda^2)
  \right]\\
  &=
  \frac{1}{2} m_{\rm em} I
  {\rm sinc}(RL\lambda^2)\\
 & \left[
    \cos(2\psi_{\rm em} + RL\lambda^2)+  i\sin(2\psi_{\rm em} + RL\lambda^2) \right]
\end{aligned}
\end{equation}
from which we can immediately read off $Q$ and $U$.  The EVPA at the top of the slab is
\begin{equation}
\tan(2\psi)
=
\frac{Q}{U}
=
\tan\left[2\psi_{\rm em} + RL(\lambda^2-\lambda_0^2)\right],
\end{equation}
from which it is apparent that the effective contribution to the compact RM is ${\rm RM}_{\rm com}=RL/2$.

The internal Faraday screen model is then defined by
\begin{equation}
  \begin{aligned}
    I^{\rm In}_{{\rm com},{\rm day},\lambda} 
    &= 
    I_{0,\rm com} \left(\lambda/\lambda_0\right)^{\alpha_{\rm com}}\\
    Q^{\rm In}_{{\rm com},{\rm day},\lambda} 
    &= 
    \frac{1}{2} m_{\rm com, day} I^{\rm In}_{{\rm com},{\rm day},\lambda}
    {\rm sinc}(2{\rm RM}_{\rm com}\lambda^2)\\
    & \cos\left[2\psi_{\rm com,day} + 2({\rm RM}_{\rm com}+{\rm RM}_{\rm ext})\lambda^2\right]\\
    U^{\rm In}_{{\rm com},{\rm day},\lambda} 
    &= 
    \frac{1}{2} m_{\rm com, day} I^{\rm In}_{{\rm com},{\rm day},\lambda}
    {\rm sinc}(2{\rm RM}_{\rm com}\lambda^2) \\
    & \sin\left[2\psi_{\rm com,day} + 2({\rm RM}_{\rm com}+{\rm RM}_{\rm ext})\lambda^2\right],
  \end{aligned}
\end{equation}
where, as with the external model, we have added the Faraday rotation from the large-scale Faraday screen.  As with the external Faraday screen model, this introduces eleven parameters, though the interpretations of the polarization fraction, and EVPA subtly differ, here referring to those of the emission process instead of derotated values.

 As with the external Faraday screen model, where useful, we will refer to $m_{\rm com,day}$ and $\psi_{\rm com,day}$, which differ among observation days, as ``dynamic'' parameters to distinguish them from the remaining ``static'' parameters.

\subsection{Markov Chain Monte Carlo Analysis}
From the models described above and the integrated EHT Stokes parameter ranges presented in Table~7 in Appendix~H2 of \citet{EHTC2020_1} ($I_{\rm EHT}$, $Q_{\rm EHT}$, $U_{\rm EHT}$) and the ALMA core Stokes parameter values for the individual SPWs in Table~\ref{tab:EHT_uvmf_spw} ($I_{\rm spw}$, $Q_{\rm spw}$, $U_{\rm spw}$), we construct a log-likelihood for each set of model parameters.
 These comprise sixty data values in total: three ($I_{\rm EHT}$, $Q_{\rm EHT}$, $U_{\rm EHT}$)  on each of four days from the EHT observations ($3\times4$ data points), three ($I_{\rm spw}$, $Q_{\rm spw}$, $U_{\rm spw}$) in four SPWs on each of four days presented here ($3\times4\times4$  data points).

We assume that the integrated EHT Stokes parameters are distributed normally with means and standard deviations  set by the centers and half-widths of the ranges; this likely over-estimates the true uncertainty on the $I_{\rm EHT}$, $Q_{\rm EHT}$, and $U_{\rm EHT}$.
The resulting log-likelihood is
\begin{multline}
  \mathcal{L} =
  \sum_{\rm day} \Bigg[
  -\frac{(I_{\rm EHT} - I_{\rm com,day,\lambda_0})^2}{2\sigma_{I_{\rm EHT}}^2}
  -\frac{(Q_{\rm EHT} 
  - Q_{\rm com,day,\lambda_0})^2}{2\sigma_{Q_{\rm EHT}}^2} \\
  -\frac{(U_{\rm EHT} - U_{\rm com,day,\lambda_0})^2}{2\sigma_{U_{\rm EHT}}^2}
  -\sum_{\rm spw=1}^{4}\frac{(I_{\rm spw} - I_{\rm tot,day,\lambda_{\rm spw}})^2}{2\sigma_{I_{\rm spw}}^2}\\
  -\sum_{\rm spw=1}^{4}\frac{(Q_{\rm spw} - Q_{\rm tot,day\lambda_{\rm spw}})^2}{2\sigma_{Q_{\rm spw}}^2} 
  -\sum_{\rm spw=1}^{4}\frac{(U_{\rm spw} - U_{\rm tot,day,\lambda_{\rm spw}})^2}{2\sigma_{U_{\rm spw}}^2} \Bigg].
  \label{eq:twocomp_loglikelihood}
\end{multline}

Linear or "uniform" priors in the natural ranges are imposed on all parameters with the exception of the compact and extended component intensity normalizations, for which logarithmic priors are chosen.  See Table~\ref{tab:twocomp_fits} for details.

The likelihood is sampled with the ensemble Markov Chain Monte Carlo (MCMC) method provided by the EMCEE python package \citep{emcee2013}.  We use 256 independent walkers, and run for $10^5$ steps, discarding the first half of the chains.  Explorations with fewer walkers and steps indicate that by this time the MCMC chains are well converged.

In addition to the models described in Appendix~\ref{app:twocomp:models}, we also considered versions of the two-component model applied to each day independently, i.e., keeping only one day in the sum in Equation~\ref{eq:twocomp_loglikelihood}.  
 In these, on each day the five parameters of the external screen and five parameters of the internal screen (a single $m_{\rm com}$ and $\psi_{com}$) are independently fit on each observation day.  Effectively, this corresponds to a forty-parameter model across the four observation days, permitting Faraday screens and emission from both the compact and extended Faraday screens to vary independently across the four observation days.

  On any given observation day, the parameters are less well constrained in this case and, with the notable exceptions of the compact component polarization fraction and EVPA,  are consistent with a single set of values across all days.  
The variable polarization properties of the compact component  matches the expectation from the EHT measurements in \citet{EHTC2020_1}.  The consistency with a single set of values for the remaining model parameters serves as a partial motivation for the more restricted variability in the models presented in Appendix~\ref{app:twocomp:models}.

\subsection{Two-component model results}
Excellent fits are found for both the external and internal Faraday screen models.  For 44 degrees of freedom, the best-fit external and internal screen models have reduced $\chi^2=3.96$   and $\chi^2=2.54$,  
respectively, both modestly small and possibly indicating that the uncertainties on the integrated EHT Stokes parameters are indeed over estimated by their half-range values.  Thus, it is possible to reproduce the variable polarimetric properties observed by ALMA with a model in which only the compact emission evolves.

\begin{figure*}
\begin{center}
\includegraphics[width=\textwidth]{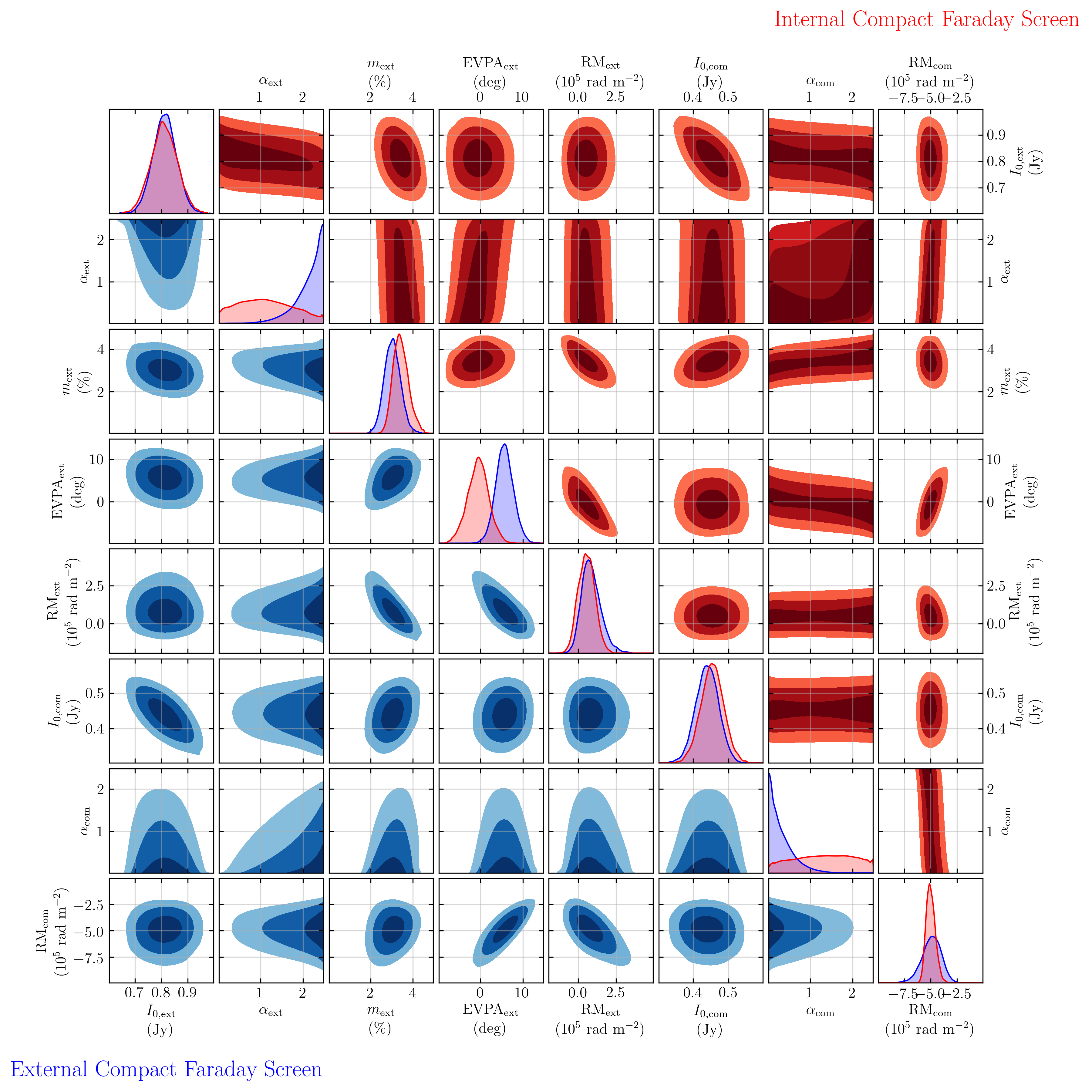}
\end{center}
\caption{Joint posteriors for static parameters of two-component polarimetric models with external (lower left; blue) and internal (upper right; red) small-scale Faraday screens.  Contours indicate 50\%, 90\%, and 95\% quantiles.}
\label{fig:twocomp_static_triangle}
\end{figure*}

\begin{figure*}
\begin{center}
\includegraphics[width=\textwidth]{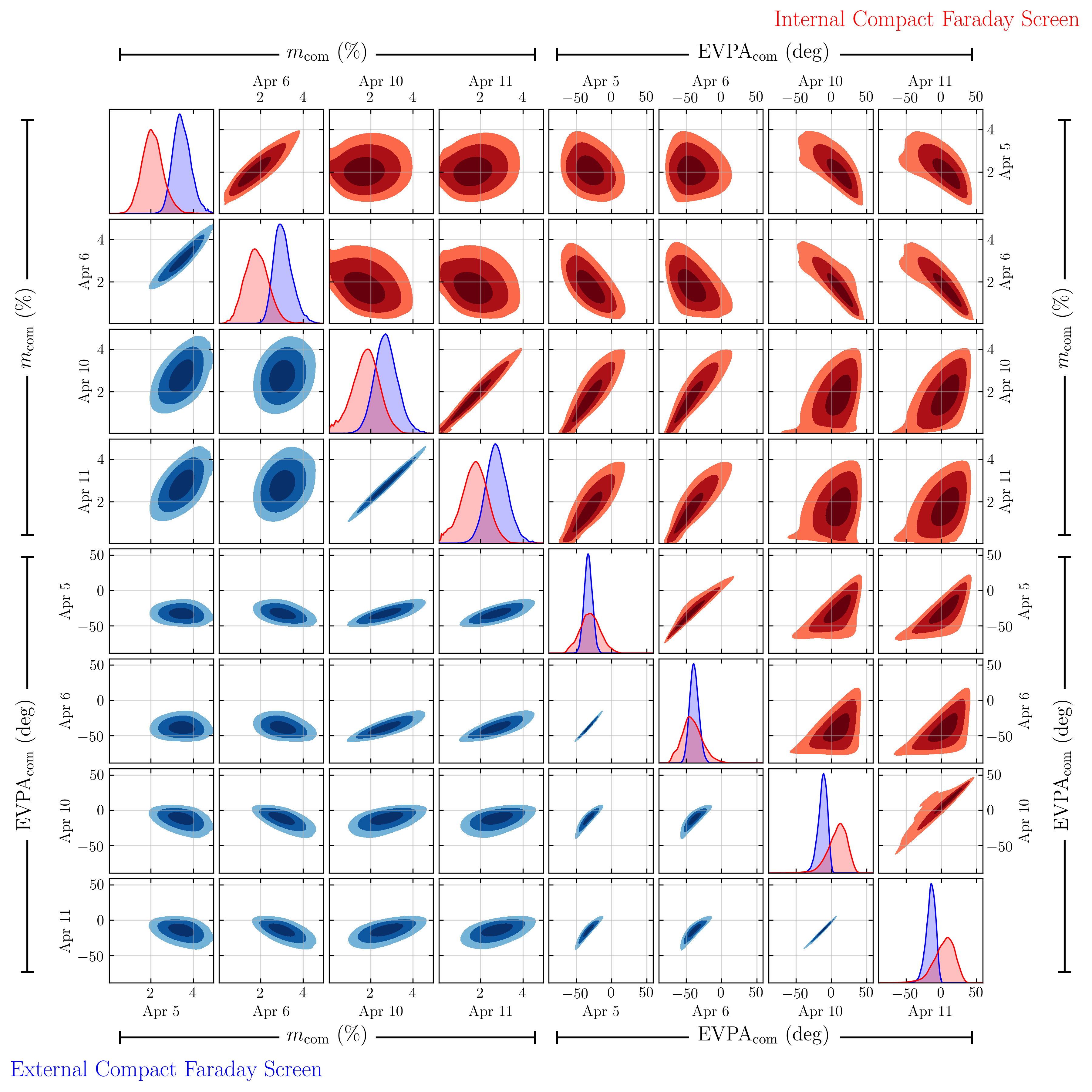}
\end{center}
\caption{Joint posteriors for dynamic parameters of two-component polarimetric models with external (lower left; blue) and internal (upper right; red) small-scale Faraday screens. For the internal Faraday screen model, the polarization fractions have been depolarized by a factor of ${\rm sinc}(2{\rm RM}_{\rm com}\lambda_0^2)$ and the EVPAs have been rotated by $({\rm RM}_{\rm com}+{\rm RM}_{\rm ext})\lambda_0^2$, corresponding to observed values, concordant with the definition for the external Faraday screen model. Contours indicate 50\%, 90\%, and 95\% quantiles.}
\label{fig:twocomp_dynamic_triangle}
\end{figure*}

\begin{table*}
\caption{Two-component model MCMC analysis priors and results.}
\label{tab:twocomp_fits}
\centering
\begin{tabular}{ccccc}
\hline\hline
Parameter & Unit & Prior\tablenotemark{a} & External Screen\tablenotemark{b} & Internal Screen\tablenotemark{b,c}   \\
\hline\hline
$I_{0,\rm ext}$        & Jy                      & $\mathcal{L}(0,\infty)$ & $0.813\pm0.044$ & $0.811\pm0.049$ \\
$\alpha_{\rm ext}$     & --                      & $\mathcal{U}[0,2.5]$    & $2.14\pm0.32$ & $1.12\pm0.62$ \\
$m_{\rm ext}$          & \%                     & $\mathcal{U}[0,100]$    & $3.03\pm0.37$ & $3.43\pm0.36$ \\
${\rm EVPA}_{\rm ext}$ & deg                     & $\mathcal{U}[-90,90]$   & $5.7\pm2.1$ & $-0.5\pm2.4$ \\
${\rm RM}_{\rm ext}$   & $10^5~{\rm rad~m^{-2}}$ & $\mathcal{U}[-100,100]$ & $0.87\pm0.64$ & $0.55\pm0.53$ \\
$I_{0,\rm com}$        & Jy                      & $\mathcal{L}(0,\infty)$ & $0.439\pm0.032$ & $0.454\pm0.032$ \\
$\alpha_{\rm com}$     & --                      & $\mathcal{U}[0,2.5]$    & $0.31\pm0.27$ & $1.31\pm0.66$ \\
${\rm RM}_{\rm com}$   & $10^5~{\rm rad~m^{-2}}$ & $\mathcal{U}[-100,100]$ & $-4.92\pm0.91$ & $-4.98\pm0.43$ \\
$m_{\rm com, Apr~5~}$  & \%                     & $\mathcal{U}[0,100]$    & $3.48\pm0.41$ & $3.66\pm0.66$ \\
$m_{\rm com, Apr~6~}$  & \%                     & $\mathcal{U}[0,100]$    & $3.06\pm0.43$ & $3.17\pm0.82$ \\
$m_{\rm com, Apr~10}$  & \%                     & $\mathcal{U}[0,100]$    & $2.73\pm0.54$ & $3.2\pm1.2$ \\
$m_{\rm com, Apr~11}$  & \%                     & $\mathcal{U}[0,100]$    & $2.78\pm0.51$ & $3.0\pm1.1$ \\
${\rm EVPA}_{\rm com, Apr~5~}$ & deg             & $\mathcal{U}[-90,90]$   & $-32.9\pm5.6$ & $14\pm10$ \\
${\rm EVPA}_{\rm com, Apr~6~}$ & deg             & $\mathcal{U}[-90,90]$   & $-38.5\pm6.3$ & $3.0\pm10.0$ \\
${\rm EVPA}_{\rm com, Apr~10}$ & deg             & $\mathcal{U}[-90,90]$   & $-12.4\pm6.5$ & $53\pm13$ \\
${\rm EVPA}_{\rm com, Apr~11}$ & deg             & $\mathcal{U}[-90,90]$   & $-14.8\pm6.6$ & $48\pm13$ \\
\noalign{\smallskip}
\hline\hline
\end{tabular}
\tablenotetext{a}{Priors types are logarithmic ($\mathcal{L}$) and uniform ($\mathcal{U}$), with ranges indicated afterward.}
\tablenotetext{b}{Means and standard deviations of parameter values are provided.}
\tablenotetext{c}{Note that LP ({\it m}) and EVPA refer to those of the emission process, not observed at the surface of the emission region.}
\end{table*}

Figures~\ref{fig:twocomp_static_triangle} and \ref{fig:twocomp_dynamic_triangle} show the joint posteriors for the external (lower left) and internal (upper right) Faraday screen models for the static and dynamic model components, respectively.  To facilitate a direct comparison with the external Faraday screen model, in Figure~\ref{fig:twocomp_dynamic_triangle}, the polarization fractions and EVPAs of the internal Faraday screen model have been depolarized and rotated to show the corresponding posteriors on their observed analogs.  Model parameter estimates, marginalized over all other parameters, are contained in Table~\ref{tab:twocomp_fits}.

After adjusting the polarization fraction and EVPA of the compact component, the properties of the two-component models are consistent among the external and internal Faraday screen models.   Strong correlations exists between many of the compact component features.  These are very strong for the polarization fractions and EVPAs   on neighboring observation days, i.e., April 5 and 6, and April 10 and 11.  This is anticipated by the similarities in the integrated polarimetric properties reported in \citet{EHTC2020_1}  between neighboring observation days, which naturally constrain the two-component polarization models accordingly.

The flux normalizations, polarization fractions, and EVPAs are well constrained.  ${\rm RM}_{\rm ext}$ is restricted to small magnitudes in both cases, typically less than $1.5\times10^5~{\rm rad~m^{-2}}$, and remains consistent with vanishing altogether.  In contrast, ${\rm RM}_{\rm com}$ is significantly non-zero, and typically of order $-5\times10^5~{\rm rad~m^{-2}}$, factors of 3-10 larger than those reported in Table~\ref{tab:EHT_uvmf_RM}, implying a significant degree of competition between the spectral variations in the two components and the residual Faraday rotation.  In neither model are the spectral indexes of the components well constrained.

\begin{figure*}
\begin{center}
\includegraphics[width=\textwidth]{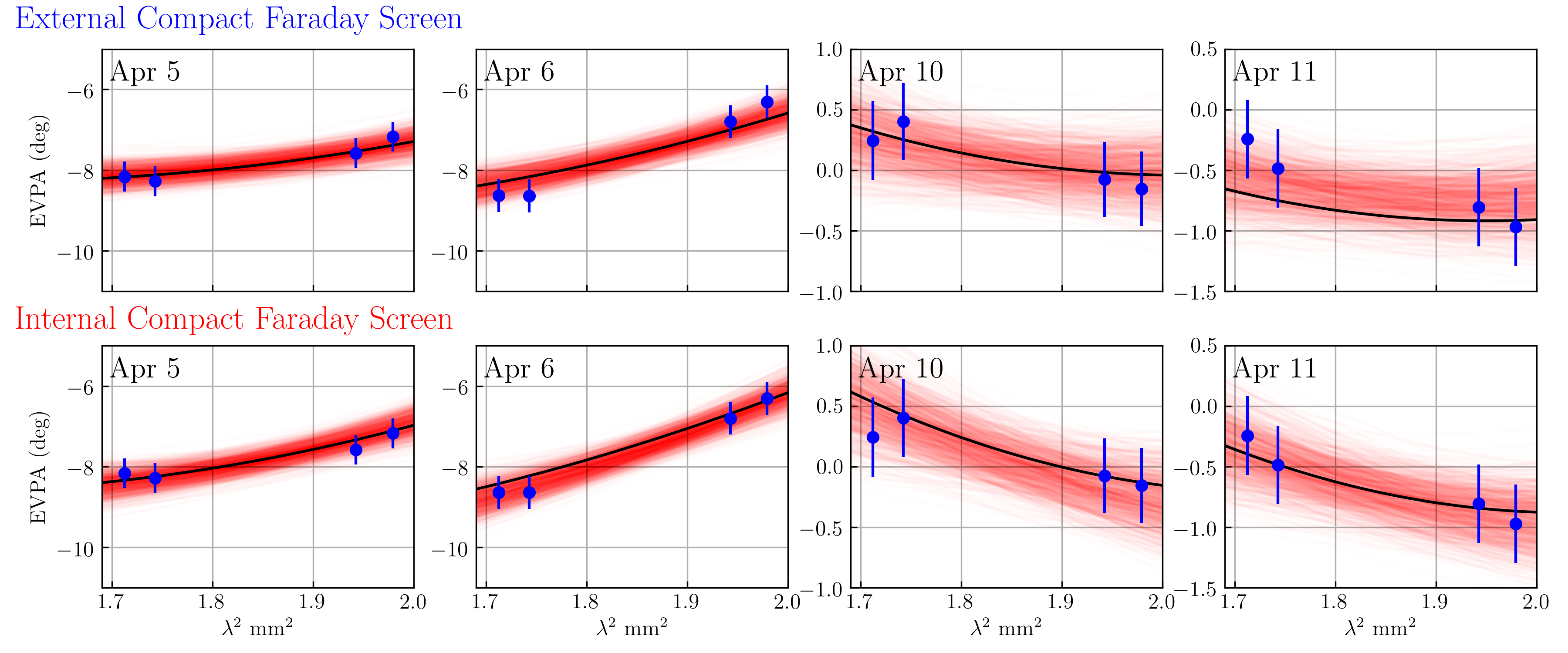}
\end{center}
\caption{EVPA as a function of $\lambda^2$ for $10^3$ draws from the two-component model posterior of the external (top) and internal (bottom) Faraday screen models.  The highest-likelihood models are shown by the black lines. 
The blue data points are the ALMA measurements (same as  Fig.~\ref{fig:RM1}).}
\label{fig:twocomp_EVPAs}
\end{figure*}

By construction, these are necessarily consistent with the results from ALMA-only RM measurements reported in Table~\ref{tab:EHT_uvmf_RM}.  Figure \ref{fig:twocomp_EVPAs} shows a number of realizations drawn from the posteriors for the wavelength-dependence of the EVPAs on each of the observation days in comparison to the measured values listed in Table~\ref{tab:EHT_uvmf_spw}.  The interday evolution in the ALMA RMs are well reproduced, despite restricting the variable elements of the model to the compact component.  Within the ALMA SPWs the EVPAs only weakly depart from the linear dependence on $\lambda^2$ indicative of Faraday rotation; outside of the ALMA SPWs this divergence can become considerably larger, implying that additional coincident polarimetric measurements at longer (and shorter) wavelengths will significantly improve constraints on the elements of the two-component model.

%
\bibliography{biblio.bib}
\end{document}